\documentclass[prx,aps,amssymb,twocolumn,superscriptaddress]{revtex4-2}
\usepackage{amsmath}
\usepackage{amssymb}
\usepackage{amsthm}
\usepackage{amsfonts}
\usepackage{listings}
\usepackage{longtable}
\usepackage{enumerate}
\usepackage{latexsym}
\usepackage{xcolor}
\usepackage{setspace} 
\usepackage{blindtext}
\usepackage{chemformula}
\usepackage{dcolumn}
\usepackage{braket}

\usepackage{multirow}

\def\ie{{\it i.e.},\ }
\def\eg{{\it e.g.}\ }
\def\ea{{\it et al.}}

\newcommand{\beq}{\begin{equation}}
\newcommand{\eneq}{\end{equation}}
\extrafloats{52}

\setcitestyle{super}

\usepackage{array,etoolbox}
\usepackage[titles]{tocloft}
\preto\tabular{\setcounter{magicrownumbers}{0}}
\newcounter{magicrownumbers}

\usepackage[english]{babel}

\newcommand{\crystal}{crystalline}

\usepackage{hyperref}
\hypersetup{
 pdfnewwindow=true, colorlinks=true,
 linkcolor=blue, anchorcolor=blue,
 citecolor=blue, filecolor=blue,
 menucolor=blue, urlcolor=blue}

\newcommand{\sirefappflatbandtopology}{Appendix~\ref{app:flatbandtopology}}
\newcommand{\sirefappkagomepyrochlore}{Appendix~\ref{app:kagomepyrochlore}}
\newcommand{\sirefappbipartite}{Appendix~\ref{app:bipartite}}
\newcommand{\sirefapptheoryexplanation}{Appendix~\ref{app:theoryexplanation}}
\newcommand{\sirefappmagflat}{Appendix~\ref{app:magflat}}
\newcommand{\sirefappallflatbands}{Appendix~\ref{app:allflatbands}}
\newcommand{\sirefapptqcdboverview}{Appendix~\ref{app:tqcdboverview}}
\newcommand{\sirefappflatbandwebsite}{Appendix~\ref{app:flatbandwebsite}}
\newcommand{\sirefappbsflatsegment}{Appendix~\ref{app:bsflatsegment}}
\newcommand{\sirefappdospeak}{Appendix~\ref{app:dospeak}}
\newcommand{\sirefappsublatticesdb}{Appendix~\ref{app:sublatticesdb}}
\newcommand{\sirefappcryssymm}{Appendix~\ref{app:crys_symm}}
\newcommand{\sirefappgeometry}{Appendix~\ref{app:geometry}}
\newcommand{\sirefappalgorithms}{Appendix~\ref{app:algorithms}}
\newcommand{\sirefappsmatrix}{Appendix~\ref{app:smatrix}}
\newcommand{\sirefCaNCl}{Appendix~\ref{CaNCl}}
\newcommand{\sirefapplistcurated}{Appendix~\ref{app:listcurated}}
\newcommand{\sirefapplistatomic}{Appendix~\ref{app:listatomic}}
\newcommand{\sirefappbestflatbands}{Appendix~\ref{app:bestflatbands}}

\newcommand{\sireftabkagomepyrochlorewyckoff}{\ref{tab:kagomepyrochlorewyckoff}}
\newcommand{\sireftabliebwyckoff}{\ref{tab:liebwyckoff}}
           
\begin{document}

%\input{flatband_macros.tex}
%TQC COMMANDS

% ICSD link to the topologicalquantumchemistry website (short version)
%\newcommand{\icsdwebshort}[1]{\href{https://www.topologicalquantumchemistry.org/\#/detail/#1}{#1}}
\newcommand{\icsdwebshort}[1]{\href{https://www.topologicalquantumchemistry.com/\#/detail/#1}{#1}}
% ICSD link to the topologicalquantumchemistry website (long version)
%\newcommand{\icsdweb}[1]{\href{https://www.topologicalquantumchemistry.org/\#/detail/#1}{ICSD #1}}
\newcommand{\icsdweb}[1]{\href{https://www.topologicalquantumchemistry.com/\#/detail/#1}{ICSD #1}}
%link to the topologicalquantumchemistry website with no ICSD:
\newcommand{\webNoICSD}{\url{https://www.topologicalquantumchemistry.org/}}
%link to the topologicalquantumchemistry website with its full name
%\newcommand{\webTQC}{\href{https://www.topologicalquantumchemistry.org/}{Topological Quantum Chemistry website}}
\newcommand{\webTQC}{\href{https://www.topologicalquantumchemistry.com/}{Topological Quantum Chemistry website}}
\newcommand{\webTQCacronym}{\href{https://www.topologicalquantumchemistry.com/}{TQCDB}}
% link to the  {Check Topological Mat website
\newcommand{\webchecktopmat}{\href{https://www.cryst.ehu.es/cryst/checktopologicalmagmat}{Check Topological Mat}}
\newcommand{\identify}{\href{www.cryst.ehu.es/cryst/identify_group}{IDENTIFY GROUP}}
%link to the topologicalquantumchemistry/flatband website with its full name
\newcommand{\webflatband}{\href{https://www.topologicalquantumchemistry.fr/flatbands}{Materials Flatband Database website}}
\newcommand{\webflatbandacronym}{\href{https://www.topologicalquantumchemistry.fr/flatbands}{MFBDB}}
%link to an ICSD on the topologicalquantumchemistry/flatband website
%\newcommand{\flatwebdirectlink}[2]{\href{http://www.nick-ux.org/flattopoofficial/index.html?ICSD=#1}{#2}}
\newcommand{\flatwebdirectlink}[2]{\href{https://www.topologicalquantumchemistry.fr/flatbands/index.html?ICSD=#1}{#2}}

% BCSID link to the topologicalquantumchemistry magnetic website (short version)
\newcommand{\bcsidwebshort}[1]{\href{https://www.topologicalquantumchemistry.fr/magnetic/index.html?BCSID=#1}{#1}}
% BCSID link to the topologicalquantumchemistry magnetic website (long version)
\newcommand{\bcsidweblong}[1]{\href{https://www.topologicalquantumchemistry.fr/magnetic/index.html?BCSID=#1}{BCSID #1}}

% link to the materials project website with its full name
\newcommand{\webmaterialsproject}{\href{https://materialsproject.org}{Materials Project}}

% link to the 
\newcommand{\webscnims}{\href{https://mits.nims.go.jp/en}{NIMS Materials Database}}

% link to the BCS website (long version) 
\newcommand{\bcslong}{\href{https://www.cryst.ehu.es/}{\emph{Bilbao Crystallographic Server}}}

%% Total entries in the ICSD
\newcommand{\TQCDTotICSDs}{181,221}
%% Wrong ICSD
\newcommand{\TQCDWrongICSDs}{184}
%% ICSD stoichiometric
\newcommand{\TQCDstoichiometric}{85,700}
%% ICSD No stoichiometric
\newcommand{\TQCDNostoichiometric}{95,521}

%%thresholds in the algorithm for lattice structure
\newcommand{\FlatBandLiebDistanceThreshold}{20\%}

\newcommand{\FlatBandLiebAngleThreshold}{10}

\newcommand{\FlatBandKagomeDistanceThreshold}{20\%}

\newcommand{\FlatBandKagomeAngleThreshold}{2}

\newcommand{\FlatBandSplitDistanceThreshold}{20\%}

%\input{sg_macros.tex}
%% Macros for the space groups

\newcommand{\sgsymb}[1]{\ifnum#1=1
$P1$\else
\ifnum#1=2
$P\bar{1}$\else
\ifnum#1=3
$P2$\else
\ifnum#1=4
$P2_1$\else
\ifnum#1=5
$C2$\else
\ifnum#1=6
$Pm$\else
\ifnum#1=7
$Pc$\else
\ifnum#1=8
$Cm$\else
\ifnum#1=9
$Cc$\else
\ifnum#1=10
$P2/m$\else
\ifnum#1=11
$P2_1/m$\else
\ifnum#1=12
$C2/m$\else
\ifnum#1=13
$P2/c$\else
\ifnum#1=14
$P2_1/c$\else
\ifnum#1=15
$C2/c$\else
\ifnum#1=16
$P222$\else
\ifnum#1=17
$P222_1$\else
\ifnum#1=18
$P2_12_12$\else
\ifnum#1=19
$P2_12_12_1$\else
\ifnum#1=20
$C222_1$\else
\ifnum#1=21
$C222$\else
\ifnum#1=22
$F222$\else
\ifnum#1=23
$I222$\else
\ifnum#1=24
$I2_12_12_1$\else
\ifnum#1=25
$Pmm2$\else
\ifnum#1=26
$Pmc2_1$\else
\ifnum#1=27
$Pcc2$\else
\ifnum#1=28
$Pma2$\else
\ifnum#1=29
$Pca2_1$\else
\ifnum#1=30
$Pnc2$\else
\ifnum#1=31
$Pmn2_1$\else
\ifnum#1=32
$Pba2$\else
\ifnum#1=33
$Pna2_1$\else
\ifnum#1=34
$Pnn2$\else
\ifnum#1=35
$Cmm2$\else
\ifnum#1=36
$Cmc2_1$\else
\ifnum#1=37
$Ccc2$\else
\ifnum#1=38
$Amm2$\else
\ifnum#1=39
$Aem2$\else
\ifnum#1=40
$Ama2$\else
\ifnum#1=41
$Aea2$\else
\ifnum#1=42
$Fmm2$\else
\ifnum#1=43
$Fdd2$\else
\ifnum#1=44
$Imm2$\else
\ifnum#1=45
$Iba2$\else
\ifnum#1=46
$Ima2$\else
\ifnum#1=47
$Pmmm$\else
\ifnum#1=48
$Pnnn$\else
\ifnum#1=49
$Pccm$\else
\ifnum#1=50
$Pban$\else
\ifnum#1=51
$Pmma$\else
\ifnum#1=52
$Pnna$\else
\ifnum#1=53
$Pmna$\else
\ifnum#1=54
$Pcca$\else
\ifnum#1=55
$Pbam$\else
\ifnum#1=56
$Pccn$\else
\ifnum#1=57
$Pbcm$\else
\ifnum#1=58
$Pnnm$\else
\ifnum#1=59
$Pmmn$\else
\ifnum#1=60
$Pbcn$\else
\ifnum#1=61
$Pbca$\else
\ifnum#1=62
$Pnma$\else
\ifnum#1=63
$Cmcm$\else
\ifnum#1=64
$Cmce$\else
\ifnum#1=65
$Cmmm$\else
\ifnum#1=66
$Cccm$\else
\ifnum#1=67
$Cmme$\else
\ifnum#1=68
$Ccce$\else
\ifnum#1=69
$Fmmm$\else
\ifnum#1=70
$Fddd$\else
\ifnum#1=71
$Immm$\else
\ifnum#1=72
$Ibam$\else
\ifnum#1=73
$Ibca$\else
\ifnum#1=74
$Imma$\else
\ifnum#1=75
$P4$\else
\ifnum#1=76
$P4_1$\else
\ifnum#1=77
$P4_2$\else
\ifnum#1=78
$P4_3$\else
\ifnum#1=79
$I4$\else
\ifnum#1=80
$I4_1$\else
\ifnum#1=81
$P\bar{4}$\else
\ifnum#1=82
$I\bar{4}$\else
\ifnum#1=83
$P4/m$\else
\ifnum#1=84
$P4_2/m$\else
\ifnum#1=85
$P4/n$\else
\ifnum#1=86
$P4_2/n$\else
\ifnum#1=87
$I4/m$\else
\ifnum#1=88
$I4_1/a$\else
\ifnum#1=89
$P422$\else
\ifnum#1=90
$P42_12$\else
\ifnum#1=91
$P4_122$\else
\ifnum#1=92
$P4_12_12$\else
\ifnum#1=93
$P4_222$\else
\ifnum#1=94
$P4_22_12$\else
\ifnum#1=95
$P4_322$\else
\ifnum#1=96
$P4_32_12$\else
\ifnum#1=97
$I422$\else
\ifnum#1=98
$I4_122$\else
\ifnum#1=99
$P4mm$\else
\ifnum#1=100
$P4bm$\else
\ifnum#1=101
$P4_2cm$\else
\ifnum#1=102
$P4_2nm$\else
\ifnum#1=103
$P4cc$\else
\ifnum#1=104
$P4nc$\else
\ifnum#1=105
$P4_2mc$\else
\ifnum#1=106
$P4_2bc$\else
\ifnum#1=107
$I4mm$\else
\ifnum#1=108
$I4cm$\else
\ifnum#1=109
$I4_1md$\else
\ifnum#1=110
$I4_1cd$\else
\ifnum#1=111
$P\bar{4}2m$\else
\ifnum#1=112
$P\bar{4}2c$\else
\ifnum#1=113
$P\bar{4}2_1m$\else
\ifnum#1=114
$P\bar{4}2_1c$\else
\ifnum#1=115
$P\bar{4}m2$\else
\ifnum#1=116
$P\bar{4}c2$\else
\ifnum#1=117
$P\bar{4}b2$\else
\ifnum#1=118
$P\bar{4}n2$\else
\ifnum#1=119
$I\bar{4}m2$\else
\ifnum#1=120
$I\bar{4}c2$\else
\ifnum#1=121
$I\bar{4}2m$\else
\ifnum#1=122
$I\bar{4}2d$\else
\ifnum#1=123
$P4/mmm$\else
\ifnum#1=124
$P4/mcc$\else
\ifnum#1=125
$P4/nbm$\else
\ifnum#1=126
$P4/nnc$\else
\ifnum#1=127
$P4/mbm$\else
\ifnum#1=128
$P4/mnc$\else
\ifnum#1=129
$P4/nmm$\else
\ifnum#1=130
$P4/ncc$\else
\ifnum#1=131
$P4_2/mmc$\else
\ifnum#1=132
$P4_2/mcm$\else
\ifnum#1=133
$P4_2/nbc$\else
\ifnum#1=134
$P4_2/nnm$\else
\ifnum#1=135
$P4_2/mbc$\else
\ifnum#1=136
$P4_2/mnm$\else
\ifnum#1=137
$P4_2/nmc$\else
\ifnum#1=138
$P4_2/ncm$\else
\ifnum#1=139
$I4/mmm$\else
\ifnum#1=140
$I4/mcm$\else
\ifnum#1=141
$I4_1/amd$\else
\ifnum#1=142
$I4_1/acd$\else
\ifnum#1=143
$P3$\else
\ifnum#1=144
$P3_1$\else
\ifnum#1=145
$P3_2$\else
\ifnum#1=146
$R3$\else
\ifnum#1=147
$P\bar{3}$\else
\ifnum#1=148
$R\bar{3}$\else
\ifnum#1=149
$P312$\else
\ifnum#1=150
$P321$\else
\ifnum#1=151
$P3_112$\else
\ifnum#1=152
$P3_121$\else
\ifnum#1=153
$P3_212$\else
\ifnum#1=154
$P3_221$\else
\ifnum#1=155
$R32$\else
\ifnum#1=156
$P3m1$\else
\ifnum#1=157
$P31m$\else
\ifnum#1=158
$P3c1$\else
\ifnum#1=159
$P31c$\else
\ifnum#1=160
$R3m$\else
\ifnum#1=161
$R3c$\else
\ifnum#1=162
$P\bar{3}1m$\else
\ifnum#1=163
$P\bar{3}1c$\else
\ifnum#1=164
$P\bar{3}m1$\else
\ifnum#1=165
$P\bar{3}c1$\else
\ifnum#1=166
$R\bar{3}m$\else
\ifnum#1=167
$R\bar{3}c$\else
\ifnum#1=168
$P6$\else
\ifnum#1=169
$P6_1$\else
\ifnum#1=170
$P6_5$\else
\ifnum#1=171
$P6_2$\else
\ifnum#1=172
$P6_4$\else
\ifnum#1=173
$P6_3$\else
\ifnum#1=174
$P\bar{6}$\else
\ifnum#1=175
$P6/m$\else
\ifnum#1=176
$P6_3/m$\else
\ifnum#1=177
$P622$\else
\ifnum#1=178
$P6_122$\else
\ifnum#1=179
$P6_522$\else
\ifnum#1=180
$P6_222$\else
\ifnum#1=181
$P6_422$\else
\ifnum#1=182
$P6_322$\else
\ifnum#1=183
$P6mm$\else
\ifnum#1=184
$P6cc$\else
\ifnum#1=185
$P6_3cm$\else
\ifnum#1=186
$P6_3mc$\else
\ifnum#1=187
$P\bar{6}m2$\else
\ifnum#1=188
$P\bar{6}c2$\else
\ifnum#1=189
$P\bar{6}2m$\else
\ifnum#1=190
$P\bar{6}2c$\else
\ifnum#1=191
$P6/mmm$\else
\ifnum#1=192
$P6/mcc$\else
\ifnum#1=193
$P6_3/mcm$\else
\ifnum#1=194
$P6_3/mmc$\else
\ifnum#1=195
$P23$\else
\ifnum#1=196
$F23$\else
\ifnum#1=197
$I23$\else
\ifnum#1=198
$P2_13$\else
\ifnum#1=199
$I2_13$\else
\ifnum#1=200
$Pm\bar{3}$\else
\ifnum#1=201
$Pn\bar{3}$\else
\ifnum#1=202
$Fm\bar{3}$\else
\ifnum#1=203
$Fd\bar{3}$\else
\ifnum#1=204
$Im\bar{3}$\else
\ifnum#1=205
$Pa\bar{3}$\else
\ifnum#1=206
$Ia\bar{3}$\else
\ifnum#1=207
$P432$\else
\ifnum#1=208
$P4_232$\else
\ifnum#1=209
$F432$\else
\ifnum#1=210
$F4_132$\else
\ifnum#1=211
$I432$\else
\ifnum#1=212
$P4_332$\else
\ifnum#1=213
$P4_132$\else
\ifnum#1=214
$I4_132$\else
\ifnum#1=215
$P\bar{4}3m$\else
\ifnum#1=216
$F\bar{4}3m$\else
\ifnum#1=217
$I\bar{4}3m$\else
\ifnum#1=218
$P\bar{4}3n$\else
\ifnum#1=219
$F\bar{4}3c$\else
\ifnum#1=220
$I\bar{4}3d$\else
\ifnum#1=221
$Pm\bar{3}m$\else
\ifnum#1=222
$Pn\bar{3}n$\else
\ifnum#1=223
$Pm\bar{3}n$\else
\ifnum#1=224
$Pn\bar{3}m$\else
\ifnum#1=225
$Fm\bar{3}m$\else
\ifnum#1=226
$Fm\bar{3}c$\else
\ifnum#1=227
$Fd\bar{3}m$\else
\ifnum#1=228
$Fd\bar{3}c$\else
\ifnum#1=229
$Im\bar{3}m$\else
\ifnum#1=230
$Ia\bar{3}d$\else
{\color{red}{Invalid SG number}}
\fi
\fi
\fi
\fi
\fi
\fi
\fi
\fi
\fi
\fi
\fi
\fi
\fi
\fi
\fi
\fi
\fi
\fi
\fi
\fi
\fi
\fi
\fi
\fi
\fi
\fi
\fi
\fi
\fi
\fi
\fi
\fi
\fi
\fi
\fi
\fi
\fi
\fi
\fi
\fi
\fi
\fi
\fi
\fi
\fi
\fi
\fi
\fi
\fi
\fi
\fi
\fi
\fi
\fi
\fi
\fi
\fi
\fi
\fi
\fi
\fi
\fi
\fi
\fi
\fi
\fi
\fi
\fi
\fi
\fi
\fi
\fi
\fi
\fi
\fi
\fi
\fi
\fi
\fi
\fi
\fi
\fi
\fi
\fi
\fi
\fi
\fi
\fi
\fi
\fi
\fi
\fi
\fi
\fi
\fi
\fi
\fi
\fi
\fi
\fi
\fi
\fi
\fi
\fi
\fi
\fi
\fi
\fi
\fi
\fi
\fi
\fi
\fi
\fi
\fi
\fi
\fi
\fi
\fi
\fi
\fi
\fi
\fi
\fi
\fi
\fi
\fi
\fi
\fi
\fi
\fi
\fi
\fi
\fi
\fi
\fi
\fi
\fi
\fi
\fi
\fi
\fi
\fi
\fi
\fi
\fi
\fi
\fi
\fi
\fi
\fi
\fi
\fi
\fi
\fi
\fi
\fi
\fi
\fi
\fi
\fi
\fi
\fi
\fi
\fi
\fi
\fi
\fi
\fi
\fi
\fi
\fi
\fi
\fi
\fi
\fi
\fi
\fi
\fi
\fi
\fi
\fi
\fi
\fi
\fi
\fi
\fi
\fi
\fi
\fi
\fi
\fi
\fi
\fi
\fi
\fi
\fi
\fi
\fi
\fi
\fi
\fi
\fi
\fi
\fi
\fi
\fi
\fi
\fi
\fi
\fi
\fi
\fi
\fi
\fi
\fi
\fi
\fi
\fi
\fi
\fi
\fi
\fi
\fi
\fi
\fi
\fi
\fi
\fi
\fi}

\newcommand{\sgsymbnum}[1]{SG #1 (\sgsymb{#1})}

%%\input{general_statistics_20201110.tex}

%% Macros for the TQC Db statistics
%% nbr of ICSDs successfully computed
\newcommand{\TQCDBNbrICSDs}{73,234}
%% nbr of ICSDs w/o SOC successfully computed
\newcommand{\TQCDBNbrNoSOCICSDs}{69,730}
%% percentage of ICSDs w/o SOC successfully computed
\newcommand{\TQCDBPercentNoSOCICSDs}{95.22\%}
%% nbr of ICSDs w/o SOC unsuccessfully computed
\newcommand{\TQCDBNbrFailedNoSOCICSDs}{3,504}
%% percentage of ICSDs w/o SOC unsuccessfully computed
\newcommand{\TQCDBPercentFailedNoSOCICSDs}{4.78\%}
%% nbr of unique materials
\newcommand{\TQCDBNbrUniqueMaterials}{38,298}
%% nbr of unique materials w/o SOC
\newcommand{\TQCDBNbrNoSOCUniqueMaterials}{36,163}
%% percentage of unique materials w/o SOC
\newcommand{\TQCDBPercentageNoSOCUniqueMaterials}{94.43\%}
%% nbr of unique materials with failed w/o SOC calculation
\newcommand{\TQCDBNbrFailedNoSOCUniqueMaterials}{2,135}
%% percentage of unique materials with failed w/o SOC calculation
\newcommand{\TQCDBPercentageFailedNoSOCUniqueMaterials}{5.57\%}

%% nbr of unique materials with f-electrons
\newcommand{\TQCDBNbrMaterialsFElectrons}{10,987}
%% percentage of unique materials with f-electrons
\newcommand{\TQCDBNbrMaterialsFElectronsPercent}{28.69\%}
%% nbr of  ICSDs  w/o SOC and without f-electrons
\newcommand{\TQCDBNbrNoSOCICSDsNoFElectrons}{52,517}
%% percentage of  ICSDs  w/o SOC and without f-electrons
\newcommand{\TQCDBNbrNoSOCICSDsNoFElectronsPercent}{75.31\%}
%% nbr of unique materials tagged as magnetic by MP
\newcommand{\TQCDBNbrMaterialsMagneticMP}{7,124}
%% percentage of unique materials tagged as magnetic by MP
\newcommand{\TQCDBNbrMaterialsMagneticMPPercent}{18.60\%}
%% nbr of unique materials tagged as magnetic by VASP or MP
\newcommand{\TQCDBNbrMaterialsMagneticMPVASP}{13,718}
%% percentage of unique materials tagged as magnetic by VASP or MP
\newcommand{\TQCDBNbrMaterialsMagneticMPVASPPercent}{35.82\%}
%% nbr of unique materials tagged as magnetic by VASP or MP, or having f-electrons
\newcommand{\TQCDBNbrMaterialsMagneticMPVASPFElectrons}{19,987}
%% percentage of unique materials tagged as magnetic by VASP or MP, or having f-electrons
\newcommand{\TQCDBNbrMaterialsMagneticMPVASPFElectronsPercent}{52.19\%}

%% nbr of unique materials with topological classification TI
\newcommand{\TQCDBNbrMaterialsTI}{6,128}
%% percentage of unique materials with topological classification TI
\newcommand{\TQCDBNbrMaterialsTIPercent}{16.00\%}
%% nbr of unique materials with topological classification SM
\newcommand{\TQCDBNbrMaterialsSM}{14,037}
%% percentage of unique materials with topological classification SM
\newcommand{\TQCDBNbrMaterialsSMPercent}{36.65\%}
%% nbr of unique materials with topological classification trivial
\newcommand{\TQCDBNbrMaterialstrivial}{18,133}
%% percentage of unique materials with topological classification trivial
\newcommand{\TQCDBNbrMaterialstrivialPercent}{47.35\%}

%% nbr of unique materials with topological subclassification NLC
\newcommand{\TQCDBNbrMaterialsNLC}{3,000}
%% percentage of unique materials with topological subclassification NLC
\newcommand{\TQCDBNbrMaterialsNLCPercent}{7.83\%}
%% nbr of unique materials with topological subclassification SEBR
\newcommand{\TQCDBNbrMaterialsSEBR}{3,128}
%% percentage of unique materials with topological subclassification SEBR
\newcommand{\TQCDBNbrMaterialsSEBRPercent}{8.17\%}
%% nbr of unique materials with topological subclassification ES
\newcommand{\TQCDBNbrMaterialsES}{4,102}
%% percentage of unique materials with topological subclassification ES
\newcommand{\TQCDBNbrMaterialsESPercent}{10.71\%}
%% nbr of unique materials with topological subclassification ESFD
\newcommand{\TQCDBNbrMaterialsESFD}{9,935}
%% percentage of unique materials with topological subclassification ESFD
\newcommand{\TQCDBNbrMaterialsESFDPercent}{25.94\%}
%% nbr of unique materials with topological subclassification LCEBR
\newcommand{\TQCDBNbrMaterialsLCEBR}{18,133}
%% percentage of unique materials with topological subclassification LCEBR
\newcommand{\TQCDBNbrMaterialsLCEBRPercent}{47.35\%}
%% nbr of topological unique materials at the Fermi level
\newcommand{\TQCDBNbrTopologicalMaterials}{20,165}
%% percentage of topological unique materials at the Fermi level
\newcommand{\TQCDBNbrTopologicalMaterialsPercent}{52.65\%}

%% nbr of unique materials w/o SOC with topological classification SM
\newcommand{\TQCDBNbrNoSOCMaterialsSM}{20,298}
%% percentage of unique materials w/o SOC with topological classification SM
\newcommand{\TQCDBNbrNoSOCMaterialsSMPercent}{56.13\%}
%% nbr of unique materials w/o SOC with topological classification trivial
\newcommand{\TQCDBNbrNoSOCMaterialstrivial}{15,865}
%% percentage of unique materials w/o SOC with topological classification trivial
\newcommand{\TQCDBNbrNoSOCMaterialstrivialPercent}{43.87\%}
%% nbr of unique materials w/o SOC with topological subclassification ES
\newcommand{\TQCDBNbrNoSOCMaterialsES}{6,006}
%% percentage of unique materials w/o SOC with topological subclassification ES
\newcommand{\TQCDBNbrNoSOCMaterialsESPercent}{16.61\%}
%% nbr of unique materials w/o SOC with topological subclassification ESFD
\newcommand{\TQCDBNbrNoSOCMaterialsESFD}{13,997}
%% percentage of unique materials w/o SOC with topological subclassification ESFD
\newcommand{\TQCDBNbrNoSOCMaterialsESFDPercent}{38.71\%}
%% nbr of unique materials w/o SOC with topological subclassification LCEBR
\newcommand{\TQCDBNbrNoSOCMaterialsLCEBR}{15,865}
%% percentage of unique materials w/o SOC with topological subclassification LCEBR
\newcommand{\TQCDBNbrNoSOCMaterialsLCEBRPercent}{43.87\%}
%% nbr of unique materials w/o SOC with topological subclassification NLCSM
\newcommand{\TQCDBNbrNoSOCMaterialsNLCSM}{251}
%% percentage of unique materials w/o SOC with topological subclassification NLCSM
\newcommand{\TQCDBNbrNoSOCMaterialsNLCSMPercent}{0.69\%}
%% nbr of unique materials w/o SOC with topological subclassification SEBRSM
\newcommand{\TQCDBNbrNoSOCMaterialsSEBRSM}{44}
%% percentage of unique materials w/o SOC with topological subclassification SEBRSM
\newcommand{\TQCDBNbrNoSOCMaterialsSEBRSMPercent}{0.12\%}
%% nbr of unique materials w/o SOC with topological subclassification NLCSM and ES
\newcommand{\TQCDBNbrNoSOCMaterialsNLCSMES}{6,257}
%% nbr of unique materials w/o SOC with topological subclassification SEBRSM and ESFD
\newcommand{\TQCDBNbrNoSOCMaterialsSEBRSMESFD}{14,041}
%% nbr of unique materials w/o SOC with topological subclassification NLCSM and SEBRSM
\newcommand{\TQCDBNbrNoSOCMaterialsNLCSMSEBRSM}{295}

%% nbr of unique materials with topological classification TI and where both SOC and w/o SOC data are available
\newcommand{\TQCDBNbrMaterialsWithSOCNoSOCTI}{5,382}
%% percentage of unique materials with topological classification TI and where both SOC and w/o SOC data are available
\newcommand{\TQCDBNbrMaterialsWithSOCNoSOCTIPercent}{14.88\%}
%% nbr of unique materials with topological classification SM and where both SOC and w/o SOC data are available
\newcommand{\TQCDBNbrMaterialsWithSOCNoSOCSM}{13,270}
%% percentage of unique materials with topological classification SM and where both SOC and w/o SOC data are available
\newcommand{\TQCDBNbrMaterialsWithSOCNoSOCSMPercent}{36.69\%}
%% nbr of unique materials with topological classification trivial and where both SOC and w/o SOC data are available
\newcommand{\TQCDBNbrMaterialsWithSOCNoSOCtrivial}{17,511}
%% percentage of unique materials with topological classification trivial and where both SOC and w/o SOC data are available
\newcommand{\TQCDBNbrMaterialsWithSOCNoSOCtrivialPercent}{48.42\%}

%% nbr of unique materials with topological subclassification NLC and where both SOC and w/o SOC data are available
\newcommand{\TQCDBNbrMaterialsWithSOCNoSOCNLC}{2,568}
%% percentage of unique materials with topological subclassification NLC and where both SOC and w/o SOC data are available
\newcommand{\TQCDBNbrMaterialsWithSOCNoSOCNLCPercent}{7.10\%}
%% nbr of unique materials with topological subclassification SEBR and where both SOC and w/o SOC data are available
\newcommand{\TQCDBNbrMaterialsWithSOCNoSOCSEBR}{2,814}
%% percentage of unique materials with topological subclassification SEBR and where both SOC and w/o SOC data are available
\newcommand{\TQCDBNbrMaterialsWithSOCNoSOCSEBRPercent}{7.78\%}
%% nbr of unique materials with topological subclassification ES and where both SOC and w/o SOC data are available
\newcommand{\TQCDBNbrMaterialsWithSOCNoSOCES}{3,785}
%% percentage of unique materials with topological subclassification ES and where both SOC and w/o SOC data are available
\newcommand{\TQCDBNbrMaterialsWithSOCNoSOCESPercent}{10.47\%}
%% nbr of unique materials with topological subclassification ESFD and where both SOC and w/o SOC data are available
\newcommand{\TQCDBNbrMaterialsWithSOCNoSOCESFD}{9,485}
%% percentage of unique materials with topological subclassification ESFD and where both SOC and w/o SOC data are available
\newcommand{\TQCDBNbrMaterialsWithSOCNoSOCESFDPercent}{26.23\%}
%% nbr of unique materials with topological subclassification LCEBR and where both SOC and w/o SOC data are available
\newcommand{\TQCDBNbrMaterialsWithSOCNoSOCLCEBR}{17,511}
%% percentage of unique materials with topological subclassification LCEBR and where both SOC and w/o SOC data are available
\newcommand{\TQCDBNbrMaterialsWithSOCNoSOCLCEBRPercent}{48.42\%}
%% nbr of unique materials with topological subclassification NLC and SEBR, where both SOC and w/o SOC data are available
\newcommand{\TQCDBNbrMaterialsWithSOCNoSOCNLCSEBR}{5,382}
%% nbr of unique materials with topological subclassification SEBR and ESFD, where both SOC and w/o SOC data are available
\newcommand{\TQCDBNbrMaterialsWithSOCNoSOCSEBRESFD}{12,299}
%% nbr of unique materials with topological subclassification NLC and ES, where both SOC and w/o SOC data are available
\newcommand{\TQCDBNbrMaterialsWithSOCNoSOCNLCES}{6,353}

%% nbr of unique supertopological materials
\newcommand{\TQCDBNbrSTopo}{879}
%% percentage of unique supertopological materials
\newcommand{\TQCDBNbrSTopoPercentage}{2.30\%}
%% nbr of unique supertopological materials with topological subclassification NLC at the Fermi level
\newcommand{\TQCDBNbrSTopoNLC}{37}
%% percentage of unique supertopological materials with topological subclassification NLC at the Fermi level
\newcommand{\TQCDBNbrSTopoNLCPercent}{0.10\%}
%% nbr of unique supertopological materials with topological subclassification SEBR at the Fermi level
\newcommand{\TQCDBNbrSTopoSEBR}{116}
%% percentage of unique supertopological materials with topological subclassification SEBR at the Fermi level
\newcommand{\TQCDBNbrSTopoSEBRPercent}{0.30\%}
%% nbr of unique supertopological materials with topological subclassification ES at the Fermi level
\newcommand{\TQCDBNbrSTopoES}{208}
%% percentage of unique supertopological materials with topological subclassification ES at the Fermi level
\newcommand{\TQCDBNbrSTopoESPercent}{0.54\%}
%% nbr of unique supertopological materials with topological subclassification ESFD at the Fermi level
\newcommand{\TQCDBNbrSTopoESFD}{407}
%% percentage of unique supertopological materials with topological subclassification ESFD at the Fermi level
\newcommand{\TQCDBNbrSTopoESFDPercent}{1.06\%}
%% nbr of unique supertopological materials with topological subclassification LCEBR at the Fermi level
\newcommand{\TQCDBNbrSTopoLCEBR}{111}
%% percentage of unique supertopological materials with topological subclassification LCEBR at the Fermi level
\newcommand{\TQCDBNbrSTopoLCEBRPercent}{0.29\%}

%% nbr of unique materials with at least one topological band
\newcommand{\TQCDBNbrTopoBand}{33,698}
%% percentage of unique materials with at least one topological band
\newcommand{\TQCDBNbrTopoBandPercent}{87.99\%}

%% nbr of band sets for unique materials
\newcommand{\TQCDBBandSetsUniqueMaterials}{1,996,728}
%% nbr of LCEBR band sets for unique materials
\newcommand{\TQCDBNbrLCEBRBandSetsUniqueMaterials}{750,504}
%% percentage of LCEBR band sets for unique materials
\newcommand{\TQCDBPercentLCEBRBandSetsUniqueMaterials}{37.59\%}
%% nbr of NLC band sets for unique materials
\newcommand{\TQCDBNbrNLCBandSetsUniqueMaterials}{859,606}
%% percentage of NLC band sets for unique materials
\newcommand{\TQCDBPercentNLCBandSetsUniqueMaterials}{43.05\%}
%% nbr of SEBR band sets for unique materials
\newcommand{\TQCDBNbrSEBRBandSetsUniqueMaterials}{379,321}
%% percentage of SEBR band sets for unique materials
\newcommand{\TQCDBPercentSEBRBandSetsUniqueMaterials}{19.00\%}
%% nbr of strong topological band sets for unique materials
\newcommand{\TQCDBNbrStrongBandSetsUniqueMaterials}{1,238,927}
%% percentage of strong topological band sets for unique materials
\newcommand{\TQCDBPercentStrongBandSetsUniqueMaterials}{62.05\%}
%% nbr of fragile band sets for unique materials
\newcommand{\TQCDBNbrFragileBandSetsUniqueMaterials}{7,297}
%% percentage of fragile band sets for unique materials
\newcommand{\TQCDBPercentFragileBandSetsUniqueMaterials}{0.37\%}

%% Total CPU time for SOC calculations, VASP1 pass
\newcommand{\TQCDBVASPICPUhSOC}{9,500,801.70}
%% Total CPU time for SOC calculations, VASP2 pass
\newcommand{\TQCDBVASPIICPUhSOC}{170,633.80}
%% Total CPU time for SOC calculations, VASP3 pass
\newcommand{\TQCDBVASPIIICPUhSOC}{2,560,882.60}
%% Total CPU time for SOC calculations, VASP4 pass
\newcommand{\TQCDBVASPIVCPUhSOC}{5,804,948.80}
%% Total CPU time for SOC calculations
\newcommand{\TQCDBVASPTotalCPUhSOC}{18,037,266.90}

%% Total CPU time for noSOC calculations, VASP1 pass
\newcommand{\TQCDBVASPICPUhNoSOC}{2,298,583.30}
%% Total CPU time for noSOC calculations, VASP2 pass
\newcommand{\TQCDBVASPIICPUhNoSOC}{56,520.70}
%% Total CPU time for noSOC calculations, VASP3 pass
\newcommand{\TQCDBVASPIIICPUhNoSOC}{773,019.50}
%% Total CPU time for noSOC calculations, VASP4 pass
\newcommand{\TQCDBVASPIVCPUhNoSOC}{1,397,550.30}
%% Total CPU time for noSOC calculations
\newcommand{\TQCDBVASPTotalCPUhNoSOC}{4,525,673.80}

%% Total CPU time for SOC and noSOC calculations (in hours)
\newcommand{\TQCDBVASPTotalCPUTimeAllHours}{22,562,940.70}
%% Total CPU time for SOC and noSOC calculations (in million of hours)
\newcommand{\TQCDBVASPTotalCPUTimeAllMillionHours}{22.60}

%% Total (SOC + w/o SOC) storage
\newcommand{\TQCDBTotalStorage}{2,037.80Gb}

%% Total w/o SOC storage
\newcommand{\TQCDBNoSOCTotalStorage}{343.30Gb}
%% percentage of storage for the w/o SOC data
\newcommand{\TQCDBPercentNoSOCTotalStorage}{16.85\%}
%% Total w/o SOC PROCAR_11.bz2 (i.e. DOS data) storage
\newcommand{\TQCDBNoSOCPROCARStorage}{173.60Gb}
%% Total w/o SOC CHGCAR.bz2 (i.e. SCC data) storage
\newcommand{\TQCDBNoSOCCHGCARStorage}{119.00Gb}

%% Total SOC storage
\newcommand{\TQCDBTotalSOCStorage}{1,694.50Gb}
%% percentage of storage for the SOC data
\newcommand{\TQCDBPercentSOCTotalStorage}{83.15\%}
%% Total SOC PROCAR_11.bz2 (i.e. DOS data) storage
\newcommand{\TQCDBPROCARSOCStorage}{1,013.80Gb}
%% percentage of storage for the SOC PROCAR_11.bz2 data
\newcommand{\TQCDBPercentPROCARSOCStorage}{49.75\%}
%% Total SOC CHGCAR.bz2 (i.e. SCC data) storage
\newcommand{\TQCDBCHGCARSOCStorage}{559.30Gb}
%% percentage of storage for the SOC CHGCAR.bz2 data
\newcommand{\TQCDBPercentCHGCARSOCStorage}{27.45\%}

%%\input{flatband_general_statistics_20211024.tex}
%% Macros for the flat band Db statistics

%% nbr of ICSDs with f-electrons
\newcommand{\TQCDBNbrICSDsFElectrons}{19,798}
%% nbr of ICSDs with f-electrons
\newcommand{\TQCDBNbrICSDsFElectronsPercent}{27.03\%}

%% nbr of ICSDs w/o SOC and with valid f-electrons results
\newcommand{\TQCDBNbrNoSOCICSDsValidFElectrons}{2,689}
%% percentage of ICSDs w/o SOC and with valid f-electrons results
\newcommand{\TQCDBNbrNoSOCICSDsValidFElectronsPercent}{3.67\%}

%% nbr of  ICSDs  w/o SOC, including those with valid f-electrons
\newcommand{\TQCDBNbrNoSOCICSDsIncludingValidFElectrons}{55,206}
%% percentage of  ICSDs  w/o SOC, including those with valid f-electrons
\newcommand{\TQCDBNbrNoSOCICSDsIncludingValidFElectronsPercent}{79.17\%}

%% nbr of unique materials w/o SOC and without f-electrons
\newcommand{\TQCDBNbrNoSOCMaterialsNoFElectrons}{26,802}
%% percentage of unique materials w/o SOC and without f-electrons
\newcommand{\TQCDBNbrNoSOCMaterialsNoFElectronsPercent}{69.98\%}

%% nbr of unique materials w/o SOC and with valid f-electrons results
\newcommand{\TQCDBNbrNoSOCMaterialsWithValidFElectrons}{1,367}
%% percentage of unique materials w/o SOC and with valid f-electrons results
\newcommand{\TQCDBNbrNoSOCMaterialsWithValidFElectronsPercent}{3.57\%}

%% nbr of unique materials w/o SOC, including those valid f-electrons results
\newcommand{\TQCDBNbrNoSOCMaterialsIncludingValidFElectrons}{28,169}
%% percentage of unique materials w/o SOC, including those with valid f-electrons results
\newcommand{\TQCDBNbrNoSOCMaterialsIncludingValidFElectronsPercent}{73.55\%}

%% nbr of curated ICSDs tagged as flat band
\newcommand{\FlatBandNbrCuratedICSDs}{6,338}
%% percentage of curated ICSDs tagged as flat band
\newcommand{\FlatBandNbrCuratedICSDsPercent}{11.48\%}
%% nbr of curated unique materials tagged as flat band
\newcommand{\FlatBandNbrCuratedMaterials}{2,379}
%% percentage of curated unique materials tagged as flat band
\newcommand{\FlatBandNbrCuratedMaterialsPercent}{8.88\%}
%% nbr of ICSDs tagged as atomic flat band
\newcommand{\FlatBandNbrAtomicFlatBandICSDs}{1,830}
%% percentage of ICSDs tagged as atomic flat band
\newcommand{\FlatBandNbrAtomicFlatBandICSDsPercent}{3.31\%}
%% nbr of unique materials tagged as atomic flat band
\newcommand{\FlatBandNbrAtomicFlatBandMaterials}{1,102}
%% percentage of unique materials tagged as atomic flat band
\newcommand{\FlatBandNbrAtomicFlatBandMaterialsPercent}{4.11\%}
%% nbr of best ICSDs tagged as flat band
\newcommand{\FlatBandNbrBestICSDs}{949}
%% percentage of best ICSDs tagged as flat band
\newcommand{\FlatBandNbrBestICSDsPercent}{1.72\%}
%% nbr of best unique materials tagged as flat band
\newcommand{\FlatBandNbrBestMaterials}{345}
%% percentage of best unique materials tagged as flat band
\newcommand{\FlatBandNbrBestMaterialsPercent}{1.22\%}

%% nbr of ICSDs with at least one sublattice
\newcommand{\FlatBandNbrICSDsOneSublattice}{24,052}
%% percentage of ICSDs with at least one sublattice
\newcommand{\FlatBandNbrICSDsOneSublatticePercent}{43.57\%}
%% nbr of ICSDs with a kagome sublattice
\newcommand{\FlatBandNbrICSDsKagome}{6,120}
%% percentage of ICSDs with a kagome sublattice
\newcommand{\FlatBandNbrICSDsKagomePercent}{11.09\%}
%% nbr of ICSDs with a kagome and a bipartite sublattices
\newcommand{\FlatBandNbrICSDsKagomeAndBipartite}{3,244}
%% percentage of ICSDs with a kagome and a bipartite sublattices
\newcommand{\FlatBandNbrICSDsKagomeAndBipartitePercent}{5.88\%}
%% nbr of ICSDs with a kagome and a lieb sublattices
\newcommand{\FlatBandNbrICSDsKagomeAndLieb}{916}
%% percentage of ICSDs with a kagome and a lieb sublattices
\newcommand{\FlatBandNbrICSDsKagomeAndLiebPercent}{1.66\%}
%% nbr of ICSDs with a kagome and a split sublattices
\newcommand{\FlatBandNbrICSDsKagomeAndSplit}{1,974}
%% percentage of ICSDs with a kagome and a split sublattices
\newcommand{\FlatBandNbrICSDsKagomeAndSplitPercent}{3.58\%}
%% nbr of ICSDs with a rigorous kagome sublattice
\newcommand{\FlatBandNbrICSDsRigorousKagome}{4,192}
%% percentage of ICSDs with a rigorous kagome sublattice
\newcommand{\FlatBandNbrICSDsRigorousKagomePercent}{7.59\%}
%% nbr of ICSDs with a approximate kagome sublattice
\newcommand{\FlatBandNbrICSDsApproximateKagome}{2,100}
%% percentage of ICSDs with a approximate kagome sublattice
\newcommand{\FlatBandNbrICSDsApproximateKagomePercent}{3.80\%}
%% nbr of curated flat band ICSDs with a kagome sublattice
\newcommand{\FlatBandNbrCuratedICSDsKagome}{1,699}
%% percentage of curated flat band ICSDs with a kagome sublattice
\newcommand{\FlatBandNbrCuratedICSDsKagomePercent}{26.81\%}
%% nbr of curated flat band ICSDs with a kagome and a bipartite sublattices
\newcommand{\FlatBandNbrCuratedICSDsKagomeAndBipartite}{1,081}
%% percentage of curated flat band ICSDs with a kagome and a bipartite sublattices
\newcommand{\FlatBandNbrCuratedICSDsKagomeAndBipartitePercent}{17.06\%}
%% nbr of curated flat band ICSDs with a kagome and a lieb sublattices
\newcommand{\FlatBandNbrCuratedICSDsKagomeAndLieb}{537}
%% percentage of curated flat band ICSDs with a kagome and a lieb sublattices
\newcommand{\FlatBandNbrCuratedICSDsKagomeAndLiebPercent}{8.47\%}
%% nbr of curated flat band ICSDs with a kagome and a split sublattices
\newcommand{\FlatBandNbrCuratedICSDsKagomeAndSplit}{904}
%% percentage of curated flat band ICSDs with a kagome and a split sublattices
\newcommand{\FlatBandNbrCuratedICSDsKagomeAndSplitPercent}{14.26\%}
%% nbr of best flat band ICSDs with a kagome sublattice
\newcommand{\FlatBandNbrBestICSDsKagome}{516}
%% percentage of best flat band ICSDs with a kagome sublattice
\newcommand{\FlatBandNbrBestICSDsKagomePercent}{54.37\%}
%% nbr of best flat band ICSDs with a kagome and a bipartite sublattices
\newcommand{\FlatBandNbrBestICSDsKagomeAndBipartite}{247}
%% percentage of best flat band ICSDs with a kagome and a bipartite sublattices
\newcommand{\FlatBandNbrBestICSDsKagomeAndBipartitePercent}{26.03\%}
%% nbr of best flat band ICSDs with a kagome and a lieb sublattices
\newcommand{\FlatBandNbrBestICSDsKagomeAndLieb}{135}
%% percentage of best flat band ICSDs with a kagome and a lieb sublattices
\newcommand{\FlatBandNbrBestICSDsKagomeAndLiebPercent}{14.23\%}
%% nbr of best flat band ICSDs with a kagome and a split sublattices
\newcommand{\FlatBandNbrBestICSDsKagomeAndSplit}{228}
%% percentage of best flat band ICSDs with a kagome and a split sublattices
\newcommand{\FlatBandNbrBestICSDsKagomeAndSplitPercent}{24.03\%}
%% nbr of ICSDs with a pyrochlore sublattice
\newcommand{\FlatBandNbrICSDsPyrochlore}{1,666}
%% percentage of ICSDs with a pyrochlore sublattice
\newcommand{\FlatBandNbrICSDsPyrochlorePercent}{3.02\%}
%% nbr of ICSDs with a pyrochlore and a bipartite sublattices
\newcommand{\FlatBandNbrICSDsPyrochloreAndBipartite}{968}
%% percentage of ICSDs with a pyrochlore and a bipartite sublattices
\newcommand{\FlatBandNbrICSDsPyrochloreAndBipartitePercent}{1.75\%}
%% nbr of ICSDs with a pyrochlore and a lieb sublattices
\newcommand{\FlatBandNbrICSDsPyrochloreAndLieb}{0}
%% percentage of ICSDs with a pyrochlore and a lieb sublattices
\newcommand{\FlatBandNbrICSDsPyrochloreAndLiebPercent}{0.00\%}
%% nbr of ICSDs with a pyrochlore and a split sublattices
\newcommand{\FlatBandNbrICSDsPyrochloreAndSplit}{338}
%% percentage of ICSDs with a pyrochlore and a split sublattices
\newcommand{\FlatBandNbrICSDsPyrochloreAndSplitPercent}{0.61\%}
%% nbr of ICSDs with a rigorous pyrochlore sublattice
\newcommand{\FlatBandNbrICSDsRigorousPyrochlore}{1,541}
%% percentage of ICSDs with a rigorous pyrochlore sublattice
\newcommand{\FlatBandNbrICSDsRigorousPyrochlorePercent}{2.79\%}
%% nbr of ICSDs with a approximate pyrochlore sublattice
\newcommand{\FlatBandNbrICSDsApproximatePyrochlore}{125}
%% percentage of ICSDs with a approximate pyrochlore sublattice
\newcommand{\FlatBandNbrICSDsApproximatePyrochlorePercent}{0.23\%}
%% nbr of curated flat band ICSDs with a pyrochlore sublattice
\newcommand{\FlatBandNbrCuratedICSDsPyrochlore}{296}
%% percentage of curated flat band ICSDs with a pyrochlore sublattice
\newcommand{\FlatBandNbrCuratedICSDsPyrochlorePercent}{4.67\%}
%% nbr of curated flat band ICSDs with a pyrochlore and a bipartite sublattices
\newcommand{\FlatBandNbrCuratedICSDsPyrochloreAndBipartite}{172}
%% percentage of curated flat band ICSDs with a pyrochlore and a bipartite sublattices
\newcommand{\FlatBandNbrCuratedICSDsPyrochloreAndBipartitePercent}{2.71\%}
%% nbr of curated flat band ICSDs with a pyrochlore and a lieb sublattices
\newcommand{\FlatBandNbrCuratedICSDsPyrochloreAndLieb}{0}
%% percentage of curated flat band ICSDs with a pyrochlore and a lieb sublattices
\newcommand{\FlatBandNbrCuratedICSDsPyrochloreAndLiebPercent}{0.00\%}
%% nbr of curated flat band ICSDs with a pyrochlore and a split sublattices
\newcommand{\FlatBandNbrCuratedICSDsPyrochloreAndSplit}{66}
%% percentage of curated flat band ICSDs with a pyrochlore and a split sublattices
\newcommand{\FlatBandNbrCuratedICSDsPyrochloreAndSplitPercent}{1.04\%}
%% nbr of best flat band ICSDs with a pyrochlore sublattice
\newcommand{\FlatBandNbrBestICSDsPyrochlore}{77}
%% percentage of best flat band ICSDs with a pyrochlore sublattice
\newcommand{\FlatBandNbrBestICSDsPyrochlorePercent}{8.11\%}
%% nbr of best flat band ICSDs with a pyrochlore and a bipartite sublattices
\newcommand{\FlatBandNbrBestICSDsPyrochloreAndBipartite}{16}
%% percentage of best flat band ICSDs with a pyrochlore and a bipartite sublattices
\newcommand{\FlatBandNbrBestICSDsPyrochloreAndBipartitePercent}{1.69\%}
%% nbr of best flat band ICSDs with a pyrochlore and a lieb sublattices
\newcommand{\FlatBandNbrBestICSDsPyrochloreAndLieb}{0}
%% percentage of best flat band ICSDs with a pyrochlore and a lieb sublattices
\newcommand{\FlatBandNbrBestICSDsPyrochloreAndLiebPercent}{0.00\%}
%% nbr of best flat band ICSDs with a pyrochlore and a split sublattices
\newcommand{\FlatBandNbrBestICSDsPyrochloreAndSplit}{10}
%% percentage of best flat band ICSDs with a pyrochlore and a split sublattices
\newcommand{\FlatBandNbrBestICSDsPyrochloreAndSplitPercent}{1.05\%}
%% nbr of ICSDs with a Lieb sublattice
\newcommand{\FlatBandNbrICSDsLieb}{1,590}
%% percentage of ICSDs with a Lieb sublattice
\newcommand{\FlatBandNbrICSDsLiebPercent}{2.88\%}
%% nbr of ICSDs with a rigorous lieb sublattice
\newcommand{\FlatBandNbrICSDsRigorousLieb}{1,202}
%% percentage of ICSDs with a rigorous lieb sublattice
\newcommand{\FlatBandNbrICSDsRigorousLiebPercent}{2.18\%}
%% nbr of ICSDs with a approximate lieb sublattice
\newcommand{\FlatBandNbrICSDsApproximateLieb}{438}
%% percentage of ICSDs with a approximate lieb sublattice
\newcommand{\FlatBandNbrICSDsApproximateLiebPercent}{0.79\%}
%% nbr of curated flat band ICSDs with a lieb sublattice
\newcommand{\FlatBandNbrCuratedICSDsLieb}{721}
%% percentage of curated flat band ICSDs with a lieb sublattice
\newcommand{\FlatBandNbrCuratedICSDsLiebPercent}{11.38\%}
%% nbr of best flat band ICSDs with a lieb sublattice
\newcommand{\FlatBandNbrBestICSDsLieb}{151}
%% percentage of best flat band ICSDs with a lieb sublattice
\newcommand{\FlatBandNbrBestICSDsLiebPercent}{15.91\%}
%% nbr of ICSDs with a bipartite sublattice
\newcommand{\FlatBandNbrICSDsBipartite}{21,175}
%% percentage of ICSDs with a bipartite sublattice
\newcommand{\FlatBandNbrICSDsBipartitePercent}{38.36\%}
%% nbr of curated flat band ICSDs with a bipartite sublattice
\newcommand{\FlatBandNbrCuratedICSDsBipartite}{3,138}
%% percentage of curated flat band ICSDs with a bipartite sublattice
\newcommand{\FlatBandNbrCuratedICSDsBipartitePercent}{49.51\%}
%% nbr of best flat band ICSDs with a bipartite sublattice
\newcommand{\FlatBandNbrBestICSDsBipartite}{432}
%% percentage of best flat band ICSDs with a bipartite sublattice
\newcommand{\FlatBandNbrBestICSDsBipartitePercent}{45.52\%}
%% nbr of ICSDs with a split sublattice
\newcommand{\FlatBandNbrICSDsSplit}{8,224}
%% percentage of ICSDs with a split sublattice
\newcommand{\FlatBandNbrICSDsSplitPercent}{14.90\%}
%% nbr of curated flat band ICSDs with a split sublattice
\newcommand{\FlatBandNbrCuratedICSDsSplit}{1,920}
%% percentage of curated flat band ICSDs with a split sublattice
\newcommand{\FlatBandNbrCuratedICSDsSplitPercent}{30.29\%}
%% nbr of best flat band ICSDs with a split sublattice
\newcommand{\FlatBandNbrBestICSDsSplit}{354}
%% percentage of best flat band ICSDs with a split sublattice
\newcommand{\FlatBandNbrBestICSDsSplitPercent}{37.30\%}
%% nbr of ICSDs without any sublattice
\newcommand{\FlatBandNbrICSDsNoSublattices}{31,154}
%% percentage of ICSDs without any sublattice
\newcommand{\FlatBandNbrICSDsNoSublatticesPercent}{56.43\%}
%% nbr of curated flat band ICSDs without any sublattice
\newcommand{\FlatBandNbrCuratedICSDsNoSublattices}{2,582}
%% percentage of curated flat band ICSDs without any sublattice
\newcommand{\FlatBandNbrCuratedICSDsNoSublatticesPercent}{40.74\%}
%% nbr of best flat band ICSDs without any sublattice
\newcommand{\FlatBandNbrBestICSDsNoSublattices}{248}
%% percentage of best flat band ICSDs without any sublattice
\newcommand{\FlatBandNbrBestICSDsNoSublatticesPercent}{26.13\%}
%% nbr of ICSDs with at least one sublattice
\newcommand{\FlatBandNbrICSDsAtLeastOneSublattice}{24,052}
%% percentage of ICSDs with at least one sublattice
\newcommand{\FlatBandNbrICSDsAtLeastOneSublatticePercent}{43.57\%}
%% nbr of curated flat band ICSDs with at least one sublattice
\newcommand{\FlatBandNbrCuratedICSDsAtLeastOneSublattice}{3,756}
%% percentage of curated flat band ICSDs with at least one sublattice
\newcommand{\FlatBandNbrCuratedICSDsAtLeastOneSublatticePercent}{59.26\%}
%% nbr of best flat band ICSDs with at least one sublattice
\newcommand{\FlatBandNbrBestICSDsAtLeastOneSublattice}{701}
%% percentage of best flat band ICSDs with at least one sublattice
\newcommand{\FlatBandNbrBestICSDsAtLeastOneSublatticePercent}{73.87\%}

%% nbr of ICSDs with only bipartite sublattice
\newcommand{\FlatBandNbrICSDsOnlyBipartiteSublattices}{17,258}
%% percentage of ICSDs with only bipartite sublattice
\newcommand{\FlatBandNbrICSDsOnlyBipartiteSublatticesPercent}{31.26\%}
%% nbr of curated flat band ICSDs with only bipartite sublattice
\newcommand{\FlatBandNbrCuratedICSDsOnlyBipartiteSublattices}{1,873}
%% percentage of curated flat band ICSDs with only bipartite sublattice
\newcommand{\FlatBandNbrCuratedICSDsOnlyBipartiteSublatticesPercent}{29.55\%}
%% nbr of best flat band ICSDs with only bipartite sublattice
\newcommand{\FlatBandNbrBestICSDsOnlyBipartiteSublattices}{169}
%% percentage of best flat band ICSDs with only bipartite sublattice
\newcommand{\FlatBandNbrBestICSDsOnlyBipartiteSublatticesPercent}{17.81\%}
%% nbr of ICSDs with only lieb sublattice
\newcommand{\FlatBandNbrICSDsOnlyLiebSublattices}{1}
%% percentage of ICSDs with only lieb sublattice
\newcommand{\FlatBandNbrICSDsOnlyLiebSublatticesPercent}{0.00\%}
%% nbr of curated flat band ICSDs with only lieb sublattice
\newcommand{\FlatBandNbrCuratedICSDsOnlyLiebSublattices}{0}
%% percentage of curated flat band ICSDs with only lieb sublattice
\newcommand{\FlatBandNbrCuratedICSDsOnlyLiebSublatticesPercent}{0.00\%}
%% nbr of best flat band ICSDs with only lieb sublattice
\newcommand{\FlatBandNbrBestICSDsOnlyLiebSublattices}{0}
%% percentage of best flat band ICSDs with only lieb sublattice
\newcommand{\FlatBandNbrBestICSDsOnlyLiebSublatticesPercent}{0.00\%}
%% nbr of ICSDs with only kagome sublattice
\newcommand{\FlatBandNbrICSDsOnlyKagomeSublattices}{2,178}
%% percentage of ICSDs with only kagome sublattice
\newcommand{\FlatBandNbrICSDsOnlyKagomeSublatticesPercent}{3.95\%}
%% nbr of curated flat band ICSDs with only kagome sublattice
\newcommand{\FlatBandNbrCuratedICSDsOnlyKagomeSublattices}{494}
%% percentage of curated flat band ICSDs with only kagome sublattice
\newcommand{\FlatBandNbrCuratedICSDsOnlyKagomeSublatticesPercent}{7.79\%}
%% nbr of best flat band ICSDs with only kagome sublattice
\newcommand{\FlatBandNbrBestICSDsOnlyKagomeSublattices}{208}
%% percentage of best flat band ICSDs with only kagome sublattice
\newcommand{\FlatBandNbrBestICSDsOnlyKagomeSublatticesPercent}{21.92\%}
%% nbr of ICSDs with only pyrochlore sublattice
\newcommand{\FlatBandNbrICSDsOnlyPyrochloreSublattices}{698}
%% percentage of ICSDs with only pyrochlore sublattice
\newcommand{\FlatBandNbrICSDsOnlyPyrochloreSublatticesPercent}{1.26\%}
%% nbr of curated flat band ICSDs with only pyrochlore sublattice
\newcommand{\FlatBandNbrCuratedICSDsOnlyPyrochloreSublattices}{124}
%% percentage of curated flat band ICSDs with only pyrochlore sublattice
\newcommand{\FlatBandNbrCuratedICSDsOnlyPyrochloreSublatticesPercent}{1.96\%}
%% nbr of best flat band ICSDs with only pyrochlore sublattice
\newcommand{\FlatBandNbrBestICSDsOnlyPyrochloreSublattices}{61}
%% percentage of best flat band ICSDs with only pyrochlore sublattice
\newcommand{\FlatBandNbrBestICSDsOnlyPyrochloreSublatticesPercent}{6.43\%}

%% nbr of most prominent structure types in the list of best flat band ICSDs
\newcommand{\FlatBandNbrProminentStructuresBestICSDs}{8}
%% nbr of best flat band ICSDs among the prominent structure types
\newcommand{\FlatBandNbrBestICSDsProminentStructures}{323}
%% percentage of best flat band ICSDs among the prominent structure types
\newcommand{\FlatBandNbrBestICSDsProminentStructuresPercent}{34.04\%}
%% nbr of best flat band unique materials among the prominent structure types
\newcommand{\FlatBandNbrBestMaterialsProminentStructures}{122}
%% percentage of best flat band unique materials among the prominent structure types
\newcommand{\FlatBandNbrBestMaterialsProminentStructuresPercent}{35.36\%}
%% nbr of best flat band ICSDs with at least one sublattice
\newcommand{\FlatBandNbrBestICSDsHeusler}{100}
%% percentage of best flat band ICSDs with structure type Heusler-AlCu2Mn
\newcommand{\FlatBandNbrBestICSDsHeuslerPercent}{10.54\%}
%% nbr of best flat band unique materials with at least one sublattice
\newcommand{\FlatBandNbrBestMaterialsHeusler}{35}
%% percentage of best flat band unique materials with structure type Heusler-AlCu2Mn
\newcommand{\FlatBandNbrBestMaterialsHeuslerPercent}{10.14\%}
%% nbr of best flat band ICSDs with at least one sublattice
\newcommand{\FlatBandNbrBestICSDsPerovskite}{69}
%% percentage of best flat band ICSDs with structure type Perovskite-CaTiO3
\newcommand{\FlatBandNbrBestICSDsPerovskitePercent}{7.27\%}
%% nbr of best flat band unique materials with at least one sublattice
\newcommand{\FlatBandNbrBestMaterialsPerovskite}{27}
%% percentage of best flat band unique materials with structure type Perovskite-CaTiO3
\newcommand{\FlatBandNbrBestMaterialsPerovskitePercent}{7.83\%}
%% nbr of best flat band ICSDs with at least one sublattice
\newcommand{\FlatBandNbrBestICSDsElpasolite}{44}
%% percentage of best flat band ICSDs with structure type Elpasolite-K2NaAlF6
\newcommand{\FlatBandNbrBestICSDsElpasolitePercent}{4.64\%}
%% nbr of best flat band unique materials with at least one sublattice
\newcommand{\FlatBandNbrBestMaterialsElpasolite}{20}
%% percentage of best flat band unique materials with structure type Elpasolite-K2NaAlF6
\newcommand{\FlatBandNbrBestMaterialsElpasolitePercent}{5.80\%}
%% nbr of best flat band ICSDs with at least one sublattice
\newcommand{\FlatBandNbrBestICSDsSrNiWo}{41}
%% percentage of best flat band ICSDs with structure type Sr2NiWO6
\newcommand{\FlatBandNbrBestICSDsSrNiWoPercent}{4.32\%}
%% nbr of best flat band unique materials with at least one sublattice
\newcommand{\FlatBandNbrBestMaterialsSrNiWo}{11}
%% percentage of best flat band unique materials with structure type Sr2NiWO6
\newcommand{\FlatBandNbrBestMaterialsSrNiWoPercent}{3.19\%}
%% nbr of best flat band ICSDs with at least one sublattice
\newcommand{\FlatBandNbrBestICSDsCuHgTi}{11}
%% percentage of best flat band ICSDs with structure type CuHg2Ti
\newcommand{\FlatBandNbrBestICSDsCuHgTiPercent}{1.16\%}
%% nbr of best flat band unique materials with at least one sublattice
\newcommand{\FlatBandNbrBestMaterialsCuHgTi}{10}
%% percentage of best flat band unique materials with structure type CuHg2Ti
\newcommand{\FlatBandNbrBestMaterialsCuHgTiPercent}{2.90\%}
%% nbr of best flat band ICSDs with at least one sublattice
\newcommand{\FlatBandNbrBestICSDsKPtCl}{20}
%% percentage of best flat band ICSDs with structure type K2PtCl6
\newcommand{\FlatBandNbrBestICSDsKPtClPercent}{2.11\%}
%% nbr of best flat band unique materials with at least one sublattice
\newcommand{\FlatBandNbrBestMaterialsKPtCl}{8}
%% percentage of best flat band unique materials with structure type K2PtCl6
\newcommand{\FlatBandNbrBestMaterialsKPtClPercent}{2.32\%}
%% nbr of best flat band ICSDs with at least one sublattice
\newcommand{\FlatBandNbrBestICSDsHalfHeusler}{21}
%% percentage of best flat band ICSDs with structure type Heusler-AlLiSi
\newcommand{\FlatBandNbrBestICSDsHalfHeuslerPercent}{2.21\%}
%% nbr of best flat band unique materials with at least one sublattice
\newcommand{\FlatBandNbrBestMaterialsHalfHeusler}{6}
%% percentage of best flat band unique materials with structure type Heusler-AlLiSi
\newcommand{\FlatBandNbrBestMaterialsHalfHeuslerPercent}{1.74\%}
%% nbr of best flat band ICSDs with at least one sublattice
\newcommand{\FlatBandNbrBestICSDsNiPbS}{17}
%% percentage of best flat band ICSDs with structure type Ni3Pb2S2
\newcommand{\FlatBandNbrBestICSDsNiPbSPercent}{1.79\%}
%% nbr of best flat band unique materials with at least one sublattice
\newcommand{\FlatBandNbrBestMaterialsNiPbS}{5}
%% percentage of best flat band unique materials with structure type Ni3Pb2S2
\newcommand{\FlatBandNbrBestMaterialsNiPbSPercent}{1.45\%}

%% nbr of ICSDs with a linegraph sublattice
\newcommand{\FlatBandNbrICSDsLinegraph}{4,409}
%% percentage of ICSDs with a linegraph sublattice
\newcommand{\FlatBandNbrICSDsLinegraphPercent}{7.99\%}
%% nbr of ICSDs with a 2d linegraph sublattice
\newcommand{\FlatBandNbrICSDsTwoDLinegraph}{3,761}
%% percentage of ICSDs with a 2d linegraph sublattice
\newcommand{\FlatBandNbrICSDsTwoDLinegraphPercent}{6.81\%}
%% nbr of ICSDs with a 2d linegraph sublattice different from kagome
\newcommand{\FlatBandNbrICSDsTwoDLinegraphNotKagome}{2,655}
%% percentage of ICSDs with a 2d linegraph sublattice different from kagome
\newcommand{\FlatBandNbrICSDsTwoDLinegraphPercentNotKagome}{4.81\%}
%% nbr of ICSDs with a 2d linegraph sublattice with a gap
\newcommand{\FlatBandNbrICSDsTwoDLinegraphGapped}{273}
%% percentage of ICSDs with a 2d linegraph sublattice with a gap
\newcommand{\FlatBandNbrICSDsTwoDLinegraphPercentGapped}{0.49\%}
%% nbr of ICSDs with a quasi-2d linegraph sublattice
\newcommand{\FlatBandNbrICSDsQuasiTwoDLinegraph}{131}
%% percentage of ICSDs with a quasi-2d linegraph sublattice
\newcommand{\FlatBandNbrICSDsQuasiTwoDLinegraphPercent}{0.24\%}
%% nbr of ICSDs with a quasi-2d linegraph sublattice different from kagome
\newcommand{\FlatBandNbrICSDsQuasiTwoDLinegraphNotKagome}{129}
%% percentage of ICSDs with a quasi-2d linegraph sublattice  different from kagome
\newcommand{\FlatBandNbrICSDsQuasiTwoDLinegraphNotKagomePercent}{0.23\%}
%% nbr of ICSDs with a quasi-2d linegraph sublattice with a gap
\newcommand{\FlatBandNbrICSDsQuasiTwoDLinegraphGapped}{7}
%% percentage of ICSDs with a quasi-2d linegraph sublattice with a gap
\newcommand{\FlatBandNbrICSDsQuasiTwoDLinegraphGappedPercent}{0.01\%}
%% nbr of ICSDs with a 3d linegraph sublattice
\newcommand{\FlatBandNbrICSDsThreeDLinegraph}{729}
%% percentage of ICSDs with a 3d linegraph sublattice
\newcommand{\FlatBandNbrICSDsThreeDLinegraphPercent}{1.32\%}
%% nbr of ICSDs with a 3d linegraph sublattice different from pyrochlore
\newcommand{\FlatBandNbrICSDsThreeDLinegraphNotPyrochlore}{340}
%% percentage of ICSDs with a 3d linegraph sublattice different from pyrochlore
\newcommand{\FlatBandNbrICSDsThreeDLinegraphNotPyrochlorePercent}{0.62\%}
%% nbr of ICSDs with a 3d linegraph sublattice with a gap
\newcommand{\FlatBandNbrICSDsThreeDLinegraphGapped}{120}
%% percentage of ICSDs with a 3d linegraph sublattice with a gap
\newcommand{\FlatBandNbrICSDsThreeDLinegraphPercentGapped}{0.22\%}

\tolerance 10000

\title{Catalogue of Flat-Band Stoichiometric Materials}

\author{Nicolas Regnault}\email{regnault@princeton.edu}
\thanks{These authors contributed equally} 
\affiliation{Department of Physics, Princeton University, Princeton, New Jersey 08544, USA}
\affiliation{Laboratoire de Physique de l'Ecole normale sup\'{e}rieure, ENS, Universit\'{e} PSL, CNRS, Sorbonne Universit\'{e}, Universit\'{e} Paris-Diderot, Sorbonne Paris Cit\'{e}, 75005 Paris, France}

\author{Yuanfeng Xu} \email{yfxu@mpi-halle.mpg.de}
\thanks{These authors contributed equally} 
\affiliation{Max Planck Institute of Microstructure Physics, 06120 Halle, Germany}

\author{Ming-Rui Li}
\thanks{These authors contributed equally} 
\affiliation{Department of Physics, Tsinghua University, Beijing, 100084, China}

\author{Da-Shuai Ma}
\thanks{These authors contributed equally} 
\affiliation{Beijing Key Laboratory of Nanophotonics and Ultrafine Optoelectronic Systems, School of Physics, Beijing Institute of Technology, Beijing 100081, China}

\author{Milena Jovanovic}
\affiliation{Department of Chemistry, Princeton University, Princeton, New Jersey 08544, USA}

\author{Ali Yazdani}
\affiliation{Department of Physics, Princeton University, Princeton, New Jersey 08544, USA}

\author{Stuart S. P. Parkin}
\affiliation{Max Planck Institute of Microstructure Physics, 06120 Halle, Germany}

\author{Claudia Felser}
\affiliation{Max Planck Institute for Chemical Physics of Solids, 01187 Dresden, Germany}

\author{Leslie M. Schoop}
\affiliation{Department of Chemistry, Princeton University, Princeton, New Jersey 08544, USA}

\author{N. Phuan Ong}
\affiliation{Department of Physics, Princeton University, Princeton, New Jersey 08544, USA}

\author{Robert J. Cava}
\affiliation{Department of Chemistry, Princeton University, Princeton, New Jersey 08544, USA}

\author{Luis Elcoro}
\thanks{These authors contributed equally} 
\affiliation{Department of Condensed Matter Physics, University of the Basque Country UPV/EHU, Apartado 644, 48080 Bilbao, Spain}

\author{Zhi-Da Song}
\thanks{These authors contributed equally} 
\affiliation{Department of Physics, Princeton University, Princeton, New Jersey 08544, USA}

\author{B. Andrei Bernevig}\email{bernevig@princeton.edu}
\thanks{These authors contributed equally} 
\affiliation{Department of Physics, Princeton University, Princeton, New Jersey 08544, USA}
\affiliation{Donostia International Physics Center, P. Manuel de Lardizabal 4, 20018 Donostia-San Sebastian, Spain}
\affiliation{IKERBASQUE, Basque Foundation for Science, Bilbao, Spain}

\date{\today}

\begin{abstract}

Topological electronic flatten bands near or at the Fermi level are a promising avenue towards unconventional superconductivity and correlated insulating states. However, the related experiments are mostly limited to the engineered materials, such as moir\'e systems~\cite{bistritzer2011moire,cao2018unconventional,cao2018correlated}. Here we present a catalogue of all the three-dimensional stoichiometric materials with flat bands around the Fermi level that exist in nature. We consider \TQCDBNbrNoSOCICSDsIncludingValidFElectrons\ materials from the Inorganic Crystal Structure Database catalogued using the \webTQC~\cite{Vergniory2019,Vergniory2021},  which provides their structural parameters, space group, band structure, density of states and topological characterization. We combine several direct signatures and properties of band flatness to a high-throughput analysis of all crystal structures. In particular, we identify materials hosting line-graph or bipartite sublattices - either in two or three dimensions - likely leading to flat bands. From this trove of information, we create the \webflatband, a powerful search engine for future theoretical and experimental studies. We use it to extract a curated list of \FlatBandNbrCuratedMaterials\ materials, with among them \FlatBandNbrBestMaterials\ promising candidates, potentially hosting flat bands whose  charge centers are not strongly localized on the atomic sites. We showcase five representative materials: \ch{KAg[CN]2}, \ch{Pb2Sb2O7}, \ch{Rb2CaH4}, \ch{Ca2NCl} and \ch{WO3}. We provide a theoretical explanation for the origin of their flat bands close to the Fermi energy using the $S$-matrix method introduced in a parallel work \cite{S-matrix}.
\end{abstract}

\maketitle

\addtocontents{toc}{\protect\setcounter{tocdepth}{0}}
\addtocontents{lot}{\protect\setcounter{lotdepth}{-1}}

\section{Introduction}\label{sec:introduction}

Electrons whose energy dispersion is bound within a narrow window are conjectured to exhibit a wide-range of interesting physics phenomena. Such electrons form a high density of states ``flat band'', where many-body effects dominate over the kinetic energy and where Fermi-surface physics gives way to strongly interacting, non-Fermi liquid behavior~\cite{kumar2021flat}. The archetypal - and until recently the only experimentally discovered - such system is the Fractional Quantum Hall effect\cite{tsui-82prl1559,Laughlin:1983p301} where anyonic (potentially non-Abelian\cite{Moore:1991p165}) quasi-particle excitations can appear under a fractional filling of an electronic flat band that develops in the presence of a large magnetic field. New developments in engineered solid-state materials have now shown that flat bands can exist even in the absence of a large magnetic field. In moir{\'e} materials such as (but not limited to) twisted bilayer graphene (TBG)~\cite{bistritzer2011moire,cao2018unconventional,cao2018correlated}, flat electronic bands are obtained by creating large, many nanometer-size moir{\'e} unit cell which folds and flattens the initial band structure of the material. This flatness plays a crucial role in the physics of TBG, leading to, e.g., both the correlated insulator states and of the strong-coupling superconductivity that renders the TBG phase diagram akin to that of the high-temperature cuprates. However, as the unit cell is large, the electron density in moir{\'e} samples is necessarily low, preventing new type of physics associated with high electron density \cite{drozdov2015conventional,drozdov2019superconductivity}. This renders the yet elusive prediction of flat bands in non-moir{\'e}, stoichiometric crystals of immediate importance.

In the present article and its accompanying Supplementary Information (SI), we address for the question of predicting and classifying all the flat bands in all the stoichiometric crystals currently present in nature, keeping in mind that not all flat bands are created equal. Extremely localized orbitals - or large unit cells with well-separated atoms - can easily give rise to mundane flat atomic bands (FAB), as the kinetic energy is suppressed by the vanishing overlap between atomic wavefunctions, as schematically shown in Fig.~\ref{fig:mainfig1}(a). The FAB is very common in layered and heavy fermion systems.  At the opposite side of the spectrum are the flat  topological bands (FTB) created by completely extended wavefunctions (such as is the case in TBG), as shown in the schematic Fig.~\ref{fig:mainfig1}(b). (See \sirefappflatbandtopology\ for a more detailed discussion of the FAB and FTB.) There, the quenching of the kinetic energy arises from interference effects despite large electron orbital overlaps and hopping. The latter type of bands can host many exotic quantum phenomena, including magnetism, fractional quantum Hall effect at zero field \cite{tang2011high,neupert2011fractional,sheng2011fractional,regnault2011fractional}, unconventional superconductivity \cite{cao2018unconventional,balents2020superconductivity,peri2021fragile}, non-Fermi liquid behavior \cite{kumar2021flat} and anomalous Landau level beyond the Onsager’s rule \cite{rhim2020quantum}. Such topological  bands can  enhance the superfluid weight in twisted bilayer graphene \cite{xie2020topology, peotta2015superfluidity} and could lead to high-temperature superconductivity. Ideal FTB near the Fermi level in crystalline materials has not yet been found; the only experimentally found FTBs are in the engineered TBG.
A third type, the flat obstructed atomic band (FOAB) lies at the interface between the polar opposites FAB and FTB: while the electron's symmetric Wannier function can be localized in real space, the Wannier center is pinned and centered at an empty site \cite{QuantumChemistry,xu2021filling} and hence delocalized from the atomic sites, as illustrated in Fig.~\ref{fig:mainfig1}(c). 

We present and implement algorithms for the detection and classification of flat bands near the Fermi level. Using the materials in our database \webTQC\ (\webTQCacronym)  (which contains most ICSD stoichiomentric structures) obtained in previous works \cite{QuantumChemistry,Vergniory2019,Vergniory2021}, we build the complementary \webflatband\ (\webflatbandacronym), where different algorithms  and search options for flat bands are provided to the user. First, we perform a brute-force search based on complementary ``flatness'' criteria such as bandwidth and density of states (DOS), to predict all the (thousands) flat-band materials in the ICSD database. We classify these bands based on their topologies. Second, using a theory that we have developed in Ref.~\cite{S-matrix} encompassing generic orbital systems with or without SOC, we perform a targeted search of flat-band materials based on the lattice geometry (such as Kagome, pyrochlore, Lieb, bipartite or split sublattices) of the compounds, whose information is present in the X-ray diffraction data on \webTQCacronym\ or ICSD. We show that geometry-based theoretical models based on the $S$-matrix method \cite{S-matrix} fit our ab-initio calculations of the flat bands remarkably well. Third, we also perform a manual check of thousands of materials for the best flat bands and select \FlatBandNbrCuratedMaterials~materials with high-quality flat bands near the Fermi level. In the main text, we showcase our methods, theoretical understanding and predictive power for five represented flat-band materials, while in the Supplementary Material we present thousands of others. Our classification and predictions take into account the different flat band natures, including their topological character, and our database is coupled to the \webmaterialsproject\ and \webscnims, providing information about the magnetic and superconducting properties (including high $T_c$) of the candidate materials.

\section{Database of Flat Bands Materials}\label{sec:databasemethodology}

In this work, we have used the \webTQCacronym \cite{Vergniory2019,Vergniory2021} as our materials database. We summarize its main feature in the Methods section and provide its detailed overview in \sirefapptqcdboverview. 
%The database used as an input the structural parameters of stoichiometric materials reported in the Inorganic Crystal Structure Database (ICSD)~\cite{ICSD}. For each entry, ab-initio calculations were performed using Density Functional Theory (DFT)~\cite{Hohenberg-PR64,Kohn-PR65} and its implementation in the Vienna Ab-initio Simulation Package (VASP)~\cite{PhysRevB.48.13115,vasp1}, with and without accounting for the spin-orbit coupling (SOC). The database provides the structural parameters, the band dispersion along high symmetry lines, the density of states, and the topological characterization for each set of bands in the material's band structure for each of the \TQCDBNbrNoSOCICSDs~ICSD entries. 
As we are interested in flat bands near the Fermi energy, we discard materials containing rare-earth elements (with the exception of La atom) and actinides as these elements usually lead to spurious flat bands due to $f$ electrons in the ab-initio calculations. 
In total, \TQCDBNbrNoSOCICSDsIncludingValidFElectrons~ICSD entries have been considered for our high-throughput search for flat bands. The automated search is based on two main, complementary, approaches: the detection of flat bands in the band structure and the DOS, and the identification of special sublattices that lead to band flattening. We will now detail each of them.

\subsection{High-throughput search of flat-bands}\label{sec:flatbandsegments}

For determining band flatness, we rely on calculations where the spin-orbit coupling (SOC) is neglected. While SOC plays an important role in the topological features near the Fermi level, it does not drastically alter the band structure nor the DOS around the Fermi energy $E_F$ $\pm 2{\rm eV}$ region where we focus our search. For each ICSD entry, our database provides the ab-initio paramagnetic-phase electronic band structure along paths made of straight lines in the Brillouin zone connecting high symmetry points (" high symmetry lines "). Each high symmetry line is well defined in every space group (SG) and it has been discretized, irrespective of its length, into 20 equally distant $k$-points (\ie points in the BZ). Since we are interested in low-energy physics, we focus our investigations on the flatness of the two highest/lowest occupied/empty bands around $E_F$, which in turn is reached when the occupation number of bands equals half the number of valence electrons. While paramagnetic calculations would failed to capture (anti-)ferromagnetic ground states, we discuss in \sirefappmagflat\ four representative ferromagnetic compounds. Our ferromagnetic calculations both matches the experimental results and while preserving the flat bands obtained in the paramagnetic calculations remain near the Fermi level.

As motivated in Section~\ref{sec:introduction}, we investigate bands that are flat in \emph{parts} of (but not necessarily over the entire) the BZ. Thus for each ICSD entry, we search for \emph{flat-band segments}: a series of $L$ consecutive $k$-points along the high symmetry lines of the band structure (we use $L=10, 20, 30, 40$ or $50$), where the energy band width is smaller than a tunable threshold $\omega$ (ranging from $25 {\rm meV}$ to $150 {\rm meV}$). The number of such flat segments for every band analyzed provides a convenient signature of band flatness. In \sirefappbsflatsegment, we provide a full discussion of the definition, the algorithm and the statistics of flat-band segments around $E_F$.

 The presence of a flat-band segment alone is not sufficient to predict the presence of interesting physics associated with it: a (quasi) one-dimensional system would equally exhibit a flat-band segment in the directions perpendicular to its dispersive direction. However, peaks (or their absence) in the DOS offer a simple and efficient way to filter out such pathological cases. Thus for each ICSD entry, we map the position and width of all DOS peaks in an energy region of $\pm 5 {\rm eV}$ around $E_F$. More details about the DOS peak detection are given in \sirefappdospeak. 

\subsection{Automated identification of sublattices}\label{sec:sublattices}

 Geometric frustration in (line-graph and bipartite) lattices is known to give rise to exact FTBs~\cite{Mielke_1992,10.1143/PTP.99.489,PhysRevB.78.125104, Liu_2014, ma2020, Chiu2020}. While initially predicted for $s$-orbitals, this property was recently generalized to a slew of other possible orbital/lattice combinations \cite{S-matrix}. This provides a crucial starting point to understand and predict  flat bands in crystalline materials: if a material hosts a line-graph or bipartite lattice as a part of its lattice structure (a "sublattice"), and if this sublattice is only  weakly perturbed or deformed by the remaining atoms or orbitals, we expect to observe FTBs. To explain the origin of the flat bands found in our high-throughput search, we have automated, using the structural parameters of every ICSD entry, the detection of five types of line-graph or bipartite sublattices:  the Kagome, pyrochlore, Lieb, bipartite or split sublattices as detailed in \sirefappkagomepyrochlore\ and \sirefappbipartite. 
 %In the following, we will introduce our search methods for the five types of sublattices in \crystal~materials.

%\textit{Kagome, pyrochlore and Lieb sublattices of a  \crystal~material lattice.} 
In 3D, the Kagome, pyrochlore and Lieb lattices can be mathematically characterized by special occupied Wyckoff positions in certain SGs. 
Using the crystalline structures of materials, we have developed a space-group method to detect the line-graph and Lieb sublattices in the Methods section.
%In \sirefappcryssymm, we have first identified the \emph{minimal} SGs that support these three lattices and the corresponding Wyckoff positions. Then, through group-subgroup relations and the split relations between the sets of Wyckoff positions on the \bcslong~(BCS)~\cite{Ivantchev:ks0038, Ivantchev:ks0141}, we have obtained all the SGs which host these lattices. The results are tabulated in the Table~\sireftabkagomepyrochlorewyckoff\ and Table~\sireftabliebwyckoff\ of \sirefappkagomepyrochlore. We dub the detection of sublattices using this tabulated information as the \emph{space-group method}. 
Although the space-group method provides a fast way to find the symmetric Kagome, pyrochlore and Lieb sublattices in crystalline materials, the exact sublattice might be spoiled by the presence of other atoms of the \emph{same element} on (or close to) this sublattice. Moreover, the space-group method discards approximate %(not exact)
sublattices which could also exhibit quasi-flat bands. To solve these issues, we have further developed a \emph{geometric method} which solely relies on the geometric features of these three sublattice types (%angles/distances/connectivity - 
discussed in \sirefappgeometry) and ignores the exact SG restrictions. In \sirefappalgorithms, we have provided a detailed presentation of our algorithms implementing the geometric method for each type of sublattice. Equipped with these methods, we sort all the possible sublattices in a material in two categories: the \emph{rigorous sublattices} which satisfy both methods and the \emph{approximate sublattices} which only satisfy the geometric method but capture weak distortions of rigorous sublattices.

%\textit{Bipartite and Split sublattices of a  \crystal~material lattice.} 
A bipartite lattice with chiral symmetry is formed by two sublattices $L$ and $\tilde{L}$ with the kinetic hopping \emph{only} between $L$ and $\tilde{L}$.
As proposed in Ref.~\cite{S-matrix} and briefly introduced in the Methods section and \sirefappsmatrix, a general method, namely the $S$-matrix method, can be used to explain the origin of flat bands in crystalline materials whose lattice contains a bipartite or split sublattice. We have developed an algorithm (detailed in \sirefappbipartite) to search for bipartite lattices from the structural parameters of each ICSD entry with the following necessary simplifying assumptions for a high-throughput analysis.
For each crystal structure in the \webflatbandacronym, we solely rely on the geometric distance between two atoms to infer the amplitude of kinetic hopping between them. By ignoring the small hopping terms based on  a tunable cutoff, we identify if a \crystal~material has a bipartite sublattice with a different number of atoms in its $L$ and $\tilde{L}$ sublattices. In the algorithm, a special case of bipartite sublattice, namely the split sublattice which has been proposed to host FTBs~\cite{ma2020}, is also detected and tagged. 

\section{Results}\label{sec:results}

We now summarize the main results obtained in the present high-throughput search. First we discuss the number of geometric sublattices detected by our algorithms and the public website we have developed to search for materials with flat bands based on the criterion in Section~\ref{sec:databasemethodology}. We discuss a manually curated list of \FlatBandNbrCuratedMaterials~materials potentially hosting FTBs, obtained using our toolset. Then, we showcase five best representative flat-band materials and  explain the flat-band segments in their band structures using the $S$-matrix method \cite{S-matrix}.

\subsection{Statistics and website}

By applying the automated analysis of the lattice structure to the \TQCDBNbrNoSOCICSDsIncludingValidFElectrons~ICSD entries of the \webflatbandacronym, we have performed a high-throughput search of the rigorous and approximate Kagome, Lieb, pyrochlore sublattices, the bipartite/split lattice with different number of atoms on their further sublattices.
\FlatBandNbrICSDsKagome~ICSDs with at least one Kagome (rigorous or approximate) sublattice have been found, including \FlatBandNbrICSDsRigorousKagome~ICSDs with a Kagome sublattice labeled as rigorous, \FlatBandNbrICSDsPyrochlore~ICSDs with a pyrochlore sublattice (rigorous or approximate) and \FlatBandNbrICSDsRigorousPyrochlore~ICSDs with a rigorous pyrochlore sublattice. For the Lieb lattice, there are \FlatBandNbrICSDsLieb~ICSDs hosting such a sublattice, including \FlatBandNbrICSDsRigorousLieb~ICSDs with rigorous Lieb sublattices. At least one bipartite sublattice (irrespective of the cutoffs) is found among \FlatBandNbrICSDsBipartite~ICSDs and split sublattices are found in \FlatBandNbrICSDsSplit~ICSDs. A breakdown of these statistics per SG is provided in \sirefappsublatticesdb. The brute-force scan of the band structures along the high symmetry lines was performed for several threshold parameters. The number of ICSDs exhibiting flat-band segments varies strongly on these parameters; for a detailed statistical analysis see \sirefappbsflatsegment.

The data generated through the automated algorithms discussed in Section~\ref{sec:databasemethodology} is available through our \webflatbandacronym (see \sirefappflatbandwebsite\ for an overview of the search engine). We actually used this website to perform an extensive investigation of  promising candidate materials exhibiting flat bands or large segment of flat bands close to the Fermi energy. The outcome of our search is provided as a list of \emph{curated flat-band materials} in \sirefapplistcurated. This list contains \FlatBandNbrCuratedICSDs~ICSD entries that can be regrouped into \FlatBandNbrCuratedMaterials~unique materials, \ie ICSDs sharing the same stoichiometric formula, SG and topological property at the Fermi energy (as defined in \sirefapptqcdboverview). The complete set of criteria applied to select these materials is provided in \sirefapplistcurated, and it includes the distance to the Fermi energy, the flat band width and topology, and the presence of a peak in the DOS. We have excluded cases where the flat bands were \emph{clear} FAB from the list of curated flat-band materials, and they are listed in \sirefapplistatomic. The statistics of detected sublattices among the curated materials are provided in Table~\ref{tab:statisticssublattices}.

\subsection{Flat-band material candidates}

Among the \FlatBandNbrCuratedMaterials~high-quality flat-band materials, we select \FlatBandNbrBestMaterials~best representative flat-band materials  in \sirefappbestflatbands\ for further experimental investigation. Most of the \FlatBandNbrBestMaterials~materials host one (or more than one) of the Kagome, pyrochlore, Lieb, bipartite and split sublattices in their crystal structures. 
For each of the five types of sublattice, we select one representative material, which hosts best flat-band segments on (or close to) the Fermi level, and explain its physical origin using the $S$-matrix method \cite{S-matrix}. All of the five representative flat-band materials are chemically realistic, experimentally paramagnetic and not Mott insulators, which is consistent with our paramagnetic calculations.

 The five typical materials are \ch{KAg[CN]2} [\icsdweb{30275}, SG 163 (\sgsymb{163})] with approximate Kagome sublattice formed by Ag atoms, \ch{Pb2Sb2O7} [\icsdweb{27120}, SG 227 (\sgsymb{227})] with pyrochlore sublattice formed by Pb atoms, \ch{Rb2CaH4} [\icsdweb{65196}, SG 139 (\sgsymb{139})] with Lieb sublattice formed by Ca and H atoms, \ch{Ca2NCl} [\icsdweb{62555}, SG 166 (\sgsymb{166})] with bipartite sublattice formed by Ca and N atoms, and
\ch{WO3} [\icsdweb{108651}, SG 221 (\sgsymb{221})] with split sublattice formed by W and O atoms. Their crystal structures are shown in Fig.~\ref{fig:mainfig2}(a)-(e) and the orbital characters of the flat bands in the five materials are shown in the orbital-projected band structures in Fig.~\ref{fig:mainfig2}(f)-(j).
As detailed in \sirefapptheoryexplanation, based on the crystal structure and orbital-projected bands of these materials, we have constructed effective tight-binding Hamiltonians using the $S$-matrix method \cite{S-matrix} and found that they can successfully explain the origins of flat bands. The flat bands of other materials of similar crystal structures can be found in the \webflatbandacronym.
In the Methods section, we use \ch{Ca2NCl} to showcase the application of the $S$-matrix method in explaining the origin of flat bands.

\section{Discussion}\label{sec:discussion}

We have performed, for the first time, a high-throughput search for flat electronic bands near the Fermi level and for the detection of line-graph and bipartite sublattices from the crystal structures of stoichiometric \crystal~materials.
We have further classified the flat bands by their topology, DOS, length of band flatness, and the types of lattices formed by the atoms whose orbitals contribute to the flat band. By successfully applying our algorithms to \TQCDBNbrNoSOCICSDsIncludingValidFElectrons~ICSD entries, we have found that \FlatBandNbrICSDsAtLeastOneSublattice\ (\FlatBandNbrICSDsAtLeastOneSublatticePercent) out of all the ICSD entries host \emph{at least} one of the Kagome, pyrochlore, Lieb, bipartite or split sublattices in their crystal structures. This proportion is raised to \FlatBandNbrCuratedICSDsAtLeastOneSublatticePercent\ for our \emph{manually curated list} of \FlatBandNbrCuratedICSDs\ ICSDs (\FlatBandNbrCuratedMaterials\ unique materials) and \FlatBandNbrBestICSDsAtLeastOneSublatticePercent\ for the \emph{best representative flat-band materials}. The appearance of flat-bands in  materials can be, in large but non-exhaustive part, theoretically understood using the $S$-matrix method~\cite{S-matrix}, as we have exemplified in five prototypical compounds.
All the results obtained in this work and detailed in the supplementary Appendices can be accessed on the \webflatbandacronym. Our results pave the way for future theoretical and experimental studies on flat-band materials combining topology and  interactions  and leading to exotic quantum phenomena, such as magnetism,  non-Fermi liquid behavior and superconductivity. Such flat-band investigations are, at present, confined to engineered twisted moire lattices in two dimensions. While the present work studies flat bands in paramagnetic band structures of 3D materials, our methods can be adapted to detect flat bands in magnetic band structures, 2D mono-layer materials, phonons and photonic crystals. Furthermore, the further classification of FOAB's will enlarge the set of flat bands whose center of charge is away from the atomic positions.

\clearpage

 {
\centerline{\bf Methods}
\vspace{0.15in}

\noindent \textbf{Topological Quantum Chemistry website}
The database used as an input the structural parameters of stoichiometric materials reported in the Inorganic Crystal Structure Database (ICSD)~\cite{ICSD}. For each entry, ab-initio calculations were performed using Density Functional Theory (DFT)~\cite{Hohenberg-PR64,Kohn-PR65} and its implementation in the Vienna Ab-initio Simulation Package (VASP)~\cite{PhysRevB.48.13115,vasp1}, with and without accounting for the spin-orbit coupling (SOC). The database provides the structural parameters, the band dispersion along high symmetry lines, the DOS, and the topological characterization for each set of bands in the material's band structure for each of the \TQCDBNbrNoSOCICSDs~ICSD entries. 

\noindent \textbf{Space-group method of detecting line-graph and Lieb sublattices}
To detect a Kagome, pyrochlore or Lieb sublattice in crystalline materials, we have first identified the \emph{minimal} SGs that support these three lattices and the corresponding Wyckoff positions (see details in \sirefappcryssymm). Then, through group-subgroup relations and the split relations between the sets of Wyckoff positions on the \bcslong~(BCS)~\cite{Ivantchev:ks0038, Ivantchev:ks0141}, we have obtained all the SGs which host these lattices. The results are tabulated in the Table~\sireftabkagomepyrochlorewyckoff\ and Table~\sireftabliebwyckoff\ of \sirefappkagomepyrochlore. We dub the detection of sublattices using this tabulated information as the \emph{space-group method}. 

\noindent  \textbf{A brief introduction of the $S$-matrix method.} 
Denoting $|L|$ and $|\tilde{L}|$ the number of atoms or orbitals in the $L$ and $\tilde{L}$ further sublattices of the bipartite or split sublattice of a material (assuming $|L| \geq |\tilde{L}|$), the Bloch Hamiltonian associated to a bipartite sublattice of a material reads
\begin{eqnarray}
H(\boldsymbol{k}) & = & \left(\begin{array}{cc}
0 & S^{\dagger}(\boldsymbol{k})\\
S(\boldsymbol{k}) & 0
\end{array}\right).\label{eq:HamiltonianBipartiteFromS}
\end{eqnarray}
Here $S(\boldsymbol{k})$ is a matrix with dimension $|L|\times |\tilde{L}|$ and $\boldsymbol{k}$ is the momentum in the Brillouin zone. A bipartite lattice with $|L| \neq |\tilde{L}|$ hosts flat bands in its band structure. For example, the Hamiltonian of Eq.~\ref{eq:HamiltonianBipartiteFromS} has at least $|L|+|\tilde{L}|-2\times rank(S(\boldsymbol{k}))$ zero-energy states, \ie the band structure has at least $|L|+|\tilde{L}|-2\times rank(S(\boldsymbol{k}))$ exact flat bands. It is also very likely that these bands exhibit nontrivial topology \cite{S-matrix}. While the other $2\times rank(S(\boldsymbol{k}))$ bands are dispersive, they are related by chiral symmetry. Although in real crystalline materials the chiral symmetry is generally broken by the intra-sublattice hopping, we find in Ref.~\cite{S-matrix} that this $S$-matrix method goes beyond the chiral symmetry. For a generalized bipartite lattice including the intra-sublattice coupling $A(k)$($B(k)$) of sublattice $L$($\tilde{L}$), if $A(k)$ has a momentum-independent eigenvalue $E=E_0$ with degeneracy $n_0$ and $|\tilde{L}|<n_0\leq |L|$, then $H(k)$ also has at least $n_0 - |\tilde{L}|$ perfectly flat bands at energy $E=E_0$ irrespective of $B(k)$. Moreover, the eigenstates of these $n_0 - |\tilde{L}|$ flat bands identical to those of the system with chiral symmetry \cite{S-matrix}.

\noindent  \textbf{The origin of flat bands in \ch{Ca2NCl}: application of $S$-matrix method.} 
As shown in Fig.~\ref{fig:mainfig2}(d) in the main text, the 3D crystal structure of \ch{Ca2NCl} is stacked, composed alternating Ca$_2$N and Cl layers.
In each Ca$_2$N layer, the Ca and N atoms occupy  honeycomb (with buckling) and triangular sublattices, respectively.
The Cl layer also forms a triangular lattice.
As shown in Fig.~\ref{fig:mainfig2}(i), the flat band and the lower dispersive bands next to it are mainly contributed by the $p$ orbitals of N atoms.
By constructing the maximal localized Wannier functions (MLWF)~\cite{PhysRevB.65.035109}, we extract an effective tight-binding model for these $p$ bands and find that its hoppings - which are computed from ab-intio methods without any additional theoretical input - obey a set of fine-tuned conditions, which in turn give rise to  flatness of the top $p$ band. As an example, the amplitude of $\sigma$-bond comes out to almost exactly the $-3$ times of $\pi$-bond. (See \sirefCaNCl\ for more details.)
Since similar flat bands also exist in many other materials of the same structure (\eg \ch{Ca2NBr}, \ch{Sr2NCl} \ea), and similar fine-tuned tight-binding ab-initio model, this points to a deeper reason for the fine-tuning conditions.

This deep reason is the $S$-matrix theory. We notice that the nearest neighbors of the N atoms are the Ca atoms and hence that Ca and N atoms form a bipartite sublattice if only the nearest neighbors hoppings are considered in a theoretical model of the bands.
As analyzed in Fig.~\ref{fig:mainfig2}(i), the conduction bands around $E=3$eV are mainly contributed by the $s$ and $d$ orbitals on the Ca atoms and the three valence bands in the energy window $-3\sim0$eV are mainly contributed by the $p$ orbitals on N atoms and partially contributed by the hybridized orbitals consisting of $s$ and $d_{z^2}$ orbitals on Ca. 
Since the $s$ and $d_{z^2}$ orbitals form the same representation ($A_1$) of the point group symmetry $C_{3v}$ isomorphic to the site-symmetry group of Ca sites, they hybridize with each other to form two hybridized orbitals.
It is  a reasonable simplification to only take into account the hybridized orbital with the lowest energy, which we refer to as the $s$ orbital in the following. 
A tight-binding model including both the N and Ca atoms is naturally an $S$-matrix theory of a bipartite lattice, where the $L$ sublattice of the bipartite (sub)-lattice consists of $p$ orbitals at the triangular lattice formed by N and the $\tilde{L}$ sublattice consists of (hybridized) $s$ orbitals at the honeycomb lattice (with buckling) formed by Ca.
The onsite energy of $s$ orbitals ($\Delta_s$) is about 3eV. 
Following the argument below Eq.~\ref{eq:HamiltonianBipartiteFromS}, there must be $|L|+|\tilde{L}|-2\times rank(S(\mathbf{k}))=3+2-4=1$ flat band at $E=0$. 
(See \sirefCaNCl\ for the explicit form of $S(\mathbf{k})$.)
As detailed in \sirefCaNCl, the $S$-matrix band structure matches the first-principles band structure very well.
Furthermore, the $S$-matrix theory also explains the fine-tuning conditions in the extracted tight-binding Hamiltonian from the MLWFs: the perturbative effective Hamiltonian for the $p$ bands, $-S(\mathbf{k})S^\dagger(\mathbf{k})/\Delta_s$,
perfectly reproduces the fine-tuning conditions of the ab-initio model.
Therefore, the $S$-matrix theory is a faithful explanation for the flat band in Ca$_2$NCl. 

}

\clearpage

\acknowledgments

We are grateful to Xi Dai, Dumitru Calugaru, Aaron Chew, Maia Vergniory and Christie Chiu for helpful discussions. We thank the referees for their suggestions that helped improving our manuscript. We acknowledge the computational resources Cobra/Draco in the Max Planck Computing and Data Facility (MPCDF) and Atlas in the Donostia International Physics Center (DIPC).  This research also used the resources of the National Energy Research Scientific Computing Center (NERSC), a U.S. Department of Energy Office of Science User Facility operated under Contract No. DE-AC02-05CH11231. This work is part of a project that has received funding from the European Research Council (ERC) under the European Union's Horizon 2020 research and innovation programme (grant agreement no. 101020833). B.A.B. and N.R. were also supported by the U.S. Department of Energy (Grant No. DE-SC0016239), and were partially supported by the National Science Foundation (EAGER Grant No. DMR 1643312), a Simons Investigator grant (No. 404513), the Office of Naval Research (ONR Grant No. N00014-20-1-2303), the Packard Foundation, the Schmidt Fund for Innovative Research, the BSF Israel US foundation (Grant No. 2018226), the Gordon and Betty Moore Foundation through Grant No. GBMF8685 towards the Princeton theory program, and a Guggenheim Fellowship from the John Simon Guggenheim Memorial Foundation. A.Y., N.P.O., R.J.C., L.M.S., B.A.B. and N.R. were supported by the NSF-MRSEC (Grant No. DMR-2011750). A.Y. was supported by NSF-DMR-1904442. B.A.B., L.M.S. and N.R. gratefully acknowledge financial support from the Schmidt DataX Fund at Princeton University made possible through a major gift from the Schmidt Futures Foundation. L.M.S. acknowledges financial support from the Packard and Sloan Foundation. L.E. was supported by the Government of the Basque Country (Project IT1301-19) and the Spanish Ministry of Science and Innovation (PID2019-106644GB-I00).  C.F. was supported by the European Research Council (ERC) Advanced Grant No.  742068 ``TOP-MAT'', Deutsche Forschungsgemeinschaft (DFG) through SFB 1143, and the Würzburg-Dresden Cluster of Excellence on Complexity and Topology in Quantum Matter-ct.qmat (EXC 2147, Project No. 390858490). S.S.P.P. acknowledges funding by the Deutsche Forschungsgemeinschaft (DFG, German Research Foundation) – Project number 314790414.

{\bf Author contribution}

B.A.B. and N.R. conceived this work; N.R. and M.R.L. performed the high-throughput calculations with the help from L.E. and Y.X.; Y.X., D.S.M., Z.D.S, M.R.L., L.E. and N.R. worked out the theoretical explanations for the flat-band materials detailed in \sirefapptheoryexplanation; The material lists in \sirefappallflatbands\ were manually selected by Y.X., M.R.L., Z.D.S, M.J. and N.R.; N.R. built the flat-band material database; D.S.M. performed the ab-initio ferromagnetic calculations advised by Y.X.; M.J., L.S., C.F. helped curate the list of materials to find the most experimentally relevant. All authors discussed the results and wrote the main text and the Methods; Y.X., Z.D.S, M.R.L, D.S.M, M.J., L.E. and N.R. wrote the supplementary materials.

{\bf Competing interests}

The authors declare no competing interests.

{\bf Corresponding authors}

Correspondence and requests for materials should be addressed to \href{bernevig@princeton.edu}{B. Andrei Bernevig} or \href{regnault@princeton.edu}{Nicolas Regnault} or \href{yfxu@mpi-halle.mpg.de}{Yuanfeng Xu}.

{\bf Data availability}
All data is available in the Supplementary Information and through our public website \webflatband (\href{https://www.topologicalquantumchemistry.fr/flatbands}{https://www.topologicalquantumchemistry.fr/flatbands}).

\clearpage

\bibliography{FlatBandMaterialsSearch}

\clearpage

\begin{figure*}[thbp]
\centering
\includegraphics[width=0.99\textwidth]{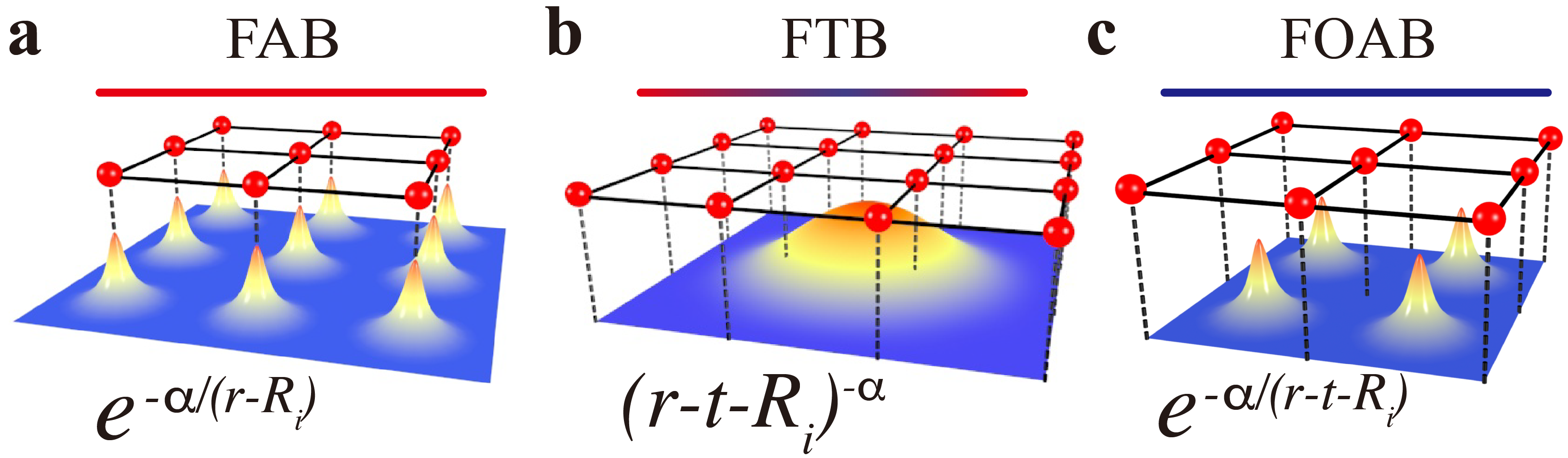}
\caption{An illustration of the three possible types of flat bands. (a) Flat atomic band (FAB): the Wannier functions associated to the flat bands are exponentially localized on the atoms' sites. (b) Flat topological band (FTB): the Bloch states are extended (with potentially a power law decay) in at least one direction of the lattice. (c) Flat obstructed atomic band (FOAB): as opposed to FAB, the corresponding Wannier functions are exponentially localized but on an empty site. In the figures, the atom sites of the 2D lattices are represented with red spheres. Below the lattices are the Bloch wave functions of the flat bands associated with their decay law, where $R_i$ is the position of atom $i$, $t$ is a fractional lattice vector and $\alpha$ is a positive coefficient.}\label{fig:mainfig1} 
\end{figure*}

\clearpage

\begin{figure*}[htbp] 
\centering
\includegraphics[width=0.99\textwidth]{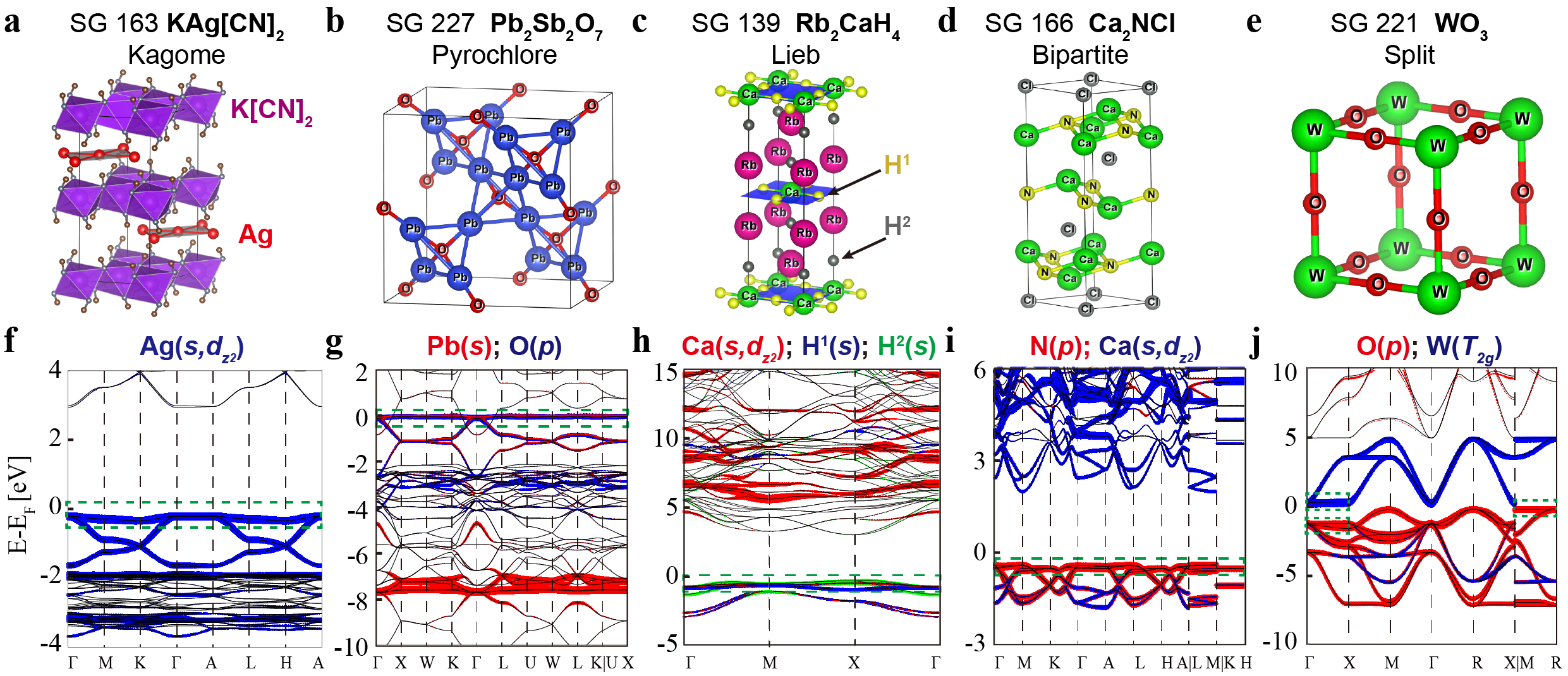}
\caption{Crystal and band structures of the representative flat-band materials. The crystal structures of (a) \ch{KAg[CN]2} which hosts an approximate Kagome sublattice formed by the Ag atoms (in red color), (b) \ch{Pb2Sb2O7} with Pb atoms at the Wyckoff position $16d$ forming a pyrochlore sublattice, (c) \ch{Rb2CaH4} with the H atoms at $4c$ (\ie the H$^1$ atoms in yellow color) and the Ca atoms at $2a$ position forming a Lieb sublattice, (d) \ch{Ca2NCl} which is stacked by alternating the \ch{Ca2N} and Cl layers, where the \ch{Ca2N} layer is identified as a bipartite sublattice in our algorithm, (e) \ch{WO3} with the W and O atoms forming a split lattice. 
For each material, its band structure and the orbital characterization of the flat bands is plotted and analyzed below its crystal structure. 
Based on the band structure analysis, the flat bands close to the Fermi level are explained by the $S$-matrix method in \sirefapptheoryexplanation.
In the crystal structure plots, SG, chemical formula and the type of sublattice host in the material are provided on the top of each panel. In the band structure plots, the flat-band segments close to the Fermi level are indicated by the dashed green lines. Orbital characters of the colored bands are provided on the top of each panel.}
\label{fig:mainfig2} 
\end{figure*} 

\clearpage

\begin{table*}
\begin{tabular}{|c||c|c|c|}
\hline
 & All ICSDs & Curated & Best  \\
\hline
\hline
\# ICSDs & \TQCDBNbrNoSOCICSDsIncludingValidFElectrons & \FlatBandNbrCuratedICSDs & \FlatBandNbrBestICSDs \\
\hline
\# Materials & \TQCDBNbrNoSOCMaterialsIncludingValidFElectrons & \FlatBandNbrCuratedMaterials & \FlatBandNbrBestMaterials \\
\hline
\hline
Kagome & \begin{tabular}{c}\FlatBandNbrICSDsKagome\\ (\FlatBandNbrICSDsKagomePercent)\end{tabular} & \begin{tabular}{c}\FlatBandNbrCuratedICSDsKagome\\ (\FlatBandNbrCuratedICSDsKagomePercent)\end{tabular} & \begin{tabular}{c}\FlatBandNbrBestICSDsKagome\\ (\FlatBandNbrBestICSDsKagomePercent)\end{tabular} \\
\hline
Pyrochlore & \begin{tabular}{c}\FlatBandNbrICSDsPyrochlore\\ (\FlatBandNbrICSDsPyrochlorePercent)\end{tabular} & \begin{tabular}{c}\FlatBandNbrCuratedICSDsPyrochlore\\ (\FlatBandNbrCuratedICSDsPyrochlorePercent)\end{tabular} & \begin{tabular}{c}\FlatBandNbrBestICSDsPyrochlore\\ (\FlatBandNbrBestICSDsPyrochlorePercent)\end{tabular} \\
\hline
Lieb & \begin{tabular}{c}\FlatBandNbrICSDsLieb\\ (\FlatBandNbrICSDsLiebPercent)\end{tabular} & \begin{tabular}{c}\FlatBandNbrCuratedICSDsLieb\\ (\FlatBandNbrCuratedICSDsLiebPercent)\end{tabular} & \begin{tabular}{c}\FlatBandNbrBestICSDsLieb\\ (\FlatBandNbrBestICSDsLiebPercent)\end{tabular} \\
\hline
Bipartite & \begin{tabular}{c}\FlatBandNbrICSDsBipartite\\ (\FlatBandNbrICSDsBipartitePercent)\end{tabular} & \begin{tabular}{c}\FlatBandNbrCuratedICSDsBipartite\\ (\FlatBandNbrCuratedICSDsBipartitePercent)\end{tabular} & \begin{tabular}{c}\FlatBandNbrBestICSDsBipartite\\ (\FlatBandNbrBestICSDsBipartitePercent)\end{tabular} \\
\hline
Split &  \begin{tabular}{c}\FlatBandNbrICSDsSplit\\ (\FlatBandNbrICSDsSplitPercent)\end{tabular}& \begin{tabular}{c}\FlatBandNbrCuratedICSDsSplit\\ (\FlatBandNbrCuratedICSDsSplitPercent)\end{tabular} & \begin{tabular}{c}\FlatBandNbrBestICSDsSplit\\ (\FlatBandNbrBestICSDsSplitPercent)\end{tabular} \\
\hline 
None & \begin{tabular}{c}\FlatBandNbrICSDsNoSublattices\\ (\FlatBandNbrICSDsNoSublatticesPercent)\end{tabular}& \begin{tabular}{c}\FlatBandNbrCuratedICSDsNoSublattices\\ (\FlatBandNbrCuratedICSDsNoSublatticesPercent)\end{tabular} & \begin{tabular}{c}\FlatBandNbrBestICSDsNoSublattices\\ (\FlatBandNbrBestICSDsNoSublatticesPercent)\end{tabular}\\
\hline
\hline
\end{tabular}
\caption[Statistics of the ICSD entries hosting sublattices.]{Statistics of the ICSD entries in the database hosting at least one sublattice for each lattice type. In the first row, we give the number of ICSDs entries for the database (first column), for the list of curated flat-band materials (second column) and for the best representative flat-band materials (third column). The second line provides the number of unique materials for the database, the list of curated flat-band materials and the best representative flat-band materials. The third, fourth, fifth, sixth and seventh rows are the statistics for the Kagome, pyrochlore, Lieb, bipartite with different number of atoms on sublattices and split lattices. An ICSD entry is considered as hosting a given type of lattice if the algorithms discussed in Section~\ref{sec:sublattices} have found at least one such a lattice irrespective of the cutoffs or being a rigorous or approximate sublattice. Note that an ICSD entry might host more than one type of sublattices. The eighth row provides the statistics for the ICSDs where no sublattices have been detected. For each column, the percentages are computed with respect to the number of ICSDs provided in the first row.}
\label{tab:statisticssublattices}
\end{table*}

\newcommand{\reffigmainfigone}{\ref{fig:mainfig1}}
\newcommand{\reftabstatisticssublattices}{\ref{tab:statisticssublattices}}
\newcommand{\refsecsublattices}{``Automated identification of line-graph and bipartite sublattices''}

%%%%%%%%%%%%%%%%%%%%%%%%%%%%%%%%%%%%%%%%%%%%%%%%%

\addtocontents{toc}{\protect\setcounter{tocdepth}{0}}
\addtocontents{lot}{\protect\setcounter{lotdepth}{-1}}

\clearpage
\onecolumngrid

\appendix

\clearpage
\begin{center}
{\bf Supplementary Information for ``Catalogue of Flat-Band Stoichiometric Materials''}
\end{center}

\tableofcontents

\clearpage

\listoftables

\clearpage

\addtocontents{toc}{\protect\setcounter{tocdepth}{3}}
\addtocontents{lot}{\protect\setcounter{lotdepth}{3}}

\section{Introduction to the Supplementary Information}\label{app:AppendixOverview_appendix}

In this work, we have presented a catalogue of stoichiometric materials hosting flat bands near the Fermi energy. In addition to the main text, we provide below Supplementary Appendices giving an in-depth discussion of our methodology and several lists of curated materials. Appendix~\ref{app:flatbandtopology} is devoted to the detailed definition of two types of flat bands: flat atomic bands and flat topological bands. In Appendix~\ref{app:database}, we give an overview of the \webTQC\ on which we rely on for our catalogue of flat-band materials. We present the \webflatband\ and the theory and algorithms for the detection of flat-band segments and density of states peaks. We also provide detailed statistics about the materials hosting the five types of sublattices leading to flat bands (namely the Kagome, pyrochlore, generic bipartite, split and Lieb lattices). 
We then present Appendices about these five types of sublattices. Appendix~\ref{app:kagomepyrochlore} focuses on Kagome, pryochlore and Lieb sublattices. We cover both the crystal symmetries of these lattices, the three dimensional space groups compatible with them and their geometrical features. We also detail the algorithms we have designed for a high throughput search in our database of materials. In Appendix~\ref{app:bipartite}, we give a short overview of the bipartite and split lattices and most specifically of our new theoretical method for finding flat bands, the $S$-matrix method, detailed in \cite{S-matrix}. We present all the technical details of the algorithms we build in order to automatically look for bipartite and split sublattices in our material database. In Appendix~\ref{app:theoryexplanation}, we provide the theoretical explanations for the presence of flat bands in seven prototypical materials: WO$_3$ [\icsdweb{108651}, SG 221 (\sgsymb{221})], Pb$_2$Sb$_2$O$_7$ [\icsdweb{27120}, SG 227 (\sgsymb{227})], CaNi$_5$ [\icsdweb{54474}, SG 191 (\sgsymb{191})], Ca$_2$NCl [\icsdweb{62555}, SG 166 (\sgsymb{166})], \ch{Rb2CaH4} [\icsdweb{65196}, SG 139 (\sgsymb{139})], \ch{KAg[CN]2} [\icsdweb{30275}, SG 163 (\sgsymb{163})]
and RbMo$_3$S$_3$ [\icsdweb{644175}, SG 176 (\sgsymb{176})]. To test the validity of our high-throughput search based on paramagnetic calculations for potentially magnetic materials, we study in Appendix~\ref{app:magflat} four materials using ab-initio calculation with on-site Hubbard interaction for $d$ and $f$ electrons, namely SrRuO$_3$[\icsdweb{69360}, SG 221 (\sgsymb{221})], Rb$_2$MnCl$_6$[\icsdweb{9347} SG 225 (\sgsymb{225})], Ba$_2$MnReO$_6$[\icsdweb{4169} SG 225 (\sgsymb{225})] and NiMnSb [\icsdweb{54255}, SG 216 (\sgsymb{216}). With magnetizations close to their experimental values, the flat band features are barely affected. Finally in  Appendix~\ref{app:allflatbands}, we give a list of \FlatBandNbrCuratedMaterials~manually curated materials with flat bands near the Fermi energy. For convenience, materials with flat atomic bands are separated from the ones with flat bands occur from wave function interference. Moreover we have selected \FlatBandNbrBestMaterials~compounds hosting the most remarkable flat bands and we provide their band structure and density of states.

\section{Flat bands with different topologies}\label{app:flatbandtopology}

In real materials, electronic bands can be flat in the entire Brillouin zone (BZ) or only be flatten along some specific high-symmetry paths. In this appendix, 
we provide a detailed description of the two types of flat bands, namely the flat atomic bands and the flat topological bands, and provide several mechanisms that could give rise to them.

\subsection{Flat atomic bands}\label{app:atomicflatband}

In crystalline materials, a set of bands that are flat in all directions or some particular directions is referred to as flat atomic bands if they form a band representation (BR) that can be induced from local orbitals.
The flat atomic bands are induced from local orbitals and hence by definition have a \emph{trivial} topology \cite{QuantumChemistry,po_symmetry-based_2017,kruthoff_topological_2017}.
However, in practice, it is difficult to fully determine whether a set of bands is a BR: the full determination requires the Bloch wave functions over the whole Brillouin zone.
Instead, we use the symmetry eigenvalues (the representations formed by the Bloch bands) to diagnose whether a set of bands is \emph{consistent} with a BR. When the symmetry eigenvalues of a set of flat bands are consistent with a BR induced from local orbitals, 
we \emph{label} the set of bands as flat atomic bands.
With this approach, it is possible that some bands labeled as flat atomic bands are nevertheless topological with some topology not diagnosed by symmetry eigenvalues.
A simple example is a 2D Chern insulator with inversion symmetry. When the Chern number is odd, the inversion eigenvalues are not consistent with any BR \cite{Turner2010Inversion,Hughes2011Inversion,Fang2012Invariants}. However, when the Chern number is even, the inversion eigenvalues can be the same as the eigenvalues of some BRs.

When the flat atomic bands are flat in {\it all} the directions, the flat atomic bands are in the atomic limit, where the hopping between some orbitals of neighboring atoms is infinitesimal, as illustrated in Fig.~\reffigmainfigone (a) of the main text. In Appendix~\ref{app:listatomic}, we provide a list of manually selected materials hosting flat atomic bands.

\begin{figure*}[ht]
\centering
\includegraphics[width=0.50\textwidth,angle=90]{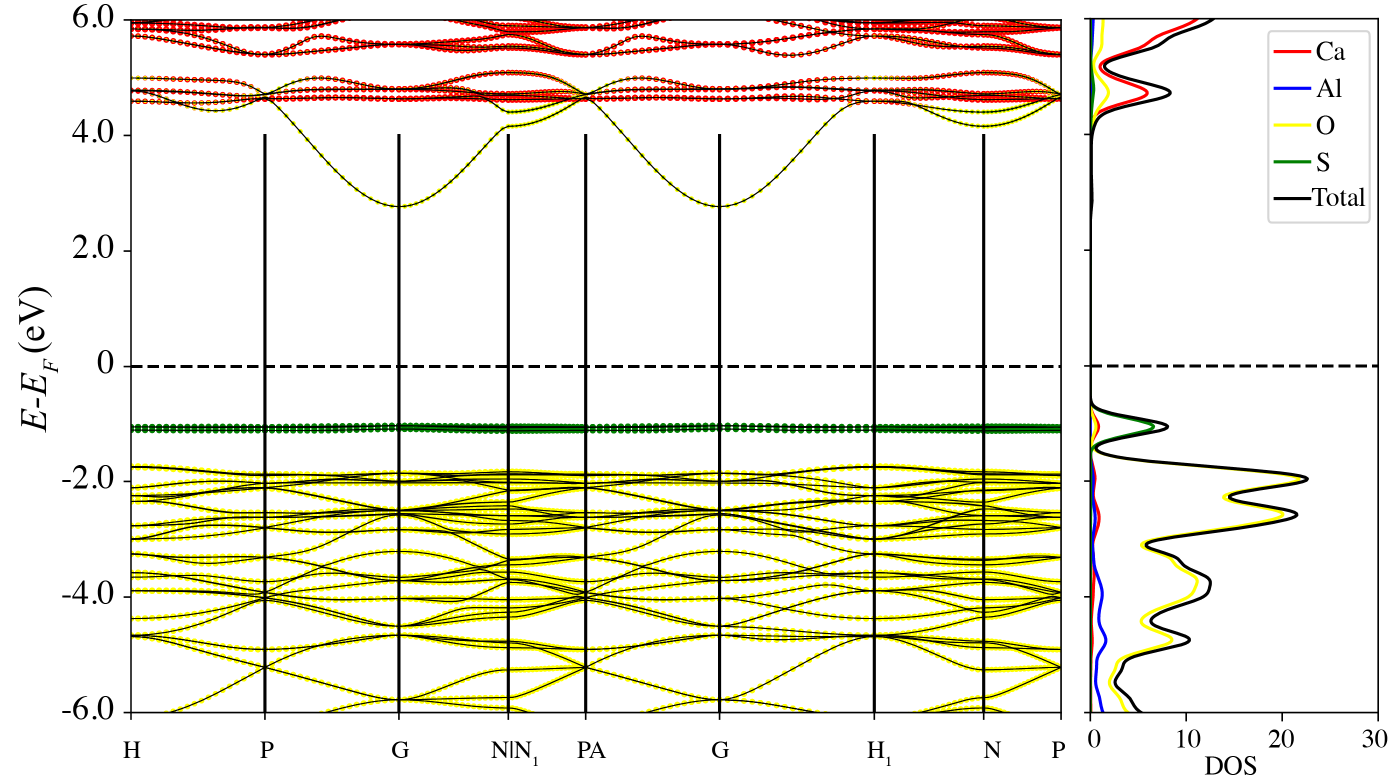}
\caption{Atom-projected band structure along the high-symmetry path (left panel) and density of states (right panel) for Al$_{12}$Ca$_8$O$_{24}$S$_2$ [\icsdweb{67589}, SG 217 (\sgsymb{217})]. The weight of orbitals on Ca, Al, O and S atoms are represented by the red, blue, yellow and green points and lines, respectively.}
\label{fig:band_67589}
\end{figure*}

As an example of flat atomic bands, we consider the compound Al$_{12}$Ca$_8$O$_{24}$S$_2$ [\icsdweb{67589}, SG 217 (\sgsymb{217})] shown in Fig.~\ref{fig:band_67589}. It has several flat bands below the Fermi level (with band index $-6\sim-1$). The BR of these bands are induced from the orbitals $\bar E_2 \oplus \bar F$ at $2a$ position, which are occupied by S atoms. In Fig.~\ref{fig:band_67589}, we plot the atom-projected band structure and density of states of the compound Al$_{12}$Ca$_8$O$_{24}$S$_2$. We find that the flat bands arise from the S atom orbitals, as indicated by the green lines. The flatness of the $p$ bands of S atoms is due to two reasons: on one hand, there is a big enough difference of the onsite energy between the $p$ orbitals on S and the orbitals on other atoms; on the other hand, the distance between two neighboring S atoms (about 8\AA) is so large that the kinetic energy between the S $p$ orbitals  is small. Hence, the flat bands contributed by the $p$ orbitals of S are flat atomic bands. 

The flat atomic bands that are flat only in one or two directions can be understood as layered systems, where the inter-layer kinetic energy is weak. In this case, all bands are flat along the inter-layer direction(s).
When the bands are flat in one (two) direction(s), each ``layer'' is a 2D plane (1D chain).

Note that heavy fermion systems may host another class of flat bands in the atomic limit. These bands are formed by localized $4f$ or $5f$ orbitals and are extremely flat. As the mean field approximation adopted in the first principle calculations fails to characterize the strong correlation of $f$ electron, we ignore the heavy fermion systems in this work as we will discuss in Appendix~\ref{app:database}.

\subsection{Flat topological bands}
Apart from the flat atomic bands defined in Appendix~\ref{app:atomicflatband}, the flat bands might be topologically nontrivial: when a set of bands that are flat in all or some particular directions do not form any BR, \ie cannot be induced from local orbitals, the set of bands forms a set of flat topological bands.
Since by the definition the topological bands are not equivalent to any local orbitals, the Wannier functions of the topological bands that preserve the symmetry of the system must be extended, \ie not exponentially localized, as shown in Fig.~\reffigmainfigone (b) of the main text.
In this work and as explained in the last subsection, we only make use of symmetry eigenvalues to diagnose the topology.
Hence a set of flat topological bands might potentially be misdiagnosed as flat atomic bands. However, all the diagnosed flat topological bands must be topological.

There are two types of flat topological bands: the stable topological bands and the fragile topological bands \cite{Po2018fragile,Jennifer2018fragile,Slager2019wilson,bradlyn_disconnected_2019,song_fragile_2020,Yang2019fragile,alexandradinata2020crystallographic}. 
The former is realized as a non-integer linear combination (NLC) of elementary BR (EBR) or split elementary BR (SEBR) \cite{QuantumChemistry,po_symmetry-based_2017,kruthoff_topological_2017,song_quantitative_2018,song_diagnosis_2018,Watanabe2018anomalous}, and the later can be realized as a difference (but not a sum) of EBRs.
Since we only rely on symmetry eigenvalues, a flat stable topological band could be labeled as fragile in this work if its symmetry eigenvalues are consistent with a difference of EBRs. On the contrary, all bands labeled as stable topological must be stable topological because by definition their symmetry eigenvalues are not consistent with any BR or difference of BRs. 
Exact flat fragile bands have been found in several lattice models \cite{Yang2019flat,Po2019faithful,Chiu2020}.
The $S$-matrix method introduced in Appendix~\ref{app:bipartite} and in Ref.~\cite{S-matrix} is a general mechanism to generate exact flat fragile bands.
On the other hand, no examples of rigorously flat (\ie dispersionless) stable topological bands have yet been found with compact support hopping. 
Nevertheless, quasi-flat stable topological bands were recently found in line-graph and split-line-graph lattice models with spin-orbit coupling \cite{leykam2018artificial,ma2020}.

\section{Database of materials with flat bands selected from TQC database}\label{app:database}

In this Appendix, we will introduce the database of materials that we are relying on for this high-throughput search of flat-band materials. We will also present the \webflatband\ and the different algorithms and search options that are provided to the user. The tools that we publicly provide through this website have been used to obtain the curated list of flat-band materials that we will present in Appendix~\ref{app:allflatbands}.

\subsection{An overview of the Topological material database}\label{app:tqcdboverview}

For our high-throughput search of materials with flat bands, we have relied on the data and database of the \webTQC. It was originally built for Ref.~\onlinecite{Vergniory2019} and upgraded both in terms of materials and features in Ref.~\onlinecite{Vergniory2021}. Beyond the two original publications related to this database, this database was also used to look for the best candidate materials with fragile bands \cite{ZhidaFragile} and to develop machine learning approach to topological feature predictions\cite{NicolasML}. We review here some important properties of this database.

The ab-initio calculations of the database were performed using Density Functional Theory (DFT)~\cite{Hohenberg-PR64,Kohn-PR65} and its implementation in the Vienna Ab-initio Simulation Package (VASP)~\cite{vasp1,PhysRevB.48.13115}.  For each material calculation, we used as input the structural parameters reported on the Inorganic Crystal Structure Database (ICSD)~\cite{ICSD}. We treated the interaction between the ion cores and the valence electrons using the projector augmented-wave method~\cite{paw1}. For the exchange-correlation potential, we used the generalized gradient approximation (GGA) with the Perdew-Burke-Ernzerhof parameterization for solids~\cite{PhysRevLett.77.3865}. Effects of spin-orbit coupling were accounted by using the second variation method~\cite{PhysRevB.62.11556}. For the plane-wave expansion, we employed a $\Gamma$-centered $\boldsymbol{k}$-point grid of size (11$\times$11$\times$11) for reciprocal space integration and a 550 eV energy cutoff. The density of states (DOS) was evaluated on a $\Gamma$-centered $\boldsymbol{k}$-point grid of size (11$\times$11$\times$11) and using the Methfessel-Paxton algorithm of order 1. We emphasize that all of the first-principle calculations are paramagnetic, \ie without the onsite spin polarization. The magnetic properties at low temperature reported on \webTQC\ are extracted from the \webmaterialsproject. 

The \webTQC\ is built upon the systematic ab-initio calculation of \TQCDstoichiometric~ICSD entries of stoichiometric materials. Among these entries, the calculation with spin-orbit coupling was performed successfully for \TQCDBNbrICSDs~materials. All of them are available through the website. Calculations without spin-orbit coupling are also available although a small fraction (\TQCDBPercentFailedNoSOCICSDs~of the successfully processed ICSD entries with spin-orbit coupling) did not converge. \TQCDBNbrICSDsFElectrons~of the available materials (\ie \TQCDBNbrICSDsFElectronsPercent) contain rare-earth elements (La, Ce, Pr, Nd, Pm, Sm, Eu, Gd, Tb, Dy, Ho, Er, Tm, and Yb) and actinides (Ac, Th, Pa, U, Np, Pu, Am, Cm, Bk, Cf, Es, Fm, Md, No and Lr) which may exhibit strong correlation effects and usually lead to spurious flat bands in the DFT calculations. For these reasons, we have discarded these materials from our flat-bandsurvey with one exception, compounds containing La as rare-earth elements. Indeed for the ab-initio calculation, La is always a cation in such compounds and the bands originated from these $f$ orbits are usually away from the Fermi energy. In our database, there are \TQCDBNbrNoSOCICSDsValidFElectrons~ICSD entries (\ie  \TQCDBNbrNoSOCICSDsValidFElectronsPercent) corresponding to this situation. There, the $f$ bands originated from the La atoms are located at least $1{\rm eV}$ above $E_f$ (typically around $2{\rm eV}$). We have highlighted some of these materials in Appendix~\ref{app:bestflatbands}.

Many ICSD entries correspond to the same stoichiometric formula and the same space group (SG), leading most of the time (but not always) to, e.g., the same topological properties at the Fermi level. For this reason the \webTQC~defines the notion of {\it unique materials}, including all the ICSD entries having the same stoichiometric formula, SG and topological classification at the Fermi level. This database contains \TQCDBNbrUniqueMaterials~such unique materials. While in some cases, all ICSD entries within a given unique materials have band structures and densities of states barely distinguishable by eye, they might in general exhibit some subtle differences for these two quantities. Automated search of flat-band features from the band structure or the density of states should thus be performed on every ICSD belonging to all unique materials.

\subsection{The flat-band material database}\label{app:flatbandwebsite}

The website \webflatband~relies on the database provided in \webTQC. It offers a dedicated search interface to look for materials with flat-band features close to the Fermi energy. As discussed above, we have filtered out those that contain any rare-earth elements except La. Moreover, to search for flat bands or flat bands along some high symmetry lines, we only focus on the DFT calculations performed without spin-orbit coupling: the band structures and density of states are usually similar with or without spin-orbit coupling, thus only considering the later simplifies the automated search without introducing any bias. Overall, the user has access to \TQCDBNbrNoSOCICSDsIncludingValidFElectrons~ICSD entries (\TQCDBNbrNoSOCMaterialsIncludingValidFElectrons~unique materials). The three main tools that we provide to search for flat bands are based on flat segments in the electronic band structure or peaks in the density of states or sublattices. We will now detail these two former, the later will be discussed in Appendix~\ref{app:sublatticesdb}.

\subsubsection{Band structure and flat-band segments}\label{app:bsflatsegment}

For each ICSD entry, we have access to the electronic band structure computed along high symmetry lines and without spin-orbit coupling. These band structures include all bands up to two times of the number of valence electrons, thus an equal number of bands above and below the last filled one. The notions of filled and empty are defined with respect to the number of electrons and the band index. Each band is evaluated along a path specific to each point group. This path is made of straight lines in the Brillouin zone (BZ) connecting two high symmetry points. The choice of these straight lines is designed to cover all cases required by band connectivity\cite{QuantumChemistry,Bandrep2} while staying connected with the related works and databases. Each straight line is discretized into 20 equally spaced $\boldsymbol{k}$ points in the BZ, including the two end high symmetry points. Note that the distance between two consecutive discretized $\boldsymbol{k}$ points might differ from one straight line to another: we use a regular mesh with the same number (20) of points for the discretization of each line but the actual distances between two pairs of high symmetry points of the BZ do not have to be equal.

\begin{figure}[ht]
\centering
\includegraphics[width=0.75\textwidth,angle=0]{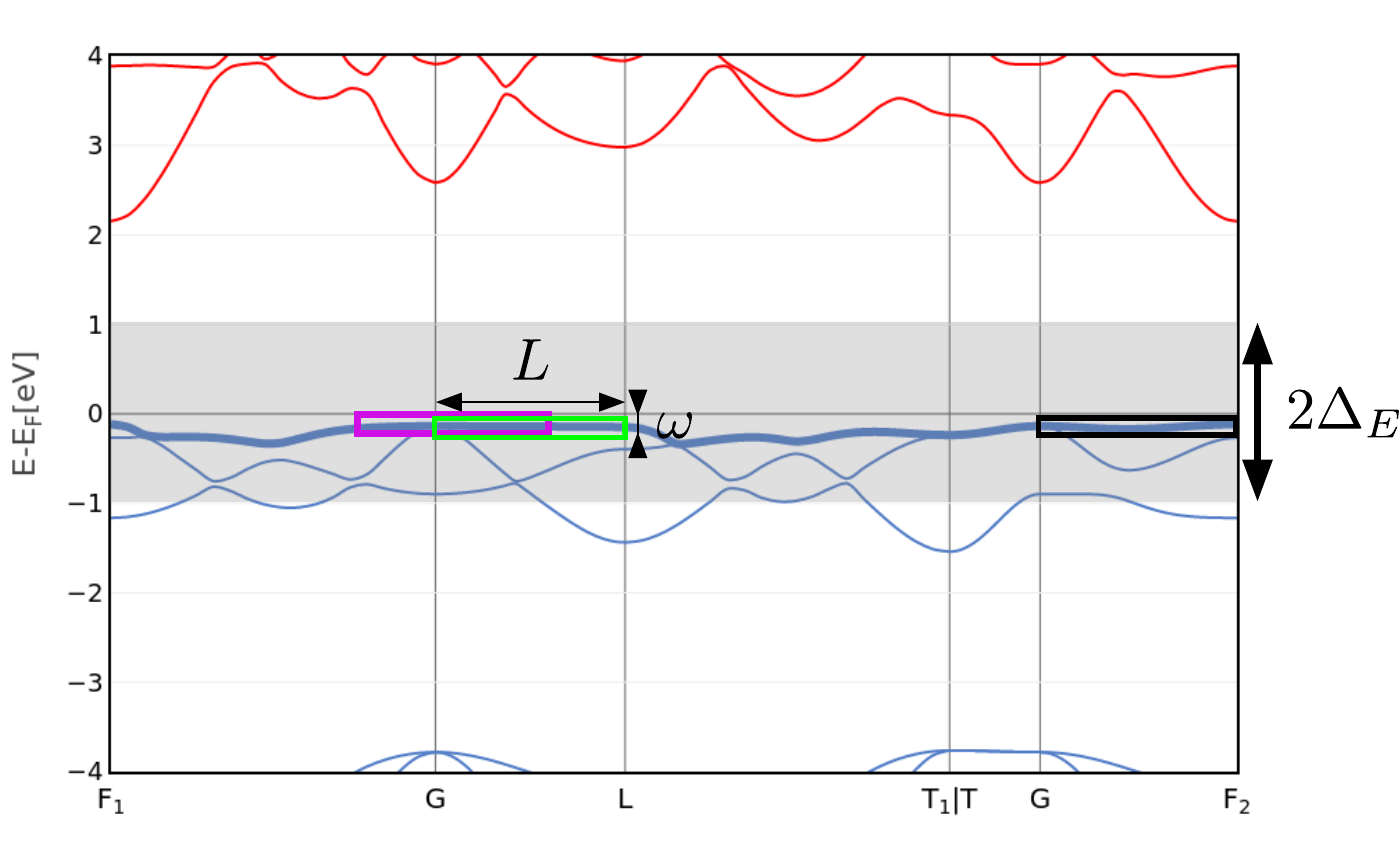}
\caption{Example of flat-band segments found in the band structure of ${\rm Ca}_2 {\rm N} {\rm Cl}$ [\icsdweb{62555}, SG 166 (\sgsymb{166})] in the space group 166 ($R\bar{3}m$), one of the best flat bands given in Appendix~\ref{app:bestflatbands}. We have set the parameters $L=20$, $\omega=150 {\rm meV}$ and $\Delta_E=1 {\rm eV}$. The best flat band closest to the Fermi level is highlight in bold blue and has a score of $0.642$. The purple, green and black boxes show three segments satisfying the constraints of the algorithm discussed in Appendix~\ref{app:bsflatsegment}.}
\label{fig:exampleflatbandsearch}
\end{figure}

To automatically search for flat segments in a given band structure of an ICSD, we apply the following procedure:

\begin{itemize}
\item We only consider at most the last 2 filled bands and 2 empty bands to focus on features near the Fermi level. We require that at least one $k$ point with an energy $E$ in the energy window parameterized by $\Delta_E$ such that $\left|E-E_F\right|< \Delta_E$. $E_F$ is the Fermi energy as computed from the VASP self consistent calculation\cite{Vergniory2021}. If any of the last 2 filled bands and 2 empty bands does not satisfy this constraint, it is discarded.
\item We define a flat segment as a series of $L$ consecutive $k$ points along the path in the BZ, whose bandwidth, \ie the difference between the highest energy and the lowest energy associated to these $L$ points, is lower than some chosen value $\omega$.
\item For each band, we test the presence of flat segments starting from every possible discretized $k$ points that define the path in the BZ. Notice that segments can overlap (see the purple and green boxes of Fig.~\ref{fig:exampleflatbandsearch}). Denoting $N_{\rm BZ}$ the total number of discretized $k$ points that define the path in the BZ, the number of flat segments $N_{\rm fl}$ is at most $N_{\rm BZ} - L + 1$. A score is then defined as the ratio
\begin{equation}
{\rm score} = \frac{N_{\rm fl}}{N_{\rm BZ} - L + 1}.\label{eq:flatbandscore}
\end{equation}
For example, a perfectly flat band would have a score of $1$.
\item In addition, we also evaluate $N_{\rm w}$ the number of bands that overlap with the energy window defined by  $\left|E-E_F\right|< \Delta_E$. It is used as a proxy to quantify how dense is the band structure in this energy window.
\end{itemize}

We have applied this procedure to all ICSDs using several discrete values of $L$ (10, 20, 30, 40 and 50), $\Delta_E$ ($0.5{\rm eV}$, $1.0{\rm eV}$, $1.5{\rm eV}$ and $2.0{\rm eV}$), and $\omega$ ($25{\rm meV}$, $50{\rm meV}$, $75{\rm meV}$, $100{\rm meV}$, $125{\rm meV}$ and $150{\rm meV}$). We exemplify the search procedure in Fig.~\ref{fig:exampleflatbandsearch} for \icsdweb{62555}. In Table~\ref{tab:flatbandstat}, we provide the number of ICSDs hosting flat-band segments for all values of $L$ and $\omega$ that we have considered. Statistics is split into different minimum threshold for the best score (as defined in Eq.~(\ref{eq:flatbandscore}) for an ICSD. A score greater than 0 means there is at least one flat-band segment. A score of $1$, \ie the maximum score means that an ICSD has one band out of the (at most) last 2 filled bands and 2 empty bands, where every sequence of $L$ consecutive $k$ points along the path in the BZ with a bandwidth smaller than $\omega$. We stress that such a score does not imply that the band has a global bandwidth smaller $\omega$.

All the above-mentioned parameters can be selected as search criteria on the \webflatband. In addition, a cap on $N_{\rm w}$, the number of bands in the energy window, can be set to look for well separated flat-band segments from other bands. Moreover, the user can require the flat band with the highest score to be topological, either as part of a topological single set of bands or all the bands up to (and including) the best flat band being topological (see Ref.~\onlinecite{Vergniory2021} for more details about the difference between these two definitions). The best score as defined in Eq.~(\ref{eq:flatbandscore}) is used both as a search criterion and to sort the search results. When selecting a given result, the band having the flat segment closest to the Fermi energy is displayed in bold for convenience (see Fig.~\ref{fig:exampleflatbandsearch}). 

\begin{longtable*}{|c|c|c|c|c|c|c|c|}
\caption[Flat-band segment statistics]{Statistics about the flat-band segments of all the ICSD entries. Note that we used the energy window $\Delta_E=0.5 {\rm eV}$ but the numbers provided here almost never depend on this parameters. The first column is the minimal best score (as defined in Eq.~(\ref{eq:flatbandscore})) and the second column is the minimum segment length $L$. The other columns are the different values for the maximum bandwidth $\omega$. We provide both the number of ICSDs hosting such a flat-band segment and the percentage with respect to the total number of ICSDs that we have considered, namely \TQCDBNbrNoSOCICSDsIncludingValidFElectrons.\label{tab:flatbandstat}}\\
\hline
Min. score & $L$  & $\omega=25 {\rm meV}$ & $\omega=50 {\rm meV}$ & $\omega=75 {\rm meV}$ & $\omega=100 {\rm meV}$ & $\omega=125 {\rm meV}$ & $\omega=150 {\rm meV}$ \\
\hline  
0.01 & 10 & 48312 {\tiny(87.5\%)} & 52693 {\tiny(95.5\%)} & 54160 {\tiny(98.1\%)} & 54534 {\tiny(98.8\%)} & 54752 {\tiny(99.2\%)} & 54876 {\tiny(99.4\%)} \\ 
0.01 & 20 & 29987 {\tiny(54.3\%)} & 38914 {\tiny(70.5\%)} & 43327 {\tiny(78.5\%)} & 46033 {\tiny(83.4\%)} & 47939 {\tiny(86.8\%)} & 49490 {\tiny(89.7\%)} \\ 
0.01 & 30 & 13610 {\tiny(24.6\%)} & 22700 {\tiny(41.1\%)} & 29037 {\tiny(52.6\%)} & 33354 {\tiny(60.4\%)} & 36537 {\tiny(66.2\%)} & 38936 {\tiny(70.5\%)} \\ 
0.01 & 40 & 7622 {\tiny(13.8\%)} & 14472 {\tiny(26.2\%)} & 19818 {\tiny(35.9\%)} & 24146 {\tiny(43.7\%)} & 27701 {\tiny(50.2\%)} & 30491 {\tiny(55.2\%)} \\ 
0.01 & 50 & 4712 {\tiny(8.5\%)} & 10179 {\tiny(18.4\%)} & 14787 {\tiny(26.8\%)} & 18583 {\tiny(33.7\%)} & 21508 {\tiny(39.0\%)} & 24372 {\tiny(44.1\%)} \\ 
\hline
0.25 & 10 & 18685 {\tiny(33.9\%)} & 29936 {\tiny(54.2\%)} & 36000 {\tiny(65.2\%)} & 40178 {\tiny(72.8\%)} & 43345 {\tiny(78.5\%)} & 45577 {\tiny(82.6\%)} \\ 
0.25 & 20 & 7223 {\tiny(13.1\%)} & 14525 {\tiny(26.3\%)} & 20081 {\tiny(36.4\%)} & 24794 {\tiny(44.9\%)} & 28636 {\tiny(51.9\%)} & 31452 {\tiny(57.0\%)} \\ 
0.25 & 30 & 3787 {\tiny(6.9\%)} & 9079 {\tiny(16.4\%)} & 13775 {\tiny(24.9\%)} & 17612 {\tiny(31.9\%)} & 20739 {\tiny(37.6\%)} & 23681 {\tiny(42.9\%)} \\ 
0.25 & 40 & 2222 {\tiny(4.0\%)} & 6160 {\tiny(11.2\%)} & 9980 {\tiny(18.1\%)} & 13432 {\tiny(24.3\%)} & 16390 {\tiny(29.7\%)} & 18866 {\tiny(34.2\%)} \\ 
0.25 & 50 & 1577 {\tiny(2.9\%)} & 4815 {\tiny(8.7\%)} & 8058 {\tiny(14.6\%)} & 11175 {\tiny(20.2\%)} & 13989 {\tiny(25.3\%)} & 16335 {\tiny(29.6\%)} \\ 
\hline
0.50 & 10 & 8416 {\tiny(15.2\%)} & 16617 {\tiny(30.1\%)} & 22871 {\tiny(41.4\%)} & 27745 {\tiny(50.3\%)} & 31376 {\tiny(56.8\%)} & 34172 {\tiny(61.9\%)} \\ 
0.50 & 20 & 2889 {\tiny(5.2\%)} & 7691 {\tiny(13.9\%)} & 12151 {\tiny(22.0\%)} & 15940 {\tiny(28.9\%)} & 19095 {\tiny(34.6\%)} & 22113 {\tiny(40.1\%)} \\ 
0.50 & 30 & 1569 {\tiny(2.8\%)} & 4866 {\tiny(8.8\%)} & 8175 {\tiny(14.8\%)} & 11433 {\tiny(20.7\%)} & 14286 {\tiny(25.9\%)} & 16781 {\tiny(30.4\%)} \\ 
0.50 & 40 & 1099 {\tiny(2.0\%)} & 3686 {\tiny(6.7\%)} & 6418 {\tiny(11.6\%)} & 9195 {\tiny(16.7\%)} & 11768 {\tiny(21.3\%)} & 14187 {\tiny(25.7\%)} \\ 
0.50 & 50 & 882 {\tiny(1.6\%)} & 2989 {\tiny(5.4\%)} & 5467 {\tiny(9.9\%)} & 7952 {\tiny(14.4\%)} & 10324 {\tiny(18.7\%)} & 12541 {\tiny(22.7\%)} \\ 
\hline
0.75 & 10 & 3415 {\tiny(6.2\%)} & 8637 {\tiny(15.7\%)} & 13323 {\tiny(24.1\%)} & 17409 {\tiny(31.5\%)} & 20847 {\tiny(37.8\%)} & 23861 {\tiny(43.2\%)} \\ 
0.75 & 20 & 1288 {\tiny(2.3\%)} & 4275 {\tiny(7.7\%)} & 7345 {\tiny(13.3\%)} & 10408 {\tiny(18.9\%)} & 13066 {\tiny(23.7\%)} & 15519 {\tiny(28.1\%)} \\ 
0.75 & 30 & 868 {\tiny(1.6\%)} & 2980 {\tiny(5.4\%)} & 5434 {\tiny(9.8\%)} & 8006 {\tiny(14.5\%)} & 10370 {\tiny(18.8\%)} & 12612 {\tiny(22.9\%)} \\ 
0.75 & 40 & 733 {\tiny(1.3\%)} & 2490 {\tiny(4.5\%)} & 4734 {\tiny(8.6\%)} & 6965 {\tiny(12.6\%)} & 9202 {\tiny(16.7\%)} & 11378 {\tiny(20.6\%)} \\ 
0.75 & 50 & 657 {\tiny(1.2\%)} & 2250 {\tiny(4.1\%)} & 4330 {\tiny(7.8\%)} & 6398 {\tiny(11.6\%)} & 8548 {\tiny(15.5\%)} & 10634 {\tiny(19.3\%)} \\ 
\hline
1.00 & 10 & 895 {\tiny(1.6\%)} & 3073 {\tiny(5.6\%)} & 5526 {\tiny(10.0\%)} & 8005 {\tiny(14.5\%)} & 10344 {\tiny(18.7\%)} & 12638 {\tiny(22.9\%)} \\ 
1.00 & 20 & 627 {\tiny(1.1\%)} & 2162 {\tiny(3.9\%)} & 4176 {\tiny(7.6\%)} & 6234 {\tiny(11.3\%)} & 8342 {\tiny(15.1\%)} & 10337 {\tiny(18.7\%)} \\ 
1.00 & 30 & 587 {\tiny(1.1\%)} & 1994 {\tiny(3.6\%)} & 3922 {\tiny(7.1\%)} & 5817 {\tiny(10.5\%)} & 7907 {\tiny(14.3\%)} & 9873 {\tiny(17.9\%)} \\ 
1.00 & 40 & 552 {\tiny(1.0\%)} & 1872 {\tiny(3.4\%)} & 3752 {\tiny(6.8\%)} & 5568 {\tiny(10.1\%)} & 7642 {\tiny(13.8\%)} & 9560 {\tiny(17.3\%)} \\ 
1.00 & 50 & 541 {\tiny(1.0\%)} & 1813 {\tiny(3.3\%)} & 3676 {\tiny(6.7\%)} & 5463 {\tiny(9.9\%)} & 7508 {\tiny(13.6\%)} & 9436 {\tiny(17.1\%)} \\
\hline
\end{longtable*}

Looking for flat-band segments along the high symmetry lines does not fully take into consideration the 3D structure of a material, whose bands could be dispersive away from high-symmetry paths (as already mentioned in Appendix~\ref{app:atomicflatband}). A typical example would be a (quasi-)one dimensional system. Any high symmetry line in the BZ perpendicular to the dispersive momentum direction of the one dimensional system would appear flat. For that reason, we also rely on a second indicator: peaks in the density of states.

\subsubsection{Peaks in the density of states}\label{app:dospeak}

The density of states (DOS) calculation is part of the processing applied to each ICSD entry displayed in the \webTQC~(see Ref.~\onlinecite{Vergniory2021} for in-depth discussion). Similarly with the search for the band structure flat-band segments, we focus on the DOS computed without spin orbit coupling. Note that we do not search for peaks in the DOS directly from the PROCAR output of VASP. Rather, we rely on the output of the python script pydos/pyband which processes the PROCAR files to generate, e.g., the DOS plots on the \webTQC.

To look for peaks in the DOS, we consider the energy range $[-5{\rm eV},5{\rm eV}]$. We compute the average DOS over this range; this value sets the minimum value to define a peak. We discretize the DOS with a step of $0.02 {\rm eV}$ to avoid spurious peaks from fast variations. We then apply the following algorithm:

\begin{enumerate}
\item We look for the largest value among the discretized DOS. It should be larger than the average DOS. The energy $E_{\rm peak}$ associated to this largest value sets the DOS peak position.
\item We compute $W_{\rm peak}$ the peak width as its Gaussian width, \ie the full width at half maximum of the peak.
\item All points of the discretized DOS within the peak width are discarded.
\item All points around the peak width where the DOS continues to decrease away from the maximum are also discarded (to avoid considering the tails of the peak as other peaks).
\item The procedure is then repeated from the first step until all discretized DOS values have been discarded or are lower than the average DOS.
\end{enumerate}

\begin{figure}[ht]
\centering
\includegraphics[width=0.9\textwidth,angle=0]{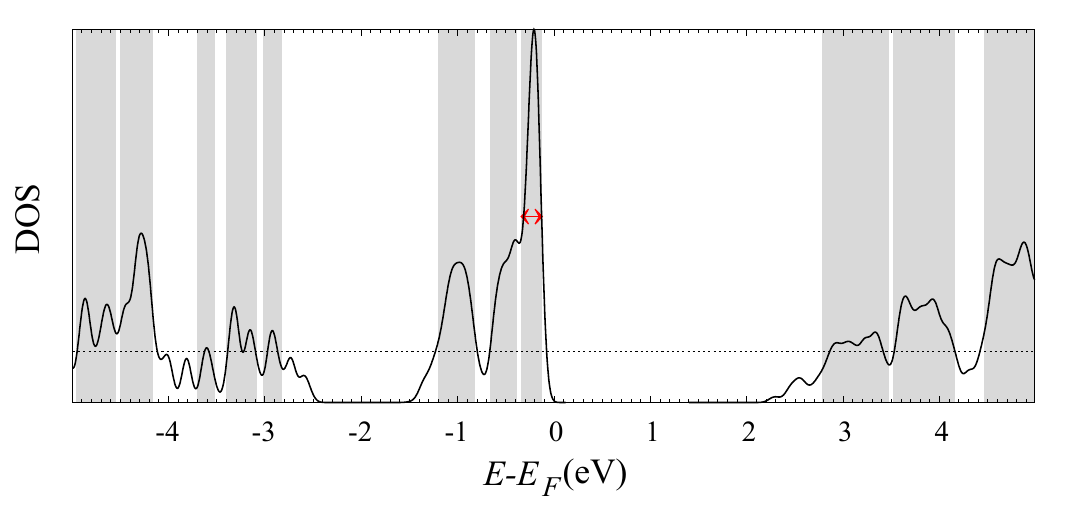}
\caption{Example of DOS peak information based on the DFT calculations for ${\rm Ca}_2 {\rm N} {\rm Br}$ [\icsdweb{153105}, SG 166 (\sgsymb{166})] in the space group 166 ($R\bar{3}m$), one of the best flat bands given in Appendix~\ref{app:bestflatbands}. The most prominent peak is located at $E_{\rm peak}-E_F \simeq -0.22 {\rm eV}$. The red double arrows indicates the Gaussian width $W_{\rm peak} \simeq 0.22 {\rm eV}$ of this peak. The dotted line is the average density of states over the energy range $[-5{\rm eV},5{\rm eV}]$. The gray areas indicate the extension of each peak.}
\label{fig:exampledospeaksearch}
\end{figure}

This crude algorithm offers a reliable and simple way to find DOS peaks. In Fig.~\ref{fig:exampledospeaksearch}, we provide an example of DOS peak processed information by this procedure. 

All ICSD DOSes available on \webflatband~have been processed with this algorithm. While not directly accessible, the DOS peak information is used by the search engine when activating ``Require a DOS peak near Ef''. Two search parameters can be tuned: the peak maximum distance to the Fermi level, \ie the maximal value for $|E_{\rm peak}-E_F|$ where $E_{\rm peak}$ is the peak position and the maximal Gaussian width $W_{\rm peak}$ for the peak closest to Fermi level. These search parameters can be used as the unique criteria or combined with other criteria such as the sublattice features or the existence of flat-band segments along the high symmetry lines.

\subsection{Sublattices and database}\label{app:sublatticesdb}

For all the ICSDs provided in \webflatband , we have performed an automated search for five types of sublattices, namely Kagome, pyrochlore, generic bipartite (with a different number of atoms in the two subsets), split and Lieb sublattices. This automated search will be detailed in Appendices~\ref{app:kagomepyrochlore} and~\ref{app:bipartite}. The full list of sublattices for each ICSD are given on their respective webpage on \webflatband. Moreover, the presence of sublattices can be used as a search criterion when looking for flat-band materials.

\begin{longtable*}{|c||c|c|c|}
\caption[Statistics of the ICSD entries hosting mutiple sublattices.]{{\it Upper part}: Statistics of the ICSD entries in the database hosting at least one Kagome or Pyrochlore sublattice and an additional sublattice: Lieb, bipartite or split. In the first row, we give the number of ICSDs entries for the database (first column), for the list of curated flat-band materials (second column) and for the best representative flat-band materials (third column). The second and third rows provide the number of ICSD entries with a Kagome (second row) or pyrochlore (third row) for each category (all ICSDs, curated flat-band materials and best representative flat-band materials). The fourth, fifth, and sixth rows are the statistics for ICSD with a Kagome sublattice and either a Lieb, bipartite with different number of atoms on sublattices and split lattices. An ICSD entry is considered as hosting a given type of lattice if the algorithms discussed in Section~\refsecsublattices of the main text have found at least one such a lattice irrespective of the cutoffs or being a rigorous or approximate sublattice. The seventh, eighth, and ninth rows are similar to the fourth, fifth, and sixth rows but for pyrochlore rather than Kagome sublattices. {\it Lower part}: Statistics of the ICSD entries in the database hosting one type of sublattice.  The tenth, eleventh, twelfth and thirteenth rows give the number of ICSDs hosting only a Kagome, pryrochlore, Lieb or bipartite sublattice respectively. All percentage are expressed with respect to the numbers in the second row. \label{tab:statisticsmultiplesublattices}}\\
\hline
 & All ICSDs & Curated & Best  \\
\hline
\hline
\# ICSDs & \TQCDBNbrNoSOCICSDsIncludingValidFElectrons & \FlatBandNbrCuratedICSDs & \FlatBandNbrBestICSDs \\
\hline
\hline
Kagome & \begin{tabular}{c}\FlatBandNbrICSDsKagome\\ (\FlatBandNbrICSDsKagomePercent)\end{tabular} & \begin{tabular}{c}\FlatBandNbrCuratedICSDsKagome\\ (\FlatBandNbrCuratedICSDsKagomePercent)\end{tabular} & \begin{tabular}{c}\FlatBandNbrBestICSDsKagome\\ (\FlatBandNbrBestICSDsKagomePercent)\end{tabular} \\
\hline
Pyrochlore & \begin{tabular}{c}\FlatBandNbrICSDsPyrochlore\\ (\FlatBandNbrICSDsPyrochlorePercent)\end{tabular} & \begin{tabular}{c}\FlatBandNbrCuratedICSDsPyrochlore\\ (\FlatBandNbrCuratedICSDsPyrochlorePercent)\end{tabular} & \begin{tabular}{c}\FlatBandNbrBestICSDsPyrochlore\\ (\FlatBandNbrBestICSDsPyrochlorePercent)\end{tabular} \\
\hline
Kagome and Lieb & \begin{tabular}{c}\FlatBandNbrICSDsKagomeAndLieb\\ (\FlatBandNbrICSDsKagomeAndLiebPercent)\end{tabular} & \begin{tabular}{c}\FlatBandNbrCuratedICSDsKagomeAndLieb\\ (\FlatBandNbrCuratedICSDsKagomeAndLiebPercent)\end{tabular} & \begin{tabular}{c}\FlatBandNbrBestICSDsKagomeAndLieb\\ (\FlatBandNbrBestICSDsKagomeAndLiebPercent)\end{tabular} \\
\hline
Kagome and Bipartite & \begin{tabular}{c}\FlatBandNbrICSDsKagomeAndBipartite\\ (\FlatBandNbrICSDsKagomeAndBipartitePercent)\end{tabular} & \begin{tabular}{c}\FlatBandNbrCuratedICSDsKagomeAndBipartite\\ (\FlatBandNbrCuratedICSDsKagomeAndBipartitePercent)\end{tabular} & \begin{tabular}{c}\FlatBandNbrBestICSDsKagomeAndBipartite\\ (\FlatBandNbrBestICSDsKagomeAndBipartitePercent)\end{tabular} \\
\hline
Kagome and Split &  \begin{tabular}{c}\FlatBandNbrICSDsKagomeAndSplit\\ (\FlatBandNbrICSDsKagomeAndSplitPercent)\end{tabular}& \begin{tabular}{c}\FlatBandNbrCuratedICSDsKagomeAndSplit\\ (\FlatBandNbrCuratedICSDsKagomeAndSplitPercent)\end{tabular} & \begin{tabular}{c}\FlatBandNbrBestICSDsKagomeAndSplit\\ (\FlatBandNbrBestICSDsKagomeAndSplitPercent)\end{tabular} \\
\hline 
Pyrochlore and Lieb & \begin{tabular}{c}\FlatBandNbrICSDsPyrochloreAndLieb\\ (\FlatBandNbrICSDsPyrochloreAndLiebPercent)\end{tabular} & \begin{tabular}{c}\FlatBandNbrCuratedICSDsPyrochloreAndLieb\\ (\FlatBandNbrCuratedICSDsPyrochloreAndLiebPercent)\end{tabular} & \begin{tabular}{c}\FlatBandNbrBestICSDsPyrochloreAndLieb\\ (\FlatBandNbrBestICSDsPyrochloreAndLiebPercent)\end{tabular} \\
\hline
Pyrochlore and Bipartite & \begin{tabular}{c}\FlatBandNbrICSDsPyrochloreAndBipartite\\ (\FlatBandNbrICSDsPyrochloreAndBipartitePercent)\end{tabular} & \begin{tabular}{c}\FlatBandNbrCuratedICSDsPyrochloreAndBipartite\\ (\FlatBandNbrCuratedICSDsPyrochloreAndBipartitePercent)\end{tabular} & \begin{tabular}{c}\FlatBandNbrBestICSDsPyrochloreAndBipartite\\ (\FlatBandNbrBestICSDsPyrochloreAndBipartitePercent)\end{tabular} \\
\hline
Pyrochlore and Split &  \begin{tabular}{c}\FlatBandNbrICSDsPyrochloreAndSplit\\ (\FlatBandNbrICSDsPyrochloreAndSplitPercent)\end{tabular}& \begin{tabular}{c}\FlatBandNbrCuratedICSDsPyrochloreAndSplit\\ (\FlatBandNbrCuratedICSDsPyrochloreAndSplitPercent)\end{tabular} & \begin{tabular}{c}\FlatBandNbrBestICSDsPyrochloreAndSplit\\ (\FlatBandNbrBestICSDsPyrochloreAndSplitPercent)\end{tabular} \\
\hline
\hline 
Only Kagome  &  \begin{tabular}{c}\FlatBandNbrICSDsOnlyKagomeSublattices\\ (\FlatBandNbrICSDsOnlyKagomeSublatticesPercent)\end{tabular}& \begin{tabular}{c}\FlatBandNbrCuratedICSDsOnlyKagomeSublattices\\ (\FlatBandNbrCuratedICSDsOnlyKagomeSublatticesPercent)\end{tabular} & \begin{tabular}{c}\FlatBandNbrBestICSDsOnlyKagomeSublattices\\ (\FlatBandNbrBestICSDsOnlyKagomeSublatticesPercent)\end{tabular} \\
\hline 
Only Pyrochlore  &  \begin{tabular}{c}\FlatBandNbrICSDsOnlyPyrochloreSublattices\\ (\FlatBandNbrICSDsOnlyPyrochloreSublatticesPercent)\end{tabular}& \begin{tabular}{c}\FlatBandNbrCuratedICSDsOnlyPyrochloreSublattices\\ (\FlatBandNbrCuratedICSDsOnlyPyrochloreSublatticesPercent)\end{tabular} & \begin{tabular}{c}\FlatBandNbrBestICSDsOnlyPyrochloreSublattices\\ (\FlatBandNbrBestICSDsOnlyPyrochloreSublatticesPercent)\end{tabular} \\
\hline 
Only Lieb  &  \begin{tabular}{c}\FlatBandNbrICSDsOnlyLiebSublattices\\ (\FlatBandNbrICSDsOnlyLiebSublatticesPercent)\end{tabular}& \begin{tabular}{c}\FlatBandNbrCuratedICSDsOnlyLiebSublattices\\ (\FlatBandNbrCuratedICSDsOnlyLiebSublatticesPercent)\end{tabular} & \begin{tabular}{c}\FlatBandNbrBestICSDsOnlyLiebSublattices\\ (\FlatBandNbrBestICSDsOnlyLiebSublatticesPercent)\end{tabular} \\
\hline 
Only Bipartite  &  \begin{tabular}{c}\FlatBandNbrICSDsOnlyBipartiteSublattices\\ (\FlatBandNbrICSDsOnlyBipartiteSublatticesPercent)\end{tabular}& \begin{tabular}{c}\FlatBandNbrCuratedICSDsOnlyBipartiteSublattices\\ (\FlatBandNbrCuratedICSDsOnlyBipartiteSublatticesPercent)\end{tabular} & \begin{tabular}{c}\FlatBandNbrBestICSDsOnlyBipartiteSublattices\\ (\FlatBandNbrBestICSDsOnlyBipartiteSublatticesPercent)\end{tabular} \\
\hline 
\hline
\end{longtable*}

For sake of completeness, we now discuss the frequency of appearance for the sublattice types among the \TQCDBNbrNoSOCICSDsIncludingValidFElectrons~ICSD entries in our database. Note that our figures here are based on ICSDs rather than unique materials. Indeed, tiny differences in the atom positions from one ICSD to another one associated to the same unique material might lead to threshold effect: a sublattice close to the maximum error bar of the automated search algorithm for one ICSD might be slightly beyond the error bar for another ICSD. From our survey, we have found that \FlatBandNbrICSDsOneSublattice~ICSDs (\FlatBandNbrICSDsOneSublatticePercent) have at least one sublattice. A large fraction of the ICSDs, \FlatBandNbrICSDsBipartitePercent~(\ie \FlatBandNbrICSDsBipartite~ICSDs), have at least one bipartite sublattice. For the Kagome sublattices, \FlatBandNbrICSDsKagome~ICSDs (\FlatBandNbrICSDsKagomePercent), the majority are rigorous Kagome sublattices (\FlatBandNbrICSDsRigorousKagome~ICSDs, \FlatBandNbrICSDsRigorousKagomePercent) as defined in Appendix~\ref{app:kagomepyrochlore}. The approximate Kagome sublattices are found in  \FlatBandNbrICSDsApproximateKagome~ICSDs (\FlatBandNbrICSDsApproximateKagomePercent). Note that an ICSD can have both a rigorous and an approximate Kagome distinct sublattice. A similar analysis can be performed for the Lieb sublattices: there are \FlatBandNbrICSDsLieb~ICSDs (\FlatBandNbrICSDsLiebPercent) with Lieb sublattices, including \FlatBandNbrICSDsRigorousLieb~ICSDs (\FlatBandNbrICSDsRigorousLiebPercent) with rigorous Lieb sublattices and \FlatBandNbrICSDsApproximateLieb~ICSDs (\FlatBandNbrICSDsApproximateLiebPercent) with approximate Lieb sublattices, as defined in Appendix~\ref{app:kagomepyrochlore}. Pyrochlore sublattices have been found in \FlatBandNbrICSDsPyrochlore~ICSDs (\FlatBandNbrICSDsPyrochlorePercent), including  \FlatBandNbrICSDsRigorousPyrochlore~ICSDs hosting (\FlatBandNbrICSDsRigorousPyrochlorePercent) rigorous sublattices and \FlatBandNbrICSDsApproximatePyrochlore~ICSDs hosting (\FlatBandNbrICSDsApproximatePyrochlorePercent). Finally, split sublattices were detected in \FlatBandNbrICSDsSplit~ICSDs (\FlatBandNbrICSDsSplitPercent). As mentioned in the caption of Table~\reftabstatisticssublattices\ in the main text, an ICSD entry might host more than one type of sublattices. A trivial example is a pyrochlore sublattice that implies the existence of Kagome sublattices. More interestingly, we provide in Table~\ref{tab:statisticsmultiplesublattices} the number of ICSDs hosting either a Kagome or a pyrochlore sublattice and any of the three other types of sublattices we have considered: Lieb, bipartite and split sublattices. We also provide in this table, the number of ICSDSs hosting a single type of sublattice. Note that the flat bands in a compound hosting solely a bipartite lattice may only be understood in the S-matrix formalism.\cite{S-matrix} Additionally, a breakdown of the number of ICSDs per space group that have a given sublattice type among the 5 types we have searched for is available in Table~\ref{tab:sublatticestat}.

{\tiny
\begin{longtable*}{|c|c|c|c|c|c|c|}
\caption[Sublattice statistics per SG]{Statistics about ICSD entries with either Kagome, pyrochlore, bipartite, split or Lieb sublattices. The first column is the space group number, the second column is the number of ICSDs that have been processed per SG. The third (seventh) column provides the number of ICSDs per SG having at least one rigorous or approximate Kagome (Lieb) sublattice. The fourth, fifth and sixth columns give the number of ICSDs per SG with at least one pyrochlore, bipartite and split sublattice respectively. Note that we count an ICSD entry as having a bipartite or split sublattice, if it hosts such a sublattice irrespective of the cutoffs defined in Appendix~\ref{app:bipartite}. Note that not all the 2D Lieb-sublattice materials are labeled as a 3D bipartite/split sublattice in the results of our algorithm and further discussion can be referred to Appendix~\ref{app:algorithmofbipartite}. \label{tab:sublatticestat}}\\
\hline
SG  & \# ICSDs & \begin{tabular}{c}\# ICSDs with\\ Kagome\end{tabular} & \begin{tabular}{c}\# ICSDs with\\ pyrochlore \end{tabular} & \begin{tabular}{c}\# ICSDs with\\ bipartite \end{tabular} & \begin{tabular}{c}\# ICSDs with\\ split \end{tabular} & \begin{tabular}{c}\# ICSDs with\\ Lieb \end{tabular}  \\
\hline  
1 & 221 & 2 {\tiny(0.9\%)} & ---  & 104 {\tiny(47.1\%)} & 59 {\tiny(26.7\%)} & ---  \\ 
2 & 1677 & 8 {\tiny(0.5\%)} & ---  & 701 {\tiny(41.8\%)} & 218 {\tiny(13.0\%)} & ---  \\ 
3 & 14 & ---  & ---  & 7 {\tiny(50.0\%)} & 3 {\tiny(21.4\%)} & ---  \\ 
4 & 275 & ---  & ---  & 122 {\tiny(44.4\%)} & 43 {\tiny(15.6\%)} & ---  \\ 
5 & 182 & 1 {\tiny(0.6\%)} & ---  & 91 {\tiny(50.0\%)} & 37 {\tiny(20.3\%)} & ---  \\ 
6 & 38 & 2 {\tiny(5.3\%)} & ---  & 13 {\tiny(34.2\%)} & 6 {\tiny(15.8\%)} & ---  \\ 
7 & 141 & ---  & ---  & 77 {\tiny(54.6\%)} & 45 {\tiny(31.9\%)} & ---  \\ 
8 & 164 & 8 {\tiny(4.9\%)} & ---  & 73 {\tiny(44.5\%)} & 22 {\tiny(13.4\%)} & 1 {\tiny(0.6\%)} \\ 
9 & 272 & 3 {\tiny(1.1\%)} & 1 {\tiny(0.4\%)} & 145 {\tiny(53.3\%)} & 44 {\tiny(16.2\%)} & ---  \\ 
10 & 38 & 2 {\tiny(5.3\%)} & ---  & 13 {\tiny(34.2\%)} & 4 {\tiny(10.5\%)} & ---  \\ 
11 & 677 & 1 {\tiny(0.1\%)} & ---  & 260 {\tiny(38.4\%)} & 51 {\tiny(7.5\%)} & ---  \\ 
12 & 1528 & 24 {\tiny(1.6\%)} & ---  & 656 {\tiny(42.9\%)} & 165 {\tiny(10.8\%)} & 1 {\tiny(0.1\%)} \\ 
13 & 246 & 1 {\tiny(0.4\%)} & ---  & 164 {\tiny(66.7\%)} & 54 {\tiny(21.9\%)} & ---  \\ 
14 & 3370 & 2 {\tiny(0.1\%)} & ---  & 1494 {\tiny(44.3\%)} & 473 {\tiny(14.0\%)} & ---  \\ 
15 & 2270 & 5 {\tiny(0.2\%)} & ---  & 1303 {\tiny(57.4\%)} & 449 {\tiny(19.8\%)} & ---  \\ 
16 & 2 & ---  & ---  & 2 {\tiny(100\%)} & 1 {\tiny(50.0\%)} & ---  \\ 
17 & 8 & ---  & ---  & 3 {\tiny(37.5\%)} & 3 {\tiny(37.5\%)} & 1 {\tiny(12.5\%)} \\ 
18 & 41 & ---  & ---  & 20 {\tiny(48.8\%)} & 12 {\tiny(29.3\%)} & ---  \\ 
19 & 327 & ---  & ---  & 116 {\tiny(35.5\%)} & 25 {\tiny(7.7\%)} & ---  \\ 
20 & 72 & 1 {\tiny(1.4\%)} & ---  & 43 {\tiny(59.7\%)} & 33 {\tiny(45.8\%)} & ---  \\ 
21 & 10 & ---  & ---  & 4 {\tiny(40.0\%)} & 4 {\tiny(40.0\%)} & ---  \\ 
22 & 7 & ---  & ---  & 3 {\tiny(42.9\%)} & 1 {\tiny(14.3\%)} & ---  \\ 
23 & 15 & ---  & ---  & 8 {\tiny(53.3\%)} & 3 {\tiny(20.0\%)} & ---  \\ 
24 & 3 & 2 {\tiny(66.7\%)} & ---  & 3 {\tiny(100\%)} & 3 {\tiny(100\%)} & ---  \\ 
25 & 56 & ---  & ---  & 27 {\tiny(48.2\%)} & 18 {\tiny(32.1\%)} & ---  \\ 
26 & 74 & 1 {\tiny(1.4\%)} & ---  & 41 {\tiny(55.4\%)} & 18 {\tiny(24.3\%)} & 1 {\tiny(1.4\%)} \\ 
27 & 1 & ---  & ---  & 1 {\tiny(100\%)} & 1 {\tiny(100\%)} & ---  \\ 
28 & 17 & ---  & ---  & 1 {\tiny(5.9\%)} & ---  & ---  \\ 
29 & 85 & ---  & ---  & 27 {\tiny(31.8\%)} & 9 {\tiny(10.6\%)} & ---  \\ 
30 & 6 & ---  & ---  & 5 {\tiny(83.3\%)} & 1 {\tiny(16.7\%)} & 1 {\tiny(16.7\%)} \\ 
31 & 240 & 8 {\tiny(3.3\%)} & ---  & 89 {\tiny(37.1\%)} & 13 {\tiny(5.4\%)} & ---  \\ 
32 & 12 & ---  & ---  & 9 {\tiny(75.0\%)} & 2 {\tiny(16.7\%)} & ---  \\ 
33 & 379 & 2 {\tiny(0.5\%)} & ---  & 200 {\tiny(52.8\%)} & 70 {\tiny(18.5\%)} & ---  \\ 
34 & 28 & 2 {\tiny(7.1\%)} & ---  & 12 {\tiny(42.9\%)} & 10 {\tiny(35.7\%)} & ---  \\ 
35 & 8 & ---  & ---  & 6 {\tiny(75.0\%)} & 3 {\tiny(37.5\%)} & ---  \\ 
36 & 372 & 1 {\tiny(0.3\%)} & ---  & 227 {\tiny(61.0\%)} & 94 {\tiny(25.3\%)} & ---  \\ 
37 & 7 & ---  & ---  & 7 {\tiny(100\%)} & 2 {\tiny(28.6\%)} & ---  \\ 
38 & 81 & 10 {\tiny(12.3\%)} & ---  & 33 {\tiny(40.7\%)} & 19 {\tiny(23.5\%)} & 13 {\tiny(16.1\%)} \\ 
39 & 13 & ---  & ---  & 6 {\tiny(46.1\%)} & 1 {\tiny(7.7\%)} & 1 {\tiny(7.7\%)} \\ 
40 & 60 & ---  & ---  & 24 {\tiny(40.0\%)} & 8 {\tiny(13.3\%)} & ---  \\ 
41 & 43 & ---  & ---  & 23 {\tiny(53.5\%)} & 20 {\tiny(46.5\%)} & ---  \\ 
42 & 4 & ---  & ---  & ---  & ---  & ---  \\ 
43 & 121 & ---  & ---  & 52 {\tiny(43.0\%)} & 29 {\tiny(24.0\%)} & ---  \\ 
44 & 80 & 4 {\tiny(5.0\%)} & 2 {\tiny(2.5\%)} & 21 {\tiny(26.2\%)} & 5 {\tiny(6.2\%)} & ---  \\ 
45 & 11 & ---  & ---  & 8 {\tiny(72.7\%)} & 5 {\tiny(45.5\%)} & 2 {\tiny(18.2\%)} \\ 
46 & 81 & 5 {\tiny(6.2\%)} & 2 {\tiny(2.5\%)} & 52 {\tiny(64.2\%)} & 42 {\tiny(51.9\%)} & 20 {\tiny(24.7\%)} \\ 
47 & 62 & 1 {\tiny(1.6\%)} & ---  & 52 {\tiny(83.9\%)} & 50 {\tiny(80.7\%)} & 3 {\tiny(4.8\%)} \\ 
48 & 2 & ---  & ---  & 1 {\tiny(50.0\%)} & 1 {\tiny(50.0\%)} & ---  \\ 
50 & 7 & ---  & ---  & 2 {\tiny(28.6\%)} & 2 {\tiny(28.6\%)} & 1 {\tiny(14.3\%)} \\ 
51 & 115 & ---  & ---  & 46 {\tiny(40.0\%)} & 16 {\tiny(13.9\%)} & 2 {\tiny(1.7\%)} \\ 
52 & 46 & 1 {\tiny(2.2\%)} & ---  & 37 {\tiny(80.4\%)} & 15 {\tiny(32.6\%)} & ---  \\ 
53 & 26 & ---  & ---  & 3 {\tiny(11.5\%)} & ---  & ---  \\ 
54 & 22 & ---  & ---  & 3 {\tiny(13.6\%)} & 3 {\tiny(13.6\%)} & ---  \\ 
55 & 337 & ---  & ---  & 164 {\tiny(48.7\%)} & 35 {\tiny(10.4\%)} & ---  \\ 
56 & 29 & ---  & ---  & 21 {\tiny(72.4\%)} & 14 {\tiny(48.3\%)} & ---  \\ 
57 & 143 & 1 {\tiny(0.7\%)} & ---  & 54 {\tiny(37.8\%)} & 27 {\tiny(18.9\%)} & ---  \\ 
58 & 461 & ---  & ---  & 162 {\tiny(35.1\%)} & 27 {\tiny(5.9\%)} & ---  \\ 
59 & 278 & ---  & ---  & 98 {\tiny(35.2\%)} & 28 {\tiny(10.1\%)} & ---  \\ 
60 & 222 & ---  & ---  & 150 {\tiny(67.6\%)} & 22 {\tiny(9.9\%)} & ---  \\ 
61 & 211 & ---  & ---  & 84 {\tiny(39.8\%)} & 25 {\tiny(11.8\%)} & ---  \\ 
62 & 4313 & 15 {\tiny(0.3\%)} & ---  & 1956 {\tiny(45.4\%)} & 503 {\tiny(11.7\%)} & 8 {\tiny(0.2\%)} \\ 
63 & 1307 & 9 {\tiny(0.7\%)} & ---  & 417 {\tiny(31.9\%)} & 132 {\tiny(10.1\%)} & 2 {\tiny(0.1\%)} \\ 
64 & 395 & 8 {\tiny(2.0\%)} & ---  & 149 {\tiny(37.7\%)} & 95 {\tiny(24.1\%)} & 57 {\tiny(14.4\%)} \\ 
65 & 192 & 4 {\tiny(2.1\%)} & ---  & 92 {\tiny(47.9\%)} & 52 {\tiny(27.1\%)} & 4 {\tiny(2.1\%)} \\ 
66 & 37 & ---  & ---  & 30 {\tiny(81.1\%)} & 27 {\tiny(73.0\%)} & 1 {\tiny(2.7\%)} \\ 
67 & 40 & ---  & ---  & 5 {\tiny(12.5\%)} & 3 {\tiny(7.5\%)} & ---  \\ 
68 & 19 & ---  & ---  & 9 {\tiny(47.4\%)} & 6 {\tiny(31.6\%)} & ---  \\ 
69 & 59 & 11 {\tiny(18.6\%)} & ---  & 26 {\tiny(44.1\%)} & 8 {\tiny(13.6\%)} & ---  \\ 
70 & 174 & 5 {\tiny(2.9\%)} & 5 {\tiny(2.9\%)} & 81 {\tiny(46.5\%)} & 38 {\tiny(21.8\%)} & ---  \\ 
71 & 355 & 5 {\tiny(1.4\%)} & ---  & 76 {\tiny(21.4\%)} & 45 {\tiny(12.7\%)} & 1 {\tiny(0.3\%)} \\ 
72 & 183 & ---  & ---  & 100 {\tiny(54.6\%)} & 63 {\tiny(34.4\%)} & ---  \\ 
73 & 18 & ---  & ---  & 5 {\tiny(27.8\%)} & ---  & ---  \\ 
74 & 221 & 33 {\tiny(14.9\%)} & 4 {\tiny(1.8\%)} & 115 {\tiny(52.0\%)} & 58 {\tiny(26.2\%)} & 4 {\tiny(1.8\%)} \\ 
75 & 5 & ---  & ---  & 4 {\tiny(80.0\%)} & 4 {\tiny(80.0\%)} & ---  \\ 
76 & 11 & ---  & ---  & 4 {\tiny(36.4\%)} & 2 {\tiny(18.2\%)} & ---  \\ 
77 & 2 & ---  & ---  & ---  & ---  & ---  \\ 
78 & 1 & ---  & ---  & ---  & ---  & ---  \\ 
79 & 13 & ---  & ---  & 3 {\tiny(23.1\%)} & 1 {\tiny(7.7\%)} & ---  \\ 
80 & 3 & ---  & ---  & 3 {\tiny(100\%)} & 2 {\tiny(66.7\%)} & ---  \\ 
81 & 18 & ---  & ---  & 5 {\tiny(27.8\%)} & 1 {\tiny(5.6\%)} & ---  \\ 
82 & 261 & ---  & ---  & 174 {\tiny(66.7\%)} & 50 {\tiny(19.2\%)} & ---  \\ 
83 & 6 & ---  & ---  & 1 {\tiny(16.7\%)} & 1 {\tiny(16.7\%)} & ---  \\ 
84 & 30 & ---  & ---  & 9 {\tiny(30.0\%)} & 4 {\tiny(13.3\%)} & ---  \\ 
85 & 49 & ---  & ---  & 27 {\tiny(55.1\%)} & 3 {\tiny(6.1\%)} & ---  \\ 
86 & 43 & ---  & ---  & 5 {\tiny(11.6\%)} & 2 {\tiny(4.7\%)} & ---  \\ 
87 & 256 & 8 {\tiny(3.1\%)} & ---  & 140 {\tiny(54.7\%)} & 115 {\tiny(44.9\%)} & ---  \\ 
88 & 275 & ---  & ---  & 187 {\tiny(68.0\%)} & 16 {\tiny(5.8\%)} & ---  \\ 
90 & 5 & ---  & ---  & 5 {\tiny(100\%)} & 2 {\tiny(40.0\%)} & ---  \\ 
91 & 8 & ---  & ---  & 8 {\tiny(100\%)} & 2 {\tiny(25.0\%)} & ---  \\ 
92 & 131 & ---  & ---  & 83 {\tiny(63.4\%)} & 71 {\tiny(54.2\%)} & ---  \\ 
94 & 1 & ---  & ---  & 1 {\tiny(100\%)} & ---  & ---  \\ 
95 & 3 & ---  & ---  & 3 {\tiny(100\%)} & 1 {\tiny(33.3\%)} & ---  \\ 
96 & 37 & 1 {\tiny(2.7\%)} & 1 {\tiny(2.7\%)} & 11 {\tiny(29.7\%)} & 6 {\tiny(16.2\%)} & ---  \\ 
97 & 1 & ---  & ---  & ---  & ---  & ---  \\ 
98 & 7 & ---  & ---  & 3 {\tiny(42.9\%)} & 2 {\tiny(28.6\%)} & ---  \\ 
99 & 182 & 48 {\tiny(26.4\%)} & ---  & 158 {\tiny(86.8\%)} & 152 {\tiny(83.5\%)} & 45 {\tiny(24.7\%)} \\ 
100 & 37 & ---  & ---  & 32 {\tiny(86.5\%)} & 17 {\tiny(46.0\%)} & ---  \\ 
102 & 12 & ---  & ---  & 5 {\tiny(41.7\%)} & 1 {\tiny(8.3\%)} & ---  \\ 
103 & 9 & ---  & ---  & ---  & ---  & ---  \\ 
104 & 4 & ---  & ---  & 2 {\tiny(50.0\%)} & 1 {\tiny(25.0\%)} & ---  \\ 
105 & 4 & ---  & ---  & ---  & ---  & ---  \\ 
106 & 1 & ---  & ---  & 1 {\tiny(100\%)} & ---  & ---  \\ 
107 & 84 & ---  & ---  & 13 {\tiny(15.5\%)} & 9 {\tiny(10.7\%)} & ---  \\ 
108 & 14 & ---  & ---  & 8 {\tiny(57.1\%)} & 4 {\tiny(28.6\%)} & ---  \\ 
109 & 33 & ---  & ---  & 6 {\tiny(18.2\%)} & ---  & ---  \\ 
110 & 36 & ---  & ---  & 33 {\tiny(91.7\%)} & 25 {\tiny(69.4\%)} & ---  \\ 
111 & 24 & 5 {\tiny(20.8\%)} & ---  & 19 {\tiny(79.2\%)} & ---  & ---  \\ 
112 & 4 & ---  & ---  & 3 {\tiny(75.0\%)} & 3 {\tiny(75.0\%)} & ---  \\ 
113 & 143 & ---  & ---  & 97 {\tiny(67.8\%)} & 77 {\tiny(53.9\%)} & ---  \\ 
114 & 41 & ---  & ---  & 19 {\tiny(46.3\%)} & 11 {\tiny(26.8\%)} & ---  \\ 
115 & 23 & 2 {\tiny(8.7\%)} & ---  & 6 {\tiny(26.1\%)} & 2 {\tiny(8.7\%)} & ---  \\ 
116 & 25 & ---  & ---  & 9 {\tiny(36.0\%)} & 4 {\tiny(16.0\%)} & ---  \\ 
117 & 9 & ---  & ---  & 5 {\tiny(55.6\%)} & 3 {\tiny(33.3\%)} & ---  \\ 
118 & 11 & ---  & ---  & 3 {\tiny(27.3\%)} & 2 {\tiny(18.2\%)} & ---  \\ 
119 & 49 & 1 {\tiny(2.0\%)} & ---  & 24 {\tiny(49.0\%)} & 10 {\tiny(20.4\%)} & ---  \\ 
120 & 22 & ---  & ---  & 8 {\tiny(36.4\%)} & 6 {\tiny(27.3\%)} & ---  \\ 
121 & 193 & ---  & ---  & 50 {\tiny(25.9\%)} & 17 {\tiny(8.8\%)} & ---  \\ 
122 & 502 & ---  & ---  & 78 {\tiny(15.5\%)} & 53 {\tiny(10.6\%)} & ---  \\ 
123 & 521 & 51 {\tiny(9.8\%)} & ---  & 177 {\tiny(34.0\%)} & 146 {\tiny(28.0\%)} & 119 {\tiny(22.8\%)} \\ 
124 & 33 & ---  & ---  & 6 {\tiny(18.2\%)} & ---  & ---  \\ 
125 & 36 & ---  & ---  & 15 {\tiny(41.7\%)} & 6 {\tiny(16.7\%)} & ---  \\ 
126 & 9 & ---  & ---  & 5 {\tiny(55.6\%)} & 4 {\tiny(44.4\%)} & ---  \\ 
127 & 271 & ---  & ---  & 72 {\tiny(26.6\%)} & 47 {\tiny(17.3\%)} & 8 {\tiny(3.0\%)} \\ 
128 & 56 & ---  & ---  & 9 {\tiny(16.1\%)} & 6 {\tiny(10.7\%)} & ---  \\ 
129 & 889 & 3 {\tiny(0.3\%)} & ---  & 108 {\tiny(12.2\%)} & 68 {\tiny(7.7\%)} & 3 {\tiny(0.3\%)} \\ 
130 & 55 & ---  & ---  & 32 {\tiny(58.2\%)} & 19 {\tiny(34.5\%)} & ---  \\ 
131 & 36 & ---  & ---  & 7 {\tiny(19.4\%)} & ---  & ---  \\ 
132 & 10 & 1 {\tiny(10.0\%)} & ---  & 5 {\tiny(50.0\%)} & 1 {\tiny(10.0\%)} & ---  \\ 
133 & 5 & ---  & ---  & 1 {\tiny(20.0\%)} & 1 {\tiny(20.0\%)} & ---  \\ 
134 & 9 & 2 {\tiny(22.2\%)} & ---  & 2 {\tiny(22.2\%)} & 2 {\tiny(22.2\%)} & ---  \\ 
135 & 76 & ---  & ---  & 66 {\tiny(86.8\%)} & 2 {\tiny(2.6\%)} & ---  \\ 
136 & 710 & ---  & ---  & 552 {\tiny(77.8\%)} & 14 {\tiny(2.0\%)} & 2 {\tiny(0.3\%)} \\ 
137 & 126 & ---  & ---  & 104 {\tiny(82.5\%)} & 34 {\tiny(27.0\%)} & ---  \\ 
138 & 17 & ---  & ---  & 9 {\tiny(52.9\%)} & 7 {\tiny(41.2\%)} & ---  \\ 
139 & 1772 & 21 {\tiny(1.2\%)} & ---  & 502 {\tiny(28.3\%)} & 406 {\tiny(22.9\%)} & 338 {\tiny(19.1\%)} \\ 
140 & 777 & ---  & ---  & 228 {\tiny(29.3\%)} & 145 {\tiny(18.7\%)} & 40 {\tiny(5.2\%)} \\ 
141 & 403 & 25 {\tiny(6.2\%)} & 20 {\tiny(5.0\%)} & 200 {\tiny(49.6\%)} & 15 {\tiny(3.7\%)} & 2 {\tiny(0.5\%)} \\ 
142 & 73 & ---  & ---  & 26 {\tiny(35.6\%)} & 16 {\tiny(21.9\%)} & ---  \\ 
143 & 18 & 2 {\tiny(11.1\%)} & ---  & 13 {\tiny(72.2\%)} & 3 {\tiny(16.7\%)} & ---  \\ 
144 & 31 & ---  & ---  & 13 {\tiny(41.9\%)} & 12 {\tiny(38.7\%)} & ---  \\ 
145 & 5 & ---  & ---  & 2 {\tiny(40.0\%)} & 2 {\tiny(40.0\%)} & ---  \\ 
146 & 105 & 3 {\tiny(2.9\%)} & ---  & 59 {\tiny(56.2\%)} & 10 {\tiny(9.5\%)} & ---  \\ 
147 & 74 & 1 {\tiny(1.4\%)} & ---  & 43 {\tiny(58.1\%)} & 11 {\tiny(14.9\%)} & ---  \\ 
148 & 740 & 10 {\tiny(1.4\%)} & ---  & 434 {\tiny(58.6\%)} & 122 {\tiny(16.5\%)} & ---  \\ 
149 & 15 & 9 {\tiny(60.0\%)} & ---  & 15 {\tiny(100\%)} & 6 {\tiny(40.0\%)} & ---  \\ 
150 & 73 & 3 {\tiny(4.1\%)} & ---  & 57 {\tiny(78.1\%)} & 27 {\tiny(37.0\%)} & ---  \\ 
151 & 4 & ---  & ---  & 3 {\tiny(75.0\%)} & ---  & ---  \\ 
152 & 269 & ---  & ---  & 194 {\tiny(72.1\%)} & 179 {\tiny(66.5\%)} & ---  \\ 
153 & 1 & ---  & ---  & ---  & ---  & ---  \\ 
154 & 138 & ---  & ---  & 109 {\tiny(79.0\%)} & 108 {\tiny(78.3\%)} & ---  \\ 
155 & 57 & 2 {\tiny(3.5\%)} & ---  & 25 {\tiny(43.9\%)} & 7 {\tiny(12.3\%)} & ---  \\ 
156 & 221 & 5 {\tiny(2.3\%)} & ---  & 116 {\tiny(52.5\%)} & 1 {\tiny(0.5\%)} & ---  \\ 
157 & 26 & 4 {\tiny(15.4\%)} & ---  & 21 {\tiny(80.8\%)} & 18 {\tiny(69.2\%)} & ---  \\ 
158 & 3 & ---  & ---  & 3 {\tiny(100\%)} & 1 {\tiny(33.3\%)} & ---  \\ 
159 & 57 & 7 {\tiny(12.3\%)} & ---  & 45 {\tiny(79.0\%)} & 6 {\tiny(10.5\%)} & ---  \\ 
160 & 291 & 51 {\tiny(17.5\%)} & ---  & 143 {\tiny(49.1\%)} & 21 {\tiny(7.2\%)} & 11 {\tiny(3.8\%)} \\ 
161 & 246 & 4 {\tiny(1.6\%)} & ---  & 200 {\tiny(81.3\%)} & 32 {\tiny(13.0\%)} & ---  \\ 
162 & 65 & 12 {\tiny(18.5\%)} & ---  & 60 {\tiny(92.3\%)} & 27 {\tiny(41.5\%)} & ---  \\ 
163 & 73 & 6 {\tiny(8.2\%)} & ---  & 45 {\tiny(61.6\%)} & 25 {\tiny(34.2\%)} & ---  \\ 
164 & 789 & 127 {\tiny(16.1\%)} & ---  & 390 {\tiny(49.4\%)} & 70 {\tiny(8.9\%)} & ---  \\ 
165 & 56 & ---  & ---  & 17 {\tiny(30.4\%)} & 9 {\tiny(16.1\%)} & ---  \\ 
166 & 1173 & 292 {\tiny(24.9\%)} & 2 {\tiny(0.2\%)} & 392 {\tiny(33.4\%)} & 125 {\tiny(10.7\%)} & ---  \\ 
167 & 845 & 7 {\tiny(0.8\%)} & ---  & 664 {\tiny(78.6\%)} & 263 {\tiny(31.1\%)} & 53 {\tiny(6.3\%)} \\ 
169 & 1 & ---  & ---  & 1 {\tiny(100\%)} & ---  & ---  \\ 
173 & 163 & 7 {\tiny(4.3\%)} & ---  & 115 {\tiny(70.5\%)} & 46 {\tiny(28.2\%)} & ---  \\ 
174 & 46 & ---  & ---  & 17 {\tiny(37.0\%)} & ---  & ---  \\ 
175 & 2 & ---  & ---  & 1 {\tiny(50.0\%)} & 1 {\tiny(50.0\%)} & ---  \\ 
176 & 312 & 8 {\tiny(2.6\%)} & ---  & 145 {\tiny(46.5\%)} & 15 {\tiny(4.8\%)} & ---  \\ 
177 & 3 & ---  & ---  & 3 {\tiny(100\%)} & 3 {\tiny(100\%)} & ---  \\ 
180 & 92 & ---  & ---  & 24 {\tiny(26.1\%)} & 20 {\tiny(21.7\%)} & ---  \\ 
181 & 15 & ---  & ---  & 6 {\tiny(40.0\%)} & 6 {\tiny(40.0\%)} & ---  \\ 
182 & 48 & 1 {\tiny(2.1\%)} & ---  & 44 {\tiny(91.7\%)} & 25 {\tiny(52.1\%)} & ---  \\ 
183 & 3 & ---  & ---  & ---  & ---  & ---  \\ 
185 & 73 & 1 {\tiny(1.4\%)} & ---  & 41 {\tiny(56.2\%)} & 9 {\tiny(12.3\%)} & ---  \\ 
186 & 716 & 89 {\tiny(12.4\%)} & ---  & 191 {\tiny(26.7\%)} & 18 {\tiny(2.5\%)} & ---  \\ 
187 & 227 & 10 {\tiny(4.4\%)} & ---  & 22 {\tiny(9.7\%)} & 1 {\tiny(0.4\%)} & ---  \\ 
188 & 27 & ---  & ---  & 24 {\tiny(88.9\%)} & 11 {\tiny(40.7\%)} & ---  \\ 
189 & 383 & 1 {\tiny(0.3\%)} & ---  & 80 {\tiny(20.9\%)} & 14 {\tiny(3.7\%)} & ---  \\ 
190 & 63 & ---  & ---  & 16 {\tiny(25.4\%)} & 7 {\tiny(11.1\%)} & ---  \\ 
191 & 662 & 374 {\tiny(56.5\%)} & ---  & 59 {\tiny(8.9\%)} & 25 {\tiny(3.8\%)} & ---  \\ 
192 & 14 & 13 {\tiny(92.9\%)} & ---  & 13 {\tiny(92.9\%)} & 6 {\tiny(42.9\%)} & ---  \\ 
193 & 362 & 3 {\tiny(0.8\%)} & ---  & 46 {\tiny(12.7\%)} & 15 {\tiny(4.1\%)} & ---  \\ 
194 & 2643 & 681 {\tiny(25.8\%)} & ---  & 544 {\tiny(20.6\%)} & 172 {\tiny(6.5\%)} & ---  \\ 
195 & 1 & ---  & ---  & 1 {\tiny(100\%)} & 1 {\tiny(100\%)} & ---  \\ 
196 & 1 & ---  & ---  & ---  & ---  & ---  \\ 
197 & 36 & ---  & ---  & 35 {\tiny(97.2\%)} & 9 {\tiny(25.0\%)} & ---  \\ 
198 & 315 & ---  & ---  & 64 {\tiny(20.3\%)} & 17 {\tiny(5.4\%)} & ---  \\ 
199 & 37 & ---  & ---  & 27 {\tiny(73.0\%)} & 12 {\tiny(32.4\%)} & ---  \\ 
200 & 25 & 6 {\tiny(24.0\%)} & ---  & 3 {\tiny(12.0\%)} & 3 {\tiny(12.0\%)} & ---  \\ 
201 & 6 & 2 {\tiny(33.3\%)} & ---  & 6 {\tiny(100\%)} & 3 {\tiny(50.0\%)} & ---  \\ 
202 & 26 & ---  & ---  & ---  & ---  & ---  \\ 
203 & 8 & 8 {\tiny(100\%)} & 6 {\tiny(75.0\%)} & 6 {\tiny(75.0\%)} & ---  & ---  \\ 
204 & 235 & 98 {\tiny(41.7\%)} & ---  & 126 {\tiny(53.6\%)} & 26 {\tiny(11.1\%)} & 8 {\tiny(3.4\%)} \\ 
205 & 357 & ---  & ---  & 62 {\tiny(17.4\%)} & 3 {\tiny(0.8\%)} & ---  \\ 
206 & 154 & ---  & ---  & 131 {\tiny(85.1\%)} & 1 {\tiny(0.7\%)} & ---  \\ 
208 & 2 & ---  & ---  & ---  & ---  & ---  \\ 
210 & 1 & ---  & ---  & 1 {\tiny(100\%)} & 1 {\tiny(100\%)} & ---  \\ 
211 & 1 & ---  & ---  & 1 {\tiny(100\%)} & 1 {\tiny(100\%)} & ---  \\ 
212 & 29 & 6 {\tiny(20.7\%)} & ---  & 21 {\tiny(72.4\%)} & 7 {\tiny(24.1\%)} & ---  \\ 
213 & 41 & 1 {\tiny(2.4\%)} & ---  & 22 {\tiny(53.7\%)} & 21 {\tiny(51.2\%)} & ---  \\ 
214 & 17 & ---  & ---  & 11 {\tiny(64.7\%)} & ---  & ---  \\ 
215 & 93 & 54 {\tiny(58.1\%)} & ---  & 12 {\tiny(12.9\%)} & 1 {\tiny(1.1\%)} & ---  \\ 
216 & 1561 & 108 {\tiny(6.9\%)} & 79 {\tiny(5.1\%)} & 58 {\tiny(3.7\%)} & 7 {\tiny(0.5\%)} & ---  \\ 
217 & 150 & 37 {\tiny(24.7\%)} & ---  & 51 {\tiny(34.0\%)} & 18 {\tiny(12.0\%)} & ---  \\ 
218 & 98 & ---  & ---  & 85 {\tiny(86.7\%)} & 81 {\tiny(82.7\%)} & ---  \\ 
219 & 13 & 12 {\tiny(92.3\%)} & ---  & 12 {\tiny(92.3\%)} & 1 {\tiny(7.7\%)} & ---  \\ 
220 & 117 & ---  & ---  & 38 {\tiny(32.5\%)} & 1 {\tiny(0.8\%)} & ---  \\ 
221 & 2271 & 1344 {\tiny(59.2\%)} & ---  & 841 {\tiny(37.0\%)} & 831 {\tiny(36.6\%)} & 828 {\tiny(36.5\%)} \\ 
223 & 451 & ---  & ---  & 13 {\tiny(2.9\%)} & ---  & ---  \\ 
224 & 41 & 5 {\tiny(12.2\%)} & ---  & 36 {\tiny(87.8\%)} & 35 {\tiny(85.4\%)} & ---  \\ 
225 & 4221 & 647 {\tiny(15.3\%)} & 9 {\tiny(0.2\%)} & 320 {\tiny(7.6\%)} & 245 {\tiny(5.8\%)} & 8 {\tiny(0.2\%)} \\ 
226 & 57 & 1 {\tiny(1.8\%)} & ---  & 1 {\tiny(1.8\%)} & ---  & ---  \\ 
227 & 1971 & 1688 {\tiny(85.6\%)} & 1535 {\tiny(77.9\%)} & 1101 {\tiny(55.9\%)} & 369 {\tiny(18.7\%)} & ---  \\ 
229 & 471 & 14 {\tiny(3.0\%)} & ---  & 19 {\tiny(4.0\%)} & 6 {\tiny(1.3\%)} & 1 {\tiny(0.2\%)} \\ 
230 & 6 & ---  & ---  & 1 {\tiny(16.7\%)} & 1 {\tiny(16.7\%)} & ---  \\ 
\hline
Total & 55206 & 6120 {\tiny(11.1\%)} & 1666 {\tiny(3.0\%)} & 21175 {\tiny(38.4\%)} & 8224 {\tiny(14.9\%)} & 1590 {\tiny(2.9\%)} \\
\hline
\end{longtable*}
}

\section{Automated search for materials with Kagome, pyrochlore or Lieb sublattice}\label{app:kagomepyrochlore}

In Appendix \ref{app:database}, we have provided statistics about the crystal sublattices found in the materials of \webTQC . These sublattices fall into five classes, namely the Kagome, pyrochlore, bipartite, split or Lieb sublattices. The two line-graph lattices, Kagome and  pyrochlore lattices, and Lieb lattice exhibit flat bands in their respective band structure \emph{only} when the related tight-binding Hamiltonian satisfies the following strict condition: the hopping amplitudes between any two nearest-neighbor sites are non-zero and equal.
In real crystalline materials, equality between two hopping amplitudes can be guaranteed only when they are related by symmetries. Otherwise such a fine tuning is never achieved exactly. For these reasons, in real crystalline materials, only the Kagome and pyrochlore lattices among the line-graph lattices can be physically realized.
For bipartite and split lattices, the condition is: the hopping amplitude is zero between any two unconnected sites.

In this appendix, we first analyze the crystal symmetries and the geometrical features of Kagome, pyrochlore and Lieb lattices. Then, by using their geometrical features, we have designed an algorithm to systematically search for the rigorous and approximate Kagome, pyrochlore or Lieb sublattices in real materials.   

\subsection{Crystal symmetries of Kagome, pyrochlore and Lieb lattices}\label{app:crys_symm}

\begin{figure}[h]
\centering
\includegraphics[width=17cm]{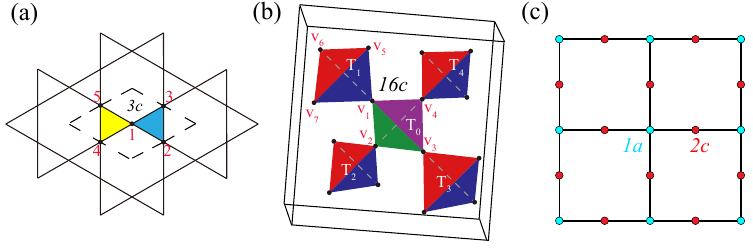}
\caption{Crystal structures of Kagome, pyrochlore and Lieb lattices. (a) The Wyckoff position $3c$ of the wallpaper group $p6mm$ form a Kagome Lattice in 2D. 
The two equilateral triangles formed by the respective set of sites \{1,2,3\} and \{1,4,5\} are related by a vertical two-fold rotational symmetry centered at the site 1.
 (b) The Wyckoff position $16c$ in space group $Fd\bar3m$ form a pyrochlore lattice in 3D. The pyrochlore lattice is formed by the regular tetrahedrons which share vertices with each others. For example, the two regular tetrahedrons $T_0$ and $T_1$ that formed by the sets of vertices $\{v_1, v_2, v_3, v_4\}$ and $\{v_1, v_5, v_6, v_7\}$, respectively, share the vertex $v_1$. Moreover, each site of the pyrochlore sublattice is also shared by three Kagome planes. For example, the triangle formed by the vertices $\{v_1, v_2, v_3\}$ in $T_0$ is co-planar with the triangle formed by the vertices $\{v_1, v_5, v_6\}$ in $T_1$. There are four equivalent Kagome sublattices related by the point group symmetry $(\bar3m)$ of the pyrochlore lattice. (c) The Wyckoff positions $1a$ and $2c$ of the wallpaper group $p4mm$ form a Lieb lattice in 2D. The 2c sites (red) are at the middle point of their two NN 1a sites (blue). The 1a sites have four $C_4$ symmetry related 2c sites NNs.}\label{fig:kagome_pyrochlore}
\end{figure}

As shown in Fig.~\ref{fig:kagome_pyrochlore}(a), the Kagome lattice in 2D consists of the $3c$ Wyckoff position of the wallpaper group $p6mm$. 
Among the 3D space groups, there are three different minimal space groups rigorously compatible with the symmetry of the Kagome lattice: \sgsymb{147} (SG 147), \sgsymb{168} (SG 168) and \sgsymb{197} (SG 197), of which the Wyckoff positions $3e$ or $3f$, $3c$ and $6b$ form a Kagome sublattice, respectively.

Using the group-subgroup relations on the \bcslong~(BCS)\cite{Ivantchev:ks0038, Ivantchev:ks0141}, we have obtained all the space groups that are super-groups of the three minimal space groups for Kagome lattices. We analyzed the subduction of the Wyckoff positions from super-groups to the minimal groups. The Wyckoff positions that form a Kagome sublattice in the super-groups have been obtained and are tabulated in Table~\ref{tab:kagomepyrochlorewyckoff}. There, we show all the crystal lattices of the 40 3D space groups that have Kagome sublattices once their corresponding Wyckoff positions provided in this table are occupied by atoms.
Note that in the above subduction of Wyckoff positions, if a Wyckoff position of a super-group splits into a Wyckoff position that forms a Kagome sublattice and some other types of Wyckoff positions coplanar to this Kagome sublattice, then the redundant Wyckoff positions spoil the Kagome lattice. Such situation happens, e.g., with the Wyckoff position $12c$ of \sgsymb{229}(SG 229). Thus, we have discarded such cases in Table~\ref{tab:kagomepyrochlorewyckoff}. 

We now turn to the pyrochlore lattice. As shown in Fig.~\ref{fig:kagome_pyrochlore}(b), a pyrochlore lattice is a 3D lattice. The minimal space groups hosting a pyrochlore lattice are \sgsymb{203} (SG 203) and \sgsymb{210} (SG 210), of which the Wyckoff positions $16c$ or $16d$ with site-symmetry group isomorphic to point group symmetry $\bar3m$ form the pyrochlore lattice. Using the same method as of the Kagome lattice, we have obtained all the space groups that are super-groups of the two minimal space groups of the pyrochlore lattice. They are tabulated in Table~\ref{tab:kagomepyrochlorewyckoff}, including the corresponding Wyckoff positions that should be occupied to get a pyrochlore lattice. Note that since pyrochlore lattice is a 3D lattice, only the Wyckoff positions tabulated in Table~\ref{tab:kagomepyrochlorewyckoff} can be occupied. Any other Wyckoff position occupied by the same element will spoil the exact pyrochlore lattice.

Finally, the Lieb lattice shown in Fig.~\ref{fig:kagome_pyrochlore}(c),  is a 2D lattice associated to the  wallpaper group $p4mm$. It is made of two different Wyckoff positions, $1a$ and $2c$. In 3D, the minimal space groups hosting Lieb lattices are \sgsymb{83} (SG 83), \sgsymb{84} (SG 84) and \sgsymb{85} (SG 85).
In Table~\ref{tab:liebwyckoff}, we tabulate these three minimal space groups and their related super-groups having Lieb sublattice, providing the related Wyckoff positions. We have also discarded the Wyckoff positions of any super-group that split into a Wyckoff position that forms a Lieb sublattice and some other types of Wyckoff positions coplanar to this Lieb sublattice, just like what we have done for the Kagome lattice.

\LTcapwidth=1.0\textwidth
\renewcommand\arraystretch{1.2}
\begin{longtable*}{|l|l|l|l|l||l|l|l|l|l|}
\caption[List of Wyckoff positions forming Kagome and pyrochlore sublattices.]{The Wyckoff positions (WPs) of 3D space groups forming Kagome or pyrochlore sublattice on special Miller plane. For each case of each SG, the column 'WP' provides the WP which form a Kagome or pyrochlore sublattice when it's occupied. The column '$\overline{WP}$' provides all the Wyckoff positions which are defined on the same plane with the Kagome lattice formed by 'WP'. When one (or more than one) of the Wyckoff positions in '$\overline{WP}$' is occupied by the atoms of the same element with 'WP', the Kagome lattice formed by the Wyckoff position in 'WP' will be spoiled. For each case, the column 'Plane' provides the Miller indices of the Kagome sublattice formed by 'WP'. Miller indices are given in the basis of the lattice vector of the conventional unit-cell (using the standard convention of the \bcslong). The column 'Lattice' indicates the type of sublattice formed by the corresponding Wyckoff positions. 'K' and 'P' stand for the Kagome and pyrochlore sublattice, respectively.}\label{tab:kagomepyrochlorewyckoff}
%\begin{tabular}{|l|l|l|l|l|l||l|l|l|l|}
\\ \hline
SG & WP & $\overline{WP}$ & Plane & Lattice & SG & WP & $\overline{WP}$ & Plane & Lattice \\ \hline
147 \sgsymb{147} & 3e &1a & [0 0 1] & K &147 \sgsymb{147} & 3f &1b & [0 0 1] & K \\ \hline 
148 \sgsymb{148} & 9d &3b & [0 0 1] & K &148 \sgsymb{148} & 9e &3a & [0 0 1] & K \\ \hline 
162 \sgsymb{162} & 3f &1a, 2c, 6i & [0 0 1] & K &162 \sgsymb{162} & 3g &1b, 2d, 6j & [0 0 1] & K \\ \hline 
163 \sgsymb{163} & 6g &2b & [0 0 1] & K &164 \sgsymb{164} & 3e &1a, 6g & [0 0 1] & K \\ \hline 
164 \sgsymb{164} & 3f &1b, 6h & [0 0 1] & K &165 \sgsymb{165} & 6e &2b & [0 0 1] & K \\ \hline 
166 \sgsymb{166} & 9d &3b, 18g & [0 0 1] & K &166 \sgsymb{166} & 9e &3a, 18f & [0 0 1] & K \\ \hline 
167 \sgsymb{167} & 18d &6b & [0 0 1] & K &168 \sgsymb{168} & 3c &1a, 2b, 6d & [0 0 1] & K \\ \hline 
175 \sgsymb{175} & 3f &1a, 2c, 6j & [0 0 1] & K &175 \sgsymb{175} & 3g &1b, 2d, 6k & [0 0 1] & K \\ \hline 
175 \sgsymb{175} & 6i &2e, 4h, 12l & [0 0 1] & K &176 \sgsymb{176} & 6g &2b & [0 0 1] & K \\ \hline 
177 \sgsymb{177} & 3f &1a, 2c, 6j, 6l & [0 0 1] & K &177 \sgsymb{177} & 3g &1b, 2d, 6k, 6m & [0 0 1] & K \\ \hline 
177 \sgsymb{177} & 6i &2e, 4h, 12n & [0 0 1] & K &183 \sgsymb{183} & 3c &1a, 2b, 6d, 6e, 12f & [0 0 1] & K \\ \hline 
184 \sgsymb{184} & 6c &2a, 4b, 12d & [0 0 1] & K &191 \sgsymb{191} & 3f &1a, 2c, 6j, 6l, 12p & [0 0 1] & K \\ \hline 
191 \sgsymb{191} & 3g &1b, 2d, 6k, 6m, 12q & [0 0 1] & K &191 \sgsymb{191} & 6i &2e, 4h, 12n, 12o, 24r & [0 0 1] & K \\ \hline 
192 \sgsymb{192} & 6f &2a, 4c, 12j, 12k & [0 0 1] & K &192 \sgsymb{192} & 6g &2b, 4d, 12l & [0 0 1] & K \\ \hline 
192 \sgsymb{192} & 12i &4e, 8h, 24m & [0 0 1] & K &193 \sgsymb{193} & 6f &2b, 4d, 12i & [0 0 1] & K \\ \hline 
194 \sgsymb{194} & 6g &2a, 12i & [0 0 1] & K &195 \sgsymb{195} & 3c &1a & [1 1 1] & K \\ \hline 
195 \sgsymb{195} & 3d &1b & [1 1 1] & K &197 \sgsymb{197} & 6b &2a & [1 1 1] & K \\ \hline 
200 \sgsymb{200} & 3c &1a & [1 1 1] & K &200 \sgsymb{200} & 3d &1b & [1 1 1] & K \\ \hline 
201 \sgsymb{201} & 6d &2a & [1 1 1] & K &202 \sgsymb{202} & 24d &4b & [1 1 1] & K \\ \hline 
202 \sgsymb{202} & 24d &4a & [1 1 1] & K &203 \sgsymb{203} & 16c &16d & [1 1 1] & K \\ \hline 
203 \sgsymb{203} & 16d &16c & [1 1 1] & K &204 \sgsymb{204} & 6b &2a & [1 1 1] & K \\ \hline 
207 \sgsymb{207} & 3c &1a, 12i & [1 1 1] & K &207 \sgsymb{207} & 3d &1b, 12j & [1 1 1] & K \\ \hline 
208 \sgsymb{208} & 6d &2a & [1 1 1] & K &209 \sgsymb{209} & 24d &4b, 48h & [1 1 1] & K \\ \hline 
209 \sgsymb{209} & 24d &4a, 48g & [1 1 1] & K &210 \sgsymb{210} & 16c &16d, 48g & [1 1 1] & K \\ \hline 
210 \sgsymb{210} & 16d &16c, 48g & [1 1 1] & K &211 \sgsymb{211} & 6b &2a, 24h & [1 1 1] & K \\ \hline 
215 \sgsymb{215} & 3c &1a & [1 1 1] & K &215 \sgsymb{215} & 3d &1b & [1 1 1] & K \\ \hline 
217 \sgsymb{217} & 6b &2a & [1 1 1] & K &218 \sgsymb{218} & 6b &2a & [1 1 1] & K \\ \hline 
219 \sgsymb{219} & 24c &8a & [1 1 1] & K &219 \sgsymb{219} & 24d &8b & [1 1 1] & K \\ \hline 
221 \sgsymb{221} & 3c &1a, 12i & [1 1 1] & K &221 \sgsymb{221} & 3d &1b, 12j & [1 1 1] & K \\ \hline 
222 \sgsymb{222} & 6b &2a, 24h & [1 1 1] & K &223 \sgsymb{223} & 6b &2a & [1 1 1] & K \\ \hline 
224 \sgsymb{224} & 6d &2a & [1 1 1] & K &225 \sgsymb{225} & 24d &4b, 48i & [1 1 1] & K \\ \hline 
225 \sgsymb{225} & 24d &4a, 48h & [1 1 1] & K &226 \sgsymb{226} & 24c &8a, 96h & [1 1 1] & K \\ \hline 
226 \sgsymb{226} & 24d &8b & [1 1 1] & K &227 \sgsymb{227} & 16c &16d, 96h & [1 1 1] & K \\ \hline 
227 \sgsymb{227} & 16d &16c, 96h & [1 1 1] & K &228 \sgsymb{228} & 48d &16a & [1 1 1] & K \\ \hline 
229 \sgsymb{229} & 6b &2a, 24h & [1 1 1] & K & 203 \sgsymb{203} & 16c & - & - & P \\ \hline
203 \sgsymb{203} & 16d & - & - & P & 210 \sgsymb{210} & 16c & - & - & P \\ \hline
210 \sgsymb{210} & 16d & - & - & P & 227 \sgsymb{227} & 16c & - & - & P \\ \hline
227 \sgsymb{227} & 16d & - & - & P & & & & & \\ \hline
% \end{tabular}
\end{longtable*}

\LTcapwidth=1.0\textwidth
\renewcommand\arraystretch{1.2}
\begin{longtable*}{|l|l|l|l||l|l|l|l|}
\caption[List of Wyckoff positions forming Lieb sublattice.]{
The Wyckoff positions (WP) of 3D space groups forming Lieb sublattice on special Miller plane.  For each case of each SG, the column 'WP(1a)' is the WP which forms the '1a' site of the Lieb lattice in Figure~\ref{fig:kagome_pyrochlore}(c).  The column 'WP(2c)' provides the list of WPs which form the 2c sites of the Lieb lattice in Figure~\ref{fig:kagome_pyrochlore}(c). The column 'WP(empty)' provides all the WPs that are defined on the same plane with the WPs in 'WP(1a)' and 'WP(2c)'. If one (or more than one) of the WP in 'WP(empty)' is occupied by any atoms, the Lieb lattice will be spoiled. For each case, the Miller indices are always [0 0 1], given in the basis of the lattice vector of the conventional unit-cell (using the standard convention of the \bcslong).}\label{tab:liebwyckoff}\\
\hline
SG & WP(1a) & WP(2c) & WP(empty) & SG & WP(1a) & WP(2c) & WP(empty)\\ \hline
83 \sgsymb{83} & 1c &2e &1a, 4j &83 \sgsymb{83} & 1a &2e &1c, 4j \\ \hline 
83 \sgsymb{83} & 1d &2f &1b, 4k &83 \sgsymb{83} & 1b &2f &1d, 4k \\ \hline 
84 \sgsymb{84} & 2d &2a, 2b &2c, 4j &84 \sgsymb{84} & 2c &2a, 2b &2d, 4j \\ \hline 
84 \sgsymb{84} & 2b &2c, 2d &2a, 4j &84 \sgsymb{84} & 2a &2c, 2d &2b, 4j \\ \hline 
85 \sgsymb{85} & 2a &4d & &85 \sgsymb{85} & 2b &4e & \\ \hline 
87 \sgsymb{87} & 2b &4c &2a, 8h &87 \sgsymb{87} & 2a &4c &2b, 8h \\ \hline 
87 \sgsymb{87} & 4d &8f & &89 \sgsymb{89} & 1c &2e &1a, 4j, 4l, 4o \\ \hline 
89 \sgsymb{89} & 1a &2e &1c, 4j, 4l, 4o &89 \sgsymb{89} & 1d &2f &1b, 4k, 4m, 4n \\ \hline 
89 \sgsymb{89} & 1b &2f &1d, 4k, 4m, 4n &93 \sgsymb{93} & 2d &2a, 2b &2c, 4j, 4k, 4l, 4m \\ \hline 
93 \sgsymb{93} & 2c &2a, 2b &2d, 4j, 4k, 4l, 4m &93 \sgsymb{93} & 2b &2c, 2d &2a, 4j, 4k, 4l, 4m \\ \hline 
93 \sgsymb{93} & 2a &2c, 2d &2b, 4j, 4k, 4l, 4m &97 \sgsymb{97} & 2b &4c &2a, 8g, 8h, 8i \\ \hline 
97 \sgsymb{97} & 2a &4c &2b, 8g, 8h, 8i &111 \sgsymb{111} & 1d &2e &1a, 4i, 4l \\ \hline 
111 \sgsymb{111} & 1a &2e &1d, 4i, 4l &111 \sgsymb{111} & 1c &2f &1b, 4j, 4k \\ \hline 
111 \sgsymb{111} & 1b &2f &1c, 4j, 4k &112 \sgsymb{112} & 2d &2a, 2c &2b, 4g, 4h, 4i, 4j \\ \hline 
112 \sgsymb{112} & 2c &2b, 2d &2a, 4g, 4h, 4i, 4j &112 \sgsymb{112} & 2b &2a, 2c &2d, 4g, 4h, 4i, 4j \\ \hline 
112 \sgsymb{112} & 2a &2b, 2d &2c, 4g, 4h, 4i, 4j &121 \sgsymb{121} & 2b &4c &2a, 8f, 8g \\ \hline 
121 \sgsymb{121} & 2a &4c &2b, 8f, 8g &123 \sgsymb{123} & 1c &2f &1a, 4j, 4l, 4n, 8p \\ \hline 
123 \sgsymb{123} & 1a &2f &1c, 4j, 4l, 4n, 8p &123 \sgsymb{123} & 1d &2e &1b, 4k, 4m, 4o, 8q \\ \hline 
123 \sgsymb{123} & 1b &2e &1d, 4k, 4m, 4o, 8q &124 \sgsymb{124} & 2d &4e &2b, 8m \\ \hline 
124 \sgsymb{124} & 2b &4e &2d, 8m &124 \sgsymb{124} & 2c &4f &2a, 8j, 8k, 8l \\ \hline 
124 \sgsymb{124} & 2a &4f &2c, 8j, 8k, 8l &125 \sgsymb{125} & 2c &4e &2a, 8i, 8k \\ \hline 
125 \sgsymb{125} & 2a &4e &2c, 8i, 8k &125 \sgsymb{125} & 2d &4f &2b, 8j, 8l \\ \hline 
125 \sgsymb{125} & 2b &4f &2d, 8j, 8l &126 \sgsymb{126} & 2b &4c &2a, 8h, 8i, 8j \\ \hline 
126 \sgsymb{126} & 2a &4c &2b, 8h, 8i, 8j &126 \sgsymb{126} & 4d &8f & \\ \hline 
128 \sgsymb{128} & 2b &4c &2a, 8h &128 \sgsymb{128} & 2a &4c &2b, 8h \\ \hline 
129 \sgsymb{129} & 2a &4d &8g &129 \sgsymb{129} & 2b &4e &8h \\ \hline 
130 \sgsymb{130} & 4b &8d & &131 \sgsymb{131} & 2d &2a, 2b &2c, 4j, 4k, 4l, 4m, 8q \\ \hline 
131 \sgsymb{131} & 2c &2a, 2b &2d, 4j, 4k, 4l, 4m, 8q &131 \sgsymb{131} & 2b &2c, 2d &2a, 4j, 4k, 4l, 4m, 8q \\ \hline 
131 \sgsymb{131} & 2a &2c, 2d &2b, 4j, 4k, 4l, 4m, 8q &132 \sgsymb{132} & 2c &4f &2a, 4i, 4j, 8n \\ \hline 
132 \sgsymb{132} & 2a &4f &2c, 4i, 4j, 8n &132 \sgsymb{132} & 2d &4e &2b, 8l, 8m \\ \hline 
132 \sgsymb{132} & 2b &4e &2d, 8l, 8m &133 \sgsymb{133} & 4b &8e &4a, 8h, 8i \\ \hline 
133 \sgsymb{133} & 4a &8e &4b, 8h, 8i &134 \sgsymb{134} & 2b &4c &2a, 8i, 8j \\ \hline 
134 \sgsymb{134} & 2a &4c &2b, 8i, 8j &134 \sgsymb{134} & 4d &4e, 4f &8k, 8l \\ \hline 
136 \sgsymb{136} & 2b &4c &2a, 4f, 4g, 8i &136 \sgsymb{136} & 2a &4c &2b, 4f, 4g, 8i \\ \hline 
138 \sgsymb{138} & 4a &4c, 4d &8g, 8h &139 \sgsymb{139} & 2b &4c &2a, 8h, 8i, 8j, 16l \\ \hline 
139 \sgsymb{139} & 2a &4c &2b, 8h, 8i, 8j, 16l &139 \sgsymb{139} & 4d &8f &16k \\ \hline 
140 \sgsymb{140} & 4b &8e &4a, 16i, 16j &140 \sgsymb{140} & 4a &8e &4b, 16i, 16j \\ \hline 
195 \sgsymb{195} & 3c &3d &1a, 6f, 6g, 6h &195 \sgsymb{195} & 3d &3c &1a, 6f, 6g, 6h \\ \hline 
195 \sgsymb{195} & 1a &3d &3c, 6f, 6g, 6h &195 \sgsymb{195} & 3c &3d &1b, 6g, 6h, 6i \\ \hline 
195 \sgsymb{195} & 3d &3c &1b, 6g, 6h, 6i &195 \sgsymb{195} & 1b &3c &3d, 6g, 6h, 6i \\ \hline 
197 \sgsymb{197} & 6b &6b &2a, 12d, 12e &200 \sgsymb{200} & 3c &3d &1a, 6e, 6f, 6g, 12j \\ \hline 
200 \sgsymb{200} & 3d &3c &1a, 6e, 6f, 6g, 12j &200 \sgsymb{200} & 1a &3d &3c, 6e, 6f, 6g, 12j \\ \hline 
200 \sgsymb{200} & 3c &3d &1b, 6f, 6g, 6h, 12k &200 \sgsymb{200} & 3d &3c &1b, 6f, 6g, 6h, 12k \\ \hline 
200 \sgsymb{200} & 1b &3c &3d, 6f, 6g, 6h, 12k &201 \sgsymb{201} & 6d &6d &2a, 12f, 12g \\ \hline 
202 \sgsymb{202} & 4b &24d &4a, 24e, 48h &202 \sgsymb{202} & 4a &24d &4b, 24e, 48h \\ \hline 
202 \sgsymb{202} & 8c &24d &48g &204 \sgsymb{204} & 6b &6b &2a, 12d, 12e, 24g \\ \hline 
207 \sgsymb{207} & 3c &3d &1a, 6e, 12h, 12i &207 \sgsymb{207} & 3d &3c &1a, 6e, 12h, 12i \\ \hline 
207 \sgsymb{207} & 1a &3d &3c, 6e, 12h, 12i &207 \sgsymb{207} & 3c &3d &1b, 6f, 12h, 12j \\ \hline 
207 \sgsymb{207} & 3d &3c &1b, 6f, 12h, 12j &207 \sgsymb{207} & 1b &3c &3d, 6f, 12h, 12j \\ \hline 
208 \sgsymb{208} & 6d &6d &2a, 6e, 6f, 12h, 12i, 12j &208 \sgsymb{208} & 6e &4b, 4c &12k, 12l \\ \hline 
209 \sgsymb{209} & 4b &24d &4a, 24e, 48g, 48h &209 \sgsymb{209} & 4a &24d &4b, 24e, 48g, 48h \\ \hline 
209 \sgsymb{209} & 8c &24d &48i &211 \sgsymb{211} & 6b &6b &2a, 12d, 12e, 24g, 24h \\ \hline 
211 \sgsymb{211} & 12d &8c &24i &215 \sgsymb{215} & 3c &3d &1a, 6f, 12h \\ \hline 
215 \sgsymb{215} & 3d &3c &1a, 6f, 12h &215 \sgsymb{215} & 1a &3d &3c, 6f, 12h \\ \hline 
215 \sgsymb{215} & 3c &3d &1b, 6g, 12h &215 \sgsymb{215} & 3d &3c &1b, 6g, 12h \\ \hline 
215 \sgsymb{215} & 1b &3c &3d, 6g, 12h &217 \sgsymb{217} & 6b &6b &2a, 12d, 12e, 24f \\ \hline 
218 \sgsymb{218} & 6b &6b &2a, 6c, 6d, 12f, 12g, 12h &219 \sgsymb{219} & 24c &24d &8a, 48f \\ \hline 
219 \sgsymb{219} & 24d &24c &8a, 48f &219 \sgsymb{219} & 8a &24d &24c, 48f \\ \hline 
219 \sgsymb{219} & 24c &24d &8b, 48g &219 \sgsymb{219} & 24d &24c &8b, 48g \\ \hline 
219 \sgsymb{219} & 8b &24c &24d, 48g &221 \sgsymb{221} & 3c &3d &1a, 6e, 12h, 12i, 24k \\ \hline 
221 \sgsymb{221} & 3d &3c &1a, 6e, 12h, 12i, 24k &221 \sgsymb{221} & 1a &3d &3c, 6e, 12h, 12i, 24k \\ \hline 
221 \sgsymb{221} & 3c &3d &1b, 6f, 12h, 12j, 24l &221 \sgsymb{221} & 3d &3c &1b, 6f, 12h, 12j, 24l \\ \hline 
221 \sgsymb{221} & 1b &3c &3d, 6f, 12h, 12j, 24l &222 \sgsymb{222} & 6b &6b &2a, 12d, 12e, 24g, 24h \\ \hline 
222 \sgsymb{222} & 12d &8c & &223 \sgsymb{223} & 6b &6b &2a, 6c, 6d, 12f, 12g, 12h, 24k \\ \hline 
223 \sgsymb{223} & 6c &6d, 8e &24j &224 \sgsymb{224} & 6d &6d &2a, 12f, 12g, 24h \\ \hline 
224 \sgsymb{224} & 12f &4b, 4c &24i, 24j &225 \sgsymb{225} & 4b &24d &4a, 24e, 48h, 48i, 96j \\ \hline 
225 \sgsymb{225} & 4a &24d &4b, 24e, 48h, 48i, 96j &225 \sgsymb{225} & 8c &24d &48g \\ \hline 
226 \sgsymb{226} & 24c &24d &8b, 48e, 96i &226 \sgsymb{226} & 24d &24c &8b, 48e, 96i \\ \hline 
226 \sgsymb{226} & 8b &24c &24d, 48e, 96i &226 \sgsymb{226} & 24c &24d &8a, 48f, 96h \\ \hline 
226 \sgsymb{226} & 24d &24c &8a, 48f, 96h &226 \sgsymb{226} & 8a &24d &24c, 48f, 96h \\ \hline 
228 \sgsymb{228} & 48d &48d &16a, 96f &229 \sgsymb{229} & 6b &6b &2a, 12d, 12e, 24g, 24h, 48j \\ \hline 
229 \sgsymb{229} & 12d &8c &48i & & & & \\ \hline
\end{longtable*}

\subsection{Geometrical features of standard Kagome, pyrochlore and Lieb lattices}\label{app:geometry}

Using the above symmetry analysis, it is straightforward to systematically filter out the crystalline materials exhibiting \emph{rigorous} Kagome, pyrochlore or Lieb sublattices by checking the occupied Wyckoff positions. However, in many of the 3D crystalline materials, there are no rigorous line-graph or Lieb sublattices, only approximate ones. These \emph{approximate} Kagome, pyrochlore or Lieb sublattices are not part of Tables~\ref{tab:kagomepyrochlorewyckoff} or~\ref{tab:liebwyckoff}. Still, they can be obtained from these rigorous sublattices by weak lattice distortion. When this distortion is weak enough, the lattice might still hosts (almost) flat bands. As the \emph{approximate} Kagome, pyrochlore and Lieb sublattices do not satisfy the minimal symmetry groups of the rigorous ones, the only efficient method to identify them is comparing their geometrical structures with those of the rigorous sublattices. In this subsection, we will analyze the key geometrical features of the rigorous Kagome, pyrochlore and Lieb lattices, which are schematically shown in Fig.~\ref{fig:kagome_pyrochlore}. In the next subsection, we will rely on these features to design numerical algorithms for searching approximate line-graph and Lieb sublattices in real crystalline materials.

\textbf{Kagome Lattice} The rigorous Kagome lattice in 2D, as shown in Fig.~\ref{fig:kagome_pyrochlore}(a), is made  of three equivalent sites corresponding to the Wyckoff position $3c$ in one unit cell with wallpaper group $p6mm$. For convenience, we label the five sites of $3c$ as 1, 2, 3, 4 and 5, where the sites 4 and 5 differ from sites 3 and 2, respectively, by a lattice translation. The 2D point group of $3c$ is $2mm$, which includes a vertical two-fold rotational symmetry and two vertical mirror symmetries. So the two regular triangles formed by sites $\{1, 2, 3\}$ and sites $\{1, 4, 5\}$, respectively, are related by the two-fold rotational symmetry.

Thus, there are two key geometrical features for the rigorous Kagome lattice in Fig.~\ref{fig:kagome_pyrochlore}(a):
\begin{enumerate}
\item For each site on the 2D lattice, it has only four co-planar nearest neighbor (NN) sites on the same plane.
\item Each site and its four NN sites form two equilateral triangles, which are related by a vertical $C_2$ symmetry.
\end{enumerate}
If a 2D sublattice occupied by the same chemical element of a given material can be characterized by the above two geometrical features within a certain threshold (as will be discussed in the next subsection), then we label it as an approximate Kagome sublattice.

\textbf{Pyrochlore Lattice} As shown in Fig.~\ref{fig:kagome_pyrochlore}(b), the rigorous pyrochlore lattice is a 3D line-graph lattice. Its geometry has the two following features:
\begin{enumerate}
    \item The unit cell of the 3D lattice is constructed by five regular tetrahedrons, each two of which share one vertex. For example, in Fig.~\ref{fig:kagome_pyrochlore}(b), the regular tetrahedron $T_0$ formed by the vertices $\{v_1, v_2, v_3, v_4\}$ and another regular tetrahedron $T_1$ formed by the vertices $\{v_1, v_5, v_6, v_7\}$ share the vertex $v_1$.
    \item Each of the four triangle surfaces in one tetrahedron is always co-planar with one triangle surface of its NN tetrahedron and the two co-planar triangles form a Kagome sublattice. For example, the triangle formed by the vertices $\{v_1, v_2, v_3\}$ in $T_0$ is co-planar with the triangle formed by the vertices $\{v_1, v_5, v_6\}$ in $T_1$ and these vertices form a Kagome sublattice. There are four equivalent Kagome sublattices which are related by the point group symmetry of the pyrochlore lattice $(\bar3m)$.
\end{enumerate}
Thus, for a 3D sublattice which is formed by the same chemical element of a given material, if it can be characterized by the above two geometrical features within a certain threshold, we consider it as an approximate pyrochlore sublattice.

\textbf{Lieb lattice}
The rigorous Lieb lattice is a 2D lattice formed by two non-equivalent atom sites, the blue and red sites as indicated in Fig.~\ref{fig:kagome_pyrochlore}(c). The blue sites form a square lattice and the red sites are on the middle of the bonds between two adjacent blue sites.

Lieb lattice has the following three key geometrical features:
\begin{enumerate}
\item It has two non-equivalent atom sites, $\{A_1\}$ and $\{A_2\}$, which are co-planar. Each atom in $\{A_1\}$ always has two NN sites, which belong to $\{A_2\}$. Each atom site in $\{A_2\}$ always has four NN sites, which belong to $\{A_1\}$. 
\item Each of the atom sites in $\{A_1\}$ is in the middle of its two NN sites. 
\item The four bonds which connect the atom in $\{A_2\}$ and its four NN sites are related by a four-fold rotational symmetry.
\end{enumerate}
A 2D sublattice occupied by one or two different chemical elements of a given material and satisfying the two above geometrical features within a certain threshold, is labelled as an approximate Lieb sublattice.

\subsection{Algorithms to search for Kagome, pyrochlore and Lieb sublattices}\label{app:algorithms}

The symmetry properties and the key geometrical features of rigorous Kagome, pyrochlore and Lieb lattices enable the high-throughput search for both rigorous and approximate line-graph sublattices in real materials. Using the key geometrical features as discussed in Appendix~\ref{app:geometry}, we have designed numerical algorithms to search for Kagome, pyrochlore and Lieb sublattices from the crystal lattice of real materials. Whether or not the line-graph and Lieb sublattices also satisfy the symmetry analysis of Appendix~\ref{app:crys_symm}, we classify them into rigorous or approximate.

\subsubsection{Kagome and pyrochlore sublattices} 

There are four steps to identify a Kagome or pyrochlore sublattice from the crystal lattice of a material.
\begin{enumerate}
    \item[Step 1] For each material, its crystal structure is parsed from the \emph{POSCAR} file, which was used in the first-principle calculations\cite{Vergniory2021}. The crystal symmetry and the list of occupied Wyckoff letters of the non-equivalent atom sites are analyzed by the $Pymatgen$ package.
    \item[Step 2] By grouping together the atom sites of the same chemical element in one unit cell, the crystal lattice is divided into several sets. For convenience, a set will be denoted by $\{A_i\}$ where $A_i$ denotes a single atom. The two following steps are applied to each set until we exhaust all possible sublattices.
    \item[Step 3] For each atom site in a certain set $\{A_i\}$, we check if it satisfies the following four criteria sequentially:
    \begin{enumerate}
        \item[C1] The atom site has and only has four co-planar nearest neighbor (NN) or next nearest neighbor (NNN) sites, which also belong to $\{A_i\}$, \emph{on certain Miller planes}. If true, we refer this atom site as a representative site and obtain the corresponding Miller indices $[hkl]$.
        \item[C2] By calculating the length of the bonds between the representative atom site and its four NN/NNN sites and the angles between the four bonds, we check whether the five sites form two regular triangles related by a vertical $C_2$ symmetry, which is centered at the representative atom site.
        \item[C3] Each of the four NN/NNN atom sites also satisfies the first two criteria.
        \item[C4] Each atom site in the set $\{A_i\}$ satisfies the criteria C1,C2 and C3 on three different Miller planes. 
    \end{enumerate}
    \item[Step 4] If a set (or part of a set) satisfies the criteria C1, C2 and C3, it is classified as a Kagome sublattice. If it moreover satisfies C4, it is also tagged as a pyrochlore sublattice.
\end{enumerate}

\begin{figure}[h]
\includegraphics[width=0.95\columnwidth]{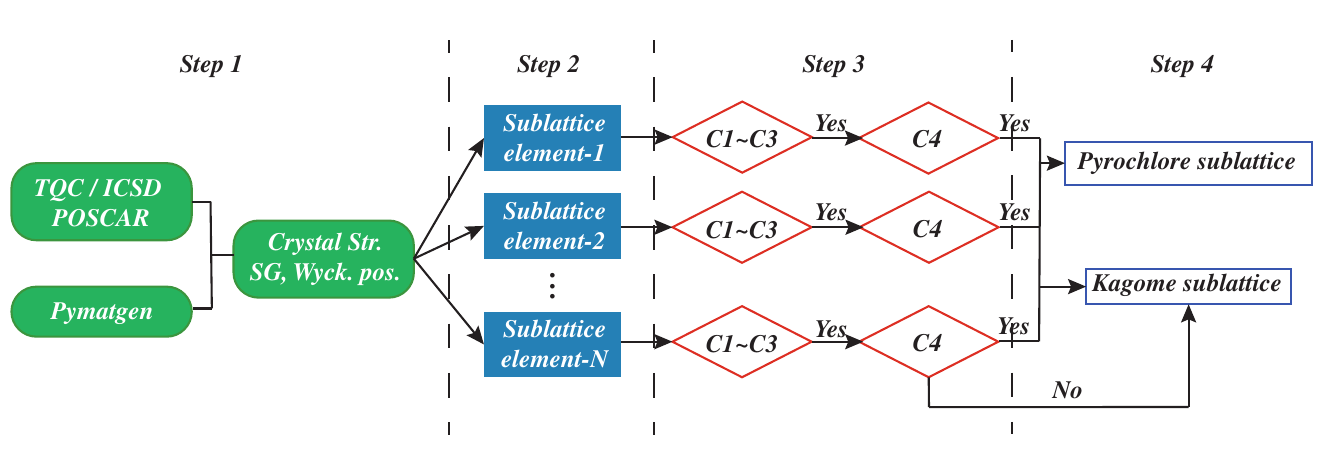}
\caption{The flow chart of the Kagome or pyrochlore sublattice identification from a 3D crystal lattice. Starting from the crystal structure of material provided as a \emph{POSCAR} file, the crystal symmetry and the Wyckoff positions of each distinct element are analyzed by the \emph{Pymatgen} package. For each sublattice, the criteria C1, C2 and C3 are applied to identify if the sublattice is a Kagome plane. If at this stage, a 2D Kagome sublattice has been identified, the criterion C4 is applied to test if the crystal structure also hosts a 3D pyrochlore sublattice.
\label{fig:workflowkagome}}
\end{figure}

The flow chart of this algorithm is shown in Fig.~\ref{fig:workflowkagome}. Based on this algorithm, we have developed a $Python$-based code to identify if the crystal lattice of a material has any Kagome or pyrochlore sublattice. In this code, we rely on several functions of the $Pymatgen$ package \cite{kirklin2015open}. The function $Pymatgen.core.structure.get\_neighbors$ is used to get the neighboring sites of an atom and the function $pymatgen.core.structure.get\_angle$ is used to calculate the angle between two bonds. The Miller indices are obtained using the function $pymatgen.core.lattice.get\_miller\_index\_from\_coords$. Our algorithm requires a set of threshold parameters to distinguish a regular triangle. More precisely, the threshold is set to $\pm\FlatBandKagomeDistanceThreshold$ of the NN/NNN bonds' length when identifying two equivalent bonds. A threshold of $\pm\FlatBandKagomeAngleThreshold^{\circ}$ is applied to discriminate two identical angles.

\subsubsection{Lieb sublattice} 

We apply a five-step process to identify a Lieb sublattice from a 3D crystal lattice of material.
\begin{enumerate}
    \item[Step 1]For each material, its crystal structure is parsed from the \emph{POSCAR} file, which was used in the first-principle calculations\cite{Vergniory2021}. The crystal symmetry and the list of Wyckoff letters of the non-equivalent atom sites are analyzed by the $Pymatgen$ package.
    \item[Step 2] We obtain all the NN sites for each atom site in the lattice.
    \item[Step 3] We select all the atom sites that have  only two NN sites of identical element and are at the equal distance of its NN sites. Such atom sites are referred as 'middle~sites' and labeled as $A_1$. The NN sites of $A_1$ are labeled as $A_2$.
    \item[Step 4] For each of the 'middle~sites' obtained in Step 3, we check sequentially the following four criteria:
    \begin{enumerate}
        \item[C1] For a given 'middle~site'  $A_1$, each of its two NN sites $A_2$ has and only has four NN sites, which are \emph{co-planar on certain Miller planes} and are also labeled as $A_1$ in Step 3.
        \item[C2] The four NNs of $A_2$ are of identical element. And the four bonds connecting $A_2$ and its four NN sites are of equal length and the angles between any two neighbor bonds equals to $90^\circ$.
        \item[C3] Both of the two NN sites $A_2$ of a 'middle~site' $A_1$ satisfy the criterion C2 on the same plane.
        \item[C4] All the 'middle~sites' on the same plane satisfy simultaneously criteria C1, C2 and C3.
    \end{enumerate}
    \item[Step 5] If there is a 2D sublattice, of which the atom sites satisfy the criteria C1 to C4 in step 4, it is classified as a Lieb lattice and the Miller indices of the Lieb sublattice are recorded.
\end{enumerate}

\begin{figure}[h]
\includegraphics[width=0.95\columnwidth]{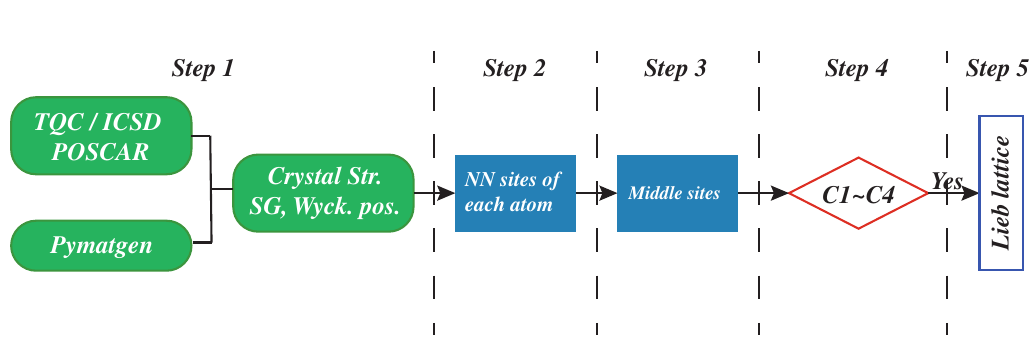}
    \caption{The flow chart of the Lieb sublattice identification from a 3D crystal lattice. Starting from the crystal structure of materials provided as a \emph{POSCAR} file, the crystal symmetry and the Wyckoff positions of each distinct element are analyzed by the \emph{Pymatgen} package. 
    The second step identifies the nearest neighbor(NN) sites of each atom. If an atom is at the middle of its two NN sites, it is referred to as the 'middle-site' (Step 3). Then at Step 4, the criteria from C1 to C4 are applied to check if the $middle~sites$ and their NN sites form a Lieb sublattice.
\label{fig:workflowlieb}}
\end{figure}

The algorithm flow chart is provided in Fig.~\ref{fig:workflowlieb}. We have implemented this algorithm in a $Python$-based code. It detects if the crystal lattice of a material has a Lieb sublattice. Our code uses a set of threshold parameters to identify whether two bonds are equal or an angle matches the expected value: a threshold is set to $\pm\FlatBandLiebDistanceThreshold$ of the NN bonds' length to identify two equivalent bonds and a threshold of $\pm\FlatBandLiebAngleThreshold^{\circ}$ is applied to discriminate two identical angles. In real crystalline materials in the \webflatband, Lieb sublattices  tend to be spoiled by lattice distortion. In this section, a larger threshold is used to identify an approximate Lieb lattice.

\subsubsection{Rigorous and approximate sublattices}

\begin{figure}[h]
\begin{tabular}{c c}
(a) \scriptsize{$\rm{Co}_{3} \rm{Sn}_{2} \rm{S}_{2}$ - \icsdweb{5435} - SG 166 ($R\bar{3}m$) - ESFD} & (b) \scriptsize{${\rm Pt}_2 {\rm Hg} {\rm Se}_3$ - \icsdweb{185808} - SG 164 (\sgsymb{164}) - SEBR} \\
\includegraphics[width=0.334\textwidth,angle=0]{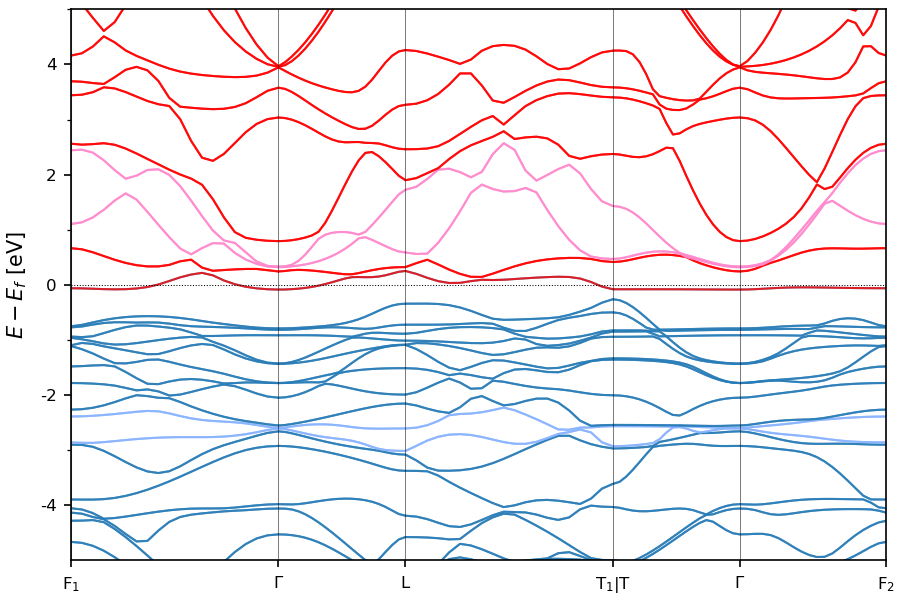}\includegraphics[width=0.14\textwidth,angle=0]{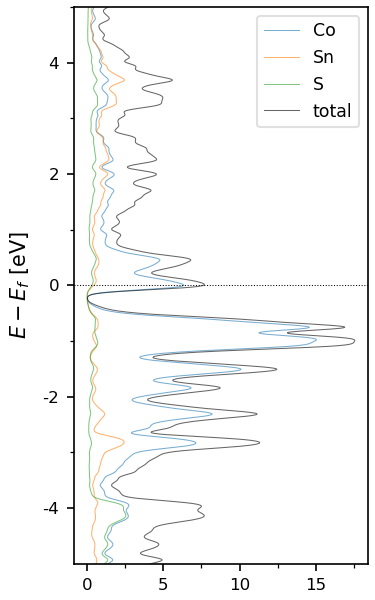} & \includegraphics[width=0.334\textwidth,angle=0]{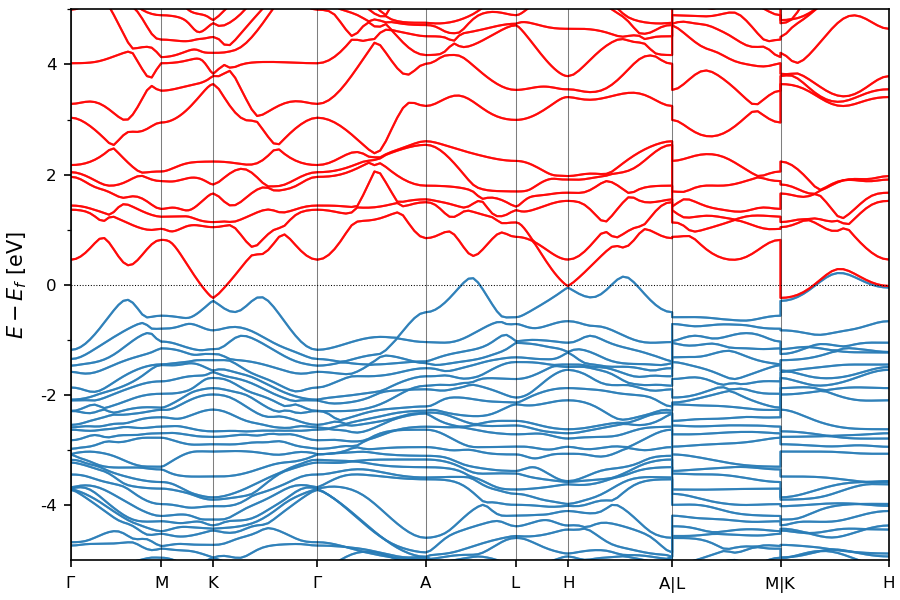}\includegraphics[width=0.14\textwidth,angle=0]{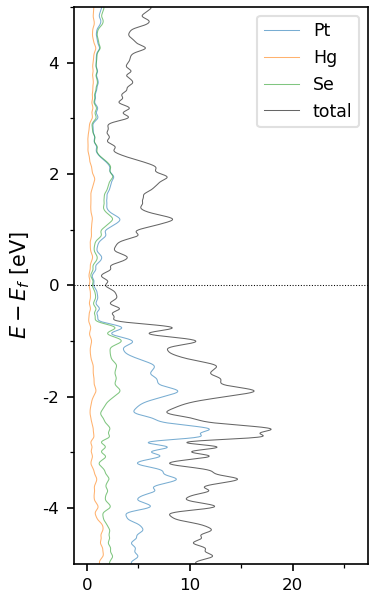} \\ 
\end{tabular}
    \caption{(a) Band structure of ${\rm Co}_3 {\rm Sn}_2 {\rm S}_2$ [\icsdweb{5435}, SG 166 (\sgsymb{166})]. (b) Band structure of ${\rm Pt}_2 {\rm Hg} {\rm Se}_3$ [\icsdweb{185808}, SG 164 (\sgsymb{164})].
\label{app:fig:spoiledkagome}}
\end{figure}

As discussed in Appendix~\ref{app:crys_symm}, rigorous sublattices can be identified from the space group and the corresponding occupied Wyckoff positions in the Table~\ref{tab:kagomepyrochlorewyckoff} or~\ref{tab:liebwyckoff}. Although these two features are analyzed for each crystal structure in the database using $Pymatgen$ package\cite{ONG2013314, togo2018textttspglib}, they are not enough to confirm that a given material hosts a rigorous sublattice. Indeed, the sublattice might be spoiled by the presence of other in-plane atoms of the \emph{same element} on (or close to) this sublattice. For example, in SG 147 (\sgsymb{147}), the Kagome sublattice formed by the Wyckoff position $3e$ with atom A occupied will be spoiled by the Wyckoff position $1a$ if $1a$ is also occupied by the identical atom. Note that the site of $1a$ is exactly on the Kagome plane formed by $3e$. The two algorithms described in Appendix~\ref{app:algorithms}(a), were designed to detect such situations. Indeed in the geometric method algorithm described above, such cases are discarded when applying criterion C1. Note that if the sublattice has as a nearest neighbor other atoms of \emph{different type}, the flat bands emerging from the sublattice are usually preserved. For example, ${\rm Co}_3 {\rm Sn}_2 {\rm S}_2$ [\icsdweb{5435}, SG 166 (\sgsymb{166})], the ${\rm Sn}$ atoms sit at the center of the ${\rm Co}$ Kagome sublattice hexagons. Still flat bands are observed in the band structure (shown in Fig.~\ref{app:fig:spoiledkagome}a). Conversely, ${\rm Pt}_2 {\rm Hg} {\rm Se}_3$ [\icsdweb{185808}, SG 164 (\sgsymb{164})] has a Kagome lattice made of ${\rm Pt}$ atoms at Wyckoff positions $3e$. But the ${\rm Pt}$ atoms at Wyckoff positions $1a$ spoils the Kagome sublattice and the band structure (shown in Fig.~\ref{app:fig:spoiledkagome}b) does not exhibit any flat band. 
 Rather than discarding cases where atoms of a different type are nearest neighbors, we have added on the \webflatband\ for each sublattice description, the location and type of any additional nearest neighbor atoms, (see for example the Kagome sublattice description for \flatwebdirectlink{5435}{${\rm Co}_3 {\rm Sn}_2 {\rm S}_2$}). 

For that reason, a material is tagged as having a \emph{rigorous} Kagome, pyrochlore or Lieb sublattice if, within its space group, its occupied Wyckoff positions belong to the corresponding set of Wyckoff positions in the Table~\ref{tab:kagomepyrochlorewyckoff} or \ref{tab:liebwyckoff} \emph{and} if the material successfully passes the corresponding test set by our algorithms. If a material does not satisfy the symmetry features of Appendix~\ref{app:crys_symm} but still successfully passes the corresponding test set by our algorithms, then the material is tagged as having an \emph{approximate} Kagome, pyrochlore or Lieb sublattice.

Among the \TQCDBNbrNoSOCICSDsIncludingValidFElectrons~ISCD entries that we have considered, \FlatBandNbrICSDsRigorousKagome~ICSDs (\FlatBandNbrICSDsRigorousKagomePercent) host at least one rigorous Kagome sublattice, while \FlatBandNbrICSDsApproximateKagome~ICSDs (\FlatBandNbrICSDsApproximateKagomePercent) host at least one approximate Kagome sublattice. Similarly, we found \FlatBandNbrICSDsRigorousPyrochlore~ICSDs (\FlatBandNbrICSDsRigorousPyrochlorePercent) with at least one rigorous pyrochlore sublattice,  \FlatBandNbrICSDsApproximatePyrochlore~ICSDs (\FlatBandNbrICSDsApproximatePyrochlorePercent) with at least one approximate pyrochlore sublattice, \FlatBandNbrICSDsRigorousLieb~ICSDs (\FlatBandNbrICSDsRigorousLiebPercent) with at least one rigorous Lieb sublattice,  \FlatBandNbrICSDsApproximateLieb~ICSDs (\FlatBandNbrICSDsApproximateLiebPercent) with at least one approximate Lieb sublattice. Note that an entry might have both a rigorous and an approximate sublattices.

\section{Automated search for materials with bipartite or split sublattices}\label{app:bipartite}

Beyond the line-graph lattices, \ie Kagome and pyrochlore lattices, and the Lieb lattice that were introduced in Appendix~\ref{app:kagomepyrochlore}, split and bipartite lattices could also host exactly flat bands~\cite{S-matrix}. Recently, these two types of lattices were shown to host topologically non-trivial bands\cite{ma2020}, which indicates that the the Wannier states cannot be localized.

A bipartite lattice is a lattice such that all its sites can be divided into two sublattices, $L$ and $\tilde{L}$. Hopping bonds \emph{only} exist in between sites in $L$ and sites in $\tilde{L}$, but do not exist within sites of the same sublattice.
Considering the translational symmetry of crystalline materials in 3D, the graph formed by the sites in $L$ and $\tilde{L}$ and the bonds connecting them can be either mathematically connected in \emph{at~least} one dimension or disconnected in all the dimensions. If the graph of a bipartite sublattice is disconnected, we refer to it as a \emph{disconnected bipartite sublattice}. For example, the three green atoms in Figure~\ref{fig:RemoveMolecularClusters}(a) is typically a group of isolated atoms in the crystal lattice and hence they form a disconnected bipartite sublattice. 
Otherwise, when the graph of a bipartite sublattice is connected in \emph{at~least} one dimension, we refer it to as the \emph{connected bipartite sublattice}. For example, a collection of the one dimensional bipartite chains is considered as a connected bipartite sublattice. 
In this work, we mainly focus on the bipartite lattice with connected graph and refer the bipartite lattice with disconnected graph as a molecular lattice. A molecular lattice also hosts flat bands but they are in the atomic limit and is thus disregarded as discussed in Appendix~\ref{app:atomicflatband}. A split lattice is a special case of bipartite lattice: each site $i$ in sublattice $\tilde{L}$ has the same hopping term to its two (and exactly two) connected sites in sublattice $L$.
For example, the Lieb lattice shown in Fig.~\ref{fig:kagome_pyrochlore}(c) is a two dimensional case of both a bipartite lattice and a split lattice. The red sites correspond to the sublattice $\tilde{L}$ and the sublattice ${L}$ is made of the blue sites. In the Lieb lattice,  bonds only connect the blue sites in $\tilde{L}$ to the red sites in ${L}$, and each site in $L$ is at the center of its two connected neighbors.

 The flat bands that exist in bipartite or split lattice can be explained with the $S$-matrix method \cite{S-matrix}. In this appendix, we provide a brief introduction to the $S$-matrix method that we will use in Appendix~\ref{app:theoryexplanation}. More details about this method can be found in Ref.\cite{S-matrix}. We then detail a numerical algorithm based on geometrical features that allows for a systematic search for crystal structures with bipartite or split sublattices among the stoichiometric materials in \webTQC.

\subsection{$S$-matrix method} \label{app:smatrix}

Given a bipartite lattice with its two sublattices labeled as $L$ and $\tilde L$, we assume the number of orbitals per unit cell in $L$ (denoted $|L|$) is always larger than that in $\tilde{L}$ (denoted $|\tilde{L}|$) , \ie $|L|\textgreater|\tilde{L}|$. Since hopping terms only occur between sites of $L$ and $\tilde{L}$, the tight binding Hamiltonian can be expressed in momentum space using the $S$-matrix as

\begin{eqnarray}
H(\boldsymbol{k}) & = & \left(\begin{array}{cc}
0 & S^{\dagger}(\boldsymbol{k})\\
S(\boldsymbol{k}) & 0
\end{array}\right),\label{app:eq:HamiltonianBipartiteFromS}
\end{eqnarray}
where $S(\boldsymbol{k})$ is a matrix with dimension $|L|\times |\tilde{L}|$ and $\boldsymbol{k}$ is the momentum in the Brillouin zone.
Thus, the square of the eigenvalues of $H(\boldsymbol{k})$ are given by
the eigenvalues of $\left(H(\boldsymbol{k})\right)^{2}=\left(\begin{array}{cc}
\tilde{T}(\boldsymbol{k}) & 0\\
0 & T(\boldsymbol{k})
\end{array}\right)$, where $\tilde{T}(\boldsymbol{k})=S^\dagger(\boldsymbol{k}) S(\boldsymbol{k})$, $T(\boldsymbol{k}) = S(\boldsymbol{k}) S^\dagger(\boldsymbol{k})$. Both the matrices $T(\boldsymbol{k})$ and $\tilde{T}(\boldsymbol{k})$ are positive semidefinite. 
As shown in the example in the end of this subsection, $T$ and $\tilde T$ serve as the effective Hamiltonians of $L$ and $\tilde{L}$ (after a rescale of energy), respectively, if $L$ and $\tilde{L}$ have different on-site energies. 

We first prove that $T(\boldsymbol{k})$ and $\tilde{T}(\boldsymbol{k})$ have the same positive eigenvalues with identical multiplicities. For any eigenstate of $\tilde{T}(\boldsymbol{k})$, $\tilde{\phi}=\tilde{\phi}(\boldsymbol{k})$, that satisfies  $\tilde{T}(\boldsymbol{k})\tilde{\phi} = E\tilde{\phi}$ and $E>0$, we can construct $\phi=S(\boldsymbol{k}) \tilde{\phi}$. Then we have $T(\boldsymbol{k}) \phi = S(\boldsymbol{k}) S^\dagger(\boldsymbol{k}) S(\boldsymbol{k}) \tilde{\phi} = S(\boldsymbol{k}) E\tilde{\phi}= E S(\boldsymbol{k})\tilde{\phi}= E\phi$, meaning  $\phi$ is an eigenstate of $T(\boldsymbol{k})$ with the same eigenvalue $E$. Considering two eigenstates of $\tilde{T}$ satisfying $\tilde{T}(\boldsymbol{k})\tilde{\phi_{i}} = E_i \tilde{\phi_{i}}$ and $\tilde{T}(\boldsymbol{k})\tilde{\phi_{j}} = E_j \tilde{\phi_{j}}$, we have $< \tilde{\phi_{i}}|\tilde{\phi_{j}}> = <\phi_{i}| S^\dagger(\boldsymbol{k}) S(\boldsymbol{k}) |\phi_{j}>= <\phi_{i}|\tilde{T}(\boldsymbol{k})|\phi_{j}>= E_j <\phi_{i}|\phi_{j}> 	\propto \delta_{i,j} $. Thus $T(\boldsymbol{k})$ and $\tilde{T}(\boldsymbol{k})$ have identical positive eigenvalues, including their multiplicities. From this property, we deduce that $T(\boldsymbol{k})$ and $\tilde{T}(\boldsymbol{k})$ lead to the same dispersive bands in the spectrum of $H(\boldsymbol{k})$. Moreover it results in a set of perfectly flat
bands at $E=0$ with degeneracy $D=\left|L\right|+\left|\tilde{L}\right|-2\times{\rm rank}\left(S(\boldsymbol{k})\right)$. (As a special case, if $S(\boldsymbol{k}) S^\dagger(\boldsymbol{k})$ has full rank, there are $\left|L\right|-\left|\tilde{L}\right|$ flat bands in the spectrum of both $H(\boldsymbol{k})$ and $T(\boldsymbol{k})$.) 
%Denoting $|L|$ and $|\tilde{L}|$ the number of atoms or orbitals in the $L$ and $\tilde{L}$ further sublattices of the bipartite or split sublattice of a material (assuming $|L| \geq |\tilde{L}|$), the Bloch Hamiltonian associated to a bipartite sublattice of a material reads
%\begin{eqnarray}
%H(\boldsymbol{k}) & = & \left(\begin{array}{cc}
%0 & S^{\dagger}(\boldsymbol{k})\\
%S(\boldsymbol{k}) & 0
%\end{array}\right).\label{eq:HamiltonianBipartiteFromS}
%\end{eqnarray}
%where $S(\boldsymbol{k})$ is a matrix with dimension $|L|\times |\tilde{L}|$ and $\boldsymbol{k}$ is the momentum in the Brillouin zone. A bipartite lattice with $|L| \neq |\tilde{L}|$ hosts flat bands in its band structure, because the Hamiltonian of Eq.~\ref{eq:HamiltonianBipartiteFromS} has at least $|L|+|\tilde{L}|-2\times rank(S(\boldsymbol{k}))$ zero-energy states, \ie %the band structure has at least 
%$|L|+|\tilde{L}|-2\times rank(S(\boldsymbol{k}))$ exact flat bands. It is also very likely that these bands exhibit topology \cite{S-matrix}. 

\begin{figure}[h]
\includegraphics[width=8.6cm]{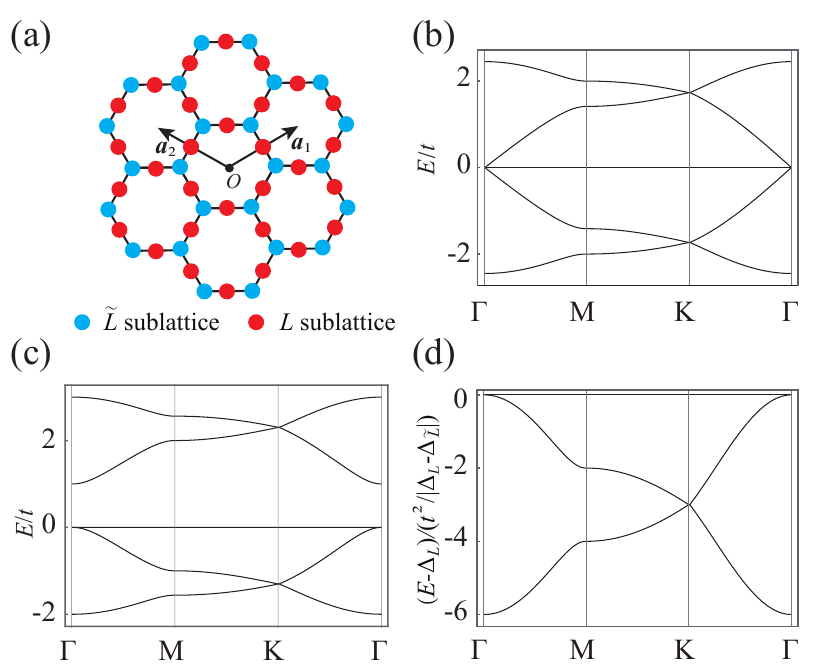}
\caption{(a) An example of bipartite lattice consists of honeycomb lattice ($\tilde{L}$ sublattice in blue) and Kagome lattice ($L$ sublattice in red). (b) Band structure of Hamiltonian $H_{L\oplus\tilde{L}}$ in Eq.~(\ref{app:eq:SplitHoneycombHAall}) without any on-site energy, \ie $\Delta_L=\Delta_{\tilde{L}}=0$. (c) Band structure of Hamiltonian $H_{L\oplus\tilde{L}}$ in Eq.~(\ref{app:eq:SplitHoneycombHAall}) and $\Delta_{L}=0$,  $\Delta_{\tilde{L}}=|t|$. (d) The band structure of the effective Hamiltonian $H_{L}$ in Eq.~(\ref{eq:hamilL}) and $\Delta_{L}=0$,  $\Delta_{\tilde{L}}=|t|$.
}
\label{fig:bipartiteexample}
\end{figure}

As an example, we consider a two dimensional bipartite lattice formed by $s$ orbitals at the Kagome sublattice and the honeycomb sublattice, as shown in Fig.~\ref{fig:bipartiteexample}(a).
We refer to the honeycomb sublattice as $\tilde{L}$ and the Kagome sublattice as $L$. 
Only hopping terms between nearest neighbors are taken into account and all of them are assumed to have the same amplitude. 
The lattice vectors $\boldsymbol{a}_{i}(i=1,2)$ are shown in Fig.~\ref{fig:bipartiteexample}(a). In this basis and focusing on the unit cell, the atoms in $\tilde L$ are located at $(\frac{1}{3},\frac{2}{3})$, $(\frac{2}{3},\frac{1}{3})$ and the atoms in ${L}$ are at $(\frac{1}{2},0)$, $(\frac{1}{2},\frac{1}{2})$, $(0,\frac{1}{2})$. We can write the explicit form of the $S$ matrix as
\begin{eqnarray}
S^{\dagger}(\boldsymbol{k}) & = & t\left(\begin{array}{ccc}
1 & 1 & e^{ik_{1}}\\
e^{ik_{2}} & 1 & 1
\end{array}\right),\label{eq:smatrix}
\end{eqnarray}
with $t$ being the strength of the nearest neighbor coupling. In the absence of on-site energy, this model is just the split graph lattice of the honeycomb lattice. The Hamiltonian can then be built using Eq.~(\ref{app:eq:HamiltonianBipartiteFromS}) and the corresponding band structure is shown in Fig.~\ref{fig:bipartiteexample}(b).

As shown in, \eg  Ref.~\cite{S-matrix}, flat bands in line-graph lattices might also be understood within the $S$-matrix formalism. For pedagogical purposes, we solely focus on the Kagome lattice which is the line-graph lattice of the honeycomb lattice. Starting from the previous example, we further add the on-site energies  $\Delta_{\tilde{L}}$ and $\Delta_{L}$ on the sublattices $\tilde{L}$ and $L$. 
Then the total Hamiltonian of the bipartite lattice $L\oplus\tilde{L}$ reads
\begin{equation}
    H_{L\oplus\tilde{L}}(\boldsymbol{k})=\left(\begin{array}{cc}
    \Delta_{\tilde{L}} \cdot \mathbb{I}_{2\times 2} & S^{\dagger}(\boldsymbol{k}) \\
    S(\boldsymbol{k}) & \Delta_{L} \cdot \mathbb{I}_{3\times 3} \\
    \end{array}\right)\;.\label{app:eq:SplitHoneycombHAall}
\end{equation}
A typical example of band structure associated to this Hamiltonian is shown in Fig.~\ref{fig:bipartiteexample}(c). 
Moreover, one has 
\begin{eqnarray}
\left(H_{L\oplus\tilde{L}}(\boldsymbol{k})-\Delta_{\tilde{L}}\cdot \mathbb{I}_{5\times 5}\right)\left(H_{L\oplus\tilde{L}}(\boldsymbol{k})-\Delta_{L}\cdot \mathbb{I}_{5\times 5}\right) & = & \left(\begin{array}{cc}
S^{\dagger}(\boldsymbol{k})S(\boldsymbol{k}) & 0\\
0 & S(\boldsymbol{k})S^{\dagger}(\boldsymbol{k})
\end{array}\right).
\end{eqnarray}
We assume the eigenvalues of $S^{\dagger}S$ are $\lambda_{i}$($i=1,2$).
Then, the eigenvalues of $H_{L\oplus\tilde{L}}$ are given by $\varepsilon^{2}-\left(\Delta_{L}+\Delta_{\tilde{L}}\right)\varepsilon+\Delta_{L}\Delta_{\tilde{L}}-\lambda_{i}=0$,
\begin{eqnarray}
\varepsilon&=&\frac{\left(\Delta_{L}+\Delta_{\tilde{L}}\right)\pm\sqrt{\left(\Delta_{L}-\Delta_{\tilde{L}}\right)^{2}+4\lambda_{i}}}{2}.\label{egen-smatrix}
\end{eqnarray}
Based on Eq. (\ref{egen-smatrix}), when $\Delta_L\neq \Delta_{\tilde{L}}$, we can derive the individual effective Hamiltonian for either $\tilde L$ or $L$ by applying the second-order perturbation theory as
\begin{equation}
    H_{\tilde L}(\boldsymbol{k})\thickapprox \Delta_{\tilde{L}}- S^{\dagger}(\boldsymbol{k})S(\boldsymbol{k})/(\Delta_{L}-\Delta_{\tilde{L}}),\label{eq:hamiLtil}
\end{equation}
and
\begin{equation}
    H_{L}(\boldsymbol{k})\thickapprox \Delta_{L} + S(\boldsymbol{k})S^{\dagger}(\boldsymbol{k})/(\Delta_{L}-\Delta_{\tilde{L}}),\label{eq:hamilL}
\end{equation} 
respectively. 
Hence, the band structures of $H_L$ and $H_{\tilde{L}}$ are given by the spectra of $T(\boldsymbol{k})$ and $\tilde{T} (\boldsymbol{k})$ (after a rescaling of energy), respectively. 
As shown in  Fig.~\ref{fig:bipartiteexample}(d), there is one flat band of degeneracy $D=|{L}|-\mathrm{rank}(S(\boldsymbol{k}))=1$ at $\Delta_L$ in the band structure of $H_{{L}}$. The effective Hamiltonian for the sublattice $L$ of Eq.~(\ref{eq:hamilL}) is nothing else than the Kagome tight binding Hamiltonian, with the nearest coupling $t^2/(\Delta_{L}-\Delta_{\tilde{L}})$ and on-site energy $\Delta_{L}$. Thus we have explained the flat band in the Kagome lattice through the $S$-matrix formalism.

\subsection{Algorithms to search for bipartite and split sublattices in crystalline materials}\label{app:algorithmofbipartite}

\begin{figure}[ht]
\includegraphics[width=0.95\columnwidth]{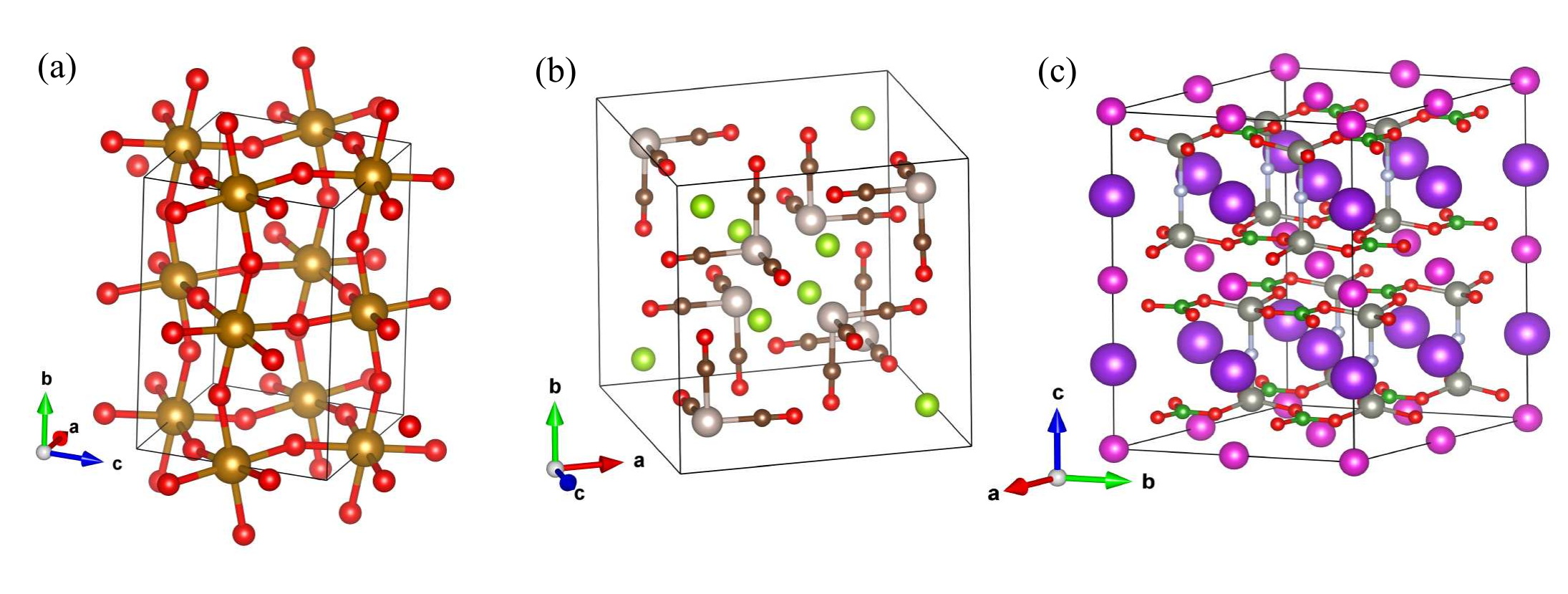}
\caption{(a) The crystal structure of $\rm{Ca}\rm{Fe}\rm{O}_3$ [\icsdweb{92330}, SG 62 (\sgsymb{62})], where $\rm{Fe}$, $\rm{O}$ are represented by the yellow, red balls and the $\rm{Ca}$ atoms are ignored. In this material, $\rm{O}$ and $\rm{Fe}$ form the sublattices $L$ and $\tilde{L}$ respectively, of a bipartite lattice and the number of $\rm{O}$ and $\rm{Fe}$ atoms are different: $|L|=3$  and $|\tilde{L}|=1$. (b)The crystal structure of $\rm{C}_{12}\rm{O}_3\rm{Ru}_4\rm{Se}_4$ [\icsdweb{92913}, SG 217 (\sgsymb{217})], where the $\rm{Ru}$, $\rm{Se}$, $\rm{C}$ and $\rm{O}$ atoms are represented by the grey, green, brown and red balls, respectively. In this material, $\rm{Ru}$, $\rm{C}$ and $\rm{O}$ form a molecular-cluster which results in the flat atomic bands. The molecular lattices like this one are excluded by the algorithm we have developed. (c)The crystal structure of a $2\times2\times1$ supercell of $\rm{Cd} \rm{Zn}_{2} \rm{K} \rm{B}_{2} \rm{O}_{6} \rm{F}$ [\icsdweb{248025}, SG 163 (\sgsymb{163})], where $\rm{Cd}$, $\rm{Zn}$, $\rm{K}$, $\rm{F}$, $\rm{B}$ and $\rm{O}$ are represented by the purple, grey, blue, white, green and red balls, respectively. In this material, there are two connected subsets along the $z$ direction. Each connected subset is a bipartite lattice with different number of atoms in its two sublattices. $\rm{F}$ and $\rm{O}$ atoms form the $L$ sublattice and $\rm{B}$ and $\rm{Zn}$ atoms form the $\tilde{L}$ sublattice ($|L|=7$  and $|\tilde{L}|=4$).
\label{fig:bipartited teexample}}
\end{figure}

In most cases, bipartite sublattices in a crystalline material are not exact if all of the possible hopping terms are considered. There usually exist hoppings between sites of the same sublattice. Nevertheless, a bipartite sublattice dressed by some additional small enough hopping terms could give rise to almost flat bands. For a systematic search of bipartite and split lattices, we solely rely on the geometric distance between two atoms to infer the hopping term amplitude between them. We can then ignore hopping terms which are likely to be small, and identify compounds that have connected bipartite sublattices. 

Our algorithm requires a cutoff for the hopping distance above which we assume two atoms are not connected by a hopping term. We base this cutoff on the shortest bond in a crystal structure we want to analyze. Thus for a crystalline material, the first step is to set the cutoff of the hopping distance.  In practice, we set the cutoff of the hopping distance $\tau$ using the following formula:

\begin{equation}
\label{equ:bondcutoff}
 \tau = \lambda \max(d,x),   
\end{equation}
where $d$ is the length of the shortest bond in the lattice. To avoid the cases where $d$ would be too small and thus excluding some non-negligible hopping terms, we also set the lower bound of $d$ as $x$. $\lambda$ ($\ge 1$) is a multiplicative coefficient to define the cutoff based on the distance $\max(d,x)$. For this article and on the \webflatband, we have considered the values $\lambda=1.2, 1.5$ and $1.7$ and $x=1.5, 1.8, 2.1, 2.4$ and $2.7 \mathrm{\AA}$.

For example in the material $\rm{Ca}\rm{Fe}\rm{O}_3$[\icsdweb{92330}, SG 62 (\sgsymb{62})] shown in Fig.~\ref{fig:bipartited teexample}(a), by setting the parameters $\lambda=1.2$ and $x=1.5 \mathrm{\AA}$, the shortest bond is the ${\rm Fe}-{\rm O}$ bond and $d=1.91 \mathrm{\AA}$. The hopping distance between ${\rm Ca}$ and ${\rm Fe}$ or ${\rm O}$ are larger than the cutoff $\tau=1.2 d=2.29 \mathrm{\AA}$. 
Hence, ${\rm Ca}$ atoms are ignored. The bonds lower than $\tau$ are assumed to be always connected, thus ${\rm Fe}$ and ${\rm O}$ form a bipartite sublattice. Indeed, the ${\rm Fe}$ atoms form the sublattice $L$ and the ${\rm O}$ atoms form the sublattice $\tilde{L}$. The difference of the atom number between $L$ and $\tilde{L}$ is 2.

We now describe the algorithm (see its flow chart in Fig.~\ref{fig:bipartitemethod}) that we have designed to identify a bipartite sublattice from a crystalline material:

\begin{enumerate}
\item[Step 1] We find the shortest bond in the 3D crystal lattice and refer its length as $d$. Then, the cutoff of the hopping distance is set as $\tau = \lambda\max(d, x)$ using the two external parameters $x$ and $\lambda$.

\item[Step 2] For each pair of atoms, if the distance between them is smaller than the cutoff $\tau$, we assume the hopping between them is nonzero and we label one as the neighbor of the other one. Otherwise, we consider there is no hopping between them. 

\item[Step 3] An atom that has less than two neighbors is considered  as a part of a molecular-like cluster, which could contribute to trivial flat atomic bands. We discard such atoms. By repeating this procedure and as shown in Fig.~\ref{fig:RemoveMolecularClusters}, in the end a molecular-like cluster is either deleted (if it was fully disconnected from any other atom) or reduced to a single atom (if the molecular-like cluster was attached to a set of connected atoms via a single atom).

\item[C1] After Step 3, if all the atoms are deleted in the unit cell, there is no bipartite sublattice. Otherwise, the following steps are applied to the remaining atoms.

\item[Step 4] For the set of atoms that are not deleted in Step 3, they can always be divided into several connected subsets, where the hopping term is non-zero between two neighbors within the same subset and is zero between any two atoms in different subsets.Using the results we obtained in Step 2, we can get all the connected subsets. For each connected subset, we apply the following steps.

\item[Step 5]
 To check if a connected subset $\mathcal{B}$ is a bipartite sublattice or not, we first define two temporary sets, $L$ and $\tilde{L}$, and a \emph{source set} $S$. Then, we choose randomly an atom from $\mathcal{B}$ and distribute it to $S$. 
 We iterate over the following Step 6 and Step 7 to populate $L$ and $\tilde{L}$ with the atoms in $\mathcal{B}$.

\item[Step 6] From the connected subset $\mathcal{B}$, we take all the atoms, which are not in $L$ or $\tilde{L}$ but are neighbors of the atoms in $S$, as the (new) \emph{target set} $T$. We distribute all the atoms in $S$ to $L$ and all the atoms in $T$ to $\tilde{L}$. Then we check if $L$ and $\tilde{L}$ satisfy the following two criteria:
\item[C2] Any atom in $L$ (or $\tilde{L}$) is not the neighbor of any other atoms in the same sublattice; Otherwise, the connected subset $\mathcal{B}$ is not a bipartite lattice and we go back to Step 5, moving to the next connected subset.
\item[C3] If all the atoms in the connected subset $\mathcal{B}$ have been distributed into either $L$ or $\tilde{L}$ then we go to C4; otherwise, we go to Step 7.

\item[Step 7] We take all the atoms, which are not in $L$ or $\tilde{L}$ but are the neighbors of the atoms in $T$, as the new \emph{source set} $S$. Then we empty the \emph{target set} $T$ and repeat Step 6.

\item[C4] If the numbers of atoms in $L$ and $\tilde{L}$ are nonzero and different, the connected subset $\mathcal{B}$ is tagged as a bipartite lattice with different numbers of atoms in its two sublattices and we go to C5. 

\item[C5] For the bipartite lattice with two sublattices $L$ and $\tilde{L}$ (up to a relabelling, we can always assume $L > \tilde{L}$), we further check if it is a split lattice. If any atom in $L$ has exactly two neighbors in $\tilde{L}$, and is at equal distance to its two neighbors, the bipartite lattice is also a split lattice. For this step, our algorithm uses a threshold corresponding to $\pm\FlatBandSplitDistanceThreshold$ of the length of the shorter bond to test if two distances are identical.
\end{enumerate}

\begin{figure}[ht]
\includegraphics[width=0.95\columnwidth]{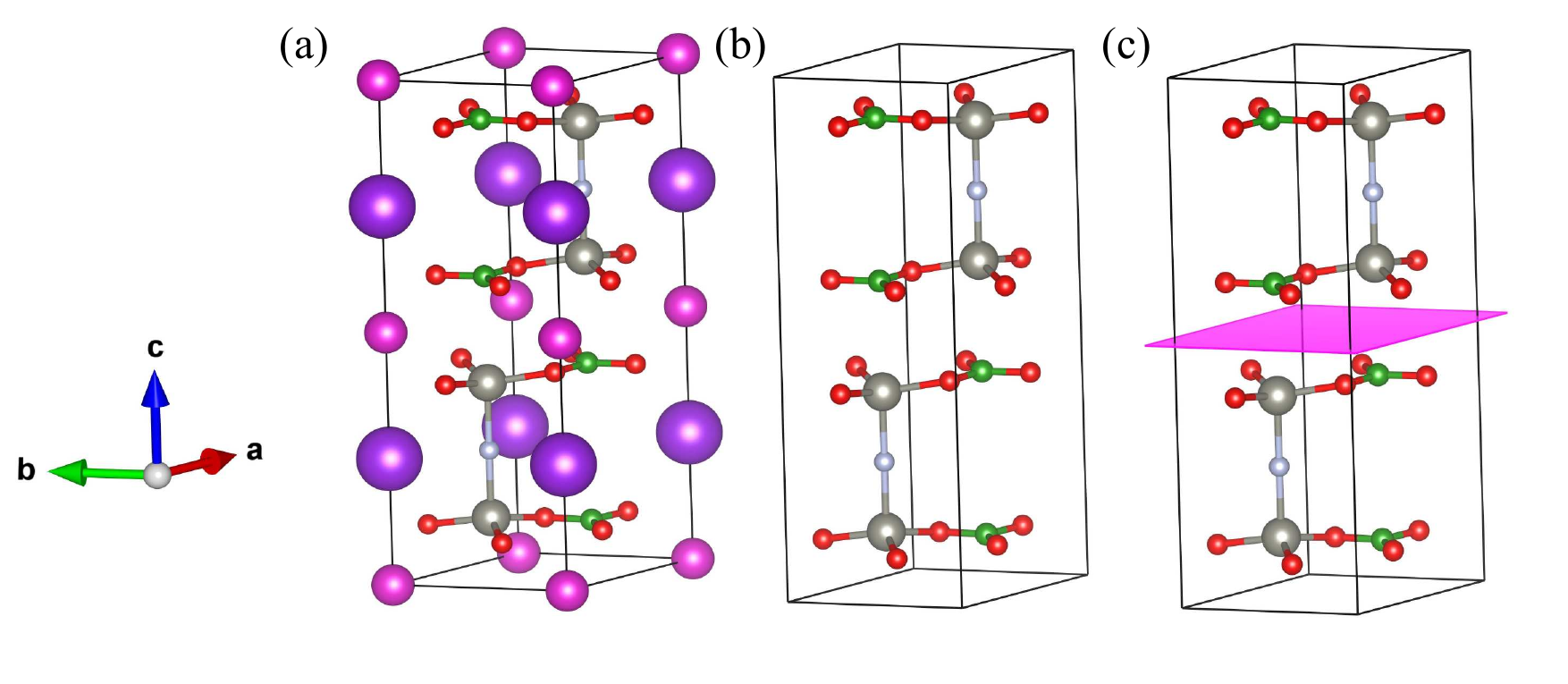}
\caption{(a)The crystal structure of a unit cell of $\rm{Cd} \rm{Zn}_{2} \rm{K} \rm{B}_{2} \rm{O}_{6} \rm{F}$ [\icsdweb{248025}, SG 163 (\sgsymb{163})], where $\rm{Cd}$, $\rm{Zn}$, $\rm{K}$, $\rm{F}$, $\rm{B}$ and $\rm{O}$ are represented by the purple, grey, blue, white, green and red balls, respectively. Only the bonds whose lengths are smaller than the cutoff $\tau=2.25\AA$ that is considered in Step 2 are shown. (b) The lattice structure after Step 3, all the Cd and K atoms are deleted and all the Zn atoms, B atoms, O atoms and F atoms are left. (c) In Step 3, we identify two connected subsets which are separated by the purple plane. In the explanation of our algorithm for this material we consider the subset above the purple plane first.\label{fig:example_of_CdZn2KB2O6F}}
\end{figure}

To illustrate how our algorithm works for the bipartite lattice with different number of atoms on sublattices, we apply it to an example of the material $\rm{Cd} \rm{Zn}_{2} \rm{K} \rm{B}_{2} \rm{O}_{6} \rm{F}$ [\icsdweb{248025}, SG 163 (\sgsymb{163})] whose crystal structure is shown in both Fig.~\ref{fig:bipartited teexample}(c) and Fig.~\ref{fig:example_of_CdZn2KB2O6F}(a). 

\begin{enumerate}
    \item[Step 1] In the first step, we identify that the bond between $\rm{O}$ (red atoms in Figure~\ref{fig:example_of_CdZn2KB2O6F}(a)) and $\rm{B}$  (green atoms in Figure~\ref{fig:example_of_CdZn2KB2O6F}(a)) is the shortest bond in $\rm{Cd} \rm{Zn}_{2} \rm{K} \rm{B}_{2} \rm{O}_{6} \rm{F}$. As the length of this shortest bond is $1.38\mathrm{\AA}$ and less  than $1.5\mathrm{\AA}$, we set the cutoff of the hopping distance as $\tau=1.5\times1.5=2.25\mathrm{\AA}$.
    \item[Step 2] Then, we discard the bonds of length larger than $\tau$ and only assume the bonds of length smaller than $\tau$ to have nonzero hopping amplitude. For example, by ignoring the bonds with zero hopping amplitude, the crystal structure of $\rm{Cd} \rm{Zn}_{2} \rm{K} \rm{B}_{2} \rm{O}_{6} \rm{F}$ is plotted in  Fig.~\ref{fig:example_of_CdZn2KB2O6F}(a), where the length of each plotted bond is less than the cutoff $\tau=2.25\mathrm{\AA}$.
    %\item Delete all the isolated atoms, i.e. the Ru atoms, of this material.
    \item[Step 3] If an atom has no, or only one, bond left, we delete it from the lattice.
    In the case of $\rm{Cd} \rm{Zn}_{2} \rm{K} \rm{B}_{2} \rm{O}_{6} \rm{F}$, all of the $\rm{Cd}$ atoms (the purple atoms in Figure~\ref{fig:example_of_CdZn2KB2O6F}(a)) and $\rm{K}$ atoms (the blue atoms in Figure~\ref{fig:example_of_CdZn2KB2O6F}(a)) have no bond left. Thus, they are removed from the lattice. And the lattice structure after this step is shown in Figure~\ref{fig:example_of_CdZn2KB2O6F}(b).
    \item[C1] As shown in Figure~\ref{fig:example_of_CdZn2KB2O6F}(b), following Step 3, all the Zn atoms (the grey atoms in Figure~\ref{fig:example_of_CdZn2KB2O6F}(a)), B atoms, O atoms and F atoms (the white atom in Figure~\ref{fig:example_of_CdZn2KB2O6F}(a)) are left in the unit cell, so continue to Step 4. 
    \item[Step 4] Using the results obtained in Step 2,  (\ie which hoppings considered  in Figure~\ref{fig:example_of_CdZn2KB2O6F}(b)), we find two \emph{connected} subsets, which are separated by the [0 0 1] Miller plane at $z=0.5c$ (where $c$ is the lattice constant along $z$ direction) in the lattice, as shown in Figure~\ref{fig:example_of_CdZn2KB2O6F}(c).
    \item[Step 5] For both of the connected subsets in Figure~\ref{fig:example_of_CdZn2KB2O6F}(c), we check whether they are bipartite sublattice separately. Now, we take the one above the purple plane in Figure~\ref{fig:example_of_CdZn2KB2O6F}(c) as an example. For convenience, we name this connected subset as $\mathcal{B}$. Considering the double counting problem of the O atoms at the unit cell's boundary, there are six O atoms, one F atom, two Zn atoms and two B atoms in $\mathcal{B}$. We first define two temporary sets, $L$ and $\tilde{L}$, and a \emph{source set} $S$. Then, we choose randomly an atom from $\mathcal{B}$ and distribute it to $S$. Here, we choose one $\rm{F}$ atom from $\mathcal{B}$.
    \item[Step 6] From the connected subset $\mathcal{B}$, we take the two $\rm{Zn}$ atoms, which are neighbors of the $\rm{F}$ atom in $S$, as the \emph{target set} $T$.
    We distribute all the atoms in $S$ (\ie one F atom) to $L$ and all the atoms in $T$ (\ie two Zn atoms) to $\tilde{L}$. Then we check if $L$ and $\tilde{L}$ satisfy the C2 and C3.
    \item[C2] As obtained in Step 2, the two Zn atoms in $\tilde{L}$ are not neighbors to each other and there is only one F atom in $L$. Thus C2 is satisfied. 
    \item[C3] In the connected subset $\mathcal{B}$, there still are six O and two B atoms left to be distributed, so C3 is not satisfied and we proceed to Step 7.
    \item[Step 7] We take the neighbors of the two $\rm{Zn}$ atoms in $T$, \ie the six $\rm{O}$ atoms, as the new \emph{source set} $S$. Then we empty the \emph{target set} $T$ and repeat Step 6.
    \item[Step 6] From the connected subset $\mathcal{B}$, we take the neighbors of the six $\rm{O}$ atoms of $S$, \ie the two $\rm{B}$ atoms which are not in $L$ or $\tilde{L}$, as the new \emph{target set} $T$. We distribute all the atoms in $S$ (\ie the six O atoms) to $L$ and all the atoms in $T$ (\ie two B atoms) to $\tilde{L}$. Then we check if $L$ and $\tilde{L}$ satisfy C2 and C3.
    \item[C2] As obtained in Step 2, the two Zn and the two B atoms in $\tilde{L}$ are not neighbors to each other. The six O and the F atom in $L$ are not neighbors to each other. So C2 is satisfied.
    \item[C3] All the atoms in $\mathcal{B}$ have been distributed to either $L$ or $\tilde{L}$. So C3 is satisfied and we move to C4.
    \item[C4] The number of atoms in $L$ is 7 which includes six O atoms and one F atom, while the number of atoms in $\tilde{L}$ is 4 which includes two Zn atoms and two B atoms. So C4 is satisfied and the connected subset $\mathcal{B}$ is a bipartite lattice with different numbers of atoms in its two sublattices. Then, we check the criteria in C5.
    \item[C5] We further check if this subset is a split lattice. Since none of the atoms in $L$ are at equal distance to their two neighbors in $\tilde{L}$ up to our threshold, the bipartite lattice is not a split lattice.
    \item[Step5] All the above steps are repeated for the lower connected subset as obtained in Step 4. 
\end{enumerate}

Within the above algorithm, we found that the subset $\mathcal{B}$ is a bipartite lattice with sublattice $L$ and $\tilde{L}$. $L$ includes six O atoms and one F atom while $\tilde{L}$ includes two Zn atoms and two B atoms.

\begin{figure}[h]
\includegraphics[width=12cm]{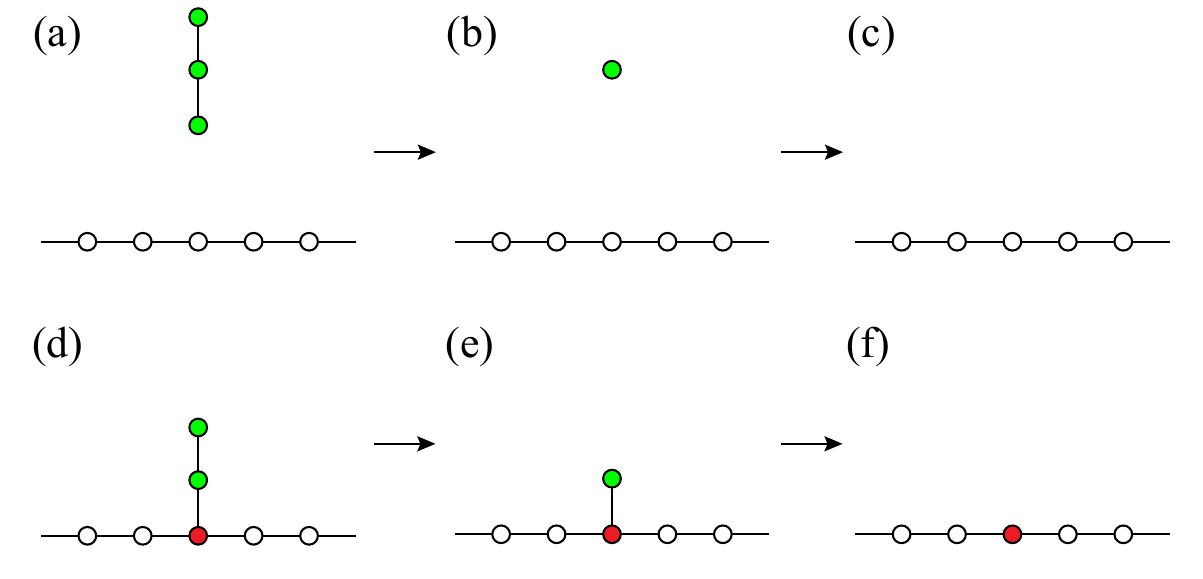}
\caption{Two of the possible cases when removing a molecular cluster. (a)-(c) The molecular cluster is made of the three green sites and is completely detached from the the fully connected sites (white circles). Acting with Step 3 once, we remove two out of the three green sites. Applying Step 3 a second time, we completely get rid of the molecular cluster. (d)-(f) The molecular cluster is made of the two green sites and one red site belonging to fully connected sites. Repetitively acting on the molecular cluster using Step 3, we end up with a single site of the molecular cluster (in red).\label{fig:RemoveMolecularClusters}}
\end{figure}

\begin{figure}[h]
\includegraphics[width=18cm]{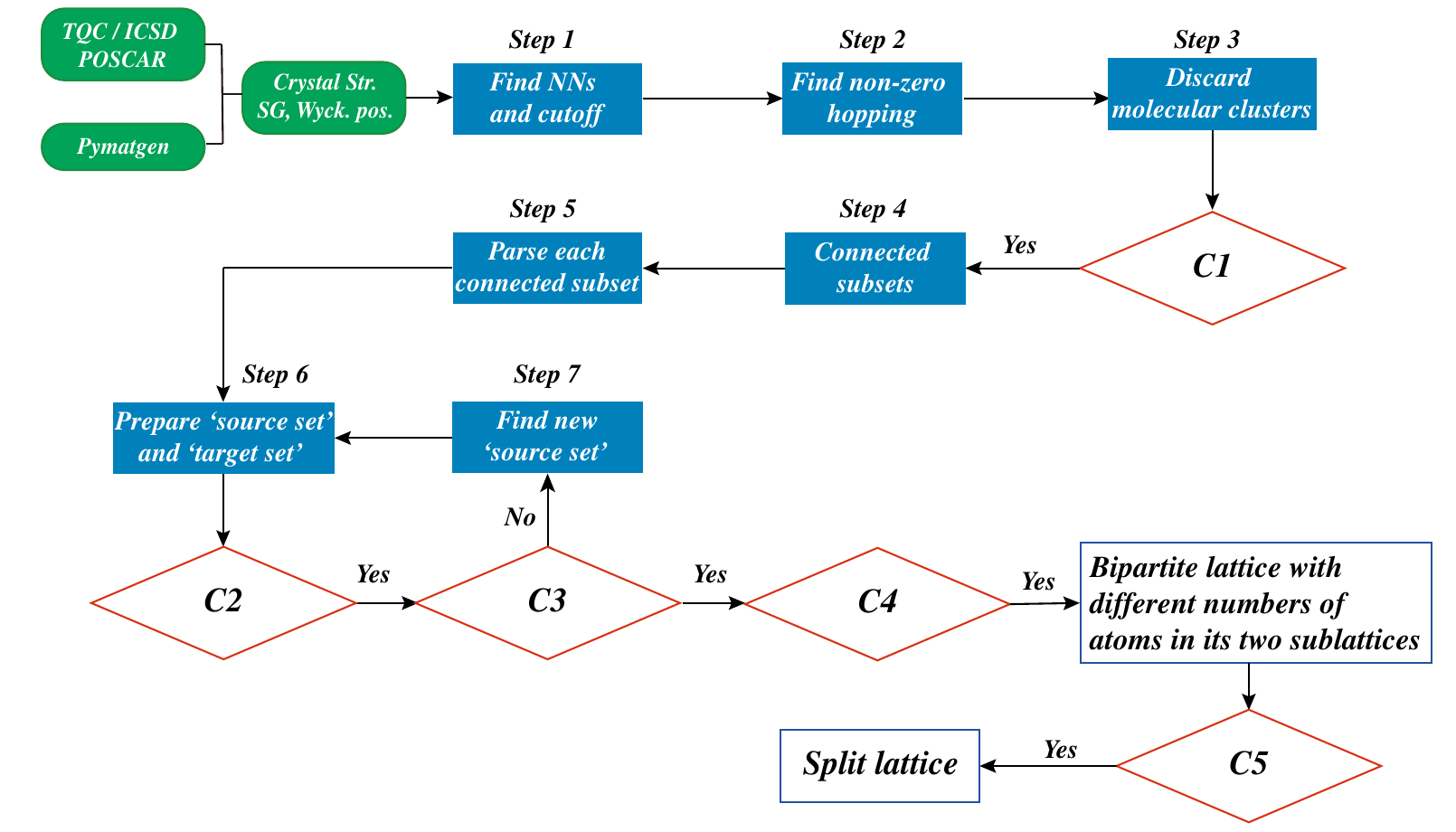}
\caption{The flow chart of the algorithm identifying bipartite sublattices from a 3D crystal lattice. Starting from the crystal structure of a material given by its \emph{POSCAR} file, we use the \emph{Pymatgen} package to analyze the crystal symmetry and the Wyckoff positions of each distinct element. Step 1 is applied to identify the nearest neighbor(NN) sites of each atom and set the cutoff of the hopping distance. In Step 2, if the hopping distance is within the cutoff, the hopping amplitude is set to non-zero; otherwise, the hopping amplitude is set to zero; In Step 3, the molecular-like clusters are removed and we exit the algorithm through C1 if there has no atom left. In Step 4, for the set of atoms that are not removed in Step 3, they are divided into several connected subsets, where the hopping amplitude is non-zero between two neighbors within the same subset and is zero between any two atoms in different connected subsets. 
The Steps 5, 6 and 7 and the criteria C2 and C3 are applied to each connected subset to identify the bipartite sublattice based on the neighbors of each atom as extracted in Step 2. The criteria C4 is applied to check whether the numbers of the atoms in the two sublattices $L$ and $\tilde{L}$ are different. If a bipartite lattice is detected, an additional test based on the criteria C5 is performed to check if the bipartite lattice is also a split lattice.
\label{fig:bipartitemethod}}
\end{figure}

To illustrate how the molecular lattices are excluded, we consider the example $\rm{C}_{12}\rm{O}_3\rm{Ru}_4\rm{Se}_4$[\icsdweb{92913}, SG 217 (\sgsymb{217})], whose crystal structure is shown in Fig.~\ref{fig:bipartited teexample}(b). Applying our algorithm to this material leads to the following sequence:
\begin{enumerate}
    \item[Step 1] In the first step, we identify that the bond between $\rm{C}$ atoms (brown atoms in Figure~\ref{fig:bipartited teexample}(b)) and $\rm{O}$ atoms (red atoms in Figure~\ref{fig:bipartited teexample}(b)) is the shortest bond in $\rm{C}_{12}\rm{O}_3\rm{Ru}_4\rm{Se}_4$. As the length of this shortest bond is smaller than $1.5\mathrm{\AA}$, we set the cutoff of the hopping distance as $\tau=\lambda1.5\mathrm{\AA}$.
    \item[Step 2] Then, we discard the bonds of length larger than $\tau$ and only the bonds of length smaller than $\tau$ are assumed to have nonzero hopping. For example, by ignoring the bonds with zero hopping, the crystal structure of $\rm{C}_{12}\rm{O}_3\rm{Ru}_4\rm{Se}_4$ is plotted in  Fig.~\ref{fig:bipartited teexample}(b) for $\tau=1.5\times1.5=2.25\mathrm{\AA}$, where the length of each plotted bond is less than the cutoff $\tau=2.25\mathrm{\AA}$.
    %\item Delete all the isolated atoms, i.e. the Ru atoms, of this material.
    \item[Step 3] If an atom has no or only one bond left, we delete it from the lattice.
    In the case of $\rm{C}_{12}\rm{O}_3\rm{Ru}_4\rm{Se}_4$, all of the $\rm{Se}$ atoms (green atoms in Figure~\ref{fig:bipartited teexample}(b)) have no bond left and every $\rm{O}$ atom has only one bond. Thus, both of them are removed from the lattice. Then, each of the left $\rm{C}$ atoms (brown atoms in Figure~\ref{fig:bipartited teexample}(b)) only has one bond and should be removed. After that, the $\rm{Ru}$ atoms (grey atoms in Figure~\ref{fig:bipartited teexample}(b)) get isolated and hence are also removed. Finally, all the atoms are removed from the lattice and this material does not belong to a \emph{connected} bipartite lattice.
\end{enumerate}

In this work, we applied our algorithm using different values for the cutoff parameters $\lambda$ and $x$ in Eq.~(\ref{equ:bondcutoff}). As we mentioned above, we adopted the values of $\lambda$ to be 1.2, 1.5 and 1.7 and the values of $x$ to be 1.5, 1.8, 2.1, 2.4 and 2.7$\mathrm{\AA}$. We have scanned the materials available in the database of Appendix~\ref{app:database} to search for bipartite sublattices potentially leading to flat bands, \ie $|L|\neq|\tilde{L}|$ and split sublattices. The individual  results are available on the \webflatband. The bipartite sublattice statistics for each setting are provided in Table~~\ref{tab:bipartitectuoffstat}. 

{\tiny
\begin{longtable*}{|c|c|c|c|}
\caption[Bipartite sublattice statistics per cutoff parameters]{Statistics for the ICSD entries with either bipartite or split sublattices with respect to the lower bound of $d$ and the multiplicative factor $\lambda$ that define the distance cutoff in Eq.~(\ref{equ:bondcutoff}). The first and second columns are the values of $x$ and $\lambda$. The third column is the number of ICSD entries hosting at least one bipartite sublattice and the fourth column is the number of ICSD entries hosting at least one split sublattice among the bipartite sublattices. All percentages are computed with respect to the total number of ICSD entries that have been considered (\TQCDBNbrNoSOCICSDsIncludingValidFElectrons~ICSDs).\label{tab:bipartitectuoffstat}}\\
\hline
$x$  & $d$ & \begin{tabular}{c}\# ICSDs with\\ bipartite \end{tabular} & \begin{tabular}{c}\# ICSDs with\\ split \end{tabular}  \\
\hline  
 $ x=1.5${\AA} & $\lambda=1.2$  & 15662 {\tiny(28.4\%)} & 7811 {\tiny(14.2\%)} \\ 
 $ x=1.5${\AA} & $\lambda=1.5$  & 6011 {\tiny(10.9\%)} & 1235 {\tiny(2.2\%)} \\ 
 $ x=1.5${\AA} & $\lambda=1.7$  & 217 {\tiny(0.4\%)} & 31 {\tiny(0.1\%)} \\ 
 $ x=1.8${\AA} & $\lambda=1.2$  & 18459 {\tiny(33.4\%)} & 7282 {\tiny(13.2\%)} \\ 
 $ x=1.8${\AA} & $\lambda=1.5$  & 1879 {\tiny(3.4\%)} & 296 {\tiny(0.5\%)} \\ 
 $ x=1.8${\AA} & $\lambda=1.7$  & 64 {\tiny(0.1\%)} & 6 {\tiny(0.0\%)} \\ 
 $ x=2.1${\AA} & $\lambda=1.2$  & 14895 {\tiny(27.0\%)} & 4721 {\tiny(8.6\%)} \\ 
 $ x=2.1${\AA} & $\lambda=1.5$  & 697 {\tiny(1.3\%)} & 102 {\tiny(0.2\%)} \\ 
 $ x=2.1${\AA} & $\lambda=1.7$  & 6 {\tiny(0.0\%)} & 1 {\tiny(0.0\%)} \\ 
 $ x=2.4${\AA} & $\lambda=1.2$  & 4704 {\tiny(8.5\%)} & 1339 {\tiny(2.4\%)} \\ 
 $ x=2.4${\AA} & $\lambda=1.5$  & 245 {\tiny(0.4\%)} & 22 {\tiny(0.0\%)} \\ 
 $ x=2.4${\AA} & $\lambda=1.7$  & ---  & ---  \\ 
 $ x=2.7${\AA} & $\lambda=1.2$  & 2863 {\tiny(5.2\%)} & 617 {\tiny(1.1\%)} \\ 
 $ x=2.7${\AA} & $\lambda=1.5$  & 37 {\tiny(0.1\%)} & 4 {\tiny(0.0\%)} \\ 
 $ x=2.7${\AA} & $\lambda=1.7$  & ---  & ---  \\
\hline
\end{longtable*}
}

{\it Discussion} At the end of this section, we comment on  the relation between Lieb, bipartite and split sublattices in \crystal~materials. As we defined before, the split sublattice is a special case of bipartite sublattice and each of the split sublattices we obtained in our results \emph{has to} be a bipartite sublattice. 
Mathematically, the Lieb lattice is a special case of both bipartite and split sublattices in 2D. 
However, in 3D \crystal~materials, the conditions in our method to identify a 2D Lieb sublattice is not stringent enough to identify if the 2D Lieb sublattice is also a bipartite or split sublattice in 3D. 
 In our geometric algorithm for Lieb sublattice, we only consider the NN bonds and the atoms on (or close to) a 2D plane. In the algorithm for bipartite sublattice, we set a cutoff $\tau$ for the hopping distance and consider all the hoppings of distance less than the cutoff $\tau$. 
 Hence more hoppings and atoms are involved and it could make a Lieb-sublattice material not a bipartite-sublattice material. Actually we have only found one such case when a Lieb lattice is not a bipartite lattice in our database, \ie $\rm{H}_{3} \rm{F}_{3} \rm{Ga} \rm{N}$ [\icsdweb{89503}, SG 39 (\sgsymb{39})].

\section{Theoretical explanations for flat bands in representative materials}\label{app:theoryexplanation}

\def\kk{\boldsymbol{k}}
\def\RR{\mathbf{R}}
\def\tt{\mathbf{t}}
\def\bb{\boldsymbol{b}}

In this appendix, we discuss seven typical materials among the \FlatBandNbrBestMaterials~materials that will be presented in Appendix~\ref{app:bestflatbands} which host clear flat bands at or near the Fermi level. Here we explain the origins of these flat bands. The seven materials are WO$_3$ [\icsdweb{108651}, SG 221 (\sgsymb{221})], Pb$_2$Sb$_2$O$_7$ [\icsdweb{27120}, SG 227 (\sgsymb{227})], CaNi$_5$ [\icsdweb{54474}, SG 191 (\sgsymb{191})],
Ca$_2$NCl [\icsdweb{62555}, SG 166 (\sgsymb{166})], $\mathrm{Rb_{2}CaH_{4}}$ [\icsdweb{65196}, SG 139 (\sgsymb{139})],
\ch{KAg[CN]2} [\icsdweb{30275}, SG 163 (\sgsymb{163})],
and RbMo$_3$S$_3$ [\icsdweb{644175}, SG 176 (\sgsymb{176})].
The first six materials have a bipartite sublattice in the crystal lattice and the flat bands can be explained using the $S$-matrix method, as introduced in Appendix~\ref{app:bipartite} and in Ref.~\cite{S-matrix}. 
For the last example, RbMo$_3$S$_3$, the flat bands are due to the quasi-1D character and the weak inter-chain couplings of this material.

\subsection{$\rm{WO_{3}}$}\label{WO3}
\begin{figure}[h]
\includegraphics[width=18cm]{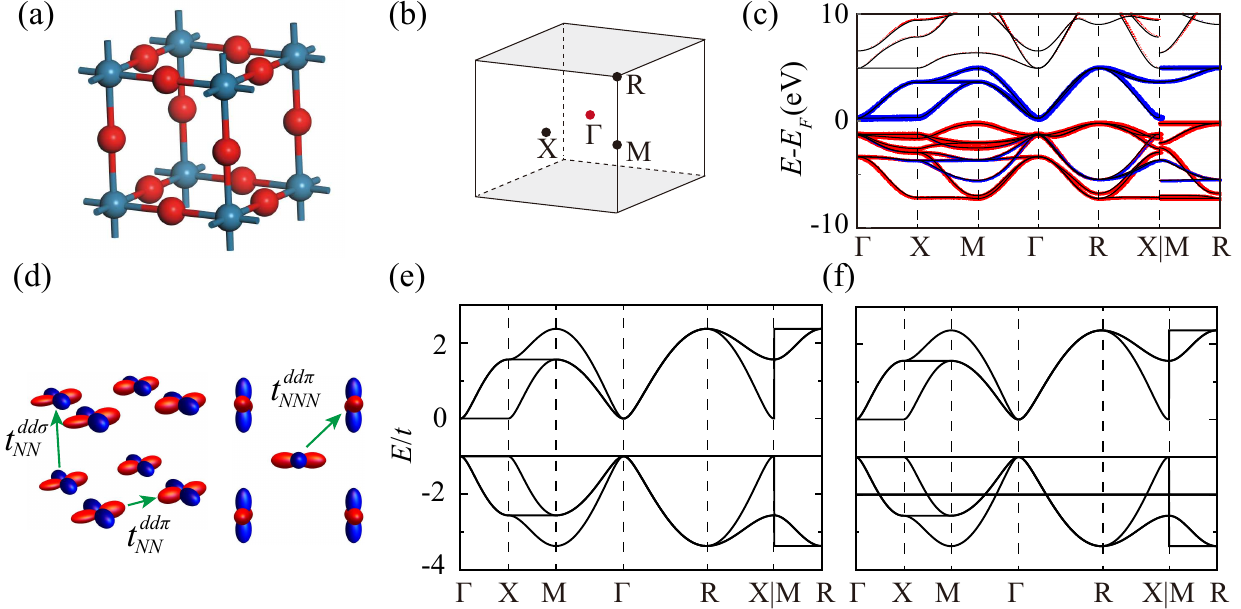}
\caption{(a) The crystal structure of $\mathrm{WO_{3}}$. $\mathrm{W}$ and $\mathrm{O}$ are represented by the blue and red balls, respectively. (b) The bulk BZ of $\mathrm{WO_{3}}$. (c) Orbital-projected band structure of $\mathrm{WO_{3}}$ without spin-orbit coupling. The weights of $d_{xy}$,$d_{yz}$ and $d_{zx}$ orbitals on W are highlighted in blue, and the weights of the $p$ orbits on O are highlighted in red. 
(d) The schematic of the three different hopping terms that are considered in Eq.~(\ref{eq:L_ReO3}). (e) The band structure of the bipartite lattice $L\oplus\tilde{L}$ constructed from the $S$ matrix in Eq.~(\ref{eq:S_ReO3}). We have chosen $\Delta=-|t|$ for our calculation. 
(f) The band structure of the bipartite lattice $L\oplus\tilde{L}$ with an additional onsite term $\delta$ that splits the $A_{2u}$ and $E_u$ orbitals at the O atoms.
\label{fig:fig1_mads}}
\end{figure}

The crystal structure of $\mathrm{WO_{3}}$ [\icsdweb{108651}, SG 221 (\sgsymb{221})] is shown in Fig.~\ref{fig:fig1_mads}(a). 
W and O occupy the $1a$ and $3c$ Wyckoff positions, respectively. From its band structure without SOC considered, as plotted in Fig.~\ref{fig:fig1_mads}(c), there are three connected bands at (and above) the Fermi level. 
Along the high-symmetry paths $\mathrm{\Gamma\rightarrow X}$, $\mathrm{X\rightarrow M}$ and $\mathrm{M\rightarrow R}$, there is always one out of the three bands that is extremely flat (but not necessarily at the Fermi level). 
By analyzing the orbital-projected band structure obtained from the first-principles calculation, the three bands come from the $d_{xy}$, $d_{yz}$ and $d_{xz}$ orbitals of W, which form the 3 dimensional single-valued irrep $T_{2g}$ of the site-symmetry-group $O_h$ of the $1a$ position. 
Below the $T_{2g}$ bands of W, there is another set of valence bands coming from $p$ orbitals of O, as indicated by the red lines in Fig.~\ref{fig:fig1_mads}(c), of which the highest bands are also flat. Similar structures of flat bands also exist in the perovskites with the chemical formulae ABO$_3$ or BO$_3$ in SG 211 (\sgsymb{221})\cite{piskunov2005hybrid,jia2017first,eglitis2018systematic}, where B is a transition element with $3d$ electrons in the outer shell sitting at the $1a$ (or $1b$) Wyckoff position and O sits at the $3c$ (or $3d$) Wyckoff position.

Both the flat conduction bands (along some paths) and the flat valence bands can be explained using the $S$-matrix method, as introduced in Appendix \ref{app:bipartite}.  For the case in WO$_3$, we refer to the sublattices of W and O as two sublattices $\tilde L$ and $L$. Then, the lattice of $L\oplus \tilde L$ is a bipartite lattice if only the bonds connecting W and its NN O atoms are taken into account. As explained in the above tight-binding method, the bands near the Fermi level come from  $d_{yz}$, $d_{xz}$, and $d_{xy}$ orbitals on W and the $p$ orbitals on O. 
We introduce the Bloch basis as 
\begin{equation} \label{eq:WO3-basis}
\ket{\phi_\alpha(\kk)} = \sum_{\RR} e^{i \RR\cdot \kk} \ket{\alpha,\RR},\qquad
\ket{\phi_a(\kk)} = \sum_{\RR} e^{i \RR  \cdot \kk} \ket{a,\RR},
\end{equation}
where $\alpha$ indexes the three $d$ orbitals, $a$ indexes the nine $p$ orbitals on the three O atoms, and $\RR$ indexes the lattice vectors, respectively. 
$\alpha=1,2,3$ correspond to the $d_{xy}$, $d_{yz}$, and $d_{xz}$ orbitals, respectively, $a=i + 3(j-1)$  correspond to the $p_x$ ($i=1$), $p_y$ ($i=2$), $p_z$ ($i=3$) orbitals at the $j$th site of the $3c$ Wyckoff position, respectively.
Here $j=1,2,3$ correspond to the sites $(\frac{1}{2},0,0)$, $(0,\frac{1}{2},0)$ and $(0,0,\frac{1}{2})$, respectively. 
Notice that we have chosen a periodic gauge for the basis in Eq.~(\ref{eq:WO3-basis}), \ie the basis is invariant if we change $\kk$ to $\kk+\mathbf{G}$ with $\mathbf{G}$ being a vector of the reciprocal lattice. 
We will use this periodic gauge throughout this section.
The $S$ matrix of this bipartite lattice is given by $S_{\alpha a}^\dagger(\boldsymbol{k}) = \bra{\phi_{\alpha}(\boldsymbol{k})} \hat{H} \ket{\phi_{a}(\boldsymbol{k})}$.
This $3\times9$-dimensional $S^\dagger(\kk)$ matrix is
\begin{equation}
S^{\dagger}(\boldsymbol{k})=t\left(\begin{array}{ccccccccc}
0 & 0 & 0 & 0 & 0 & 1-e^{-ik_{z}} & 0 & 1-e^{-ik_{y}} & 0\\
0 & 0 & 1-e^{-ik_{z}} & 0 & 0 & 0 & 1-e^{-ik_{x}} & 0 & 0\\
0 & 1-e^{-ik_{y}} & 0 & 1-e^{-ik_{x}} & 0 & 0 & 0 & 0 & 0
\end{array}\right)\;.\label{eq:S_ReO3}
\end{equation}
Here $t$ is the hopping between the $d_{xy}$ orbital at $(0,0,0)$ and $p_x$ orbital at $(\frac{1}{2},0,0)$ which can be obtained from the maximally localized Wannier function (MLWF) calculations \cite{PhysRevB.65.035109} and is $-1.654~$eV. Here, we build the MLWF by the $p$ orbitals and the $T_{2g}$ orbitals of W atom. The inter-sublattice couplings \eg the coupling between $p_z$ orbitals at $(0,0,\frac{1}{2})$ and $(0,\frac{1}{2},0)$ (0.179$~$eV), and the coupling between $d_{xy}$ orbitals at $(0,0,0)$ and $(1,0,0)$ (-0.222$~$eV) are ignored. Since these couplings are much smaller than those we consider, neglecting these couplings is reasonable.
Other NN hoppings between $T_{2g}$ and $p$ orbitals are equivalent to this hopping due to the symmetries. 
We further assume the onsite energy of $p$ orbitals as $\Delta$ ($<0$), then the total Hamiltonian reads
\begin{equation}
    H_{L\oplus\tilde{L}}(\boldsymbol{k})=\left(\begin{array}{cc}
    0_{3\times 3} & S^{\dagger}(\kk) \\
    S(\kk) & \Delta \cdot \mathbb{I}_{9\times 9} \\
    \end{array}\right)\;.\label{eq:H_ReO3}
\end{equation}
To discover the flat valence bands (over the whole Brillouin zone), we consider the SVD decomposition of the $S(\kk)$ matrix as $S(\kk) = U(\kk) \Sigma(\kk) V^\dagger(\kk)$, where $U(\kk)$ is a nine-by-nine unitary matrix, $V(\kk)$ is a three-by-three unitary matrix, and $\Sigma(\kk)$ is a nine-by-three rectangular diagonal matrix with real non-negative diagonal elements. 
We apply the unitary transformation to the Hamiltonian
\begin{equation}
    \begin{pmatrix}
    V^\dagger(\kk) & 0_{3\times 9} \\
    0_{9\times 3} & U^\dagger(\kk) 
    \end{pmatrix} \cdot
    H_{L\oplus\tilde{L}}(\boldsymbol{k}) \cdot 
    \begin{pmatrix}
    V(\kk) & 0_{3\times 9} \\
    0_{9\times 3} & U(\kk) 
    \end{pmatrix} = 
    \begin{pmatrix}
    0_{3\times 3} &  \Sigma^T(\kk)\\
    \Sigma(\kk) & \Delta \cdot \mathbb{I}_{9\times 9} 
    \end{pmatrix}.
\end{equation}
Suppose that the diagonal elements of $\Sigma(\kk)$ are $\xi_n(\kk)$ with $n=1,2,3$, then the system has six branches of dispersive bands $\frac{\Delta}2 \pm \sqrt{\xi_n^2(\kk) + \Delta^2/4}$ and six flat bands at $\Delta$.
If one branch of $\xi_n(\kk)$ is a constant $\xi_c$ over the whole Brillouin zone, we would have one more flat band at $\frac{\Delta}2 \pm \sqrt{\xi_c^2 + \Delta^2/4}$ and an additional flat band at $0$. 
For a generic Hamiltonian in form of Eq.~(\ref{eq:H_ReO3}), there are $|L| - \mathrm{rank} (S(\kk))$ states at $\Delta$ and $|\tilde{L}| - \mathrm{rank} (S(\kk))$ states at 0. 
The other $2\times \mathrm{rank}(S(\kk))$ states must have energies either below (above) $\Delta$ or above (below) 0 for negative (positive) $\Delta$. 
The particular $S(\kk)$ matrix defined in Eq.~(\ref{eq:S_ReO3}) has rank three at generic momenta, hence there are 6 flat bands at the energy $\Delta$, as shown in  Fig.~\ref{fig:fig1_mads}(e).

Since the $T_{2g}$-bands and $p$-bands are well separated in energy, we can derive the individual effective Hamiltonians for them by applying the second-order perturbation theory, already exemplified in Appendix~\ref{app:smatrix}, as 
\begin{equation}
    H_{\tilde L}(\boldsymbol{k})\thickapprox - S^{\dagger}(\kk)S(\kk)/\Delta,\label{eq:tL}
\end{equation}
and
\begin{equation}
    H_{L}(\boldsymbol{k})\thickapprox\Delta + S(\kk)S^{\dagger}(\kk)/\Delta,\label{eq:L}
\end{equation} 
respectively (The exact expressions are provided in Appendix~\ref{app:smatrix}). 
The effective Hamiltonian of $T_{2g}$-bands is a diagonal matrix
\begin{equation}
    H_{\tilde L}(\boldsymbol{k})= -2t^2/\Delta \left(\begin{array}{ccc}
    2 - \cos k_y - \cos k_z & 0 & 0 \\
    0 & 2 - \cos k_x - \cos k_z & 0 \\
    0 & 0 & 2 - \cos k_x - \cos k_y \\
    \end{array}\right)\;.\label{eq:L_ReO3}
\end{equation}
With parameters given by $t^{dd\pi}_{NN} = - t^2/|\Delta|$, $t^{dd\delta}_{NN} = t^{dd\pi}_{NNN}=0$, $\Delta_d = - \frac{4t^2}{\Delta} = \frac{4t^2}{|\Delta|}$. $\Delta_d$ is the onsite energy of $T_{2g}$ orbitals, $t_{NN}^{dd\pi}$  ($t_{NNN}^{dd\pi}$) is the amplitude of the NN (NNN) in-plane hopping while $t_{NN}^{dd\delta}$ is the amplitude of NN out-of-plane hopping, as schematically shown in Fig.~\ref{fig:fig1_mads}(d). Taking the $T_{2g}$ orbitals as basis, we built the MLWF.  The values of $\Delta_d$, $t_{NN}^{dd\pi}$, $t_{NNN}^{dd\pi}$ and $t_{NN}^{dd\delta}$  are 2.45eV, -0.62eV, 0.01eV and -0.006eV, respectively. Then, the Hamiltonian in Eq.~(\ref{eq:L_ReO3}) is a diagonal matrix with elements $2t_{NN}^{dd\pi}\left\{ \begin{array}{ccc}\cos k_{y}+\cos k_{z}, & \cos k_{x}+\cos k_{z}, & \cos k_{x}+\cos k_{y}\end{array}\right\} $ (up to the constant shift $\Delta_d$). 
We can prove that there is always a flat band along a path parallel to the direction of the $k_{i}$ ($i=x,y,z$) axis, and the energy of the flat band is $E=2t_{NN}^{dd\pi}(\cos  k_j + \cos  k_k)$, where $i\neq j \neq k$. 
\begin{itemize}
\item Along the high-symmetry path $\mathrm{\Gamma\rightarrow X}$, with $k_y=k_z=0$, the Hamiltonian in Eq.~(\ref{eq:L_ReO3}) becomes a diagonal matrix with three diagonal elements, $2t_{NN}^{dd\pi}\left\{ \begin{array}{ccc}2, & 1+\cos k_{x}, & 1+\cos k_{x}\end{array}\right\} + \left\{\Delta_d,\Delta_d,\Delta_d\right\}$, resulting in one flat band at $E=4t_{NN}^{dd\pi}+ \Delta_d$=-0.03eV. 
This flat band has a touching point at the  $\Gamma$ point with a double-degenerate dispersive band.
\item Along the high-symmetry path $\mathrm{X\rightarrow M}$, with $k_x=\pi$ and $k_z=0$, the Hamiltonian in Eq.~(\ref{eq:L_ReO3}) becomes a diagonal matrix with three diagonal elements, $2t_{NN}^{dd\pi}\left\{ \begin{array}{ccc}1+\cos k_{y}, & 0,-1+\cos k_{y} & \end{array}\right\}+ \left\{\Delta_d,\Delta_d,\Delta_d\right\} $, resulting in one flat band at $E= \Delta_d$=2.45eV. 
This flat band has touching points at $X$ and $M$ points with two dispersive bands, respectively.
\item Along the high-symmetry path $\mathrm{R\rightarrow M}$, with $k_x=k_y=\pi$, the Hamiltonian in Eq.~(\ref{eq:L_ReO3}) becomes a diagonal matrix with three diagonal elements, $2t_{NN}^{dd\pi}\left\{ \begin{array}{ccc}-1+\cos k_{z}, & -1+\cos k_{z},-2 & \end{array}\right\}+ \left\{\Delta_d,\Delta_d,\Delta_d\right\} $, resulting in a flat bands at $E=-4t_{NN}^{dd\pi}+ \Delta_d$=4.93eV.
This flat band has a touching point at $\mathrm{R}$ point with a double-degenerate dispersive band.
\end{itemize}

The dispersion of the valence bands from the first-principle calculations is more complicated than the dispersion of the valence bands in the $S$-matrix model, as can be seen by comparing Fig.~\ref{fig:fig1_mads}(c) and Fig.~\ref{fig:fig1_mads}(e). Indeed, we have omitted several symmetry-allowed terms in the $S$-matrix model. 
For example, the site-symmetry-groups of the O atoms are all $D_{4h}$.
The $p_{x},p_y, p_z$ orbitals form a two-dimensional irrep $E_u$ and a one-dimensional irrep $A_{2u}$ of the point group $D_{4h}$. The two irreps in principle should have different onsite energies due to the crystal field splitting. 
In the $S$-matrix construction we have set both onsite energies as $\Delta$ for simplicity.
Now we consider splitting the $A_{2u}$ and $E_u$ orbitals by adding an additional onsite energy $\delta$ on the $A_{2u}$ orbitals.
Notice that the $A_{2u}$ irreps at $(\frac{1}{2},0,0)$, $(0,\frac{1}{2},0)$, $(0,0,\frac{1}{2})$ are formed by $p_x$, $p_y$, and $p_z$ orbitals, respectively.
From the $S$-matrix in Eq.~(\ref{eq:S_ReO3}), one can see that these $A_{2u}$ orbitals are decoupled from the other orbitals.
Hence, for $\delta=0$, the $A_{2u}$ orbitals contribute to three  of the six flat bands at $\Delta$.
A nonzero $\delta$ just shifts the energy of these three flat bands to $\Delta + \delta$, as shown in Fig.~\ref{fig:fig1_mads}(f), where $\delta$ is set to $-|t|$ (note that $\delta$ does not have to be equal to $\Delta$). 
The band structure with nonzero $\delta$ still has perfect flat-band segments in the valence bands. 
These flat bands will have finite band widths if the (smaller) hopping terms between $p$ orbitals at different O atoms are taken into account.
We leave this discussion for future studies.

The above analysis using the $S$-matrix method could be adapted to explain the flat bands of the other perovskites, such as 
$\mathrm{ReO_{3}} $ [\icsdweb{16810}, SG 221 (\sgsymb{221})]\cite{ReO3Ong},
$\mathrm{NbF_{3}} $ [\icsdweb{25596}, SG 221 (\sgsymb{221})],
$\mathrm{BaTiO_{3}}$ [\icsdweb{27971}, SG 221 (\sgsymb{221})], 
$\mathrm{CaTiO_{3}}$ [\icsdweb{162924}, SG 221 (\sgsymb{221})],
$\mathrm{SrTiO_{3}}$ [\icsdweb{80874}, SG 221 (\sgsymb{221})],
$\mathrm{NaNbO_{3}}$ [\icsdweb{28588}, SG 221 (\sgsymb{221})],
$\mathrm{NaTaO_{3}}$ [\icsdweb{88378}, SG 221 (\sgsymb{221})],
$\mathrm{CrLaO_{3}}$ [\icsdweb{28930}, SG 221 (\sgsymb{221})],
$\mathrm{CrSrO_{3}}$ [\icsdweb{108903}, SG 221 (\sgsymb{221})],
$\mathrm{BaFeO_{3}}$ [\icsdweb{29096}, SG 221 (\sgsymb{221})],
$\mathrm{SrFeO_{3}}$ [\icsdweb{91062}, SG 221 (\sgsymb{221})],
$\mathrm{LaFeO_{3}}$ [\icsdweb{29118}, SG 221 (\sgsymb{221})],
$\mathrm{LaMnO_{3}}$ [\icsdweb{674501}, SG 221 (\sgsymb{221})],
$\mathrm{SrRuO_{3}}$ [\icsdweb{82983}, SG 221 (\sgsymb{221})],
$\mathrm{BaRuO_{3}}$ [\icsdweb{673893}, SG 221 (\sgsymb{221})],
$\mathrm{SrCoO_{3}}$ [\icsdweb{77142}, SG 221 (\sgsymb{221})],
$\mathrm{SrVO_{3}}$ [\icsdweb{88982}, SG 221 (\sgsymb{221})],
$\mathrm{CaTcO_{3}}$ [\icsdweb{671082}, SG 221 (\sgsymb{221})],
$\mathrm{SrTcO_{3}}$ [\icsdweb{109076}, SG 221 (\sgsymb{221})],
$\mathrm{CaTcO_{3}}$ [\icsdweb{671086}, SG 221 (\sgsymb{221})],
$\mathrm{TiPbO_{3}}$ [\icsdweb{187295}, SG 221 (\sgsymb{221})],
$\mathrm{CrPbO_{3}}$ [\icsdweb{160196}, SG 221 (\sgsymb{221})],
$\mathrm{CaMnO_{3}}$ [\icsdweb{168902}, SG 221 (\sgsymb{221})],
$\mathrm{SrMnO_{3}}$ [\icsdweb{188415}, SG 221 (\sgsymb{221})],
$\mathrm{VPbO_{3}}$ [\icsdweb{187637}, SG 221 (\sgsymb{221})],
$\mathrm{KNbO_{3}}$ [\icsdweb{190920}, SG 221 (\sgsymb{221})], 
and 
$\mathrm{VBaO_{3}}$ [\icsdweb{191203}, SG 221 (\sgsymb{221})]. Note that all of these materials are listed in Appendix~\ref{app:listcurated}.

\subsection{$\rm{Pb_{2}Sb_{2}O_{7}}$}\label{PbSbO}

\begin{figure}[t]
\includegraphics[width=18cm]{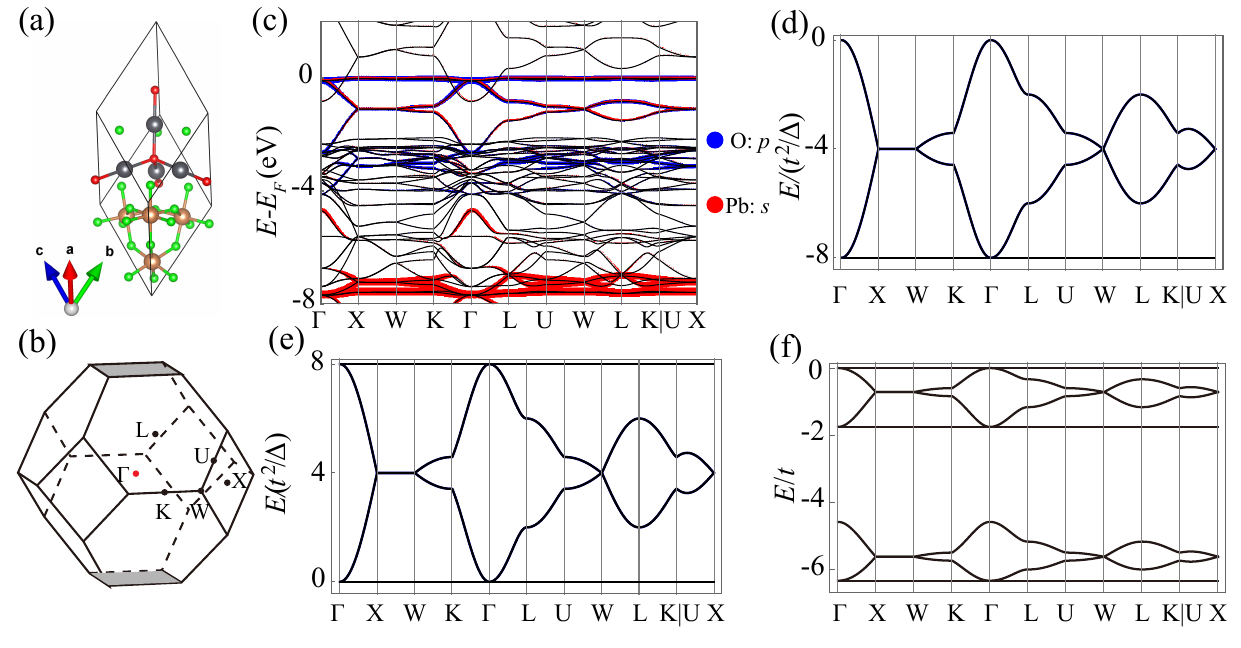}
\caption{(a) The crystal structure of $\mathrm{Pb_{2}Sb_{2}O_{7}}$. The Pb and Sb atoms are in gray and brown, respectively. The O atoms at $8b$ and $48f$ are in red and green, respectively. (b) The bulk BZ of $\mathrm{Pb_{2}Sb_{2}O_{7}}$. (c) Band structure calculated by GGA-PBE of $\mathrm{Pb_{2}Sb_{2}O_{7}}$ without spin-orbit coupling. The weights of the $s$ orbital on Pb and the $p$ orbitals on O atoms that are located at $8b$ are highlighted in red and blue, respectively. (d) The band structure of Hamiltonian $H_{\tilde{L}}$ in Eq.~(\ref{eq:ObSbO}). 
(e) The band structure of Hamiltonian $H_{L}$ in Eq.~(\ref{eq:ObSbO}). (f) The band structure of Hamiltonian $H_{L\oplus\tilde{L}}$ in Eq.~(\ref{eq:ObSbO1}).
}
\label{fig:fig7_mads}
\end{figure}

The crystal structure of $\mathrm{Pb_{2}Sb_{2}O_{7}}$ [\icsdweb{27120}, SG 227 (\sgsymb{227})(origin choice 1 as defined on the \bcslong)]
is shown in Fig.~\ref{fig:fig7_mads}(a). The $16d$ and $16c$ Wyckoff positions are occupied by the
Pb and Sb atoms, respectively. O atoms occupy two different Wyckoff positions, $8b$ and $48f$. 
As tabulated in Table~\ref{tab:kagomepyrochlorewyckoff}, each of the sublattices formed by Pb and Sb is a pyrochlore lattice, which is a 3D line-graph lattice that could host flat bands \cite{PhysRevB.78.125104}. 
As shown in the orbital-projected band structure of $\mathrm{Pb_{2}Sb_{2}O_{7}}$ in Fig.~\ref{fig:fig7_mads}(c), around the Fermi level, there are two flat bands mainly composed
by the $p$ orbitals of the O located at the $8b$ position. 
As indicated in the orbital-projected band structure, the $p$ orbitals of O are hybridized with the $s$ orbitals of Pb.
The other bands between -8eV and -2eV are mainly contributed by the O atoms at the $48f$ Wyckoff position.
The hybridization between the $s$ orbitals of O at $8b$ and these bands is tiny because the distances from the O atoms at $8b$ to the Sb atoms and O atoms at $48f$ are larger than those to the Pb atoms.
In fact, in \webflatband\ and Appendix \ref{app:listcurated}, the crystal structure of $\mathrm{Pb_{2}Sb_{2}O_{7}}$ is identified to have a bipartite lattice formed by Pb and the O at $8b$. 
As shown below, the flat bands come from the $S$-matrix of this bipartite lattice.

We refer to the two sublattices formed by O (at $8b$) and Pb (at $16d$) as $L$ and $\tilde{L}$, respectively. Then, the lattice of $L\oplus\tilde{L}$ is a bipartite lattice if only the bonds connecting Pb and its NN O atoms are taken into account. 
We construct the $S$ matrix by putting three $p$ orbitals at each of the $8b$ positions $\left(\frac{1}{2},\frac{1}{2},\frac{1}{2}\right)$, $\left(\frac{3}{4},\frac{3}{4},\frac{3}{4}\right)$ and an $s$ orbital at each of the $16d$ positions $\left(\frac{5}{8},\frac{5}{8},\frac{5}{8}\right)$,$\left(\frac{5}{8},\frac{5}{8},\frac{1}{8}\right)$,$\left(\frac{5}{8},\frac{1}{8},\frac{5}{8}\right)$ and $\left(\frac{1}{8},\frac{5}{8},\frac{5}{8}\right)$, with coordinates given in the primitive unit cell spanned by $\boldsymbol{a}_{1}=a(0,1/2,1/2)$, $\boldsymbol{a}_{2}=a(1/2,0,1/2)$, $\boldsymbol{a}_{3}=a(1/2,1/2,0)$, with $a$ being the lattice parameter.
We use $\alpha$ and $a$ to label the orbitals in the sublattices $\tilde{L}$ and $L$, respectively.
The index $\alpha=1,2,3,4$ corresponds to the four $s$ orbitals at the four sites in $\tilde{L}$.
The index $a=i + 2(j-1)$  corresponds to the $p_x$ ($j=1$), $p_y$ ($j=2$), $p_z$ ($j=3$) orbitals at the $i$'th site ($i=1,2$) in $L$.
Then the $S$ matrix can be written as 
\begin{equation}
S^{\dagger}(\kk)=t\left(\begin{array}{cccccc}
1 & -1 & 1 & -1 & 1 & -1\\
-1 & e^{-i k_{3}} & -1 & e^{-i k_{3}} & 1 & -e^{-i k_{3}}\\
-1 & e^{-i k_{2}} & 1 & -e^{-i k_{2}} & -1 & e^{-i k_{2}}\\
1 & -e^{-i k_{1}} & -1 & e^{-i k_{1}} & -1 & e^{-i k_{1}}
\end{array}\right). \label{eq:s-s-1}
\end{equation}
Here, $t$ is the strength of the coupling between $s$ orbital at $\left(\frac{5}{8},\frac{5}{8},\frac{5}{8}\right)$ and $p_x$ orbital at $\left(\frac{1}{2},\frac{1}{2},\frac{1}{2}\right)$, and $(k_1,k_2,k_3)$ is the coordinate of the momentum $\kk= \sum_{i=1}^3 k_i \bb_i$, where $\bb_i$ is the reciprocal lattice basis corresponding to the $\boldsymbol{a}_i$ given above.
We assume the onsite energies of the $p$ and $s$ orbitals to be $\Delta_s$ and $\Delta_p$, respectively, and they satisfy $\Delta_s<\Delta_p<0$. 
Although $\Delta_s$ and $\Delta_p$ are free parameters, to simplify the expression of the following Hamiltonian and the associated eigenvalues of flat bands, we choose a set of particular values
$\Delta_p=\left(\sqrt{2}-\sqrt{10}\right)|t|$,  $\Delta_s=-\left(\sqrt{2}+\sqrt{10}\right)|t|$. The total Hamiltonian is given by 
\begin{equation}
    H_{L \oplus \tilde{L}} (\kk) =
    \begin{pmatrix}
    \Delta_s \mathbb{I}_{4\times 4} & S^\dagger(\kk) \\
    S(\kk) & \Delta_p \mathbb{I}_{6\times 6}
    \end{pmatrix}.\label{eq:ObSbO1}
\end{equation}
Following the same analysis below Eq.~(\ref{eq:H_ReO3}), one can find that there will be $|L|-\mathrm{rank}(S(\kk))$ states at the energy $\Delta_p$ and $|\tilde{L}|-\mathrm{rank}(S(\kk))$ states at the energy $\Delta_s$. 
While the other $2\times\mathrm{rank}(S(\kk))$ states have the energies $\frac{\Delta_s + \Delta_p}{2} \pm \sqrt{\xi^2_n(\kk) + (\Delta_s-\Delta_p)^2/4}$, with $\xi_n(\kk)$ being the nonzero singular values of $S(\kk)$. 
Since $\mathrm{rank}(S(\kk))=|\tilde{L}|=4$ at generic momenta, there are two flat bands at $E=\Delta_p \approx -1.75|t|$, as shown in Fig.~\ref{fig:fig7_mads}(e).
The effective Hamiltonian for $L$ ($p$-bands) and $\tilde{L}$ ($s$-bands) can be derived through the second order perturbation theory as 
\begin{equation}
    H_{L}(\kk) \thickapprox \Delta_p + \frac{S(\kk) S^\dagger(\kk)}{\Delta_p-\Delta_s},\quad
    H_{\tilde{L}}(\kk) \thickapprox \Delta_s - \frac{S^\dagger(\kk) S(\kk)}{\Delta_p-\Delta_s}, \label{eq:ObSbO}
\end{equation}
respectively. 
The energy spectra of $H_{\tilde L}$ and $H_{L}$ are given in Fig.~\ref{fig:fig7_mads}(d) and (e), respectively, which  qualitatively reproduce the band structures of $H_{\tilde{L} \oplus L}(\kk)$.

One may notice that in addition to the flat bands at $\Delta_p$, there are two other sets of flat bands at the top and the bottom of the band structure of $H_{L\oplus \tilde{L}}$ which are both double-degenerate. 
They come from the double-degenerate singular value $2 \sqrt{2}t$ of $S(\kk)$ at arbitrary $\kk$.
Indeed from the above analysis, this singular value leads to flat bands at $\frac{\Delta_s+\Delta_p}{2} \pm \sqrt{8t^2 + (\Delta_s-\Delta_p)^2/4}$, \ie $0$ and $-2 \sqrt{10}|t|$, respectively. 
The flat band at zero energy corresponds to the first-principles flat band at the Fermi level (Fig.~\ref{fig:fig7_mads}(c)).
Hence, the bipartite lattice in this example give rise to the flat bands at $E=\Delta_p$, which come from the rank of $S(\kk)$, as well as to the two degenerate extra flat bands which come from the particular singular values of $S(\kk)$.

 The band structure features obtained from $S$-matrix method is consistent with the first-principle calculations of $\mathrm{Pb_{2}Sb_{2}O_{7}}$.
Such an analysis applies to some other materials with the pyrochlore structure, such as 
$\mathrm{Sn_{2}Nb_{2}O_{7}}$ [\icsdweb{163817}, SG 227 (\sgsymb{227})],
$\mathrm{Ca_{2}Nb_{2}O_{7}}$ [\icsdweb{22411}, SG 227 (\sgsymb{227})],
$\mathrm{Bi_{2}Pt_{2}O_{7}}$ [\icsdweb{202346}, SG 227 (\sgsymb{227})],
$\mathrm{Bi_{2}Ir_{2}O_{7}}$ [\icsdweb{161103}, SG 227 (\sgsymb{227})],
$\mathrm{Bi_{2}Rh_{2}O_{7}}$ [\icsdweb{161099}, SG 227 (\sgsymb{227})],
$\mathrm{Ge_{2}In_{2}O_{7}}$ [\icsdweb{239593}, SG 227 (\sgsymb{227})],
and 
$\mathrm{Sn_{2}Ta_{2}O_{7}}$ [\icsdweb{27119}, SG 227 (\sgsymb{227})].  Note that all of these materials are part of the curated list in Appendix~\ref{app:listcurated}.

\subsection{$\rm{CaNi}_{5}$}\label{app:BaPd}

\begin{figure*}[t]
\includegraphics[width=18cm]{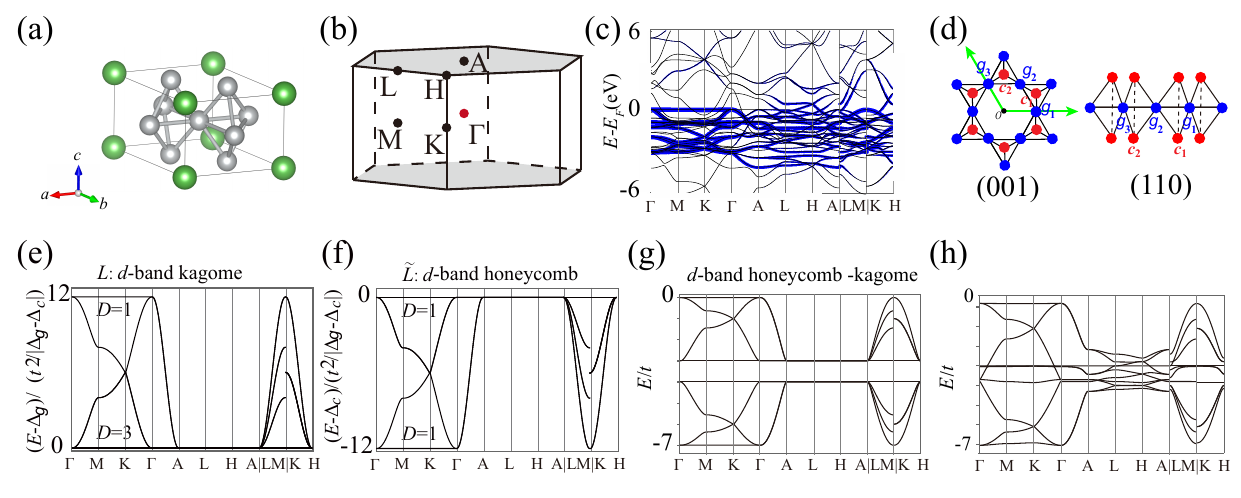}
\caption{(a) The crystal structure of $\mathrm{CaNi_{5}}$. The Ca and Ni atoms are represented by the green and gray balls, respectively. (b) The bulk BZ of $\mathrm{CaNi_{5}}$. (c) Band structure of $\mathrm{CaNi_{5}}$ without spin-orbit coupling, where the weights of  $d_{xz}$ and $d_{yz}$ orbitals on Ni are highlighted in blue. (d) The schematic pictures of the top- and side-view (\ie along the respective (001) and (110) directions) of the bipartite lattice in $\mathrm{CaNi_{5}}$.
(e) and (f)  The band structures of the effective Hamiltonians of $d_{xz}, d_{yz}$ orbitals on Kagome ($L$) and honeycomb ($\tilde{L}$) lattices, where the effective Hamiltonians are obtained from the $S$-matrix method through second order perturbation theory. The degeneracy $D$ of each segment of flat bands is indicated in the plots.
(g) The band structure of $d_{xz}, d_{yz}$ orbitals on the bipartite lattice $\widetilde{L}\oplus L$. 
(h) The band structure of $\widetilde{L}\oplus L$ with intra-sublattice couplings.
\label{fig:fig2_mads}}
\end{figure*}

The crystal structure of $\mathrm{CaNi_{5}}$ [\icsdweb{54474}, SG 191 (\sgsymb{191})] is shown in Fig.~\ref{fig:fig2_mads}(a). 
The $1a$ Wyckoff position is occupied by Ca and both the $3g$ and $2c$ Wyckoff positions are occupied by Ni.
The band structure of $\mathrm{CaNi_{5}}$ is plotted in Fig.~\ref{fig:fig2_mads}(c), which has an extremely flat band at the Fermi level in the $k_z=0$ plane.
By analyzing the orbital-projected band structure, we find that the flat bands mainly arise from the $d_{xz}$ and $d_{yz}$ orbitals on Ni.

In $\mathrm{CaNi_{5}}$, Ni form two sublattices: Ni atoms at $2c$ form a honeycomb sublattice, and Ni atoms at $3g$ form a Kagome sublattice.
In the following, to understand the origin of this flat band, we first only consider couplings between two sublattices.
We will consider couplings within each sublattice at the end of this subsection.
We refer to the sublattices formed by Ni atoms at $2c$ and $3g$ as $\tilde{L}$ and $L$, respectively.
We label the $3g$ and $2c$ positions as $g_{i}$ (${i=1,2,3}$) and $c_{i}$ (${i=1,2}$), respectively, as shown in Fig.~\ref{fig:fig2_mads} (d).
They form a bipartite lattice since we ignore the hoppings within each of them. 
We introduce the $S$ matrix formed by the $d_{xz}$ and $d_{yz}$ orbitals in $\tilde{L}$ and $L$ as $S_{\alpha a}^\dagger = \bra{\phi_{\alpha}(\kk)} H \ket{\phi_{a}(\kk)}$.
$\alpha=i + 2(j-1)$  corresponds to the $d_{yz}$ ($j=1$), $d_{xz}$ ($j=2$) orbitals at the $i$th site ($i=1,2$) in $\tilde{L}$.
$a=i + 3(j-1)$ corresponds to the $d_{yz}$ ($j=1$), $d_{xz}$ ($j=2$) orbitals at the $i$th site ($i=1,2,3$) in ${L}$.  Based on the MLWF, we consider the coupling between $d_{xz}$ orbital at $c_2$ and $d_{yz}$ orbital at $g_3$ and set its strength to $t$.
The $S$-matrix reads
\begin{equation}
S^{\dagger}(\kk)=\frac{t}{4}\left(1+e^{ik_{3}}\right)\left(\begin{array}{cccccc}
4 & 1 & e^{ik_{1}} & 0 & -\sqrt{3} & \sqrt{3}e^{ik_{1}}\\
4e^{ik_{2}} & 1 & 1 & 0 & -\sqrt{3} & \sqrt{3}\\
0 & -\sqrt{3} & \sqrt{3}e^{ik_{1}} & 0 & 3 & 3e^{ik_{1}}\\
0 & -\sqrt{3} & \sqrt{3} & 0 & 3 & 3
\end{array}\right)\;,\label{eqS-1-1}
\end{equation}
where  $(k_1,k_2,k_3)$ are the coordinates of the momentum $\kk= \sum_{i=1}^3 k_i \bb_i$, where $\bb_i$ is the reciprocal lattice basis corresponding to the primitive unit cell spanned by $\boldsymbol{a}_{1}=b(1,0,0)$, $\boldsymbol{a}_{2}=b(-1/2,\sqrt{3}/2,0)$, $\boldsymbol{a}_{3}=c(0,0,1)$, with $b$ and $c$ being the lattice parameters. Note that the values of the hopping amplitudes appearing in $S^{\dagger}_{i,j}(\kk)$ are constrained by the space group symmetries. 

The Hamiltonians of the bipartite lattice $L\oplus \tilde L$ and the $\tilde{L}$ and $L$ are given by Eqs.~(\ref{eq:ObSbO1}) and~(\ref{eq:ObSbO}) except that $\left\{ \Delta_{s},\Delta_{p}\right\}$ should be replaced by $\left\{ \Delta_{c},\Delta_{g}\right\}$. 
Here $\Delta_{c}$ is the onsite energy of d orbitals on the honeycomb lattice and $\Delta_{g}$ is the onsite energy of d orbitals on the Kagome lattice.
In the following we choose $\Delta_c=-4|t|$ and $\Delta_g=-3|t|$.
The $S(\kk)$ matrix has rank three at generic momentum.
Following the same analysis below Eq.~(\ref{eq:H_ReO3}), one can find that in the band structure of $H_{L\oplus\tilde{L}}$ there will be $|L|-\mathrm{rank}(S(\kk))=3$ states at the energy $\Delta_g$ and $|\tilde{L}|-\mathrm{rank}(S(\kk))=1$ states at the energy $\Delta_c$, as shown in Fig.~\ref{fig:fig2_mads}(g). 
In addition, since $S$ has a constant singular value $2 \sqrt{3}t |\cos\frac{k_3}2|$ in a horizontal plane with fixed $k_3$,
due to the discussion below Eq.~(\ref{eq:H_ReO3}), there is a set of 2D flat bands at $\frac{\Delta_g+\Delta_c}{2} \pm \sqrt{12t^2 \cos^2 \frac{k_3}2 + (\Delta_g-\Delta_c)^2/4}$.
For $k_3=0$, the two additional flat bands at  $E=0$ and $E=-7t$, respectively.

The effective Hamiltonian $H_{\tilde{L}} \thickapprox \Delta_c - S^\dagger(\kk) S(\kk) / ( \Delta_g - \Delta_c)$ describes a 3D system consisting of honeycomb lattice layers. 
It has single particle dimension four. 
Since $S(\kk)$ has rank three over the whole Brillouin zone (or smaller ranks at high symmetry momenta) and has a constant singular value $2\sqrt3 t |\cos\frac{k_3}2|$ in a horizontal plane with fixed $k_3$, the band structure of $H_{\tilde{L}}$ has a 3D flat band at $E=\Delta_{c}$ and a 2D flat band $E=\Delta_{c}-\frac{12t^{2}}{\Delta_{g}-\Delta_{c}} \cos^2 \frac{k_3}2$ in any fixed-$k_3$ plane (Fig.~\ref{fig:fig2_mads}(f)) \cite{Barreteau_2017}. 
Similarly, the effective Hamiltonian of the Kagome lattice ($H_{{L}} \thickapprox \Delta_g - S(\kk) S^\dagger(\kk) / ( \Delta_c - \Delta_g)$), which has dimension six, has a three-fold degenerate 3D flat band at $E=\Delta_{g}$ and a 2D flat band at  $E=\Delta_{g}-\frac{12t^{2}}{\Delta_{c}-\Delta_{g}} \cos^2 \frac{k_3}2$ in any fixed-$k_3$ plane (Fig.~\ref{fig:fig2_mads}(e)).

Comparing with the band structure from first-principle calculations in Fig.~\ref{fig:fig2_mads}(c), we identify the flat band of $L\oplus \tilde{L}$ at $E=0$ as the flat band near the Fermi level in $\mathrm{CaNi_{5}}$. 
To further check the robustness of the flat bands in Fig.~\ref{fig:fig2_mads}(g), we consider the couplings within each sublattice. 
We set the hopping between $d_{yz}$ orbital at $c_1$ and $d_{xz}$ orbital at $c_2$ ($d_{yz}$ orbital at $g_1$ and $d_{yz}$ orbital at $g_2$), as well as the equivalent hopping terms due to symmetries, to $0.1t$ ($0.2t$) (Fig.~\ref{fig:fig2_mads}(d)).
The resulting band structure is shown in Fig.~\ref{fig:fig2_mads}(h).
One can see that these intra-sublattice hopping terms barely have any effect on the flat band at $E=0$ and $E=-7t$ in the $k_3=0$ plane.

The above explanations of the extremely flat band in $\mathrm{CaNi_{5}}$ can also be applied to some other materials with $\mathrm{AB_5}$-type structure in SG $191$ (\sgsymb{191}), such as 
$\mathrm{LaNi_5}$ [\icsdweb{242027},SG 191 (\sgsymb{191})], 
$\mathrm{ScNi_5}$ [\icsdweb{646468}, SG 191 (\sgsymb{191})],
$\mathrm{CaPd_5}$ [\icsdweb{106357}, SG 191 (\sgsymb{191})],
$\mathrm{SrPd_5}$ [\icsdweb{105707}, SG 191 (\sgsymb{191})],
$\mathrm{BaPd_5}$ [\icsdweb{616030}, SG 191 (\sgsymb{191})],
$\mathrm{YNi_5}$ [\icsdweb{647075}, SG 191 (\sgsymb{191})]
and 
$\mathrm{YFe_5}$ [\icsdweb{103699}, SG 191 (\sgsymb{191})].  Note that all of these materials are listed in Appendix~\ref{app:listcurated}.

\subsection{$\rm{Ca_{2}NCl}$}\label{CaNCl}

\begin{figure}[t]
\includegraphics[width=18cm]{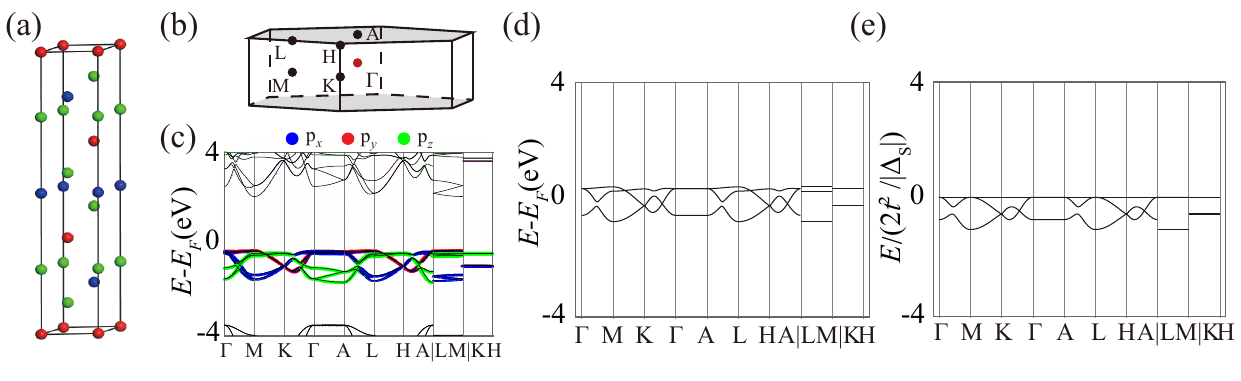}
\caption{
 (a) The crystal structure of $\mathrm{Ca_{2}NCl}$, where the N atoms are represented by the blue balls. (b) The bulk BZ of $\mathrm{Ca_{2}NCl}$. (c) Band structure of $\mathrm{Ca_{2}NCl}$ without spin-orbit coupling. The weights of the $p$ orbits N on the highest set of valence bands are highlighted in blue for the $p_x$ orbitals, red for the $p_y$ orbitals and green for the $p_z$ orbitals.  
(d) Band structure of the three band Hamiltonian in Eq.~(\ref{HamilSr1}) with the parameters extracted from the MLWF calculation. (e) Band structure given by the $S$-matrix method given in Eq.~(\ref{eq:CaNCl1ha}).
\label{fig:fig3_mads}}
\end{figure}

The crystal structure and BZ of $\mathrm{Ca_{2}NCl}$ [\icsdweb{62555}, SG 166 (\sgsymb{166})] are shown in Figs.~\ref{fig:fig3_mads}(a) and (b), respectively.
Since it is a layered material, we have chosen to use the conventional unit cell rather than the primitive unit cell for the band structure calculation.
The $3a$, $3b$ and $6c$ Wyckoff positions are occupied by the Cl, N and Ca atoms, respectively. The lattice parameters of the conventional cell are $a=b=3.67\mathring{\mathrm{A}}$ and $c=19.89\mathring{\mathrm{A}}$.
As shown in the band structure in Fig.~\ref{fig:fig3_mads}(c), the highest valence band is flat along all the high-symmetry paths. 
From the orbital-projected band structure, we see that the dominant contribution to the flat band comes from the $p$ orbitals of N. 
Since the distance between two neighboring N layers is large ($c/3=6.63\mathring{\mathrm{A}}$), in the following we will omit the inter-layer couplings.
This is a reasonable approximation because, as shown in Fig.~\ref{fig:fig3_mads}(c), the band dispersion of the non-flat bands in the $k_z$ direction, \eg paths $\Gamma$A, LM, KH, is weak compared to the in-plane dispersion.

To explain the flat bands in Fig.~\ref{fig:fig3_mads}(c), we first build a 2D tight-binding model consisting of $p_x,p_y$ and $p_z$ orbitals on the N atoms in a single N layer. 
The N atoms in a single layer form a simple triangular lattice.
The tight-binding Hamiltonian on the $p_{x},p_y,p_z$ orbitals is given by
\begin{equation}
H\left(\boldsymbol{k}\right)=2\left(\begin{array}{cc}
H_{12}\left(\boldsymbol{k}\right) & H_{3}\left(\boldsymbol{k}\right)\\
H_{3}^\dagger\left(\boldsymbol{k}\right) &\Delta+ t_{z}^{pp\pi}\left[\cos k_{1}+\cos k_{2}+\cos \left(k_{1}+k_{2}\right)\right]
\end{array}\right)\;,\label{HamilSr1}
\end{equation}
with the terms
{\footnotesize
\begin{equation}
H_{12}\left(\boldsymbol{k}\right)=\left(\begin{array}{cc}
t^{pp\sigma}\cos k_{1}+\frac{t^{pp\sigma}+3t^{pp\pi}}{4}\left[\cos k_{2}+\cos \left(k_{1}+k_{2}\right)\right] & \frac{\sqrt{3}\left(t^{pp\pi}-t^{pp\sigma}\right)}{4}\left[\cos k_{2}-\cos \left(k_{1}+k_{2}\right)\right]\\
\frac{\sqrt{3}\left(t^{pp\pi}-t^{pp\sigma}\right)}{4}\left[\cos k_{2}-\cos \left(k_{1}+k_{2}\right)\right] & t^{pp\pi}\cos k_{1}+\frac{3t^{pp\sigma}+t^{pp\pi}}{4}\left[\cos k_{2}+\cos \left(k_{1}+k_{2}\right)\right]
\end{array}\right)\;.\label{HamilSr2}
\end{equation}
}%%
and 
\begin{equation}
H_{3}^{\dagger}\left(\boldsymbol{k}\right)=\left(\begin{array}{cc}
\frac{\sqrt{3}t^p}{8}\left[\mathrm{cos}k_{2}-\mathrm{cos}\left(k_{1}+k_{2}\right)\right] & \frac{t^p}{8}\left[-2\mathrm{cos}k_{1}+\mathrm{cos}k_{2}+\mathrm{cos}\left(k_{1}+k_{2}\right)\right]\end{array}\right)
\end{equation}
Here $t^{pp\pi}$ and $t^{pp\sigma}$ are the $\pi$-bond and $\sigma$-bond strengths between $p_x$ and $p_y$ orbitals, $t_{z}^{pp\pi}$ is the $\pi$-bond strength between $p_z$ orbitals, and $t^p/8$ is the strength of coupling between a $p_y$ orbital at $(0,0,0)$ and a $p_z$ orbital at $(0,1,0)$.  As mentioned in Appendix~\ref{app:BaPd}, the hopping amplitudes appearing in Eq.~(\ref{HamilSr1}) are constrained by the space group symmetries.
$(k_1,k_2)$ is the coordinate of the momentum $\kk= \sum_{i=1}^2 k_i \bb_i$, where $\bb_i$ is the reciprocal lattice basis corresponding to the unit cell spanned by $\boldsymbol{a}_{1}=a(1,0)$, $\boldsymbol{a}_{2}=a(-\frac{1}{2},\frac{\sqrt{3}}{2})$.
We set the difference of the onsite energy of $p_z$ and $p_{x,y}$ orbitals as $\Delta$. 
From the MLWF we extract the parameters $\Delta=0.04$eV, $t^{pp\sigma}=0.21$eV, $t_z^{pp\pi}=-0.10$eV, $t^{pp\pi}=-0.08$eV, and $t^{p}=0.16$eV.
The tight-binding band structure is shown in Fig.~\ref{fig:fig3_mads}(d).
It correctly reproduces the flat band from the first-principles calculation.
We find that the flatness comes from a set of conditions for the hopping terms that are approximately satisfied in the MLWF model. As an example, 
we take the path $\rm \Gamma M$ ($k_1=0$) to show one such conditions.
In this path, the Hamiltonian is simplified to 
\begin{equation}
H(0,k_2) = 2\begin{pmatrix}
t^{pp\sigma} + \frac{t^{pp\sigma}+3t^{pp\pi}}{2} \cos k_2 & 0 & 0 \\
0 & t^{pp\sigma} + \frac{3t^{pp\sigma}+t^{pp\pi}}{2} \cos k_2 & -\frac{t^p}4 + \frac{t^p}4 \cos k_2 \\ 
0 & -\frac{t^p}4 + \frac{t^p}4 \cos k_2 & \Delta + t_z^{pp\pi} (1+ 2 \cos k_2) 
\end{pmatrix}\label{app:eq:CaNClMLWFKzero}
\end{equation}
There would be an \emph{exact} flat band at $E=2t^{pp\sigma}$ along this path if $t^{pp\sigma}+3t^{pp\pi}=0$.
This condition is approximately satisfied by the actual values extracted from MLWFs ($t^{pp\sigma}+3t^{pp\pi}=-0.03$eV). As shown in Fig.~\ref{fig:fig3_mads}(d), if the condition $t^{pp\sigma}+3t^{pp\pi}=0$ is not strictly satisfied well, the band near $E=2t^{pp\sigma}$ becomes weakly dispersive but is still relatively flat. This follows from the first-principles calculations.

The bands around 3eV in Fig.~\ref{fig:fig3_mads}(c) are mostly due to the Ca atoms. 
We notice that the nearest neighbors of N atoms are the Ca atoms.
Hence the strongest hopping terms in this system are those between N and Ca atoms.
In the following, we construct a $S$-matrix between the N and Ca atoms and derive Eq.~(\ref{HamilSr1}) as an effective Hamiltonian from the second-order perturbation of this $S$-matrix model.
The set of conditions that guarantee the flat band are automatically implied by the $S$-matrix. 
We refer to the sublattices formed by Ca atoms and N atoms as $\tilde{L}$ and $L$, respectively.
We introduce the $S$ matrix formed by the $s$ or $d_z^2$ orbital in $\tilde{L}$ and $p$ orbitals in $L$ as $S_{\alpha a}^\dagger = \bra{\phi_{\alpha}(\kk)} H \ket{\phi_{a}(\kk)}$.
The index $\alpha=(1,2)$  corresponds to the $s$ orbitals at $(\frac{1}{3},\frac{2}{3},x)$ and $(\frac{1}{3},\frac{2}{3},1-x)$ in $\tilde{L}$.
$\alpha=(1,2,3)$ corresponds to the $p_x$, $p_y$ and $p_z$ orbitals at $(0,0,1/2)$ in ${L}$.
The $S$-matrix reads
\begin{eqnarray}
S^{\dagger}(\kk) & = & \frac{t}{4}\left(\begin{array}{ccc}
-\sqrt{3}\left(-1+e^{ik_{1}}\right)e^{ik_{2}} & -e^{i\left(k_{1}+k_{2}\right)}-e^{ik_{2}}+2 & 2\lambda\left(1+e^{ik_{2}}+e^{i\left(k_{1}+k_{2}\right)}\right)\\
-\sqrt{3}\left(-1+e^{ik_{1}}\right) & -2e^{i\left(k_{1}+k_{2}\right)}+e^{ik_{1}}+1 & -2\lambda\left(1+e^{ik_{1}}+e^{i\left(k_{1}+k_{2}\right)}\right)
\end{array}\right),\label{eq:CaNCl}
\end{eqnarray}
Here, $t$ is the strength of coupling between the $s$ or $d_z^2$ orbital at $(\frac{1}{3},\frac{2}{3},x)$ and the $p_y$ orbital at $(0,0,0)$, $\lambda t$ is the strength of coupling between the $s$ orbital at $(\frac{1}{3},\frac{2}{3},x)$ and the $p_z$ orbital at $(0,0,0)$.  
The Hamiltonians of the bipartite lattice $L\oplus \tilde L$ and the $\tilde{L}$ and $L$ are given by Eq.~(\ref{eq:ObSbO1}) and  Eq.~(\ref{eq:ObSbO}), where $\Delta_p=0$ is the onsite energy of $L$ and $\Delta_s>0$ is the onsite energy of $\tilde L$. 
The effective Hamiltonian of $L$ ($p$-bands) is given by
\begin{equation}
    H_{L}(\kk) \thickapprox  - \frac{S(\kk) S^\dagger(\kk)}{\Delta_s}, \label{eq:CaNCl1}
\end{equation}
The explicit form of $H_L$ is given by 
{\footnotesize{}\begin{eqnarray}
H_{L}(\boldsymbol{k}) & \thickapprox & -\frac{S(\boldsymbol{k})S^{\dagger}(\boldsymbol{k})}{\Delta_{s}}\nonumber \\
 & = & \frac{-t^{2}}{4\Delta_{s}}\left(\begin{array}{ccc}
3-3\mathrm{cos}k_{1} & \sqrt{3}\left[\mathrm{cos}k_{2}-\mathrm{cos}\left(k_{1}+k_{2}\right)\right] & \sqrt{3}\lambda\left[\mathrm{cos}k_{2}-\mathrm{cos}\left(k_{1}+k_{2}\right)\right]\\
\sqrt{3}\left[\mathrm{cos}k_{2}-\mathrm{cos}\left(k_{1}+k_{2}\right)\right] & 3+\mathrm{cos}k_{1}-2\left[\mathrm{cos}k_{2}+\mathrm{cos}\left(k_{1}+k_{2}\right)\right] & \lambda\left[-2\mathrm{cos}k_{1}+\mathrm{cos}k_{2}+\mathrm{cos}\left(k_{1}+k_{2}\right)\right]\\
\sqrt{3}\lambda\left[\mathrm{cos}k_{2}-\mathrm{cos}\left(k_{1}+k_{2}\right)\right] & \lambda\left[-2\mathrm{cos}k_{1}+\mathrm{cos}k_{2}+\mathrm{cos}\left(k_{1}+k_{2}\right)\right] & 4\lambda^{2}\left[\frac{3}{2}+\mathrm{cos}k_{1}+\mathrm{cos}k_{2}+\mathrm{cos}\left(k_{1}+k_{2}\right)\right]
\end{array}\right).\label{eq:CaNCl1ha}
\end{eqnarray}}%%
From the MLWF we know that the strength of coupling between the $s$ orbital at $(\frac{1}{3},\frac{2}{3},x)$ and the $p_y$ orbital at $(0,0,0)$ is bigger than that  of coupling between the $s$ orbital at $(\frac{1}{3},\frac{2}{3},x)$ and the $p_z$ orbital at $(0,0,0)$. Meanwhile, they have opposite sign. Thus, we set $\lambda\simeq -0.71$ in the calculation. The band structure of the $L$ sublattice is shown in Fig.~\ref{fig:fig3_mads}(e).
Since the rank of $S(\kk)$ is two at generic momentum, there is a single flat band at $E=0$.
$H_L$ has the same form as Eq.~(\ref{HamilSr1}) with the five parameters given by $t^{pp\sigma} = 3t^2/(8\Delta_s)$, $t^{pp\pi} =-t^2/(8\Delta_s)$, $t_z^{pp\pi}=-\lambda^2t^2/(2\Delta_s)$, $t^p =-\lambda t^2/\Delta_s$, and $\Delta=-3t^2/(8\Delta_s)(2\lambda^2-1)$. 
Since the $S$-matrix only has two free parameters, the yielded five parameters will satisfy three conditions which are $t_p^2=16t^{pp\pi}t_z^{pp\pi}$, $3t_p^2=32(\Delta+2t^{pp\pi})t^{pp\pi}$, and $t^{pp\sigma} + 3 t^{pp\pi}=0$.
Note that the last condition $t^{pp\sigma} + 3 t^{pp\pi}=0$, which is responsible for the flat band in the path $\rm \Gamma M$ was also derived in Eq.~(\ref{app:eq:CaNClMLWFKzero}) from the MLWF model.

The bands obtained from first-principle calculations are more complicated than the $S$-matrix band. For example, the $p_z$-band has a non-negligible bandwidth along the $k_3$ direction, implying that the hopping terms between $p_z$ orbitals in different layers are not negligible. 
Thus, a more realistic model can be obtained by adding these hopping terms to the $S$-matrix model.

The explanations that we have developed above can be applied to some other compounds, such as 
$\mathrm{Ca_{2}NBr}$ [\icsdweb{62556}, SG 166 (\sgsymb{166})],
$\mathrm{Sr_{2}NCl}$ [\icsdweb{172595}, SG 166 (\sgsymb{166})],
$\mathrm{Sr_{2}NBr}$ [\icsdweb{172601}, SG 166 (\sgsymb{166})], 
$\mathrm{Ca_{2}N}$ [\icsdweb{90632}, SG 166 (\sgsymb{166})].  Note that all of these materials are part of the curated list provided in Appendix~\ref{app:listcurated}.

\subsection{$\rm{Rb_{2}CaH_{4}}$}\label{RbCaH}

\begin{figure}[h]
\includegraphics[width=18cm]{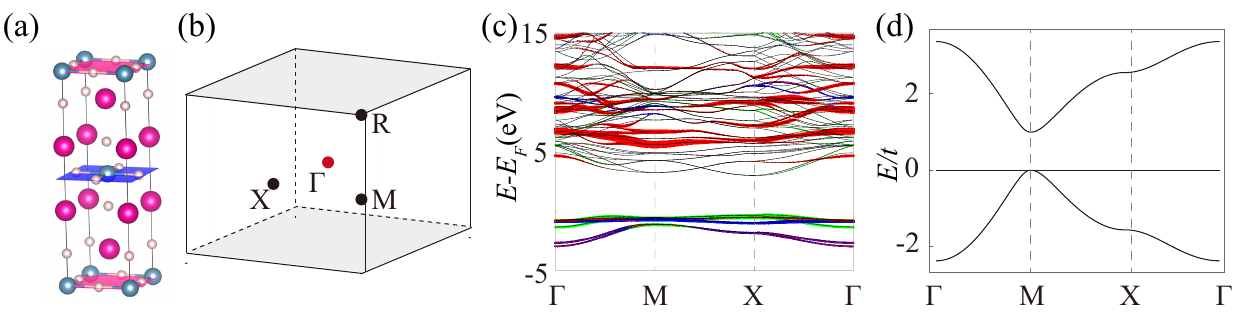}
\caption{
 (a) The crystal structure of the unit cell of $\mathrm{Rb_{2}CaH_{4}}$, where the Ca, H, and Rb atoms are represented by the cyan, pink, and purple balls, respectively. The Ca and H atoms form two Lieb lattices on the (001) Miller plane, as indicated by the pink and blue planes at $z=0$ and $z=0.5c$ (where $c$ is the lattice constant along the (001) direction of the unit cell).(b) The 3D bulk BZ of $\mathrm{Rb_{2}CaH_{4}}$. (c) Orbital-projected band structure along the in-plane high-symmetry $k$-paths of $\mathrm{Rb_{2}CaH_{4}}$ in the absence of spin-orbit coupling. The weight of $s$ orbitals on H atoms sitting at $4c$ and $4e$ are highlighted in blue and green, respectively, and the weight of $s$ and $d_z^2$orbitals on Ca atoms at $2e$ is highlighted in red. 
(d) The band structure along the in-plane high-symmetry $k$-paths of the Lieb lattice given by the $S$-matrix method with $\delta=t$.
\label{fig:fig8_mads}}
\end{figure}

The crystal structure of $\mathrm{Rb_{2}CaH_{4}}$ [\icsdweb{65196}, SG 139 (\sgsymb{139})] is shown in Fig.~\ref{fig:fig8_mads}(a). The $4e$ and $2a$ Wyckoff positions are occupied by the Rb and Ca atoms, respectively. H atoms occupy two non-equivalent Wyckoff positions, $4c$ and $4e$. Ca atoms and the H atoms at $4e$ form two Lieb sublattices in one unit cell, which could host flat topological bands \cite{ma2020}, as shown in Fig.~\ref{fig:fig8_mads}(a). 

From the band structures calculations (without SOC) in the \webflatband, there are a set of flat bands near the Fermi level.
As plotted in Fig.~\ref{fig:fig8_mads}(c), 
we have analyzed the orbital-projected band structure of $\mathrm{Rb_{2}CaH_{4}}$ and found the flat bands are mainly composed by the $s$ orbitals on H atoms sitting at the $4c$ and $4e$ positions. We also find that all the four bands contributed by the $s$ orbitals on H atoms at $4e$ are flat in the whole BZ and they are flat atomic bands, which can be explained by the weak hybridization between H atoms at $4e$ and the other atoms.
While for the four bands which are contributed by the $s$ orbitals on H atoms sitting at $4c$, only two of them are flat and the other two bands are dispersive, implying a larger kinetic energy of the $s$ electron at $4c$.
In the following, we discard the flat atomic bands contributed by the H atoms at $4e$ and explain the origin of the other flat bands contributed by the H atoms at $4c$ using $S$-matrix method.

In the crystal structure of $\mathrm{Rb_{2}CaH_{4}}$, the distance between two neighboring Lieb sublattices is so large (7.42\AA) that the coupling between them is weak. Thus, we can simplify the 3D crystal lattice in $\mathrm{Rb_{2}CaH_{4}}$ with a 2D Lieb lattice, which is indicated by the pink plane in Fig. \ref{fig:fig8_mads}(a). For the 2D lieb lattice, we refer to the two sublattices formed by putting $s$ orbitals at H sublattice \ie  \{$\left(\frac{1}{2},0\right)$, $\left(0,\frac{1}{2}\right)$\} and putting $s$ or $d_z^2$ orbitals at Ca sublattice \ie $\left(0,0\right)$ as $L$ and $\tilde{L}$, respectively. 
Then the $S$ matrix can be written as
\begin{eqnarray}
S^{\dagger}(\boldsymbol{k})	& = & t\left(\begin{array}{cc}
1+e^{-ik_{1}} & 1+e^{-ik_{2}}
\end{array}\right)
\end{eqnarray}
where $t$ is the strength of the coupling between the $s$ orbital at $\left(\frac{1}{2},0\right)$ and the $s$ or $d_z^2$ orbital at $\left(0,0\right)$, ;$(k_1,k_2)$ are the coordinates of the momentum $\kk= \sum_{i=1}^2 k_i \bb_i$, where $\bb_i$ is the reciprocal lattice vector that corresponds to the in-plane lattice vectors  $\boldsymbol{a}_1=a(1,0)$ and $\boldsymbol{a}_2=a(0,1)$ (where $a$ is the in-plane lattice constant). We assume the on-site energy of $s$ orbital on H is 0 and that of the Ca atom is $\Delta$. 
Then the total Hamiltonian of the bipartite lattice reads
\begin{equation}
    H_{L\oplus\tilde{L}}(\boldsymbol{k})=\left(\begin{array}{cc}
    \Delta & S^{\dagger}(\kk) \\
    S(\kk) &  0_{2\times 2} \\
    \end{array}\right)\;.\label{eq:Rb2CaH4}
\end{equation}

As the band structure of this Lieb lattice plotted in Fig.~\ref{fig:fig8_mads}(d), there is a flat band at 0, which corresponds to the flat bands contributed by the $s$ orbitals of H at $4c$ in the first-principles calculations. 
Note that the weak band splitting of the dispersive bands of energy range from $-5 {\rm eV}$ to $-1{\rm eV}$ comes from the weak coupling between the two Lieb lattices in the $(001)$ direction.
The explanations that we have developed above can be applied to some other compounds, such as 
$\mathrm{Cs_{2}MgH_{4}}$ [\icsdweb{162261}, SG 139 (\sgsymb{139})],
and 
$\mathrm{K_{2}MgH_{4}}$ [\icsdweb{68358}, SG 139 (\sgsymb{139})].

\subsection{$\rm{KAg[CN]_2}$}\label{KAgCN}

\begin{figure}[h]
\includegraphics[width=18cm]{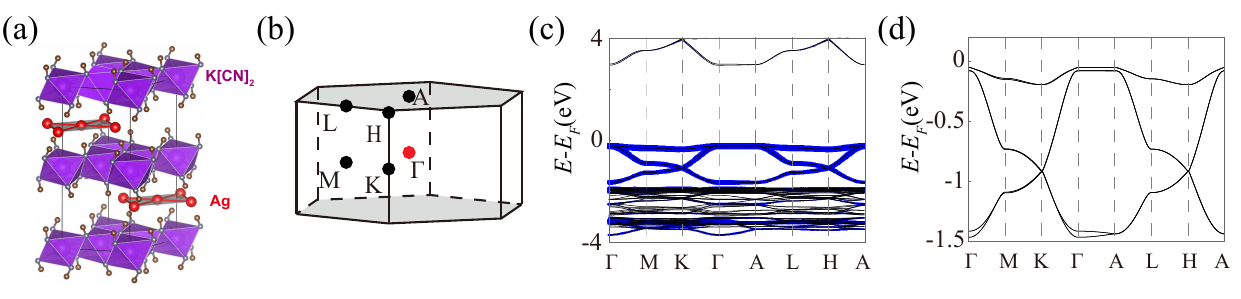}
\caption{(a) The crystal structure of $\rm{KAg[CN]_2}$ which is a stacking of alternating molecular layer of \ch{K[CN]2} and Kagome layer of Ag atoms (in red color). (b) The bulk BZ of $\rm{KAg[CN]_2}$ with its high-symmetry $k$ points. (c) The orbital-projected band structure of $\rm{KAg[CN]_2}$ along the high-symmetry lines from the first-principle calculation without SOC. The bands contributed by the $s$ and $d_{z^2}$ orbitals on Ag atoms are highlighted in blue. (d) The band structure of $\rm{KAg[CN]_2}$ from the first-principle calculation with SOC (zoomed over the energy range $-1.5{\rm eV}$ and $0.5{\rm eV}$), where the minimal direct band gap between the flat bands and the dispersive bands is about 25 meV.
\label{fig:fig9_mads}}
\end{figure}

As shown in the Fig.~\ref{fig:fig9_mads}(a), the crystal structure of \ch{KAg[CN]2} [\icsdweb{30275}, SG 163 (\sgsymb{163})] is a layered structure of alternating molecular layers of \ch{K[CN]2} and approximate Kagome layers of Ag atoms. The Ag atoms sitting at $6h$ Wyckoff position form two layers of {\it approximate} Kagome sublattices per unit cell on the $(001)$ plane as indicated by the red spheres in Fig.~\ref{fig:fig9_mads}(a).
We emphasize that generic $6h$ positions do not form a Kagome lattice guaranteed by symmetries. However, the actual positions of the Ag atoms form an approximate Kagome lattice.

As shown in the band structure without SOC in Fig.~\ref{fig:fig9_mads}(c) and on the \webflatband, there are two spinless flat bands near the Fermi level and the flat bands are mainly contributed by the hybridized orbital of the $s$ and $d_{z^2}$ orbitals on Ag atoms. 
The distance between two neighboring Kagome sublattices is so large (8.90\AA) that the coupling between them is weak and hence all the bands along the $\Gamma$A path are flat. Thus, we can simplify the 3D crystal lattice in \ch{KAg[CN]2} to a 2D Kagome lattice. 
Since the $s$ and $d_{z^2}$ orbitals form the same single-value irrep $A$ of the point group symmetry $C_2$ at the Wyckoff position $6h$, they in general hybridize with each other and form two hybridized orbitals.
We model the six bands in the energy range $-1.5\sim0$eV in Fig.~\ref{fig:fig9_mads}(c) by a single hybridized orbital - the one with lower energy - at every Ag atom.
In the following we refer to this hybridized orbital as $s$ orbital because it transforms under the symmetries as an $s$ orbital.
The flat bands of $s$-orbital Kagome lattice have been fully studied in several previous works \cite{Mielke_1992,PhysRevB.78.125104,Liu_2014,Barreteau_2017,ma2020}. In Appendix~\ref{app:smatrix}, we have also constructed the effective Hamiltonian of the $s$-orbital Kagome model and explained the origin of its flat bands using an alternative $S$-matrix method. 

In the presence of SOC \cite{ma2020}, the flat bands  of the $s$-orbital Kagome lattice in 2D are gapped from the dispersive bands in the whole BZ and have a  strong topology \cite{QuantumChemistry}, \ie the $\mathbb{Z}_2$ topological insulator phase protected by time-reversal symmetry. 
For the 3D material \ch{KAg[CN]2}, as shown in Fig.~\ref{fig:fig9_mads}(d), the SOC also fully gaps the flat bands from the dispersive bands.
A representation analysis shows that the SOC-gapped flat bands host at least a fragile topology: The irreps of the flat bands at the high-symmetry $k$ points, $\{\bar A_4\bar A_4,\bar{\Gamma}_4\bar{\Gamma}_5+\bar{\Gamma}_6\bar{\Gamma}_7,\bar H_4\bar H_5+\bar H_6, \bar K_4\bar K_5+\bar K_6,\bar L_2\bar L_2,\bar M_3\bar M_4+\bar M_5\bar M_6\}$, can only be written as a difference of two BRs, \eg  $\bar E_1@c$$\oplus$$^1\bar{E}^2\bar{E}@d$$\ominus$$\bar{E_1}@a$, but not a sum of BRs. 
The fragile topology has also been identified in Ref.~\cite{song_fragile_2020}.
However, here we argue that the topology of the SOC-gapped flat bands might be a {\it stable higher-order topology with hinge states} rather than a fragile topology.
According to Ref. \cite{ma2020}, the single-layer Kagome lattice with SOC is a 2D $\mathbb{Z}_2$ topological insulator. 
Then there will be two layers of 2D topological insulators per unit cell occupying the $6h$ positions.
Because the $6h$ positions include all the $C_{2x}$-axes (and equivalent axes due to $C_{3z}$ and inversion symmetries), all the $C_{2x}$-axes are occupied by the 2D topological insulators.
This state is nothing but a layer construction of the $C_2$ and time-reversal protected higher-order topological insulator \cite{fang2019new,song_quantitative_2018}. 

\subsection{$\rm{RbMo_{3}S_3}$}\label{RbMoS}

\begin{figure}[h]
\includegraphics[width=18cm]{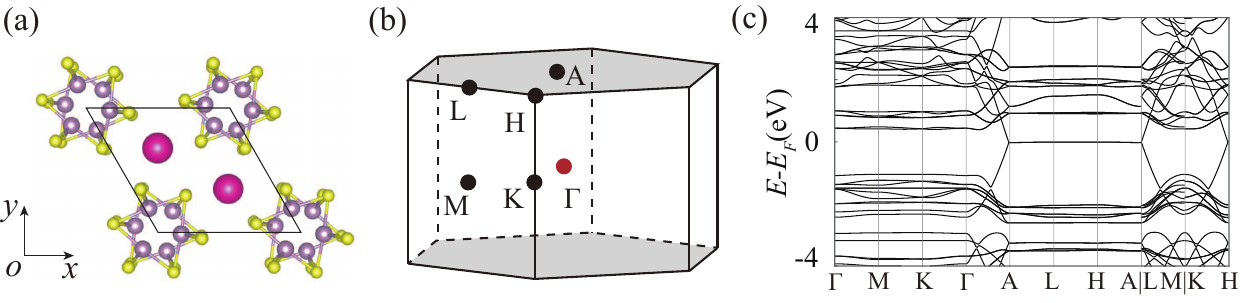}
\caption{(a) The crystal structure of $\mathrm{RbMo_{3}S_3}$. The Mo (in purple) and S atoms (in yellow) form a 1D atomic chain along the $z$ direction. (b) The bulk BZ of $\mathrm{RbMo_{3}S_3}$. (c) Band structure of $\mathrm{RbMo_{3}S_3}$ from first-principle calculations where spin-orbit coupling is ignored.
\label{fig:fig4_mads}}
\end{figure}

$\mathrm{RbMo_{3}S_3}$ [\icsdweb{644175}, SG 176 (\sgsymb{176})] possesses flat bands at the Fermi level. The crystal structure of $\mathrm{RbMo_{3}S_3}$ is shown in Fig.~\ref{fig:fig4_mads}(a), where Mo and S atoms form a quasi-1D atomic chain along the $z$ direction. The band structure of $\mathrm{RbMo_{3}S_3}$ is shown in Fig.~\ref{fig:fig4_mads}(c). There are flat bands on the $k_z=0$ and $k_z=\pi$ planes located at the Fermi level. These flat bands are mainly composed of the $E_{1g}$ orbitals of Mo atoms. By constructing the MLWF, we find that the maximal hopping strength between the $E_{1g}$ orbitals of two neighboring chains is about 0.01eV, which is much smaller than the maximal intra-chain hopping (0.76eV). 
Hence, the flatness of bands at the Fermi level is due to the weak inter-chain coupling of the quasi-1D crystal structure.

\clearpage

\section{Examples of magnetic flat-band materials}\label{app:magflat}

In the present work, flat-band materials are obtained from the \webTQC\ by analyzing their crystal and band structures. As discussed in Appendix \ref{app:database}, all the first-principles calculations performed on the \webTQC\ are forced to be paramagnetic, \ie without the on-site spin polarization. However, the paramagnetic phase might not be the ground state for all flat-band materials. In Appendix \ref{app:allflatbands}, we have tagged the materials whose ground states are likely to be magnetic, reporting  their magnetic properties found on the \webmaterialsproject. To theoretically predict the magnetic properties of materials, we need to compare the total energy of different magnetic configurations in the ab-initio calculations. 

\begin{figure*}
    \centering
    \includegraphics[width=18cm]{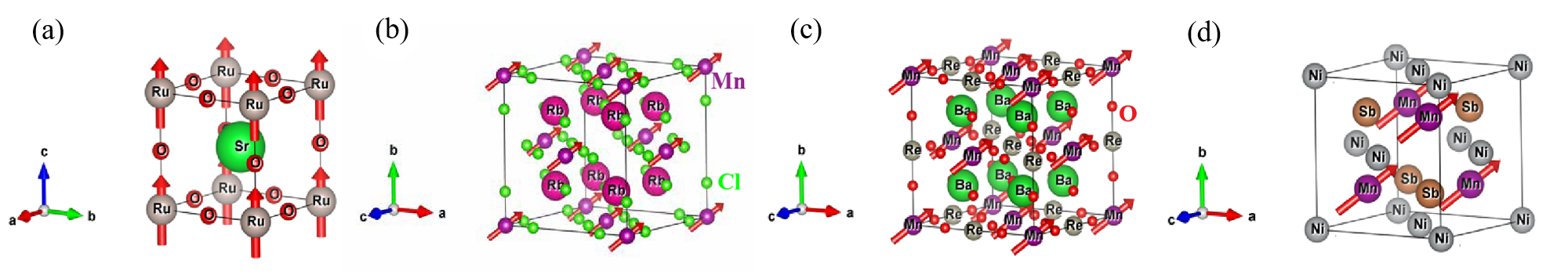}
    \caption{The experimental crystal and magnetic structures of the flat-band materials (a) SrRuO$_3$[\icsdweb{69360}, SG 221 (\sgsymb{221})], (b) Rb$_2$MnCl$_6$[\icsdweb{9347} SG 225 (\sgsymb{225})], (c) Ba$_2$MnReO$_6$[\icsdweb{4169} SG 225 (\sgsymb{225})] and (d) NiMnSb [\icsdweb{54255}, SG 216 (\sgsymb{216}).}
    \label{fig:magflat}
\end{figure*}

In this appendix, we showcase four flat-band materials with experimentally known ferromagnetic ground states and compare the total energies of the paramagnetic and ferromagnetic states obtained from the first-principles calculations. The four materials are  SrRuO$_3$[\icsdweb{69360}, SG 221 (\sgsymb{221})], Rb$_2$MnCl$_6$[\icsdweb{9347} SG 225 (\sgsymb{225})], Ba$_2$MnReO$_6$[\icsdweb{4169} SG 225 (\sgsymb{225})] and NiMnSb [\icsdweb{54255}, SG 216 (\sgsymb{216}). Their experimental crystal and magnetic structures are shown in Fig.~\ref{fig:magflat}.

\begin{figure}[h]
\centering
\begin{tabular}{c c}
\scriptsize{$\rm{Sr} \rm{Ru} \rm{O}_{3}$ - \icsdweb{69360} - SG 221 (\sgsymb{221}) - NM - $U=0$} & \scriptsize{$\rm{Sr} \rm{Ru} \rm{O}_{3}$ - \icsdweb{69360} - SG 221 (\sgsymb{221}) - NM - $U=3$}\\
\includegraphics[width=0.45\textwidth,angle=0]{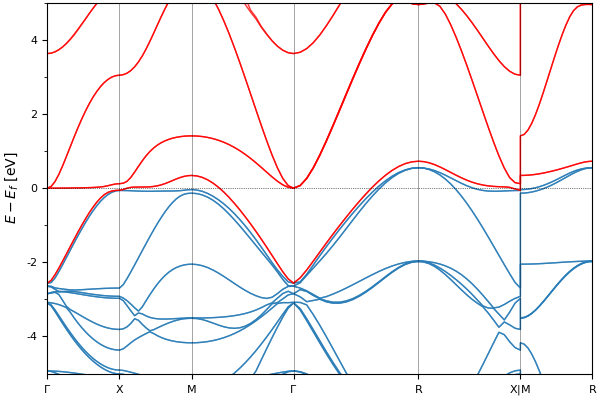} & \includegraphics[width=0.45\textwidth,angle=0]{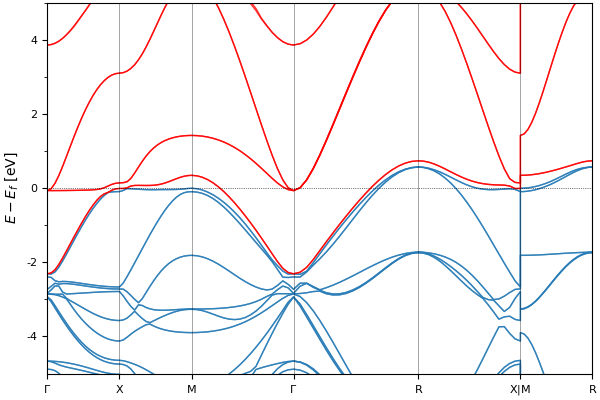}\\
\scriptsize{$\rm{Sr} \rm{Ru} \rm{O}_{3}$ - \icsdweb{69360} - SG 221 (\sgsymb{221}) - FM - $U=0$} & \scriptsize{$\rm{Sr} \rm{Ru} \rm{O}_{3}$ - \icsdweb{69360} - SG 221 (\sgsymb{221}) - FM - $U=3$}\\
\includegraphics[width=0.45\textwidth,angle=0]{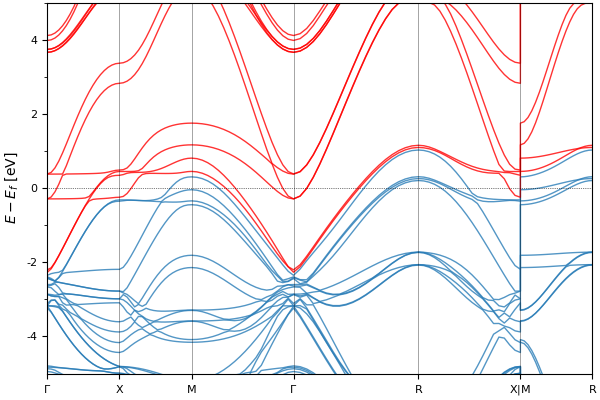} & \includegraphics[width=0.45\textwidth,angle=0]{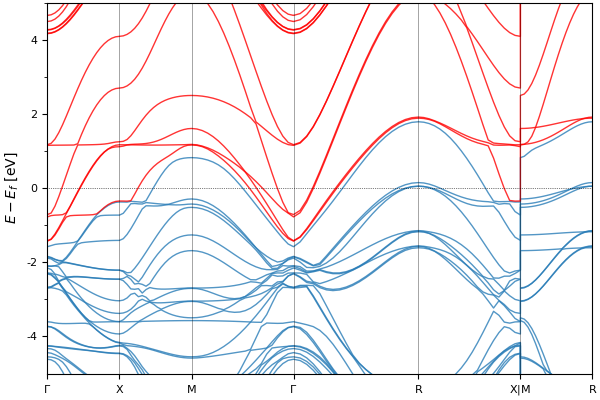}\\
\end{tabular}
\caption{Band structure for SrRuO$_3$[\icsdweb{69360}, SG 221 (\sgsymb{221})] obtained with an on-site Hubbard $U$ value for $d$ or $f$ electron in the mean-field approximation and including the spin-orbit coupling. Plots in the upper panel are for paramagnetic calculations (NM) while those of the lower panel are for ferromagnetic calculations (FM). Here we have considered both $U=0$ (left panel) and $U=3$ (right panel).}
\label{fig:app:magrusro}
\end{figure}

\begin{figure}[h]
\centering
\begin{tabular}{c c}
\scriptsize{$\rm{Rb}_{2} \rm{Mn} \rm{Cl}_{6}$ - \icsdweb{9347} - SG 225 (\sgsymb{225}) - NM - $U=0$} & \scriptsize{$\rm{Rb}_{2} \rm{Mn} \rm{Cl}_{6}$ - \icsdweb{9347} - SG 225 (\sgsymb{225}) - NM - $U=4$}\\
\includegraphics[width=0.45\textwidth,angle=0]{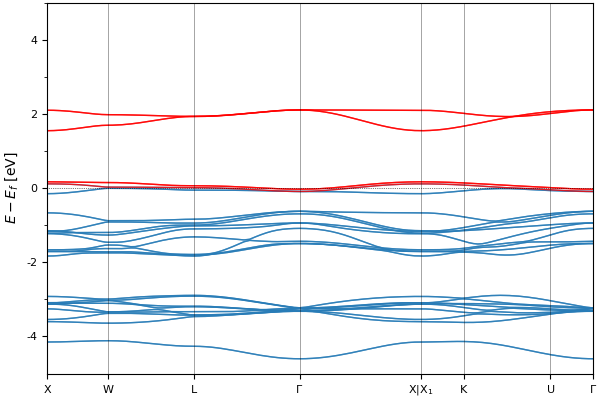} & \includegraphics[width=0.45\textwidth,angle=0]{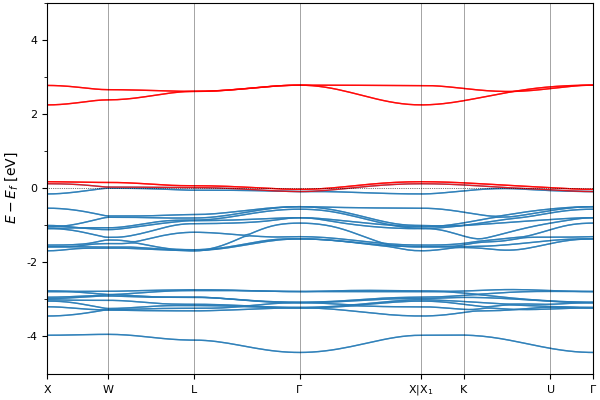}\\
\scriptsize{$\rm{Rb}_{2} \rm{Mn} \rm{Cl}_{6}$ - \icsdweb{9347} - SG 225 (\sgsymb{225}) - FM - $U=0$} & \scriptsize{$\rm{Rb}_{2} \rm{Mn} \rm{Cl}_{6}$ - \icsdweb{9347} - SG 225 (\sgsymb{225}) - FM - $U=4$}\\
\includegraphics[width=0.45\textwidth,angle=0]{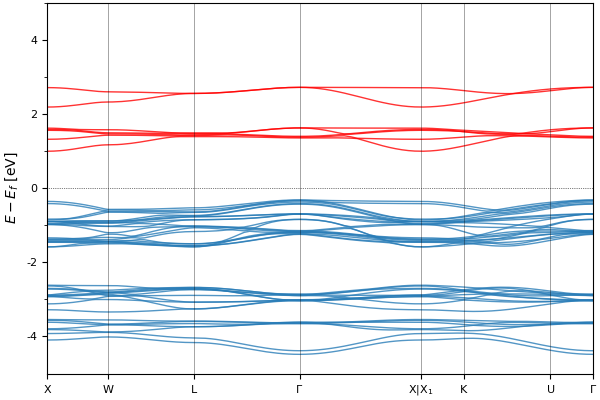} & \includegraphics[width=0.45\textwidth,angle=0]{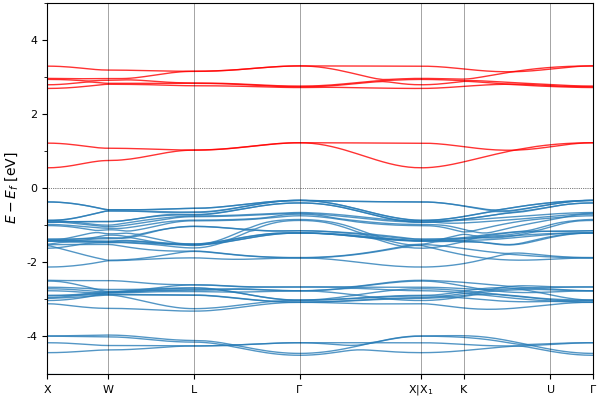}\\
\end{tabular}
\caption{Band structure for $\rm{Rb}_{2} \rm{Mn} \rm{Cl}_{6}$[\icsdweb{9347}, SG 225 (\sgsymb{225})] obtained with an on-site Hubbard $U$ value for $d$ or $f$ electron in the mean-field approximation and including the spin-orbit coupling. Plots in the upper panel are for paramagnetic calculations (NM) while those of the lower panel are for ferromagnetic calculations (FM). Here we have considered both $U=0$ (left panel) and $U=4$ (right panel).}
\label{fig:app:magrbmncl}
\end{figure}

\begin{figure}[h]
\centering \begin{tabular}{c c}
\scriptsize{$\rm{Ba}_{2} \rm{Mn}\rm{Re} \rm{O}_{6}$ - \icsdweb{4169} - SG 225 (\sgsymb{225}) - NM - $U=0$} & \scriptsize{$\rm{Ba}_{2} \rm{Mn}\rm{Re} \rm{O}_{6}$ - \icsdweb{4169} - SG 225 (\sgsymb{225}) - NM - $U=4$}\\
\includegraphics[width=0.45\textwidth,angle=0]{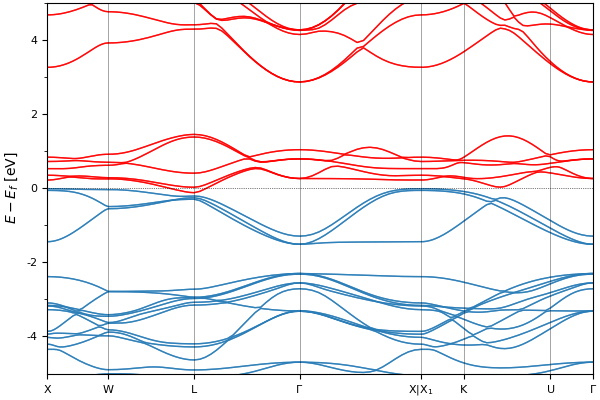} & \includegraphics[width=0.45\textwidth,angle=0]{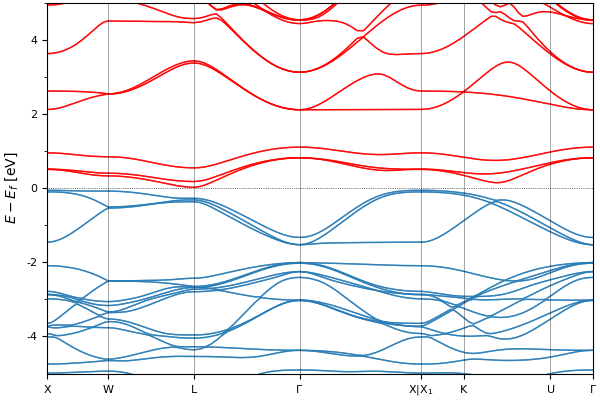}\\
\scriptsize{$\rm{Ba}_{2} \rm{Mn}\rm{Re} \rm{O}_{6}$ - \icsdweb{4169} - SG 225 (\sgsymb{225}) - FM - $U=0$} & \scriptsize{$\rm{Ba}_{2} \rm{Mn}\rm{Re} \rm{O}_{6}$ - \icsdweb{4169} - SG 225 (\sgsymb{225}) - FM - $U=4$}\\
\includegraphics[width=0.45\textwidth,angle=0]{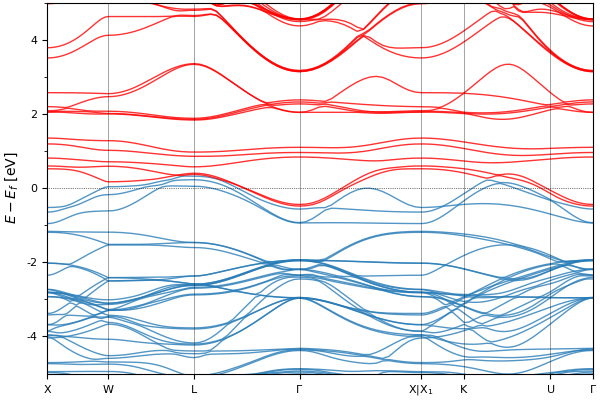} & \includegraphics[width=0.45\textwidth,angle=0]{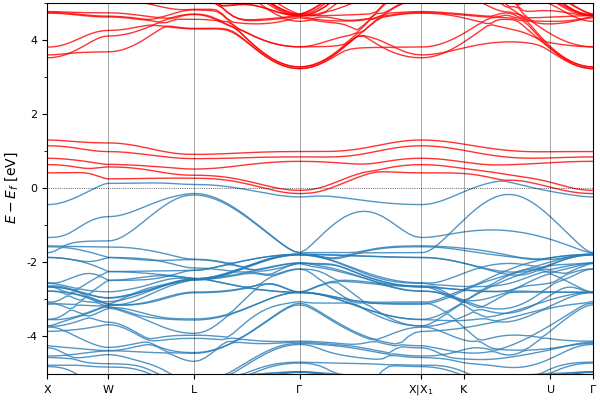}\\
\end{tabular}
\caption{Band structure for $\rm{Ba}_{2} \rm{Mn}\rm{Re} \rm{O}_{6}$[\icsdweb{4169}, SG 225 (\sgsymb{225})] obtained with an on-site Hubbard $U$ value for $d$ or $f$ electron in the mean-field approximation and including the spin-orbit coupling. Plots in the upper panel are for paramagnetic calculations (NM) while those of the lower panel are for ferromagnetic calculations (FM). Here we have considered both $U=0$ (left panel) and $U=4$ (right panel).}
\label{fig:app:magbamnreo}
\end{figure}

\begin{figure}[h]
\centering
\begin{tabular}{c c}
\scriptsize{$\rm{Ni} \rm{Mn} \rm{Sb}$ - \icsdweb{54255} - SG 216 (\sgsymb{216}) - NM - $U=0$} & \scriptsize{$\rm{Ni} \rm{Mn} \rm{Sb}$ - \icsdweb{54255} - SG 216 (\sgsymb{216}) - NM - $U=4$}\\
\includegraphics[width=0.45\textwidth,angle=0]{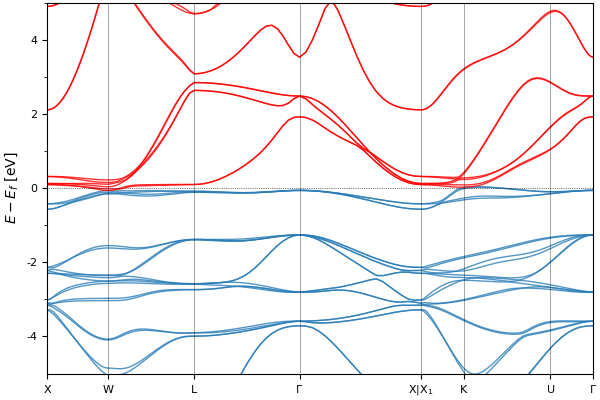} & \includegraphics[width=0.45\textwidth,angle=0]{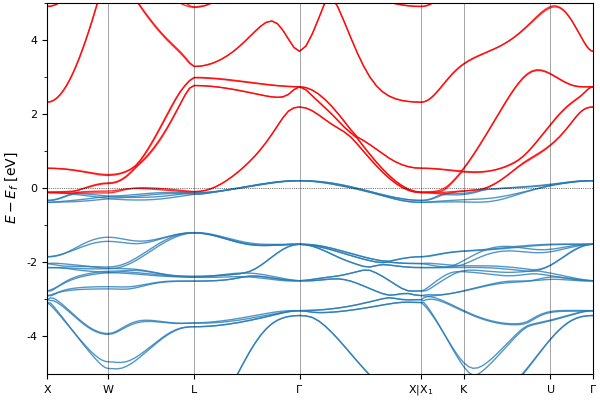}\\
\scriptsize{$\rm{Ni} \rm{Mn} \rm{Sb}$ - \icsdweb{54255} - SG 216 (\sgsymb{216}) - FM - $U=0$} & \scriptsize{$\rm{Ni} \rm{Mn} \rm{Sb}$ - \icsdweb{54255} - SG 216 (\sgsymb{216}) - FM - $U=4$}\\
\includegraphics[width=0.45\textwidth,angle=0]{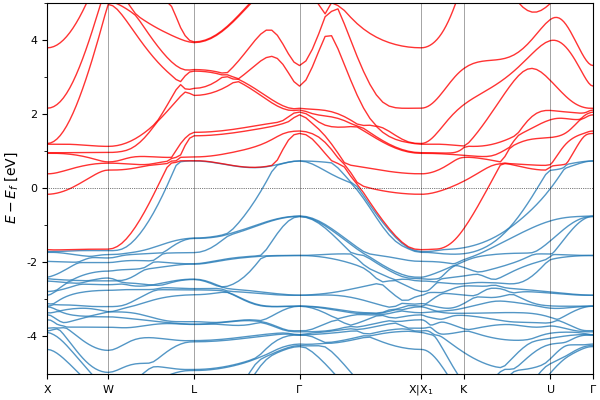} & \includegraphics[width=0.45\textwidth,angle=0]{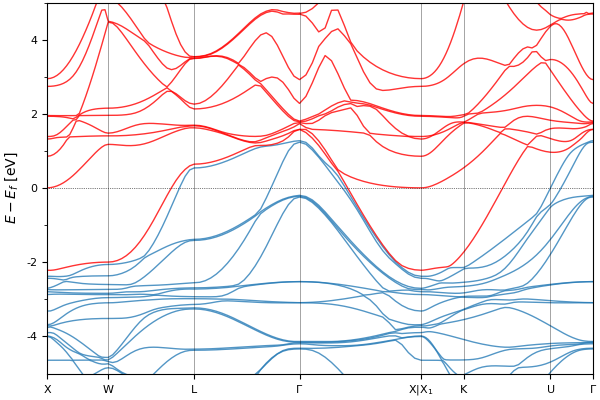}\\
\end{tabular}
\caption{Band structure for NiMnSb[\icsdweb{54255}, SG 216 (\sgsymb{216})] obtained with an on-site Hubbard $U$ value for $d$ or $f$ electron in the mean-field approximation and including the spin-orbit coupling. Plots in the upper panel are for paramagnetic calculations (NM) while those of the lower panel are for ferromagnetic calculations (FM). Here we have considered both $U=0$ (left panel) and $U=4$ (right panel).}
\label{fig:app:magnimnsb}
\end{figure}

For the ferromagnetic calculations we adopt the GGA+U method (LDAUTYPE=2, \ie the simplified (rotationally invariant) approach to the LSDA+U method \cite{LSDA+U}), which considers the on-site Hubbard $U$ value of $d$ or $f$ electron in the mean-field approximation. 
By comparing the total energy of the non-magnetic and ferromagnetic phases for different $U$ values (varying from $0$ to $5{\rm eV}$) with and without SOC, as provided in the Tables~\ref{tab:magNiMnSb}-\ref{tab:magBa2MnReO6}, we find that the ferromagnetic phases always have a lower energy and hence the paramagnetic phase is not the ground state of these materials. The theoretically obtained magnetic moments of these four materials are in \emph{good agreement with the experimental results}. 
Therefore, their magnetic ground states could be captured by the GGA+U method. Here we take $U=3~$eV for Ru and $U=4~$eV for Mn atom to meet the experimental results of magnetic structure.
As shown in Figs.~\ref{fig:app:magrusro}-\ref{fig:app:magnimnsb}, with the on-site Hubbard $U$ value of $d$ or $f$ electron in the  mean field approximation is considered, the flat bands of the selected materials in the paramagnetic calculations are preserved and remain near the Fermi level. For example, in the ferromagnetic calculations of SrRuO$_3$[\icsdweb{69360}, SG 221 (\sgsymb{221})] shown in Fig.~\ref{fig:app:magrusro} (for two different values of $U$ and providing both the paramagnetic and the ferromagnetic band structures), the flat bands along high symmetry lines $\Gamma \rm{X}$ and $\rm{M} \rm{R}$ at Fermi level are not settling at $E=0$ anymore but split into two sets of flat bands near the Fermi level. 
Similar conclusions hold for the three other ferromagnetic materials that we have considered in this Appendix.
While a high-throughput ferromagnetic calculation is out of reach, these examples provide compelling evidence that the flat bands found in the paramagnetic calculations discussed in the results section of the manuscript, will remain in the ferromagnetic calculations.

\begin{table*}[h]
\caption[The total energy comparison for paramagnetic and ferromagnetic states of SrRuO$_3$]{The total energy comparison for paramagnetic and ferromagnetic states of SrRuO$_3$[\icsdweb{69360}, SG 221 (\sgsymb{221})].  In the FM case the magnetization points along the [001] direction of the conventional unit cell. Here, we only provide the $z$-component of the magnetic moment $m_z$ in the last column. The ground atste of this material is reported to be ferromagnetic with moment 2.8 $\mu B/Ru$ \cite{longo1968magnetic,bushmeleva2006evidence,cao1997thermal}. Our calculations show that the ferromagnetic (FM) state is more stable than the nonmagnetic (NM) state.
}\label{tab:magRuSrO3}
\vspace{0.2cm}
\renewcommand\arraystretch{1.5}
\begin{tabular}{p{1.4cm}<{ \centering}p{1.4cm}<{ \centering}p{3.4cm}<{ \centering}p{1.4cm}<{ \centering}p{1.4cm}<{ \centering}p{1.4cm}<{ \centering}p{1.4cm}<{ \centering}p{1.4cm}<{ \centering}p{1.4cm}<{ \centering}}
\hline
   &&&  {\bf U=0} &  {\bf U=1} &  {\bf U=2} &  {\bf U=3} &  {\bf U=4} &  {\bf U=5} \\
\hline
\multirow{3}{*}{Non-SOC} & NM & Total energy (eV) &0 &  1.046 &  2.074 &  3.088 &  4.091 &  5.06 \\
\cline{2-9}
 & \multirow{2}{*}{FM} & Total energy (eV) &-0.079&  0.839&  1.730&  2.608&  3.464&  4.271 \\
\cline{3-9}
 & & Mag-mom ($\mu$B) &   1.594 &   2.250 &   2.254 &   2.262 &   2.629 &   2.733 \\
\hline
\multirow{3}{*}{SOC} & NM & Total energy (eV) &0.000& 1.032& 2.030& 2.980& 3.828& 4.585 \\
\cline{2-9}
 & \multirow{2}{*}{FM} & Total energy (eV) &   -0.078& 0.841& 1.731& 2.602& 3.451& 4.265 \\
\cline{3-9}
 & &  $m_z$($\mu$B) &1.6077 &   2.118 &   2.231 &   2.244 &   2.259 &   2.808 \\
\hline
\end{tabular}  
 \end{table*}

 \begin{table*}[h]
\caption[The total energy comparison for paramagnetic and ferromagnetic states of Rb$_2$MnCl$_6$]{The total energy comparison for paramagnetic and ferromagnetic states of Rb$_2$MnCl$_6$[\icsdweb{9347}, SG 225 (\sgsymb{225})].  In the FM case the magnetization points along the [1,1,$\bar1$] direction of the conventional unit cell. Here, we only provide the $x$-component of the magnetic moment $m_x$ in the last column. The lattice victors of the primary unit cell are $\boldsymbol{a_{1}}=a\left(\begin{array}{ccc} 0 & 1/2 & 1/2\end{array}\right)$, $\boldsymbol{a_{2}}=a\left(\begin{array}{ccc} 1/2 & 0 & 1/2\end{array}\right)$ and $\boldsymbol{a_{3}}=a\left(\begin{array}{ccc} 1/2 & 1/2 & 0\end{array}\right)$ with $a$ is the lattice parameter. The total magnetic moment is $3~\mu B/Mn$ which is in good agreement with the experimental $3.9~\mu B/Mn$ \cite{lalancette1972crystal,faizan2019electronic}. The results show that the ferromagnetic (FM) state is more stable than the nonmagnetic (NM) state.
}\label{tab:magRb2MnCl6}
\vspace{0.2cm}
\renewcommand\arraystretch{1.5}
\begin{tabular}{p{1.4cm}<{ \centering}p{1.4cm}<{ \centering}p{3.4cm}<{ \centering}p{1.4cm}<{ \centering}p{1.4cm}<{ \centering}p{1.4cm}<{ \centering}p{1.4cm}<{ \centering}p{1.4cm}<{ \centering}p{1.4cm}<{ \centering}}
\hline
   &&&  {\bf U=0} &  {\bf U=1} &  {\bf U=2} &  {\bf U=3} &  {\bf U=4} &  {\bf U=5} \\
\hline
\multirow{3}{*}{Non-SOC} & NM & Total energy (eV) &   0.000& 1.219& 2.433& 3.641& 4.841& 6.031 \\
\cline{2-9}
 & \multirow{2}{*}{FM} & Total energy (eV) &  -1.793& -1.092& -0.419& 0.224& 0.833& 1.407 \\
\cline{3-9}

 & & Mag-mom ($\mu$B) &  3.000  & 3.000 &3.000 &2.999 &3.000 &3.000 \\
\hline

\multirow{3}{*}{SOC} & NM & Total energy (eV) &0.000 &  1.051 &  2.040  &  2.983 &  3.845 &  4.519 \\
\cline{2-9}
 & \multirow{2}{*}{FM} & Total energy (eV) &  -1.807 & -1.102 & -0.425 &0.222 &0.837 &1.417 \\
\cline{3-9}

 & &  $m_x$($\mu$B) & 1.732  & 1.732 &1.732 &1.732 &1.732 &1.732 \\
\hline  
\end{tabular}  
 \end{table*}
 
\begin{table*}[h]
\caption[The total energy comparison for paramagnetic and ferromagnetic states of Ba$_2$MnReO$_6$]{The total energy comparison for paramagnetic and ferromagnetic states of Ba$_2$MnReO$_6$[\icsdweb{4169}, SG 225 (\sgsymb{225})].  In the FM case the magnetization points along the [1,1,$\bar1$] direction of the conventional cell. Here, we only give the $x$-component of the magnetic moment $m_x$ in the last column. The lattice victors of the primary unit cell are $\boldsymbol{a_{1}}=a\left(\begin{array}{ccc} 0 & 1/2 & 1/2\end{array}\right)$, $\boldsymbol{a_{2}}=a\left(\begin{array}{ccc} 1/2 & 0 & 1/2\end{array}\right)$ and $\boldsymbol{a_{3}}=a\left(\begin{array}{ccc} 1/2 & 1/2 & 0\end{array}\right)$ with $a$ is the lattice parameter. The theoretical magnetic moment is $4~\mu B/Mn$ which is in good agreement with the experimental $3.9~\mu B/Mn$ \cite{khattak1975magnetic}. 
It shows that the ferromagnetic (FM) state is more stable than the nonmagnetic (NM) state.
}\label{tab:magBa2MnReO6}
\vspace{0.2cm}
\renewcommand\arraystretch{1.5}
\begin{tabular}{p{1.4cm}<{ \centering}p{1.4cm}<{ \centering}p{3.4cm}<{ \centering}p{1.4cm}<{ \centering}p{1.4cm}<{ \centering}p{1.4cm}<{ \centering}p{1.4cm}<{ \centering}p{1.4cm}<{ \centering}p{1.4cm}<{ \centering}}
\hline
   &&&  {\bf U=0} &  {\bf U=1} &  {\bf U=2} &  {\bf U=3} &  {\bf U=4} &  {\bf U=5} \\
\hline
\multirow{3}{*}{Non-SOC} &         NM & Total energy (eV) &       0.000 & 0.902 & 1.760  & 2.572 &3.337 & 4.056 \\
\cline{2-9}
 & \multirow{2}{*}{FM} & Total energy (eV) &                       -1.331& -0.869& -0.502& -0.201& 0.053& 0.278 \\
\cline{3-9}

 & & Mag-mom ($\mu$B) &                                              3.124  &         3.506 &        3.8272 &        3.999 &        4.000 &        4.000 \\
\hline

\multirow{3}{*}{SOC} &         NM & Total energy (eV) &        0.000& 0.903& 1.762& 2.571& 3.331& 4.044 \\
\cline{2-9}
 & \multirow{2}{*}{FM} & Total energy (eV) &                       -1.269& -0.804& -0.431& -0.126& 0.131& 0.357 \\
\cline{3-9}

 & &  $m_x$($\mu$B) &                                             1.828  &         2.044 &        2.207 &        2.321 &        2.332 &        2.335 \\
\hline  \end{tabular}  
 \end{table*}
 
\begin{table*}[h]
\caption[The total energy comparison for paramagnetic and ferromagnetic states of NiMnSb]{The total energy comparison for paramagnetic and ferromagnetic states of NiMnSb[\icsdweb{54255}, SG 216 ($F\overline{4}3m$))].  In the FM case the magnetization points along the [1,1,$\bar{1}$] direction of the conventional cell. Here, we only provide the $x$-component of the magnetic moment $m_x$ in the last column. The lattice victors of the primary unit cell are $\boldsymbol{a_{1}}=a\left(\begin{array}{ccc} 0 & 1/2 & 1/2\end{array}\right)$, $\boldsymbol{a_{2}}=a\left(\begin{array}{ccc} 1/2 & 0 & 1/2\end{array}\right)$ and $\boldsymbol{a_{3}}=a\left(\begin{array}{ccc} 1/2 & 1/2 & 0\end{array}\right)$ with $a$ is the lattice parameter. The total magnetic moment is $4~\mu B/Mn$ and is in good agreement with the experiments ($4.2~\mu B/Mn$) \cite{otto1989half}. The results show that the ferromagnetic (FM) state is more stable than the nonmagnetic or paramagnetic (NM) state.}\label{tab:magNiMnSb}
\vspace{0.2cm}
\renewcommand\arraystretch{1.5}
\begin{tabular}{p{1.4cm}<{ \centering}p{1.4cm}<{ \centering}p{3.4cm}<{ \centering}p{1.4cm}<{ \centering}p{1.4cm}<{ \centering}p{1.4cm}<{ \centering}p{1.4cm}<{ \centering}p{1.4cm}<{ \centering}p{1.4cm}<{ \centering}}
\hline
&&& {\bf U=0} &  {\bf U=1} &  {\bf U=2} &  {\bf U=3} &  {\bf U=4} &  {\bf U=5} \\
\hline
\multirow{3}{*}{Non-SOC} & NM & Total energy (eV) & 0  & 1.246 & 2.488 &3.658 &5.003 &   6.247 \\
\cline{2-9}
 & \multirow{2}{*}{FM} & Total energy (eV) &   -1.445  &-0.842 &-0.301 &0.176 &   0.59 &   0.954 \\
\cline{3-9}

 & & Mag-mom ($\mu$B) &  4.000  & 4.000 &4.040 &4.150 &4.225 &4.289 \\
\hline

\multirow{3}{*}{SOC} & NM & Total energy (eV) &0 &1.236 &2.399 &3.405 &4.300 &5.115 \\
\cline{2-9}
 & \multirow{2}{*}{FM} & Total energy (eV) &   -1.434 & -0.833 & -0.292 & 0.185 & 0.598 & 0.965 \\
\cline{3-9}

 & &  $m_x$($\mu$B) & 2.309  & 2.310 &2.335 &2.396 &2.441 &2.481 \\
\hline  
\end{tabular}  
 \end{table*}
 
\clearpage
\section{Disconnected flat bands manually selected from flat-band database}\label{app:allflatbands}

\newcommand{\ICSDFigRef}[1]{
\ifnum#1=670051 \ref{fig:bestflatband1}\fi
\ifnum#1=670060 \ref{fig:bestflatband1}\fi
\ifnum#1=189752 \ref{fig:bestflatband1}\fi
\ifnum#1=62544 \ref{fig:bestflatband1}\fi
\ifnum#1=75222 \ref{fig:bestflatband1}\fi
\ifnum#1=154123 \ref{fig:bestflatband1}\fi
\ifnum#1=36255 \ref{fig:bestflatband1}\fi
\ifnum#1=416669 \ref{fig:bestflatband1}\fi
\ifnum#1=150279 \ref{fig:bestflatband2}\fi
\ifnum#1=79479 \ref{fig:bestflatband2}\fi
\ifnum#1=84782 \ref{fig:bestflatband2}\fi
\ifnum#1=28471 \ref{fig:bestflatband2}\fi
\ifnum#1=155847 \ref{fig:bestflatband2}\fi
\ifnum#1=99894 \ref{fig:bestflatband2}\fi
\ifnum#1=59723 \ref{fig:bestflatband2}\fi
\ifnum#1=60957 \ref{fig:bestflatband2}\fi
\ifnum#1=79341 \ref{fig:bestflatband3}\fi
\ifnum#1=431760 \ref{fig:bestflatband3}\fi
\ifnum#1=263050 \ref{fig:bestflatband3}\fi
\ifnum#1=83493 \ref{fig:bestflatband3}\fi
\ifnum#1=672719 \ref{fig:bestflatband3}\fi
\ifnum#1=281593 \ref{fig:bestflatband3}\fi
\ifnum#1=94740 \ref{fig:bestflatband3}\fi
\ifnum#1=72887 \ref{fig:bestflatband3}\fi
\ifnum#1=81167 \ref{fig:bestflatband4}\fi
\ifnum#1=643441 \ref{fig:bestflatband4}\fi
\ifnum#1=33996 \ref{fig:bestflatband4}\fi
\ifnum#1=90991 \ref{fig:bestflatband4}\fi
\ifnum#1=247564 \ref{fig:bestflatband4}\fi
\ifnum#1=9008 \ref{fig:bestflatband4}\fi
\ifnum#1=63228 \ref{fig:bestflatband4}\fi
\ifnum#1=16691 \ref{fig:bestflatband4}\fi
\ifnum#1=25700 \ref{fig:bestflatband5}\fi
\ifnum#1=22065 \ref{fig:bestflatband5}\fi
\ifnum#1=33664 \ref{fig:bestflatband5}\fi
\ifnum#1=65453 \ref{fig:bestflatband5}\fi
\ifnum#1=65454 \ref{fig:bestflatband5}\fi
\ifnum#1=65455 \ref{fig:bestflatband5}\fi
\ifnum#1=65456 \ref{fig:bestflatband5}\fi
\ifnum#1=65457 \ref{fig:bestflatband5}\fi
\ifnum#1=65458 \ref{fig:bestflatband6}\fi
\ifnum#1=65460 \ref{fig:bestflatband6}\fi
\ifnum#1=65461 \ref{fig:bestflatband6}\fi
\ifnum#1=173488 \ref{fig:bestflatband6}\fi
\ifnum#1=150702 \ref{fig:bestflatband6}\fi
\ifnum#1=153544 \ref{fig:bestflatband6}\fi
\ifnum#1=187669 \ref{fig:bestflatband6}\fi
\ifnum#1=190593 \ref{fig:bestflatband6}\fi
\ifnum#1=91791 \ref{fig:bestflatband7}\fi
\ifnum#1=95518 \ref{fig:bestflatband7}\fi
\ifnum#1=150701 \ref{fig:bestflatband7}\fi
\ifnum#1=151703 \ref{fig:bestflatband7}\fi
\ifnum#1=187662 \ref{fig:bestflatband7}\fi
\ifnum#1=192327 \ref{fig:bestflatband7}\fi
\ifnum#1=263049 \ref{fig:bestflatband7}\fi
\ifnum#1=673198 \ref{fig:bestflatband7}\fi
\ifnum#1=424067 \ref{fig:bestflatband8}\fi
\ifnum#1=76202 \ref{fig:bestflatband8}\fi
\ifnum#1=170538 \ref{fig:bestflatband8}\fi
\ifnum#1=202917 \ref{fig:bestflatband8}\fi
\ifnum#1=23563 \ref{fig:bestflatband8}\fi
\ifnum#1=89092 \ref{fig:bestflatband8}\fi
\ifnum#1=4287 \ref{fig:bestflatband8}\fi
\ifnum#1=393 \ref{fig:bestflatband8}\fi
\ifnum#1=83367 \ref{fig:bestflatband9}\fi
\ifnum#1=61198 \ref{fig:bestflatband9}\fi
\ifnum#1=58201 \ref{fig:bestflatband9}\fi
\ifnum#1=623956 \ref{fig:bestflatband9}\fi
\ifnum#1=90123 \ref{fig:bestflatband9}\fi
\ifnum#1=155174 \ref{fig:bestflatband9}\fi
\ifnum#1=191359 \ref{fig:bestflatband9}\fi
\ifnum#1=191360 \ref{fig:bestflatband9}\fi
\ifnum#1=94079 \ref{fig:bestflatband10}\fi
\ifnum#1=246541 \ref{fig:bestflatband10}\fi
\ifnum#1=16254 \ref{fig:bestflatband10}\fi
\ifnum#1=65196 \ref{fig:bestflatband10}\fi
\ifnum#1=68358 \ref{fig:bestflatband10}\fi
\ifnum#1=162261 \ref{fig:bestflatband10}\fi
\ifnum#1=105729 \ref{fig:bestflatband10}\fi
\ifnum#1=247194 \ref{fig:bestflatband10}\fi
\ifnum#1=19077 \ref{fig:bestflatband11}\fi
\ifnum#1=20294 \ref{fig:bestflatband11}\fi
\ifnum#1=50185 \ref{fig:bestflatband11}\fi
\ifnum#1=41787 \ref{fig:bestflatband11}\fi
\ifnum#1=52999 \ref{fig:bestflatband11}\fi
\ifnum#1=55926 \ref{fig:bestflatband11}\fi
\ifnum#1=607535 \ref{fig:bestflatband11}\fi
\ifnum#1=607756 \ref{fig:bestflatband11}\fi
\ifnum#1=57842 \ref{fig:bestflatband12}\fi
\ifnum#1=24570 \ref{fig:bestflatband12}\fi
\ifnum#1=42519 \ref{fig:bestflatband12}\fi
\ifnum#1=42139 \ref{fig:bestflatband12}\fi
\ifnum#1=93699 \ref{fig:bestflatband12}\fi
\ifnum#1=1522 \ref{fig:bestflatband12}\fi
\ifnum#1=153173 \ref{fig:bestflatband12}\fi
\ifnum#1=155126 \ref{fig:bestflatband12}\fi
\ifnum#1=191592 \ref{fig:bestflatband13}\fi
\ifnum#1=194588 \ref{fig:bestflatband13}\fi
\ifnum#1=51017 \ref{fig:bestflatband13}\fi
\ifnum#1=60006 \ref{fig:bestflatband13}\fi
\ifnum#1=250123 \ref{fig:bestflatband13}\fi
\ifnum#1=65699 \ref{fig:bestflatband13}\fi
\ifnum#1=71028 \ref{fig:bestflatband13}\fi
\ifnum#1=30275 \ref{fig:bestflatband13}\fi
\ifnum#1=20041 \ref{fig:bestflatband14}\fi
\ifnum#1=250498 \ref{fig:bestflatband14}\fi
\ifnum#1=27955 \ref{fig:bestflatband14}\fi
\ifnum#1=182051 \ref{fig:bestflatband14}\fi
\ifnum#1=182052 \ref{fig:bestflatband14}\fi
\ifnum#1=153864 \ref{fig:bestflatband14}\fi
\ifnum#1=245665 \ref{fig:bestflatband14}\fi
\ifnum#1=260887 \ref{fig:bestflatband14}\fi
\ifnum#1=24859 \ref{fig:bestflatband15}\fi
\ifnum#1=40062 \ref{fig:bestflatband15}\fi
\ifnum#1=25005 \ref{fig:bestflatband15}\fi
\ifnum#1=410393 \ref{fig:bestflatband15}\fi
\ifnum#1=425834 \ref{fig:bestflatband15}\fi
\ifnum#1=62555 \ref{fig:bestflatband15}\fi
\ifnum#1=62556 \ref{fig:bestflatband15}\fi
\ifnum#1=44926 \ref{fig:bestflatband15}\fi
\ifnum#1=72409 \ref{fig:bestflatband16}\fi
\ifnum#1=67261 \ref{fig:bestflatband16}\fi
\ifnum#1=67263 \ref{fig:bestflatband16}\fi
\ifnum#1=300110 \ref{fig:bestflatband16}\fi
\ifnum#1=300111 \ref{fig:bestflatband16}\fi
\ifnum#1=300190 \ref{fig:bestflatband16}\fi
\ifnum#1=402573 \ref{fig:bestflatband16}\fi
\ifnum#1=409919 \ref{fig:bestflatband16}\fi
\ifnum#1=401026 \ref{fig:bestflatband17}\fi
\ifnum#1=25766 \ref{fig:bestflatband17}\fi
\ifnum#1=25767 \ref{fig:bestflatband17}\fi
\ifnum#1=5440 \ref{fig:bestflatband17}\fi
\ifnum#1=425137 \ref{fig:bestflatband17}\fi
\ifnum#1=5435 \ref{fig:bestflatband17}\fi
\ifnum#1=420728 \ref{fig:bestflatband17}\fi
\ifnum#1=5437 \ref{fig:bestflatband17}\fi
\ifnum#1=62748 \ref{fig:bestflatband18}\fi
\ifnum#1=62749 \ref{fig:bestflatband18}\fi
\ifnum#1=68151 \ref{fig:bestflatband18}\fi
\ifnum#1=615156 \ref{fig:bestflatband18}\fi
\ifnum#1=33579 \ref{fig:bestflatband18}\fi
\ifnum#1=152588 \ref{fig:bestflatband18}\fi
\ifnum#1=608322 \ref{fig:bestflatband18}\fi
\ifnum#1=608323 \ref{fig:bestflatband18}\fi
\ifnum#1=34350 \ref{fig:bestflatband19}\fi
\ifnum#1=248042 \ref{fig:bestflatband19}\fi
\ifnum#1=4226 \ref{fig:bestflatband19}\fi
\ifnum#1=407809 \ref{fig:bestflatband19}\fi
\ifnum#1=10142 \ref{fig:bestflatband19}\fi
\ifnum#1=29343 \ref{fig:bestflatband19}\fi
\ifnum#1=2279 \ref{fig:bestflatband19}\fi
\ifnum#1=27599 \ref{fig:bestflatband19}\fi
\ifnum#1=9752 \ref{fig:bestflatband20}\fi
\ifnum#1=28603 \ref{fig:bestflatband20}\fi
\ifnum#1=42735 \ref{fig:bestflatband20}\fi
\ifnum#1=105707 \ref{fig:bestflatband20}\fi
\ifnum#1=106357 \ref{fig:bestflatband20}\fi
\ifnum#1=54474 \ref{fig:bestflatband20}\fi
\ifnum#1=58672 \ref{fig:bestflatband20}\fi
\ifnum#1=54245 \ref{fig:bestflatband20}\fi
\ifnum#1=54422 \ref{fig:bestflatband21}\fi
\ifnum#1=105336 \ref{fig:bestflatband21}\fi
\ifnum#1=59435 \ref{fig:bestflatband21}\fi
\ifnum#1=658018 \ref{fig:bestflatband21}\fi
\ifnum#1=99999 \ref{fig:bestflatband21}\fi
\ifnum#1=109145 \ref{fig:bestflatband21}\fi
\ifnum#1=410390 \ref{fig:bestflatband21}\fi
\ifnum#1=8193 \ref{fig:bestflatband21}\fi
\ifnum#1=10486 \ref{fig:bestflatband22}\fi
\ifnum#1=60847 \ref{fig:bestflatband22}\fi
\ifnum#1=41921 \ref{fig:bestflatband22}\fi
\ifnum#1=102582 \ref{fig:bestflatband22}\fi
\ifnum#1=102542 \ref{fig:bestflatband22}\fi
\ifnum#1=102722 \ref{fig:bestflatband22}\fi
\ifnum#1=151196 \ref{fig:bestflatband22}\fi
\ifnum#1=53535 \ref{fig:bestflatband22}\fi
\ifnum#1=20805 \ref{fig:bestflatband23}\fi
\ifnum#1=107419 \ref{fig:bestflatband23}\fi
\ifnum#1=601018 \ref{fig:bestflatband23}\fi
\ifnum#1=612227 \ref{fig:bestflatband23}\fi
\ifnum#1=612233 \ref{fig:bestflatband23}\fi
\ifnum#1=54985 \ref{fig:bestflatband23}\fi
\ifnum#1=672665 \ref{fig:bestflatband23}\fi
\ifnum#1=105477 \ref{fig:bestflatband23}\fi
\ifnum#1=107033 \ref{fig:bestflatband24}\fi
\ifnum#1=107420 \ref{fig:bestflatband24}\fi
\ifnum#1=185106 \ref{fig:bestflatband24}\fi
\ifnum#1=423939 \ref{fig:bestflatband24}\fi
\ifnum#1=672261 \ref{fig:bestflatband24}\fi
\ifnum#1=191648 \ref{fig:bestflatband24}\fi
\ifnum#1=191658 \ref{fig:bestflatband24}\fi
\ifnum#1=191659 \ref{fig:bestflatband24}\fi
\ifnum#1=191660 \ref{fig:bestflatband25}\fi
\ifnum#1=191661 \ref{fig:bestflatband25}\fi
\ifnum#1=191662 \ref{fig:bestflatband25}\fi
\ifnum#1=191664 \ref{fig:bestflatband25}\fi
\ifnum#1=671982 \ref{fig:bestflatband25}\fi
\ifnum#1=191657 \ref{fig:bestflatband25}\fi
\ifnum#1=54255 \ref{fig:bestflatband25}\fi
\ifnum#1=53001 \ref{fig:bestflatband25}\fi
\ifnum#1=54343 \ref{fig:bestflatband26}\fi
\ifnum#1=104955 \ref{fig:bestflatband26}\fi
\ifnum#1=672030 \ref{fig:bestflatband26}\fi
\ifnum#1=102552 \ref{fig:bestflatband26}\fi
\ifnum#1=191171 \ref{fig:bestflatband26}\fi
\ifnum#1=76848 \ref{fig:bestflatband26}\fi
\ifnum#1=674693 \ref{fig:bestflatband26}\fi
\ifnum#1=100355 \ref{fig:bestflatband26}\fi
\ifnum#1=183701 \ref{fig:bestflatband27}\fi
\ifnum#1=53745 \ref{fig:bestflatband27}\fi
\ifnum#1=76413 \ref{fig:bestflatband27}\fi
\ifnum#1=105353 \ref{fig:bestflatband27}\fi
\ifnum#1=193569 \ref{fig:bestflatband27}\fi
\ifnum#1=42600 \ref{fig:bestflatband27}\fi
\ifnum#1=58038 \ref{fig:bestflatband27}\fi
\ifnum#1=59438 \ref{fig:bestflatband27}\fi
\ifnum#1=103856 \ref{fig:bestflatband28}\fi
\ifnum#1=105020 \ref{fig:bestflatband28}\fi
\ifnum#1=105689 \ref{fig:bestflatband28}\fi
\ifnum#1=187963 \ref{fig:bestflatband28}\fi
\ifnum#1=187969 \ref{fig:bestflatband28}\fi
\ifnum#1=247193 \ref{fig:bestflatband28}\fi
\ifnum#1=670622 \ref{fig:bestflatband28}\fi
\ifnum#1=670623 \ref{fig:bestflatband28}\fi
\ifnum#1=670626 \ref{fig:bestflatband29}\fi
\ifnum#1=104859 \ref{fig:bestflatband29}\fi
\ifnum#1=191170 \ref{fig:bestflatband29}\fi
\ifnum#1=15423 \ref{fig:bestflatband29}\fi
\ifnum#1=24479 \ref{fig:bestflatband29}\fi
\ifnum#1=28145 \ref{fig:bestflatband29}\fi
\ifnum#1=28146 \ref{fig:bestflatband29}\fi
\ifnum#1=28930 \ref{fig:bestflatband29}\fi
\ifnum#1=43722 \ref{fig:bestflatband30}\fi
\ifnum#1=44568 \ref{fig:bestflatband30}\fi
\ifnum#1=60611 \ref{fig:bestflatband30}\fi
\ifnum#1=69360 \ref{fig:bestflatband30}\fi
\ifnum#1=76763 \ref{fig:bestflatband30}\fi
\ifnum#1=76790 \ref{fig:bestflatband30}\fi
\ifnum#1=76797 \ref{fig:bestflatband30}\fi
\ifnum#1=77389 \ref{fig:bestflatband30}\fi
\ifnum#1=88982 \ref{fig:bestflatband31}\fi
\ifnum#1=109076 \ref{fig:bestflatband31}\fi
\ifnum#1=168902 \ref{fig:bestflatband31}\fi
\ifnum#1=187637 \ref{fig:bestflatband31}\fi
\ifnum#1=188415 \ref{fig:bestflatband31}\fi
\ifnum#1=191203 \ref{fig:bestflatband31}\fi
\ifnum#1=247066 \ref{fig:bestflatband31}\fi
\ifnum#1=422858 \ref{fig:bestflatband31}\fi
\ifnum#1=671082 \ref{fig:bestflatband32}\fi
\ifnum#1=671086 \ref{fig:bestflatband32}\fi
\ifnum#1=673496 \ref{fig:bestflatband32}\fi
\ifnum#1=15424 \ref{fig:bestflatband32}\fi
\ifnum#1=29119 \ref{fig:bestflatband32}\fi
\ifnum#1=674060 \ref{fig:bestflatband32}\fi
\ifnum#1=30612 \ref{fig:bestflatband32}\fi
\ifnum#1=165988 \ref{fig:bestflatband32}\fi
\ifnum#1=77071 \ref{fig:bestflatband33}\fi
\ifnum#1=108651 \ref{fig:bestflatband33}\fi
\ifnum#1=181465 \ref{fig:bestflatband33}\fi
\ifnum#1=93543 \ref{fig:bestflatband33}\fi
\ifnum#1=15775 \ref{fig:bestflatband33}\fi
\ifnum#1=15776 \ref{fig:bestflatband33}\fi
\ifnum#1=27425 \ref{fig:bestflatband33}\fi
\ifnum#1=28598 \ref{fig:bestflatband33}\fi
\ifnum#1=28601 \ref{fig:bestflatband34}\fi
\ifnum#1=97028 \ref{fig:bestflatband34}\fi
\ifnum#1=109252 \ref{fig:bestflatband34}\fi
\ifnum#1=251068 \ref{fig:bestflatband34}\fi
\ifnum#1=427115 \ref{fig:bestflatband34}\fi
\ifnum#1=4169 \ref{fig:bestflatband34}\fi
\ifnum#1=6027 \ref{fig:bestflatband34}\fi
\ifnum#1=22114 \ref{fig:bestflatband34}\fi
\ifnum#1=22118 \ref{fig:bestflatband35}\fi
\ifnum#1=96688 \ref{fig:bestflatband35}\fi
\ifnum#1=99061 \ref{fig:bestflatband35}\fi
\ifnum#1=157016 \ref{fig:bestflatband35}\fi
\ifnum#1=157603 \ref{fig:bestflatband35}\fi
\ifnum#1=157886 \ref{fig:bestflatband35}\fi
\ifnum#1=246112 \ref{fig:bestflatband35}\fi
\ifnum#1=257115 \ref{fig:bestflatband35}\fi
\ifnum#1=26735 \ref{fig:bestflatband36}\fi
\ifnum#1=60440 \ref{fig:bestflatband36}\fi
\ifnum#1=60441 \ref{fig:bestflatband36}\fi
\ifnum#1=102385 \ref{fig:bestflatband36}\fi
\ifnum#1=102483 \ref{fig:bestflatband36}\fi
\ifnum#1=102682 \ref{fig:bestflatband36}\fi
\ifnum#1=102687 \ref{fig:bestflatband36}\fi
\ifnum#1=102996 \ref{fig:bestflatband36}\fi
\ifnum#1=105327 \ref{fig:bestflatband37}\fi
\ifnum#1=52994 \ref{fig:bestflatband37}\fi
\ifnum#1=25324 \ref{fig:bestflatband37}\fi
\ifnum#1=52954 \ref{fig:bestflatband37}\fi
\ifnum#1=53080 \ref{fig:bestflatband37}\fi
\ifnum#1=53086 \ref{fig:bestflatband37}\fi
\ifnum#1=53555 \ref{fig:bestflatband37}\fi
\ifnum#1=54546 \ref{fig:bestflatband37}\fi
\ifnum#1=54595 \ref{fig:bestflatband38}\fi
\ifnum#1=57654 \ref{fig:bestflatband38}\fi
\ifnum#1=57793 \ref{fig:bestflatband38}\fi
\ifnum#1=57807 \ref{fig:bestflatband38}\fi
\ifnum#1=76227 \ref{fig:bestflatband38}\fi
\ifnum#1=102433 \ref{fig:bestflatband38}\fi
\ifnum#1=102755 \ref{fig:bestflatband38}\fi
\ifnum#1=103641 \ref{fig:bestflatband38}\fi
\ifnum#1=103644 \ref{fig:bestflatband39}\fi
\ifnum#1=104926 \ref{fig:bestflatband39}\fi
\ifnum#1=105339 \ref{fig:bestflatband39}\fi
\ifnum#1=108436 \ref{fig:bestflatband39}\fi
\ifnum#1=169468 \ref{fig:bestflatband39}\fi
\ifnum#1=185966 \ref{fig:bestflatband39}\fi
\ifnum#1=185999 \ref{fig:bestflatband39}\fi
\ifnum#1=186059 \ref{fig:bestflatband39}\fi
\ifnum#1=186060 \ref{fig:bestflatband40}\fi
\ifnum#1=671340 \ref{fig:bestflatband40}\fi
\ifnum#1=675102 \ref{fig:bestflatband40}\fi
\ifnum#1=53525 \ref{fig:bestflatband40}\fi
\ifnum#1=103473 \ref{fig:bestflatband40}\fi
\ifnum#1=186057 \ref{fig:bestflatband40}\fi
\ifnum#1=673582 \ref{fig:bestflatband40}\fi
\ifnum#1=9347 \ref{fig:bestflatband40}\fi
\ifnum#1=59894 \ref{fig:bestflatband41}\fi
\ifnum#1=245747 \ref{fig:bestflatband41}\fi
\ifnum#1=604 \ref{fig:bestflatband41}\fi
\ifnum#1=605 \ref{fig:bestflatband41}\fi
\ifnum#1=9022 \ref{fig:bestflatband41}\fi
\ifnum#1=9023 \ref{fig:bestflatband41}\fi
\ifnum#1=292 \ref{fig:bestflatband41}\fi
\ifnum#1=72546 \ref{fig:bestflatband41}\fi
\ifnum#1=58675 \ref{fig:bestflatband42}\fi
\ifnum#1=52216 \ref{fig:bestflatband42}\fi
\ifnum#1=58915 \ref{fig:bestflatband42}\fi
\ifnum#1=105334 \ref{fig:bestflatband42}\fi
\ifnum#1=105458 \ref{fig:bestflatband42}\fi
\ifnum#1=106236 \ref{fig:bestflatband42}\fi
\ifnum#1=27120 \ref{fig:bestflatband42}\fi
\ifnum#1=33672 \ref{fig:bestflatband42}\fi
\ifnum#1=187534 \ref{fig:bestflatband43}\fi
\ifnum#1=27119 \ref{fig:bestflatband43}\fi
\ifnum#1=27121 \ref{fig:bestflatband43}\fi
\ifnum#1=27330 \ref{fig:bestflatband43}\fi
\ifnum#1=161104 \ref{fig:bestflatband43}\fi
\ifnum#1=163817 \ref{fig:bestflatband43}\fi
\ifnum#1=76980 \ref{fig:bestflatband43}\fi
\ifnum#1=290133 \ref{fig:bestflatband43}\fi
\ifnum#1=608815 \ref{fig:bestflatband44}\fi
}

\newcommand{\bestflatcaption}[1]{Band structure and density of states for materials with the most remarkable flat-band features near the Fermi energy. (part #1/44)}

\newcommand{\bestperstructurecaption}[1]{List of best compounds belonging to the family of #1 compounds. The first column is the chemical formula, the second column is the representative ICSD used for the band structure and projected DOS plots. The third column is the space group, the fourth column is the link to figure showing the corresponding band structure and projected DOS. The last two columns provide the magnetic and superconducting properties, following the same notations than Table.~\ref{tab:curatedflatband}.}

In this appendix, we provide a list of curated materials with flat bands. The analysis was manually performed on the database presented in Appendix~\ref{app:tqcdboverview}. For convenience, materials with flat atomic bands are handled separately. We highlight \FlatBandNbrBestMaterials~compounds hosting the most interesting flat-band features by providing their band structure and DOS. 

\subsection{List of curated materials with flat bands}\label{app:listcurated}

We start with the list of \FlatBandNbrCuratedMaterials~curated unique materials (as defined in Appendix~\ref{app:tqcdboverview}), corresponding to \FlatBandNbrCuratedICSDs~ICSD entries. This list is available in Table~\ref{tab:curatedflatband}. It was obtained by performing a manual search in our database, using the tools available on \webflatband. We have applied several criteria to obtain this list:

\begin{itemize}
\item We required that the region of $\Delta_E = 0.5 {\rm eV}$ around the Fermi level (as defined in Appendix~\ref{app:bsflatsegment}) to host only a few bands along the high symmetry lines so that the flat bands can be easily distinguished from the dispersing bands.
\item Materials with trivial flat atomic bands, of which the BRs are classified as topologically trivial (\ie LCEBR), have been excluded at this stage. The methodology to detect these materials will be detailed in Appendix~\ref{app:listatomic}, including a complete list of candidates.
\item We fixed the band width of the flat-band segment to be less than $\omega=0.1{\rm eV}$ (see Appendix~\ref{app:bsflatsegment}).
\item We requested to have at least one DOS peak in the region $[-1{\rm eV},1{\rm eV}]$ around Fermi level with a Gaussian width $W_{\rm peak}$ smaller than $1 {\rm eV}$ (see Appendix~\ref{app:dospeak}).
\end{itemize}
Note that for a given unique compound, we might have one ICSD fulfilling all these criteria and another ICSD not satisfying one of them. Indeed, we use sharp cutoffs and different ICSD entries of a unique compound might have slightly different band structure or DOS (as explained in Appendix~\ref{app:database}). One such example is ${\rm K}_2 {\rm O}_7 {\rm Zn}_6$ [\icsdweb{1120}, SG 102 (\sgsymb{102})] and [\icsdweb{2496}, SG 102 (\sgsymb{102})]. The former fulfills our conditions, the latter does not. Since the differences are barely noticeable, we include all the ICSD entries of a unique compound as soon as one of them satisfies all the criteria defined above.

For each entry in Table~\ref{tab:curatedflatband}, we provide the chemical formula, space group, all related ICSD numbers and possible sublattices that we have found using the procedures described in Appendices~\ref{app:kagomepyrochlore} and~\ref{app:bipartite}. We also provide three additional attributes. First, we rely on the \webmaterialsproject\cite{MaterialsProject} to give the magnetic properties (when available). Similarly, we indicate if a material is know to be superconductor based on the information from the \webscnims\cite{NIMS}. Finally, we tag some materials as ``high quality". These materials have a structure determination including refinement (in case of powder data including Rietveld refinement), temperature factors, pressure in the range of $0.09-0.11 {\rm MPa}$, temperature within the range of $285-300 {\rm K}$ and standard deviation given for cell parameters. These are basically compounds whose crystallographic structure has been well-determined and are easy to synthesize. 

Among the \FlatBandNbrCuratedMaterials~curated unique materials, some of them have already been theoretically analyzed or experimentally probed in the context of flat-band materials. While we cannot pretend to be exhaustive, we list below 11 such materials, their related references and the main features of their studies.

\begin{itemize}
\item ${\rm Ru} {\rm O}_2$ [\icsdweb{172178}, SG 136 (\sgsymb{136})] was studied by angle-resolved photoemission spectroscopy (ARPES) experiments in Ref.~\cite{jovic2019dirac} and was reported to host topologically trivial flat-band surface state.

\item The electronic structure properties of ${\rm Y}_2 {\rm Ni}_7$ [\icsdweb{647060}, SG 166 (\sgsymb{166})]
have been theoretically studied in Ref.~\cite{Crivello_2020}, pointing toward the presence the flat bands to explain the magnetic properties of these materials.

\item ${\rm Al} {\rm Fe}_2 {\rm B}_2$ [\icsdweb{241907}, SG 65 (\sgsymb{65})] is a layered ferromagnet studied by ARPES in Ref.~\cite{PhysRevB.101.245129} and have been shown to host flat bands near the Fermi energy.

\item Flat bands near the Fermi level have been observed in ARPES for ${\rm Ca} {\rm Cu}_3 {\rm Ru}_4 {\rm O}_{12}$ [\icsdweb{95715}, SG 204 (\sgsymb{204})] in Ref.~\cite{PhysRevB.102.035111}.

\item In Ref.~\cite{yin2019negative}, a scanning tunneling microscopy (STM) experiment was performed on the ferromagnetic ${\rm Co}_3 {\rm Sn}_2 {\rm S}_2$ [\icsdweb{624867}, SG 166 (\sgsymb{166})]. A pronounced conductance peak near the Fermi level was observed and identified as arising from the flat bands. In Ref.~\cite{Co3Sn2S2_2020}, the flat bands of ferromagnetic ${\rm Co}_3 {\rm Sn}_2 {\rm S}_2$ were also associated to a sharp peak in the optical conductivity measurement. 

\item Topological flat bands in the vicinity of the Fermi level in ${\rm Co} {\rm Sn}$ [\icsdweb{161118}, SG 191 (\sgsymb{191})] were studied theoretically and observed in the ARPES experiments in Ref.~\cite{Kang_2020, Liu2020}.
The tuning of the flat bands in doped ${\rm Co} {\rm Sn}$ was also reported experimentally in Ref.~\cite{sales2021tuning}.

\item The material ${\rm Mg} {\rm V}_2 {\rm O}_4$ [\icsdweb{674693}, SG 216 (\sgsymb{216})] in its ferromagnetic phase was predicted by first-principles calculations in Ref.~\cite{PhysRevB.102.155144} to host a three-dimensional flat band near the Fermi energy.

\item ${\rm Pd}_3 {\rm P}_2 {\rm S}_8$ [\icsdweb{16296}, SG 164 (\sgsymb{164})] with a Kagome van-der-Waals structure was experimentally studied in Ref.~\cite{park2020kagome} where the magnetic measurements were explained by the flat bands predicted in the first-principles calculations.

\item ${\rm Pd}_3 {\rm Pb}$ [\icsdweb{42600}, SG 221 (\sgsymb{221})] was studied theoretically in Ref.~\cite{PhysRevB.98.035130} and experimentally by Shubnikov–de Haas oscillations in Ref.~\cite{PhysRevB.101.245113}, pointing toward a flat-band semimetal.

\item A minimal microscopic model within a large-$S$ expansion based was built for ${\rm Sr}_2 {\rm Fe} {\rm Mo} {\rm O}_6$ [\icsdweb{181752}, SG 225 (\sgsymb{225})] in Ref.~\cite{Yamada_2020}, leading to the spin-polarized flat bands and four massive or massless Dirac dispersions.

\item ${\rm Y} {\rm Cr}_6 {\rm Ge}_6$ [\icsdweb{658018}, SG 191 (\sgsymb{191})] has been probed by ARPES in Ref.~\cite{yang2019evidence} to unveil planar flat bands associated to the Kagome sublattice.

\end{itemize}

\LTcapwidth=1.0\textwidth
\renewcommand\arraystretch{1.0}
\begin{longtable*}{|c|c|c|c|c|c|c|c|c|c|}
    \caption[Manually curated list of materials with flat bands]{Manually curated list of materials with flat bands. We provide in this table the list of the \FlatBandNbrCuratedMaterials~entries of curately chosen flat-band materials, indexed by the first column. The second column gives the chemical formula, the third column the space group symbol (and number in parentheses). The fourth column indicates the topological classification at the Fermi level: topological insulator (TI), topological semi-metal (SM) or trivial insulator (Triv.). The fifth column indicates if the material also belongs to the list of best materials given in Appendix~\ref{app:bestflatbands} and the figure displaying its band structure and DOS. The sixth column details the sublattices found for each material and common to every ICSD entries of this material. Here 'k' stands for Kagome, 'p' for pyrochlore, 'b' for bipartite, 's' for split lattice and 'l' for Lieb sublattice. A star superscript after 'k' ('l') indicate that only approximate Kagome (Lieb) sublattices were found (as defined in Appendix~\ref{app:kagomepyrochlore}). The seventh column gives the magnetic properties as obtained from \webmaterialsproject\ (NM for non-magnetic, FM for ferromagnetic, AFM for anti-ferromagnetic, FiM for ferrimagnetic, '---' when the information is not available). The eighth column indicates if the material is known to be superconductor according to the \webscnims. 'SC' stands for superconductor while 'SC$^{*}$' means that alloys based on this material are superconductor. The ninth column tells which material are "high-quality" as defined in Appendix~\ref{app:listcurated}. The last column gives all the ICSD entries related to each material. If a given ICSD has additional sublattices compared to those common to all ICSD entries, a superscript label provides the list of these additional sublattices.} 
    \label{tab:curatedflatband}\\
\hline
\end{longtable*}

\subsection{List of curated materials with flat atomic bands}\label{app:listatomic}

We have also searched for materials with flat atomic bands defined in Appendix~\ref{app:atomicflatband} and which are excluded in Appendix~\ref{app:listcurated}. Since we cannot probe directly the absence of coupling between orbitals of neighboring atoms, we rely on a series of indicators that are easier to access using the database described in Appendix~\ref{app:database}. Our search criteria are:

\begin{itemize}
    \item The bands should be extremely flat along all the directions in the BZ.
    \item The flat bands should be the first set of conduction bands or the last set of valence bands near the Fermi level.
    \item The DOS peak associated to the set of flat bands should have its center at a maximum distance of $1 {\rm eV}$ from the Fermi energy and its gaussian width lower than $1 {\rm eV}$ (see Appendix~\ref{app:dospeak}).
    \item The BRs of the flat bands should be classified by the symmetry eigenvalues as topologically trivial and non-fragile (\ie LCEBR)~ \cite{Po2018fragile,Jennifer2018fragile,Slager2019wilson,bradlyn_disconnected_2019,song_fragile_2020,Yang2019fragile,alexandradinata2020crystallographic,Yang2019flat,Po2019faithful,Chiu2020}.
\end{itemize}

For flat atomic bands, the electronic states are (exponentially) localized on one occupied Wyckoff position. In addition we could have \emph{flat molecular bands} where the electronic states are (exponentially) localized on several occupied Wyckoff positions. In some cases, these flat molecular bands could be identified by properties such as bonding within molecular unit, such as poly-atomic ions, coordination-complexes and their ions, or neutral molecules. The rigorous distinction between flat atomic bands and flat molecular bands requires a systematic evaluation of the Wannier states which is beyond the scope of this article. For that reason, we include the compounds with flat molecular bands in the list of flat atomic band compounds.

Based on these criteria, we have obtained \FlatBandNbrAtomicFlatBandMaterials~unique materials (\FlatBandNbrAtomicFlatBandICSDs~ICSD entries) exhibiting flat atomic bands near the Fermi level. These materials are listed in Table~\ref{tab:atomicflatband}.

%%Additionally, we have identified flat-band materials in which flat bands are localized on molecular or ionic clusters rather than on single atoms. The compounds were selected based on the same criteria as the compounds on the curated list (as explained in Appendix~\ref{app:listcurated}), with additional requirements that:
%%\begin{itemize}
%%    \item The BRs of the flat bands are classified by the symmetry eigenvalues as topologically trivial and non-fragile (\ie LCEBR)~
%%    \item The flat bands are the first set of conduction bands, the last set of valence bands, or both. 
%%   \item The flat bands are composed of states that:
%%    \begin{itemize}
%%        \item host lone electron pairs on atoms of one of elements in the compound.
%%        \item arise from the bonding within molecular unit, such as poly-atomic ions, coordination-complexes and their ions, or neutral molecules.
%%        \item belong to atoms that are more than $5.0 \mathrm{\AA}$ apart.
%%    \end{itemize}
%%    \item There are no covalent bonds between sub-units of a compound, and the crystal lattice is held together by ionic, van der Waals, or London interactions.
%%\end{itemize}

\LTcapwidth=1.0\textwidth
\renewcommand\arraystretch{1.0}
\begin{longtable*}{|c|c|c|c|c|c|c|c|c|}
    \caption[Manually curated list of materials with atomic flat bands]{Manually curated list of materials with atomic flat bands. We provide in this table the list of the \FlatBandNbrAtomicFlatBandMaterials~entries of atomic flat-band materials, indexed by the first column. The second column gives the chemical formula, the third column the space group symbol (and number in parentheses). The fourth column indicates the topological classification at the Fermi level: topological insulator (TI), topological semi-metal (SM) or trivial insulator (Triv.). The fifth column details the sublattices found for each material and common to every ICSD entries of this material. Here 'k' stands for Kagome, 'p' for pyrochlore, 'b' for bipartite, 's' for split lattice and 'l' for Lieb sublattice. A star superscript after 'k' ('l') indicates that only approximate Kagome (Lieb) sublattices were found (as defined in Appendix~\ref{app:kagomepyrochlore}). The sixth column gives the magnetic properties as obtained from \webmaterialsproject\ (NM for non-magnetic, FM for ferromagnetic, AFM for anti-ferromagnetic, FiM for ferrimagnetic, '---' when the information is not available). The seventh column indicates if the material is known to be superconductor according to the \webscnims. SC stands for superconductor while SC$^{*}$ means that alloys based on this material are superconductors. The eighth column tells which material are "high-quality" as defined in Appendix~\ref{app:listcurated}. The last column gives all the ICSD entries related to each material. If a given ICSD has additional sublattices compared to those common to all ICSD entries, a superscript label provides the list of these additional sublattices.} 
    \label{tab:atomicflatband}\\
\hline

\end{longtable*}

\subsection{Best flat bands near or at the Fermi level}\label{app:bestflatbands}

Among the \FlatBandNbrCuratedMaterials~curated flat-band materials given in Table~\ref{tab:curatedflatband} of non-atomic flat-band materials, we present in this subsection the \FlatBandNbrBestMaterials~best candidates (corresponding to \FlatBandNbrBestICSDs~ICSD entries). They have been selected with their experimental relevance in mind: we have targeted  materials where flat bands are located at or close to the Fermi level, without any gating/doping. For that purpose, we have applied the following two criteria:
\begin{enumerate}
\item The highest occupied band or the lowest unoccupied band
near the Fermi level is classified as a flat band according to the definition given in Appendix~\ref{app:bsflatsegment}.
\item The highest occupied or lowest unoccupied flat band is perfectly flat, \ie almost dispersionless, along at least one high-symmetry line of the BZ. Note that we did not apply here any bandwidth threshold.
\end{enumerate}

\begin{figure}[ht]
\centering
\includegraphics[width=0.9\textwidth,angle=0]{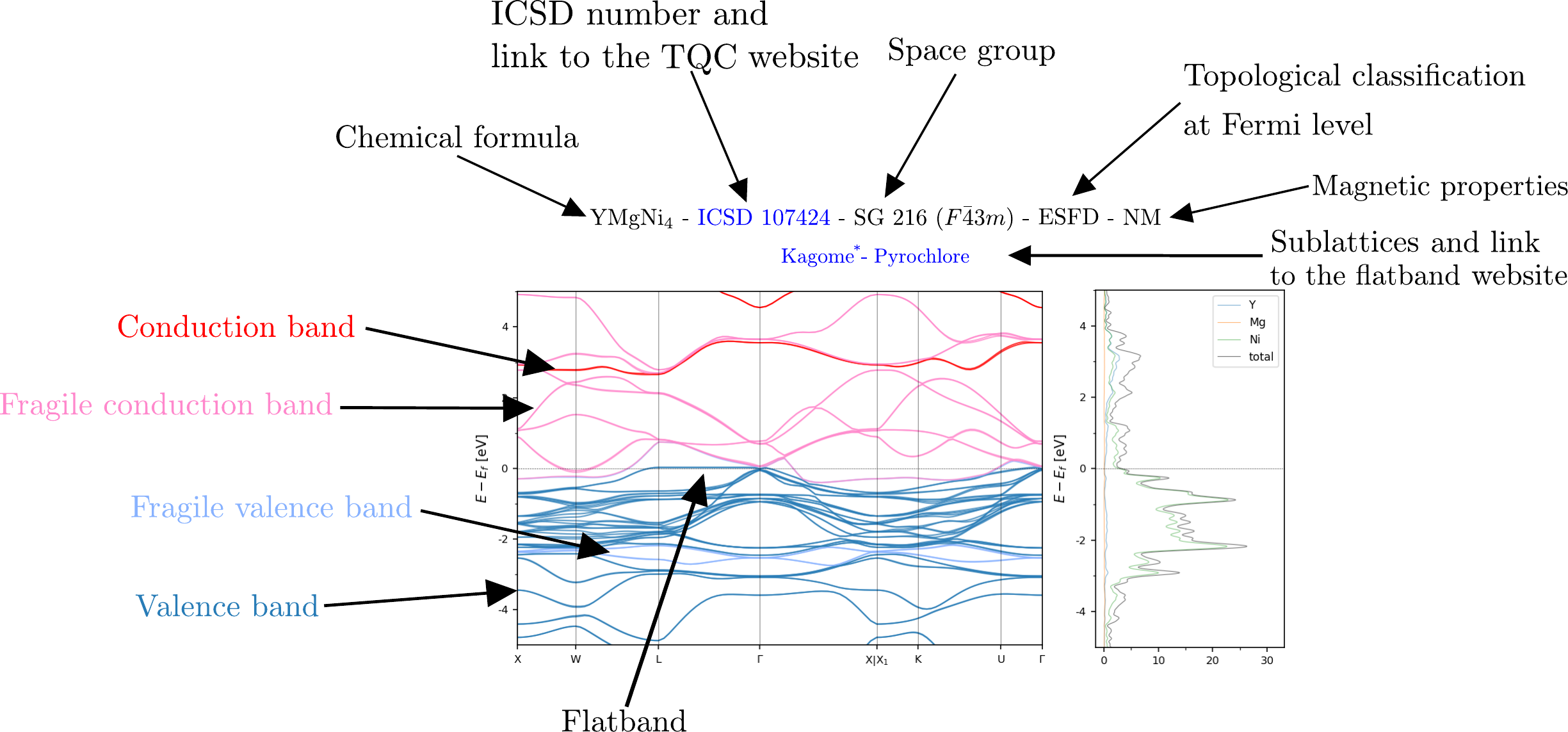}
\caption{Example band structure and projected density of states plots.  In this figure, we show the band structure and the projected density of states of ${\rm Y Mg Ni}_4$ [\icsdweb{107424}, SG 216 $F\bar{4}3m$] as an example of the labeling scheme and information contained within each of the band structure plots shown in this section.  First, at the top of each band structure plot, we provide the chemical formula, ICSD number with a hyperlink to~\webNoICSD, space group symbol and number, the topological classification at the Fermi level (LCEBR, NLC, SEBR, ES, or ESFD) and the magnetic properties as available from \webmaterialsproject\ (NM for non-magnetic, FM for ferromagnetic, AFM for anti-ferromagnetic, FiM for ferrimagnetic and n.a. when the information is not available). Then we further list the hosted sublattices: Kagome, pyrochlore, bipartite, split or Lieb. A superscript $^*$ for Kagome (Lieb) means that only approximate sublattices exist for Kagome (Lieb), as defined in Appendix~\ref{app:kagomepyrochlore}. In the band structure plot, trivial and stable topological conduction (valence) bands are labeled in red (blue), and fragile conduction (valence) bands are labeled in pink (light blue).}
\label{fig:examplebsdos}
\end{figure}

The band structure and projected density of states of each of these \FlatBandNbrBestMaterials~best candidates are given in Figs.~\ref{fig:bestflatband1},~\ref{fig:bestflatband2},~\ref{fig:bestflatband3},~\ref{fig:bestflatband4}, ~\ref{fig:bestflatband5},~\ref{fig:bestflatband6},~\ref{fig:bestflatband7},~\ref{fig:bestflatband8},~\ref{fig:bestflatband9},~\ref{fig:bestflatband10},~\ref{fig:bestflatband11},~\ref{fig:bestflatband12},~\ref{fig:bestflatband13},~\ref{fig:bestflatband14}, ~\ref{fig:bestflatband15},~\ref{fig:bestflatband16},~\ref{fig:bestflatband17},~\ref{fig:bestflatband18},~\ref{fig:bestflatband19},~\ref{fig:bestflatband20},~\ref{fig:bestflatband21},~\ref{fig:bestflatband22},~\ref{fig:bestflatband23},~\ref{fig:bestflatband24}, ~\ref{fig:bestflatband25},~\ref{fig:bestflatband26},~\ref{fig:bestflatband27},~\ref{fig:bestflatband28},~\ref{fig:bestflatband29},~\ref{fig:bestflatband30},~\ref{fig:bestflatband31},~\ref{fig:bestflatband32},~\ref{fig:bestflatband33},~\ref{fig:bestflatband34},~\ref{fig:bestflatband35},~\ref{fig:bestflatband36},~\ref{fig:bestflatband37},~\ref{fig:bestflatband38},~\ref{fig:bestflatband39},~\ref{fig:bestflatband40},~\ref{fig:bestflatband41},~\ref{fig:bestflatband42},~\ref{fig:bestflatband43} and~\ref{fig:bestflatband44}. In Fig.~\ref{fig:examplebsdos}, we detail the information provided for each material. 
Note that the materials
 \ch{KAg[CN]2} [\icsdweb{30275}, SG 163 (\sgsymb{163})], \ch{Pb2Sb2O7} [\icsdweb{27120}, SG 227 (\sgsymb{227})], \ch{Rb2CaH4} [\icsdweb{65196}, SG 139 (\sgsymb{139})], \ch{Ca2NCl} [\icsdweb{62555}, SG 166 (\sgsymb{166})] and
\ch{WO3} [\icsdweb{108651}, SG 221 (\sgsymb{221})]
have been used as prototypical cases with line-graph or bipartite sublattice in the main text. The five materials and \ch{CaNi5} [\icsdweb{54474}, SG 191 (\sgsymb{191})] appearing in this list of best candidates are discussed in details in Appendix~\ref{app:theoryexplanation}, focusing on the origin of their flat bands. As pointed out in Appendix~\ref{app:listcurated}, some candidates presented here have been reported theoretically or experimentally to host flat bands near the Fermi level: $\rm{Co}_3\rm{Sn}_2\rm{S}_2$ [\icsdweb{624867}, SG 166 (\sgsymb{166})]\cite{wang2018large,liu2018giant,Co3Sn2S2_2020}, $\rm{Sr}_2\rm{Fe}\rm{Mo}\rm{O}_6$ [\icsdweb{181752}, SG 225 (\sgsymb{225})]\cite{Yamada_2020} and $\rm{Pd}_3\rm{Pb}$ [\icsdweb{42600}, SG 221 (\sgsymb{221})]\cite{PhysRevB.101.245113}.

Compounds with the same crystal structure and varying chemical composition can have remarkably similar band structures. In the following subsections, we briefly describe structural and composition requirements for the \FlatBandNbrProminentStructuresBestICSDs\ most prominent groups of these \FlatBandNbrBestMaterialsProminentStructures\ unique materials (including \FlatBandNbrBestICSDsProminentStructures\ ICSDs) among the list of best compounds.

\subsubsection{Heusler-\ch{AlCu2Mn}}\label{app:heusler}

Heusler compounds are a class of intermetallic crystals with generic chemical formula $\rm{X}_2\rm{Y}\rm{Z}$, which crystallize in the cubic space group SG 225 (\sgsymb{225}). X and Y are transition metals and Z is a main-group element. In some cases, Y can be an alkaline earth element. Atoms of X, Y, and Z elements occupy $8c$, $4a$, and $4b$ Wyckoff positions, respectively. They are known for easily-tunable band gaps, diverse magnetic properties, such as magneto-optical, magnetocaloric, and magneto-structural characteristics, half-metallic ferromagnetism, and flat bands \cite{graf2011simple,kurosaki2020crystal}. Moreover, Heusler compounds with electron counts near 27 electrons are superconductive \cite{klimczuk2012superconductivity,winterlik2008ni, graf2011simple, winterlik2008electronic, felser2001metal}. We refer the reader to the review Ref.~\cite{graf2011simple} for a detailed discussion about Heusler compounds.

The list of best compounds contains \FlatBandNbrBestMaterialsHeusler\ unique materials (corresponding to  \FlatBandNbrBestICSDsHeusler\ ICSD entries) that are Heusler compounds. These materials are given in Table~\ref{tab:bestHeusler}.

\begin{longtable*}{|c|c|c|c|c|c|}
\caption[List of best Heusler-AlCu$_2$Mn compounds]{\bestperstructurecaption{Heusler-AlCu$_2$Mn}}
\label{tab:bestHeusler}\\
\hline
\hline
 {\scriptsize{chem. formula}} & ICSD & {\scriptsize{space group}} & {\scriptsize{figure}} & {\scriptsize{magn.}} & {\scriptsize{\begin{tabular}{c} super- \\ conduct.  \end{tabular}}} \\
\hline
%% best compounds with structure type Heusler-AlCu2Mn
 $\rm{Li} \rm{Co}_{2} \rm{Ge}$ & \icsdwebshort{25324} & SG 225 (\sgsymb{225})) & Fig.\ICSDFigRef{25324}  & NM & ---  \\
 $\rm{Co} \rm{Fe}_{2} \rm{Ge}$ & \icsdwebshort{52954} & SG 225 (\sgsymb{225})) & Fig.\ICSDFigRef{52954}  & FM & ---  \\
 $\rm{Co}_{2} \rm{Zn} \rm{Ge}$ & \icsdwebshort{52994} & SG 225 (\sgsymb{225})) & Fig.\ICSDFigRef{52994}  & FM & ---  \\
 $\rm{Co}_{2} \rm{Ti} \rm{Si}$ & \icsdwebshort{53080} & SG 225 (\sgsymb{225})) & Fig.\ICSDFigRef{53080}  & FM & ---  \\
 $\rm{Co}_{2} \rm{V} \rm{Si}$ & \icsdwebshort{53086} & SG 225 (\sgsymb{225})) & Fig.\ICSDFigRef{53086}  & FM & ---  \\
 $\rm{Fe} \rm{Ru}_{2} \rm{Si}$ & \icsdwebshort{53525} & SG 225 (\sgsymb{225})) & Fig.\ICSDFigRef{53525}  & FM & ---  \\
 $\rm{Fe}_{2} \rm{V} \rm{Si}$ & \icsdwebshort{53555} & SG 225 (\sgsymb{225})) & Fig.\ICSDFigRef{53555}  & FM & ---  \\
 $\rm{Zr} \rm{Ni}_{2} \rm{In}$ & \icsdwebshort{54546} & SG 225 (\sgsymb{225})) & Fig.\ICSDFigRef{54546}  & NM & ---  \\
 $\rm{Hf} \rm{Ni}_{2} \rm{In}$ & \icsdwebshort{54595} & SG 225 (\sgsymb{225})) & Fig.\ICSDFigRef{54595}  & NM & ---  \\
 $\rm{Al} \rm{Cr} \rm{Fe}_{2}$ & \icsdwebshort{57654} & SG 225 (\sgsymb{225})) & Fig.\ICSDFigRef{57654}  & FM & ---  \\
 $\rm{Al} \rm{Fe}_{3}$ & \icsdwebshort{57793} & SG 225 (\sgsymb{225})) & Fig.\ICSDFigRef{57793}  & FM & ---  \\
 $\rm{Al} \rm{Fe}_{2} \rm{Mo}$ & \icsdwebshort{57807} & SG 225 (\sgsymb{225})) & Fig.\ICSDFigRef{57807}  & FM & ---  \\
 $\rm{Mn}_{3} \rm{Si}$ & \icsdwebshort{76227} & SG 225 (\sgsymb{225})) & Fig.\ICSDFigRef{76227}  & FM & SC  \\
 $\rm{Co} \rm{Fe}_{2} \rm{Ga}$ & \icsdwebshort{102385} & SG 225 (\sgsymb{225})) & Fig.\ICSDFigRef{102385}  & FM & ---  \\
 $\rm{Co}_{2} \rm{Hf} \rm{Ga}$ & \icsdwebshort{102433} & SG 225 (\sgsymb{225})) & Fig.\ICSDFigRef{102433}  & FM & ---  \\
 $\rm{Co}_{2} \rm{Hf} \rm{Sn}$ & \icsdwebshort{102483} & SG 225 (\sgsymb{225})) & Fig.\ICSDFigRef{102483}  & FM & ---  \\
 $\rm{Co}_{2} \rm{Sn} \rm{Ti}$ & \icsdwebshort{102682} & SG 225 (\sgsymb{225})) & Fig.\ICSDFigRef{102682}  & FM & ---  \\
 $\rm{Co}_{2} \rm{Sn} \rm{Zr}$ & \icsdwebshort{102687} & SG 225 (\sgsymb{225})) & Fig.\ICSDFigRef{102687}  & FM & ---  \\
 $\rm{Cr} \rm{Fe}_{2} \rm{Ga}$ & \icsdwebshort{102755} & SG 225 (\sgsymb{225})) & Fig.\ICSDFigRef{102755}  & FM & ---  \\
 $\rm{Cu}_{2} \rm{Mn} \rm{In}$ & \icsdwebshort{102996} & SG 225 (\sgsymb{225})) & Fig.\ICSDFigRef{102996}  & FM & ---  \\
 $\rm{Fe}_{2} \rm{Ga} \rm{V}$ & \icsdwebshort{103473} & SG 225 (\sgsymb{225})) & Fig.\ICSDFigRef{103473}  & NM & ---  \\
 $\rm{Fe}_{2} \rm{Sn} \rm{Ti}$ & \icsdwebshort{103641} & SG 225 (\sgsymb{225})) & Fig.\ICSDFigRef{103641}  & NM & ---  \\
 $\rm{Fe}_{2} \rm{Sn} \rm{V}$ & \icsdwebshort{103644} & SG 225 (\sgsymb{225})) & Fig.\ICSDFigRef{103644}  & FM & ---  \\
 $\rm{Mn} \rm{Ni}_{2} \rm{Sn}$ & \icsdwebshort{104926} & SG 225 (\sgsymb{225})) & Fig.\ICSDFigRef{104926}  & FM & ---  \\
 $\rm{Ni} \rm{Rh}_{2} \rm{Sn}$ & \icsdwebshort{105327} & SG 225 (\sgsymb{225})) & Fig.\ICSDFigRef{105327}  & FM & ---  \\
 $\rm{Ni}_{2} \rm{Sc} \rm{Sn}$ & \icsdwebshort{105339} & SG 225 (\sgsymb{225})) & Fig.\ICSDFigRef{105339}  & NM & ---  \\
 $\rm{Fe}_{3} \rm{Ga}$ & \icsdwebshort{108436} & SG 225 (\sgsymb{225})) & Fig.\ICSDFigRef{108436}  & FM & ---  \\
 $\rm{Co}_{2} \rm{Ti} \rm{Ge}$ & \icsdwebshort{169468} & SG 225 (\sgsymb{225})) & Fig.\ICSDFigRef{169468}  & FM & ---  \\
 $\rm{Co}_{2} \rm{Ti} \rm{Al}$ & \icsdwebshort{185966} & SG 225 (\sgsymb{225})) & Fig.\ICSDFigRef{185966}  & FM & ---  \\
 $\rm{Fe}_{2} \rm{Cr} \rm{Sn}$ & \icsdwebshort{185999} & SG 225 (\sgsymb{225})) & Fig.\ICSDFigRef{185999}  & FM & ---  \\
 $\rm{Fe}_{2} \rm{Ti} \rm{Ge}$ & \icsdwebshort{186057} & SG 225 (\sgsymb{225})) & Fig.\ICSDFigRef{186057}  & NM & ---  \\
 $\rm{Fe}_{2} \rm{Ti} \rm{As}$ & \icsdwebshort{186059} & SG 225 (\sgsymb{225})) & Fig.\ICSDFigRef{186059}  & FM & ---  \\
 $\rm{Fe}_{2} \rm{Ti} \rm{Sb}$ & \icsdwebshort{186060} & SG 225 (\sgsymb{225})) & Fig.\ICSDFigRef{186060}  & FM & ---  \\
 $\rm{Ru}_{2} \rm{V} \rm{Ge}$ & \icsdwebshort{671340} & SG 225 (\sgsymb{225})) & Fig.\ICSDFigRef{671340}  & n.a. & ---  \\
 $\rm{Fe}_{2} \rm{Zr} \rm{P}$ & \icsdwebshort{675102} & SG 225 (\sgsymb{225})) & Fig.\ICSDFigRef{675102}  & n.a. & ---  
\\
\hline
\end{longtable*}

\subsubsection{Heusler-AlLiSi}\label{app:halfheusler}

Heusler-AlLiSi compounds, a sub-class of half-Heusler compounds, have the XYZ general formula and non-centrosymmetric SG 216 (\sgsymb{216}). There is a wide range of choices for X, Y, and Z elements. The compounds can contain main-group elements, transition metals, and elements whose atoms form cations, which can be an alkaline, alkaline earth, or transition-metal, or rate-earth elements. The lattice can be viewed as zinc blende-type sub-lattice ($4a$ and $4c$ Wyckoff positions) with occupied $4b$ positions. \cite{graf2011simple}. Members of half-Heusler family are known for superconductivity \cite{timm2017inflated,nakajima2015topological,pavlosiuk2016antiferromagnetism} and theromelectric properties \cite{fu2015realizing, zeier2016engineering}.

The list of best compounds contains \FlatBandNbrBestMaterialsHalfHeusler\ unique materials (corresponding to  \FlatBandNbrBestICSDsHalfHeusler\ ICSD entries) that are Heusler-AlLiSi compounds. These materials are given in Table~\ref{tab:bestHalfHeusler}.

\begin{longtable*}{|c|c|c|c|c|c|}
\caption[List of best Heusler-AlLiSi compounds]{\bestperstructurecaption{Heusler-AlLiSi}}
\label{tab:bestHalfHeusler}\\
\hline
\hline
 {\scriptsize{chem. formula}} & ICSD & {\scriptsize{space group}} & {\scriptsize{figure}} & {\scriptsize{magn.}} & {\scriptsize{\begin{tabular}{c} super- \\ conduct.  \end{tabular}}} \\
\hline
%% best compounds with structure type Heusler-AlLiSi
 $\rm{Co} \rm{Mn} \rm{Sb}$ & \icsdwebshort{53001} & SG 216 (\sgsymb{216})) & Fig.\ICSDFigRef{53001}  & FM & ---  \\
 $\rm{Ni} \rm{Mn} \rm{Sb}$ & \icsdwebshort{54255} & SG 216 (\sgsymb{216})) & Fig.\ICSDFigRef{54255}  & FM & ---  \\
 $\rm{Rh} \rm{Mn} \rm{Sb}$ & \icsdwebshort{54343} & SG 216 (\sgsymb{216})) & Fig.\ICSDFigRef{54343}  & FM & ---  \\
 $\rm{Co} \rm{Nb} \rm{Sn}$ & \icsdwebshort{102552} & SG 216 (\sgsymb{216})) & Fig.\ICSDFigRef{102552}  & FM & ---  \\
 $\rm{Mn} \rm{Pt} \rm{Sn}$ & \icsdwebshort{104955} & SG 216 (\sgsymb{216})) & Fig.\ICSDFigRef{104955}  & FM & ---  \\
 $\rm{Li} \rm{Ca} \rm{C}$ & \icsdwebshort{672030} & SG 216 (\sgsymb{216})) & Fig.\ICSDFigRef{672030}  & n.a. & ---  
\\
\hline
\end{longtable*}

\subsubsection{Heusler-\ch{CuHg2Ti}}\label{app:inverseheusler}

Intermetallic compounds in the \ch{CuHg2Ti} structure type, known as inverse Heusler structure, have a $\rm{X}_2\rm{YZ}$ generalized chemical formula. Similarly to the Heusler compounds discussed in Apppendix~\ref{app:heusler}, they are composed of transition metals X and Y, and a main-group element Z. They crystallize in the cubic space group SG 216 (\sgsymb{216}), like half-Heusler compounds (Apppendix~\ref{app:halfheusler}). However, the coordination around all sites is tetrahedral, unlike in the Heusler compounds. Two X atoms occupy non-equivalent $4a$ and $4d$ Wyckoff positions, while Y and Z atoms occupy $4b$ and $4c$ positions, respectively.  This lattice type arises in cases when Y element is more electronegative than X element. Sometimes atoms at $4a$ and $4d$ positions belong to different element leading to XX'YZ structure \cite{graf2011simple}.

The list of best compounds contains \FlatBandNbrBestMaterialsCuHgTi\ unique materials (corresponding to  \FlatBandNbrBestICSDsCuHgTi\ ICSD entries) that have \ch{CuHg2Ti} structure type. These materials are given in Table~\ref{tab:bestCuHgTi}.

\begin{longtable*}{|c|c|c|c|c|c|}
\caption[List of best CuHg$_2$Ti compounds]{\bestperstructurecaption{CuHg$_2$Ti}}
\label{tab:bestCuHgTi}\\
\hline
\hline
 {\scriptsize{chem. formula}} & ICSD & {\scriptsize{space group}} & {\scriptsize{figure}} & {\scriptsize{magn.}} & {\scriptsize{\begin{tabular}{c} super- \\ conduct.  \end{tabular}}} \\
\hline
%% best compounds with structure type CuHg2Ti
 $\rm{Mn}_{2} \rm{Co} \rm{As}$ & \icsdwebshort{191648} & SG 216 (\sgsymb{216})) & Fig.\ICSDFigRef{191648}  & FM & ---  \\
 $\rm{Co} \rm{Fe} \rm{Ti} \rm{Al}$ & \icsdwebshort{191657} & SG 216 (\sgsymb{216})) & Fig.\ICSDFigRef{191657}  & NM & ---  \\
 $\rm{Co} \rm{Fe} \rm{Ti} \rm{Ga}$ & \icsdwebshort{191658} & SG 216 (\sgsymb{216})) & Fig.\ICSDFigRef{191658}  & NM & ---  \\
 $\rm{Co} \rm{Fe} \rm{Ti} \rm{Si}$ & \icsdwebshort{191659} & SG 216 (\sgsymb{216})) & Fig.\ICSDFigRef{191659}  & FM & ---  \\
 $\rm{Co} \rm{Fe} \rm{Ti} \rm{Ge}$ & \icsdwebshort{191660} & SG 216 (\sgsymb{216})) & Fig.\ICSDFigRef{191660}  & FM & ---  \\
 $\rm{Co} \rm{Fe} \rm{Ti} \rm{As}$ & \icsdwebshort{191661} & SG 216 (\sgsymb{216})) & Fig.\ICSDFigRef{191661}  & FM & ---  \\
 $\rm{Co} \rm{Fe} \rm{Ti} \rm{Sb}$ & \icsdwebshort{191662} & SG 216 (\sgsymb{216})) & Fig.\ICSDFigRef{191662}  & FM & ---  \\
 $\rm{Co} \rm{Fe} \rm{V} \rm{Ga}$ & \icsdwebshort{191664} & SG 216 (\sgsymb{216})) & Fig.\ICSDFigRef{191664}  & FM & ---  \\
 $\rm{Fe}_{2} \rm{Mn} \rm{Ge}$ & \icsdwebshort{671982} & SG 216 (\sgsymb{216})) & Fig.\ICSDFigRef{671982}  & n.a. & ---  \\
 $\rm{Fe}_{2} \rm{Mn} \rm{Al}$ & \icsdwebshort{672261} & SG 216 (\sgsymb{216})) & Fig.\ICSDFigRef{672261}  & n.a. & ---  
\\
\hline
\end{longtable*}

\subsubsection{Perovskite-\ch{CaTiO3}}\label{app:perovskite}

Cubic perovskites which crystallize in SG 221 (\sgsymb{221}) have a generalized chemical formula $\rm{XYZ}_3$, where X and Y are cations, while Z is an anion. In traditional structures, smaller Y atoms are octahedrally coordinated by 6 Z atoms, while larger X atoms have XII-fold Z shell. X atoms are at $1a$, Y atoms are at $1b$, and Z atoms are at $3c$ Wyckoff positions \cite{pena2001chemical}. The covalent framework of these compounds is formed by Y-Z bonds, while X-Z bonds are ionic in character, and X cations provide the charge balance. Perovskites are known for magnetism \cite{eng2003investigations, atou1999structure, meng2018strain, goto2005anticorrelation, liu2017brief, filippetti2002coexistence}, and superconductivity \cite{bednorz1986possible,he2001superconductivity,cava1988superconductivity}.

The list of best compounds contains \FlatBandNbrBestMaterialsPerovskite\ unique materials (corresponding to  \FlatBandNbrBestICSDsPerovskite\ ICSD entries) that are Perovskite-\ch{CaTiO3} compounds. These materials are given in Table~\ref{tab:bestPerovskite}.

\begin{longtable*}{|c|c|c|c|c|c|}
\caption[List of best Perovskite-CaTi$O_3$ compounds]{\bestperstructurecaption{Perovskite-CaTiO$_3$}}
\label{tab:bestPerovskite}\\
\hline
\hline
 {\scriptsize{chem. formula}} & ICSD & {\scriptsize{space group}} & {\scriptsize{figure}} & {\scriptsize{magn.}} & {\scriptsize{\begin{tabular}{c} super- \\ conduct.  \end{tabular}}} \\
\hline
%% best compounds with structure type Perovskite-CaTiO3
 $\rm{K} \rm{Mn} \rm{F}_{3}$ & \icsdwebshort{15423} & SG 221 (\sgsymb{221})) & Fig.\ICSDFigRef{15423}  & FM & ---  \\
 $\rm{K} \rm{Fe} \rm{F}_{3}$ & \icsdwebshort{15424} & SG 221 (\sgsymb{221})) & Fig.\ICSDFigRef{15424}  & FM & ---  \\
 $\rm{Cs} (\rm{Hg} \rm{Br}_{3})$ & \icsdwebshort{24479} & SG 221 (\sgsymb{221})) & Fig.\ICSDFigRef{24479}  & NM & ---  \\
 $\rm{K} \rm{V} \rm{F}_{3}$ & \icsdwebshort{28145} & SG 221 (\sgsymb{221})) & Fig.\ICSDFigRef{28145}  & FM & ---  \\
 $\rm{Rb} \rm{V} \rm{F}_{3}$ & \icsdwebshort{28146} & SG 221 (\sgsymb{221})) & Fig.\ICSDFigRef{28146}  & FM & ---  \\
 $\rm{La} (\rm{Cr} \rm{O}_{3})$ & \icsdwebshort{28930} & SG 221 (\sgsymb{221})) & Fig.\ICSDFigRef{28930}  & FM & ---  \\
 $\rm{La} (\rm{Mn} \rm{O}_{3})$ & \icsdwebshort{29119} & SG 221 (\sgsymb{221})) & Fig.\ICSDFigRef{29119}  & FM & ---  \\
 $\rm{Rb} (\rm{Mn} \rm{F}_{3})$ & \icsdwebshort{43722} & SG 221 (\sgsymb{221})) & Fig.\ICSDFigRef{43722}  & FM & ---  \\
 $\rm{Rh}_{3} \rm{Y} \rm{B}$ & \icsdwebshort{44568} & SG 221 (\sgsymb{221})) & Fig.\ICSDFigRef{44568}  & NM & ---  \\
 $\rm{Na} \rm{V} \rm{F}_{3}$ & \icsdwebshort{60611} & SG 221 (\sgsymb{221})) & Fig.\ICSDFigRef{60611}  & FM & ---  \\
 $\rm{Sr} (\rm{Ru} \rm{O}_{3})$ & \icsdwebshort{69360} & SG 221 (\sgsymb{221})) & Fig.\ICSDFigRef{69360}  & FM & ---  \\
 $\rm{Fe}_{3} \rm{Zn} \rm{C}$ & \icsdwebshort{76763} & SG 221 (\sgsymb{221})) & Fig.\ICSDFigRef{76763}  & FM & ---  \\
 $\rm{Co}_{3} \rm{Mg} \rm{C}$ & \icsdwebshort{76790} & SG 221 (\sgsymb{221})) & Fig.\ICSDFigRef{76790}  & FM & ---  \\
 $\rm{Co}_{3} \rm{Zn} \rm{C}$ & \icsdwebshort{76797} & SG 221 (\sgsymb{221})) & Fig.\ICSDFigRef{76797}  & FM & ---  \\
 $\rm{Y} \rm{Rh}_{3} \rm{C}$ & \icsdwebshort{77389} & SG 221 (\sgsymb{221})) & Fig.\ICSDFigRef{77389}  & NM & ---  \\
 $\rm{Sr} (\rm{V} \rm{O}_{3})$ & \icsdwebshort{88982} & SG 221 (\sgsymb{221})) & Fig.\ICSDFigRef{88982}  & FM & ---  \\
 $\rm{Sr} (\rm{Tc} \rm{O}_{3})$ & \icsdwebshort{109076} & SG 221 (\sgsymb{221})) & Fig.\ICSDFigRef{109076}  & NM & ---  \\
 $\rm{Ca} (\rm{Mn} \rm{O}_{3})$ & \icsdwebshort{168902} & SG 221 (\sgsymb{221})) & Fig.\ICSDFigRef{168902}  & FM & ---  \\
 $\rm{Pb} (\rm{V} \rm{O}_{3})$ & \icsdwebshort{187637} & SG 221 (\sgsymb{221})) & Fig.\ICSDFigRef{187637}  & FM & ---  \\
 $\rm{Sr} (\rm{Mn} \rm{O}_{3})$ & \icsdwebshort{188415} & SG 221 (\sgsymb{221})) & Fig.\ICSDFigRef{188415}  & AFM & ---  \\
 $\rm{Ba} (\rm{V} \rm{O}_{3})$ & \icsdwebshort{191203} & SG 221 (\sgsymb{221})) & Fig.\ICSDFigRef{191203}  & FM & ---  \\
 $\rm{In} \rm{N} \rm{Co}_{3}$ & \icsdwebshort{247066} & SG 221 (\sgsymb{221})) & Fig.\ICSDFigRef{247066}  & FM & ---  \\
 $\rm{Cd} \rm{Co}_{3} \rm{N}$ & \icsdwebshort{422858} & SG 221 (\sgsymb{221})) & Fig.\ICSDFigRef{422858}  & FM & ---  \\
 $\rm{Ca} \rm{Tc} \rm{O}_{3}$ & \icsdwebshort{671082} & SG 221 (\sgsymb{221})) & Fig.\ICSDFigRef{671082}  & n.a. & ---  \\
 $\rm{Ba} \rm{Tc} \rm{O}_{3}$ & \icsdwebshort{671086} & SG 221 (\sgsymb{221})) & Fig.\ICSDFigRef{671086}  & n.a. & ---  \\
 $\rm{Zn} \rm{Sn} \rm{O}_{3}$ & \icsdwebshort{673496} & SG 221 (\sgsymb{221})) & Fig.\ICSDFigRef{673496}  & n.a. & ---  \\
 $\rm{K} \rm{Sb} \rm{O}_{3}$ & \icsdwebshort{674060} & SG 221 (\sgsymb{221})) & Fig.\ICSDFigRef{674060}  & n.a. & ---  
\\
\hline
\end{longtable*}

\subsubsection{Double perovskite crystal structure: \ch{K2PtCl6}, Elpasolite-\ch{K2NaAlF6}, and \ch{Sr2NiWO6}}

Unit cell of double perovskites is created by the doubling of perovskite unit cell, $\rm{XYZ}_3$ along all three crystallographic axes and introducing a change every other Y site. In \ch{K2PtCl6} lattice type, with a generalized formula $\rm{X}_2\rm{YZ}_6$, a vacancy is introduced. In elpasolites, with a $\rm{X}_2\rm{Y'YZ}_6$ general formula, Y atom is substitute by Y' atom. The symmetry reduces from SG 221 (\sgsymb{221}) to SG 225 (\sgsymb{225}). Finally, in compounds of \ch{Sr2NiWO6} lattice type elpasolite lattice is tetragonally distorted as the size of X cations decreases with respect to sizes of Y and Y' cations. Space group reduces farther to SG 87 (\sgsymb{87}) \cite{maughan2019perspectives,wolf2021doubling}.

Usually, X is an alkaline, alkaline earth, or rare-earth elements, Y is a transition metal or a main-group element, and Z is a halide. In elpasolites, Y' can be an alkaline or alkaline earth metal, a transition metal, or a main-group element. Moreover, Y and Y' can be the same element, but in different oxidation states \cite{schoop2013lone}. Additionally, Z element in elpasolites can be an oxide anion instead of a halide. Transition from elpasolite to the \ch{Sr2NiWO6} lattice is the most common when Y and Y' are transition metals \cite{maughan2019perspectives,wolf2021doubling, eriksson2006high, martinez2003synthesis}.

The list of best compounds contains \FlatBandNbrBestMaterialsKPtCl\ unique materials (\FlatBandNbrBestICSDsKPtCl\ ICSD entries), \FlatBandNbrBestMaterialsElpasolite\ unique materials (\FlatBandNbrBestICSDsElpasolite\ ICSD entries) and \FlatBandNbrBestMaterialsSrNiWo\ unique materials (\FlatBandNbrBestICSDsSrNiWo\ ICSD entries) that are \ch{K2PtCl6}, elpasolites, and \ch{Sr2NiWO6} compounds, respectively. These materials are given in Tables~\ref{tab:bestKPtCl}, ~\ref{tab:bestElpasolite} and~\ref{tab:bestSrNiWo}.

\begin{longtable*}{|c|c|c|c|c|c|}
\caption[List of best Elpasolite-K$_2$NaAlF$_6$ compounds]{\bestperstructurecaption{Elpasolite-\ch{K2NaAlF6}}}
\label{tab:bestElpasolite}\\
\hline
\hline
 {\scriptsize{chem. formula}} & ICSD & {\scriptsize{space group}} & {\scriptsize{figure}} & {\scriptsize{magn.}} & {\scriptsize{\begin{tabular}{c} super- \\ conduct.  \end{tabular}}} \\
\hline
%% best compounds with structure type Elpasolite-K2NaAlF6
 $\rm{Ba}_{2} \rm{Mn} (\rm{Re} \rm{O}_{6})$ & \icsdwebshort{4169} & SG 225 (\sgsymb{225})) & Fig.\ICSDFigRef{4169}  & FM & ---  \\
 $\rm{K}_{2} \rm{Na} (\rm{Al} \rm{F}_{6})$ & \icsdwebshort{6027} & SG 225 (\sgsymb{225})) & Fig.\ICSDFigRef{6027}  & NM & ---  \\
 $\rm{K} \rm{Tl}_{2} \rm{Mo} \rm{F}_{6}$ & \icsdwebshort{15775} & SG 225 (\sgsymb{225})) & Fig.\ICSDFigRef{15775}  & FM & ---  \\
 $\rm{Na} \rm{Tl}_{2} \rm{Mo} \rm{F}_{6}$ & \icsdwebshort{15776} & SG 225 (\sgsymb{225})) & Fig.\ICSDFigRef{15776}  & FM & ---  \\
 $\rm{K}_{2} \rm{Na} \rm{Tl} \rm{F}_{6}$ & \icsdwebshort{22114} & SG 225 (\sgsymb{225})) & Fig.\ICSDFigRef{22114}  & NM & ---  \\
 $\rm{Cs}_{2} \rm{Na} \rm{Tl} \rm{F}_{6}$ & \icsdwebshort{22118} & SG 225 (\sgsymb{225})) & Fig.\ICSDFigRef{22118}  & NM & ---  \\
 $\rm{Ba}_{2} \rm{Co} \rm{W} \rm{O}_{6}$ & \icsdwebshort{27425} & SG 225 (\sgsymb{225})) & Fig.\ICSDFigRef{27425}  & FM & ---  \\
 $\rm{Sr}_{2} \rm{Co} (\rm{W} \rm{O}_{6})$ & \icsdwebshort{28598} & SG 225 (\sgsymb{225})) & Fig.\ICSDFigRef{28598}  & FM & ---  \\
 $\rm{Sr}_{2} \rm{Co} (\rm{Mo} \rm{O}_{6})$ & \icsdwebshort{28601} & SG 225 (\sgsymb{225})) & Fig.\ICSDFigRef{28601}  & AFM & ---  \\
 $\rm{Ba}_{2} \rm{Fe} \rm{Mo} \rm{O}_{6}$ & \icsdwebshort{96688} & SG 225 (\sgsymb{225})) & Fig.\ICSDFigRef{96688}  & FM & ---  \\
 $\rm{Ba}_{2} (\rm{Co} \rm{Mo} \rm{O}_{6})$ & \icsdwebshort{97028} & SG 225 (\sgsymb{225})) & Fig.\ICSDFigRef{97028}  & AFM & ---  \\
 $\rm{Ba}_{2} (\rm{Fe} \rm{W} \rm{O}_{6})$ & \icsdwebshort{99061} & SG 225 (\sgsymb{225})) & Fig.\ICSDFigRef{99061}  & AFM & ---  \\
 $\rm{Ba}_{2} \rm{Fe} (\rm{Re} \rm{O}_{6})$ & \icsdwebshort{109252} & SG 225 (\sgsymb{225})) & Fig.\ICSDFigRef{109252}  & FM & ---  \\
 $\rm{Sr}_{2} (\rm{Ga} \rm{Sb} \rm{O}_{6})$ & \icsdwebshort{157016} & SG 225 (\sgsymb{225})) & Fig.\ICSDFigRef{157016}  & NM & ---  \\
 $\rm{Sr}_{2} (\rm{Fe} \rm{Mo} \rm{O}_{6})$ & \icsdwebshort{157603} & SG 225 (\sgsymb{225})) & Fig.\ICSDFigRef{157603}  & AFM & ---  \\
 $\rm{Sr}_{2} \rm{Y} (\rm{Sb} \rm{O}_{6})$ & \icsdwebshort{157886} & SG 225 (\sgsymb{225})) & Fig.\ICSDFigRef{157886}  & NM & ---  \\
 $\rm{Ba}_{2} \rm{Ca} \rm{Te} \rm{O}_{6}$ & \icsdwebshort{246112} & SG 225 (\sgsymb{225})) & Fig.\ICSDFigRef{246112}  & NM & ---  \\
 $\rm{Sr}_{2} \rm{Fe} \rm{Os} \rm{O}_{6}$ & \icsdwebshort{251068} & SG 225 (\sgsymb{225})) & Fig.\ICSDFigRef{251068}  & AFM & ---  \\
 $\rm{Cs}_{2} \rm{In} \rm{Ag} \rm{Cl}_{6}$ & \icsdwebshort{257115} & SG 225 (\sgsymb{225})) & Fig.\ICSDFigRef{257115}  & n.a. & ---  \\
 $\rm{Pb}_{2} \rm{Na} \rm{I} \rm{O}_{6}$ & \icsdwebshort{427115} & SG 225 (\sgsymb{225})) & Fig.\ICSDFigRef{427115}  & NM & ---  
\\
\hline
\end{longtable*}

\begin{longtable*}{|c|c|c|c|c|c|}
\caption[List of best \ch{K2PtCl6} compounds]{\bestperstructurecaption{\ch{K2PtCl6}}}
\label{tab:bestKPtCl}\\
\hline
\hline
 {\scriptsize{chem. formula}} & ICSD & {\scriptsize{space group}} & {\scriptsize{figure}} & {\scriptsize{magn.}} & {\scriptsize{\begin{tabular}{c} super- \\ conduct.  \end{tabular}}} \\
\hline
%% best compounds with structure type K2PtCl6
 $\rm{K}_{2} (\rm{Sn} \rm{Cl}_{6})$ & \icsdwebshort{604} & SG 225 (\sgsymb{225})) & Fig.\ICSDFigRef{604}  & NM & ---  \\
 $(\rm{N} \rm{H}_{4})_{2} (\rm{Sn} \rm{Cl}_{6})$ & \icsdwebshort{605} & SG 225 (\sgsymb{225})) & Fig.\ICSDFigRef{605}  & NM & ---  \\
 $\rm{Rb}_{2} \rm{Sn} \rm{Cl}_{6}$ & \icsdwebshort{9022} & SG 225 (\sgsymb{225})) & Fig.\ICSDFigRef{9022}  & NM & ---  \\
 $\rm{Cs}_{2} \rm{Sn} \rm{Cl}_{6}$ & \icsdwebshort{9023} & SG 225 (\sgsymb{225})) & Fig.\ICSDFigRef{9023}  & NM & ---  \\
 $\rm{Rb}_{2} \rm{Mn} \rm{Cl}_{6}$ & \icsdwebshort{9347} & SG 225 (\sgsymb{225})) & Fig.\ICSDFigRef{9347}  & FM & ---  \\
 $\rm{K}_{2} \rm{Ta} \rm{Cl}_{6}$ & \icsdwebshort{59894} & SG 225 (\sgsymb{225})) & Fig.\ICSDFigRef{59894}  & FM & ---  \\
 $\rm{Rb}_{2} (\rm{Nb} \rm{Cl}_{6})$ & \icsdwebshort{245747} & SG 225 (\sgsymb{225})) & Fig.\ICSDFigRef{245747}  & FM & ---  \\
 $\rm{Y}_{2} \rm{Ni}_{6} \rm{C}$ & \icsdwebshort{673582} & SG 225 (\sgsymb{225})) & Fig.\ICSDFigRef{673582}  & n.a. & ---  
\\
\hline
\end{longtable*}

\begin{longtable*}{|c|c|c|c|c|c|}
\caption[List of best Sr$_2$NiWO$_6$ compounds]{\bestperstructurecaption{\ch{Sr2NiWO6}}}
\label{tab:bestSrNiWo}\\
\hline
\hline
 {\scriptsize{chem. formula}} & ICSD & {\scriptsize{space group}} & {\scriptsize{figure}} & {\scriptsize{magn.}} & {\scriptsize{\begin{tabular}{c} super- \\ conduct.  \end{tabular}}} \\
\hline
%% best compounds with structure type Sr2NiWO6
 $\rm{Sr}_{2} \rm{Ni} (\rm{W} \rm{O}_{6})$ & \icsdwebshort{91791} & SG 87 (\sgsymb{87})) & Fig.\ICSDFigRef{91791}  & AFM & ---  \\
 $\rm{Ba}_{2} (\rm{Fe} \rm{W} \rm{O}_{6})$ & \icsdwebshort{95518} & SG 87 (\sgsymb{87})) & Fig.\ICSDFigRef{95518}  & AFM & ---  \\
 $\rm{Sr}_{2} (\rm{Fe} \rm{Mo} \rm{O}_{6})$ & \icsdwebshort{150701} & SG 87 (\sgsymb{87})) & Fig.\ICSDFigRef{150701}  & FM & ---  \\
 $\rm{Sr}_{2} (\rm{Fe} \rm{Re} \rm{O}_{6})$ & \icsdwebshort{150702} & SG 87 (\sgsymb{87})) & Fig.\ICSDFigRef{150702}  & FM & ---  \\
 $\rm{Sr}_{2} \rm{Mg} (\rm{W} \rm{O}_{6})$ & \icsdwebshort{151703} & SG 87 (\sgsymb{87})) & Fig.\ICSDFigRef{151703}  & NM & ---  \\
 $\rm{Sr}_{2} (\rm{Co} \rm{Mo} \rm{O}_{6})$ & \icsdwebshort{153544} & SG 87 (\sgsymb{87})) & Fig.\ICSDFigRef{153544}  & FM & ---  \\
 $\rm{Sr}_{2} \rm{Co} \rm{Re} \rm{O}_{6}$ & \icsdwebshort{173488} & SG 87 (\sgsymb{87})) & Fig.\ICSDFigRef{173488}  & AFM & ---  \\
 $\rm{Sr}_{2} \rm{Mg} \rm{Mo} \rm{O}_{6}$ & \icsdwebshort{187662} & SG 87 (\sgsymb{87})) & Fig.\ICSDFigRef{187662}  & NM & ---  \\
 $\rm{Sr}_{2} \rm{Mn} \rm{Mo} \rm{O}_{6}$ & \icsdwebshort{187669} & SG 87 (\sgsymb{87})) & Fig.\ICSDFigRef{187669}  & FM & ---  \\
 $\rm{Sr}_{2} \rm{Co} \rm{W} \rm{O}_{6}$ & \icsdwebshort{190593} & SG 87 (\sgsymb{87})) & Fig.\ICSDFigRef{190593}  & AFM & ---  \\
 $\rm{Sr}_{2} \rm{Co} \rm{Nb} \rm{O}_{6}$ & \icsdwebshort{192327} & SG 87 (\sgsymb{87})) & Fig.\ICSDFigRef{192327}  & n.a. & ---  
\\ 
\hline
\end{longtable*}

\subsubsection{Shandite-\ch{Ni3Pb2S2}}

Ni3Pb2S2 lattice type in SG 166 (\sgsymb{166}), known as shandite, is a lattice with a $\rm{X}_3\rm{Y}_2\rm{Z}_2$ general formula. But this lattice type also includes compounds with a $\rm{X}_3\rm{YY'}\rm{Z}_2$ formula. X atoms belong to a transition metal element, while Y (Y') and Z atoms are main-group elements. X atoms occupy $3a$ Wyckoff positions, Y atoms are at $3b$ positions, and Z atoms are at $6c$ positions.  The bonding in these compounds can be approximated by a Kagome Hamiltonian for $d$ orbitals, though small contributions form Y's and Z's orbitals to the bands near the Fermi level are possible \cite{jiao2019signatures, skinner2013electronic}. 

The list of best compounds contains \FlatBandNbrBestMaterialsNiPbS\ unique materials (corresponding to  \FlatBandNbrBestICSDsNiPbS\ ICSD entries) that are Shandite-\ch{Ni3Pb2S2} compounds, and are listed in Table~\ref{tab:bestNiPbS}. Note that all of them host a rigorous Kagome sublattice, as expected.

\begin{longtable*}{|c|c|c|c|c|c|}
\caption[List of best Ni$_3$Pb$_2$S$_2$ compounds]{\bestperstructurecaption{\ch{Ni3Pb2S2}}}
\label{tab:bestNiPbS}\\
\hline
\hline
 {\scriptsize{chem. formula}} & ICSD & {\scriptsize{space group}} & {\scriptsize{figure}} & {\scriptsize{magn.}} & {\scriptsize{\begin{tabular}{c} super- \\ conduct.  \end{tabular}}} \\
\hline
%% best compounds with structure type Ni3Pb2S2
 $\rm{Co}_{3} \rm{Sn}_{2} \rm{S}_{2}$ & \icsdwebshort{5435} & SG 166 (\sgsymb{166})) & Fig.\ICSDFigRef{5435}  & FM & ---  \\
 $\rm{Co}_{3} \rm{In} \rm{Sn} \rm{S}_{2}$ & \icsdwebshort{5437} & SG 166 (\sgsymb{166})) & Fig.\ICSDFigRef{5437}  & NM & ---  \\
 $\rm{Rh}_{3} \rm{In} \rm{Pb} \rm{S}_{2}$ & \icsdwebshort{5440} & SG 166 (\sgsymb{166})) & Fig.\ICSDFigRef{5440}  & NM & ---  \\
 $\rm{Rh}_{3} \rm{Sn}_{2} \rm{S}_{2}$ & \icsdwebshort{420728} & SG 166 (\sgsymb{166})) & Fig.\ICSDFigRef{420728}  & NM & ---  \\
 $\rm{In} \rm{Sn} \rm{Co}_{3} \rm{S}_{2}$ & \icsdwebshort{425137} & SG 166 (\sgsymb{166})) & Fig.\ICSDFigRef{425137}  & n.a. & ---  
\\
\hline
\end{longtable*}

\begin{figure}[h]
\centering
\input{appendices/figurelatex/bestflatband/bestflatband_1_44.tex}
\caption{\bestflatcaption{1}}
\label{fig:bestflatband1}
\end{figure}

\begin{figure}[h]
\centering
\input{appendices/figurelatex/bestflatband/bestflatband_2_44.tex}
\caption{\bestflatcaption{2}}
\label{fig:bestflatband2}
\end{figure}

\begin{figure}[h]
\centering
\input{appendices/figurelatex/bestflatband/bestflatband_3_44.tex}
\caption{\bestflatcaption{3}}
\label{fig:bestflatband3}
\end{figure}

\begin{figure}[h]
\centering
\input{appendices/figurelatex/bestflatband/bestflatband_4_44.tex}
\caption{\bestflatcaption{4}}
\label{fig:bestflatband4}
\end{figure}

\begin{figure}[h]
\centering
\input{appendices/figurelatex/bestflatband/bestflatband_5_44.tex}
\caption{\bestflatcaption{5}}
\label{fig:bestflatband5}
\end{figure}

\begin{figure}[h]
\centering
\input{appendices/figurelatex/bestflatband/bestflatband_6_44.tex}
\caption{\bestflatcaption{6}}
\label{fig:bestflatband6}
\end{figure}

\begin{figure}[h]
\centering
\input{appendices/figurelatex/bestflatband/bestflatband_7_44.tex}
\caption{\bestflatcaption{7}}
\label{fig:bestflatband7}
\end{figure}

\begin{figure}[h]
\centering
\input{appendices/figurelatex/bestflatband/bestflatband_8_44.tex}
\caption{\bestflatcaption{8}}
\label{fig:bestflatband8}
\end{figure}

\begin{figure}[h]
\centering
\input{appendices/figurelatex/bestflatband/bestflatband_9_44.tex}
\caption{\bestflatcaption{9}}
\label{fig:bestflatband9}
\end{figure}

\begin{figure}[h]
\centering
\input{appendices/figurelatex/bestflatband/bestflatband_10_44.tex}
\caption{\bestflatcaption{10}}
\label{fig:bestflatband10}
\end{figure}

\begin{figure}[h]
\centering
\input{appendices/figurelatex/bestflatband/bestflatband_11_44.tex}
\caption{\bestflatcaption{11}}
\label{fig:bestflatband11}
\end{figure}

\begin{figure}[h]
\centering
\input{appendices/figurelatex/bestflatband/bestflatband_12_44.tex}
\caption{\bestflatcaption{12}}
\label{fig:bestflatband12}
\end{figure}

\begin{figure}[h]
\centering
\input{appendices/figurelatex/bestflatband/bestflatband_13_44.tex}
\caption{\bestflatcaption{13}}
\label{fig:bestflatband13}
\end{figure}

\begin{figure}[h]
\centering
\input{appendices/figurelatex/bestflatband/bestflatband_14_44.tex}
\caption{\bestflatcaption{14}}
\label{fig:bestflatband14}
\end{figure}

\begin{figure}[h]
\centering
\input{appendices/figurelatex/bestflatband/bestflatband_15_44.tex}
\caption{\bestflatcaption{15}}
\label{fig:bestflatband15}
\end{figure}

\begin{figure}[h]
\centering
\input{appendices/figurelatex/bestflatband/bestflatband_16_44.tex}
\caption{\bestflatcaption{16}}
\label{fig:bestflatband16}
\end{figure}

\begin{figure}[h]
\centering
\input{appendices/figurelatex/bestflatband/bestflatband_17_44.tex}
\caption{\bestflatcaption{17}}
\label{fig:bestflatband17}
\end{figure}

\begin{figure}[h]
\centering
\input{appendices/figurelatex/bestflatband/bestflatband_18_44.tex}
\caption{\bestflatcaption{18}}
\label{fig:bestflatband18}
\end{figure}

\begin{figure}[h]
\centering
\input{appendices/figurelatex/bestflatband/bestflatband_19_44.tex}
\caption{\bestflatcaption{19}}
\label{fig:bestflatband19}
\end{figure}

\begin{figure}[h]
\centering
\input{appendices/figurelatex/bestflatband/bestflatband_20_44.tex}
\caption{\bestflatcaption{20}}
\label{fig:bestflatband20}
\end{figure}

\begin{figure}[h]
\centering
\input{appendices/figurelatex/bestflatband/bestflatband_21_44.tex}
\caption{\bestflatcaption{21}}
\label{fig:bestflatband21}
\end{figure}

\begin{figure}[h]
\centering
\input{appendices/figurelatex/bestflatband/bestflatband_22_44.tex}
\caption{\bestflatcaption{22}}
\label{fig:bestflatband22}
\end{figure}

\begin{figure}[h]
\centering
\input{appendices/figurelatex/bestflatband/bestflatband_23_44.tex}
\caption{\bestflatcaption{23}}
\label{fig:bestflatband23}
\end{figure}

\begin{figure}[h]
\centering
\input{appendices/figurelatex/bestflatband/bestflatband_24_44.tex}
\caption{\bestflatcaption{24}}
\label{fig:bestflatband24}
\end{figure}

\begin{figure}[h]
\centering
\input{appendices/figurelatex/bestflatband/bestflatband_25_44.tex}
\caption{\bestflatcaption{25}}
\label{fig:bestflatband25}
\end{figure}

\begin{figure}[h]
\centering
\input{appendices/figurelatex/bestflatband/bestflatband_26_44.tex}
\caption{\bestflatcaption{26}}
\label{fig:bestflatband26}
\end{figure}

\begin{figure}[h]
\centering
\input{appendices/figurelatex/bestflatband/bestflatband_27_44.tex}
\caption{\bestflatcaption{27}}
\label{fig:bestflatband27}
\end{figure}

\begin{figure}[h]
\centering
\input{appendices/figurelatex/bestflatband/bestflatband_28_44.tex}
\caption{\bestflatcaption{28}}
\label{fig:bestflatband28}
\end{figure}

\begin{figure}[h]
\centering
\input{appendices/figurelatex/bestflatband/bestflatband_29_44.tex}
\caption{\bestflatcaption{29}}
\label{fig:bestflatband29}
\end{figure}

\begin{figure}[h]
\centering
\input{appendices/figurelatex/bestflatband/bestflatband_30_44.tex}
\caption{\bestflatcaption{30}}
\label{fig:bestflatband30}
\end{figure}

\begin{figure}[h]
\centering
\input{appendices/figurelatex/bestflatband/bestflatband_31_44.tex}
\caption{\bestflatcaption{31}}
\label{fig:bestflatband31}
\end{figure}

\begin{figure}[h]
\centering
\input{appendices/figurelatex/bestflatband/bestflatband_32_44.tex}
\caption{\bestflatcaption{32}}
\label{fig:bestflatband32}
\end{figure}

\begin{figure}[h]
\centering
\input{appendices/figurelatex/bestflatband/bestflatband_33_44.tex}
\caption{\bestflatcaption{33}}
\label{fig:bestflatband33}
\end{figure}

\begin{figure}[h]
\centering
\input{appendices/figurelatex/bestflatband/bestflatband_34_44.tex}
\caption{\bestflatcaption{34}}
\label{fig:bestflatband34}
\end{figure}

\begin{figure}[h]
\centering
\input{appendices/figurelatex/bestflatband/bestflatband_35_44.tex}
\caption{\bestflatcaption{35}}
\label{fig:bestflatband35}
\end{figure}

\begin{figure}[h]
\centering
\input{appendices/figurelatex/bestflatband/bestflatband_36_44.tex}
\caption{\bestflatcaption{36}}
\label{fig:bestflatband36}
\end{figure}

\begin{figure}[h]
\centering
\input{appendices/figurelatex/bestflatband/bestflatband_37_44.tex}
\caption{\bestflatcaption{37}}
\label{fig:bestflatband37}
\end{figure}

\begin{figure}[h]
\centering
\input{appendices/figurelatex/bestflatband/bestflatband_38_44.tex}
\caption{\bestflatcaption{38}}
\label{fig:bestflatband38}
\end{figure}

\begin{figure}[h]
\centering
\input{appendices/figurelatex/bestflatband/bestflatband_39_44.tex}
\caption{\bestflatcaption{39}}
\label{fig:bestflatband39}
\end{figure}

\begin{figure}[h]
\centering
\input{appendices/figurelatex/bestflatband/bestflatband_40_44.tex}
\caption{\bestflatcaption{40}}
\label{fig:bestflatband40}
\end{figure}

\begin{figure}[h]
\centering
\input{appendices/figurelatex/bestflatband/bestflatband_41_44.tex}
\caption{\bestflatcaption{41}}
\label{fig:bestflatband41}
\end{figure}

\begin{figure}[h]
\centering
\input{appendices/figurelatex/bestflatband/bestflatband_42_44.tex}
\caption{\bestflatcaption{42}}
\label{fig:bestflatband42}
\end{figure}

\begin{figure}[h]
\centering
\input{appendices/figurelatex/bestflatband/bestflatband_43_44.tex}
\caption{\bestflatcaption{43}}
\label{fig:bestflatband43}
\end{figure}

\begin{figure}[h]
\centering
\input{appendices/figurelatex/bestflatband/bestflatband_44_44.tex}
\caption{\bestflatcaption{44}}
\label{fig:bestflatband44}
\end{figure}

\end{document}